\documentclass[preprint2]{aastex62}

\usepackage{natbib}
\usepackage{lineno}
\usepackage{rotating}
\usepackage[caption=false]{subfig}
\usepackage{array, diagbox}
\usepackage{slashbox}
\usepackage{multirow}
\usepackage{enumerate}
\usepackage{chngpage}
\usepackage{hyperref}
\usepackage{ulem} 
\def \aj {AJ}
\def \apj {ApJ}
\def \apjs {ApJS}
\def \apjl {ApJ}

\def \mnras {MNRAS}
\def \aap {A\&A}

\def \araa {ARA\&A}
\def \pasp {PASP}

\def\dg{\hbox{$^\circ$}}
\def\arcmin{\hbox{$^\prime$}}
\def\arcsec{\hbox{$^{\prime\prime}$}}
\def\utw{\smash{\rlap{\lower5pt\hbox{$\sim$}}}}
\def\udtw{\smash{\rlap{\lower6pt\hbox{$\approx$}}}}

\def\farcm{\hbox{$.\!^\prime$}}
\def\farcs{\hbox{$.\!\!^{\prime\prime}$}}

\received{\today}
\revised{\today}
\accepted{\today}
\submitjournal{ApJS}

\shorttitle{ROGUE~\textrm{I}: SDSS galaxies with FIRST core identifications}
\shortauthors{Kozie{\l}-Wierzbowska et al.}

\begin{document}

\title{ Radio sources associated with Optical Galaxies and having Unresolved or Extended morphologies (ROGUE).\\ \textrm{I}. A catalog of SDSS galaxies with FIRST core identifications}

\correspondingauthor{Dorota Kozie{\l}-Wierzbowska}
\email{dorota.koziel@uj.edu.pl}

\author[0000-0003-4323-0984]{Dorota Kozie{\l}-Wierzbowska}
\affil{Astronomical Observatory of Jagiellonian University, ul.\ Orla 171, 30-244 Krak\'ow, Poland}

\author[0000-0002-2224-6664]{Arti Goyal}
\affil{Astronomical Observatory of Jagiellonian University, ul.\ Orla 171, 30-244 Krak\'ow, Poland}

\author[0000-0003-2644-6441]{Natalia~\.{Z}ywucka}
\affil{Astronomical Observatory of Jagiellonian University, ul.\ Orla 171, 30-244 Krak\'ow, Poland}
\affil{Centre for Space Research, North-West University, Potchefstroom 2520, South Africa}

\begin{abstract}
We present the catalog of Radio sources associated with Optical Galaxies and having Unresolved or Extended morphologies I (ROGUE~I), consisting of 32,616 spectroscopically selected galaxies. It is the largest handmade catalog of this kind, obtained by cross-matching galaxies from the Sloan Digital Sky Survey (SDSS) Data Release 7 and radio sources from both the First Images of Radio Sky at Twenty Centimetre (FIRST) survey and the NRAO VLA Sky Survey \textit{without imposing a limit to the radio flux densities}. The catalog provides a \textit{visual} classification of radio and optical morphologies of galaxies presenting a FIRST core within 3\arcsec\ of the optical position. The radio morphological classification is performed by examining the radio-optical overlays of linear sizes equal to 1 Mpc at the source distance, while the 120\arcsec\ image snapshots from the SDSS database are used for optical classification. The results of our search are: (i) single-component unresolved and elongated, radio sources constitute the major group in the ROGUE~I catalog ($\sim$90\%), and $\sim$8\% exhibiting {\it extended} morphologies,
(ii) samples of 269, 730, and 115 Fanaroff-Riley (FR) type I, II, and hybrid galaxies, respectively, are presented (iii) we report 55 newly discovered giant/possible giant, 16 double--double, 9 X--shaped, and 25 Z--shaped radio sources, (iv) on the optical front, most galaxies have elliptical morphologies ($\sim$62\%) while spirals form the second major category ($\sim$17\%) followed by distorted ($\sim$ 12\%) and lenticular ($\sim$7\%) morphologies, (v) division between the FR~I and the FR~II sources in the radio--optical luminosity plane is blurred, in tune with recent studies.
\end{abstract}


\section{Introduction} 
\label{sec:intro}

It is widely accepted that the broad range in radio powers (up to 4 decades), sizes (up to Mpc scales), and radio morphologies of jetted active galactic nuclei (AGNs) stems from the complex interplay between the jet production efficiency (linked to the properties of accretion flow around the supermassive black hole (SMBH) located at the centers of galaxies), interaction of jets with the host galaxy, and the large scale environment they reside in \citep[see, for a recent review,][]{Padovani17, Blandford19}. While accretion onto the SMBH has been unequivocally accepted as the main source of AGN activity, it is not clear what fraction of it is channeled as kinetic energy of a jet as the majority of the optically selected galaxies and their high luminosity counterparts, quasars, do not show significant radio activity \citep[see][and references therein]{Balokovic12,Koziel-Wierzbowska17}. Moreover, the energy exchange between the evolving radio source and the interstellar/intergalactic medium on small as well as large scales forms an important part of the feedback that governs the co-evolution of galaxies, central SMBH, and galaxy groups/clusters over cosmological timescales \citep[see, for a review,][]{McNamara07, Fabian12}.  

The wide contrast between the two main morphological classes of radio galaxies, i.e., Fanaroff-Riley type I and II (FR~I, FR~II; \citealt{Fanaroff74}), and the extremely large number of `young' radio galaxies as compared to the `evolved' ones is already driving the need to obtain large samples of radio galaxies in order to derive statistical inferences on the physics of radio AGN \citep[e.g.,][]{Best12, Tadhunter16}. In this regard, a reliable association of radio sources with their host galaxies is crucial for identifying the conditions responsible for the radio AGN trigger from the nuclear regions (i.e., different accretion flows; \citealt{Ineson15}), host galaxies (i.e., positive/negative feedback; \citealt{Kalfountzou17}), merger history \citep{Singh15b}, and the cluster environment (i.e., rich vs. poor environment; \citealt{Mingo17}). 

\begin{table*}[!ht]
\caption{Summary of the previous/current results on samples of radio sources}
\label{AGNsamples}
\scriptsize
\begin{center}
\begin{tabular}{cp{4cm}p{3cm}ccccc}\hline
  Sample & \centering{Selection procedure}  & \centering{$z$ and flux density cuts} &  Sky coverage  & Number &  Radio  & Optical\\
    (1)  &      \centering{(2)}             &       \centering{(3)}               &      (4)       &   (5)  &    (6)      &     (7)  \\
\hline
\citet{Machalski99} & Cross-matching of LCRS galaxy sample with NVSS sample using a matching radius $\Delta\mathrm{r}=2\farcs5$ &  $z<0.2$; optical magnitude $<18$ and NVSS flux density $>2.5$ mJy & $\sim$0.2 sr & 1,157 & No & No \\ 
\citet{McMahon02} & Cross-matching of POSS~I sources with FIRST sample using $\Delta\mathrm{r}=20\arcsec$ & optical magnitude $<20$ and FIRST flux density $>1.0$ mJy & $\sim$4,150 deg$^2$ & 70,000 & No & No\\ 
\citet{Sadler02} & Cross-matching of 2dF Galaxy redshift Survey with NVSS sample using $\Delta\mathrm{r}=15\arcsec$ & $z<0.438$; optical magnitude $<19.5$ and NVSS flux density $>2.5$ mJy &  $\sim$325 deg$^2$ & 757 & No & No \\
\citet{Best05} & Cross-matching of SDSS DR\,2 spectroscopic sample with NVSS using $\Delta\mathrm{r}=3\arcmin$ and then matching with FIRST & $0.01<z<0.56$; optical magnitude $<17.77$ and NVSS flux density $>5.0$ mJy & $\sim$3,324 deg$^2$   & 2,712  &  No & No\\  
\citet{Gendre08} &  Cross-identification of FIRST and NVSS samples and then search of optical counterparts in SuperCosmos Sky Survey \citep{Hambly01}& $0.003<z<3.5$;  NVSS flux density $>1.3$ Jy & $\sim$4,924 deg$^2$   & 274  &  Yes & No\\ 
\citet{Kimball08}$^\dag$ &  Cross-identification of FIRST, NVSS, WENSS samples with SDSS sample within $\Delta\mathrm{r}=2\arcsec$ & optical magnitude $<17.77$ & $\sim$3,000 deg$^2$  & 2,885  &  No & No\\ 
\citet{Gendre10} &  Cross-identification of FIRST and NVSS samples and then search of optical counterparts in SDSS and 2\,MASS \citep{Skrutskie06} & $0.003<z<3.5$;  NVSS flux density $>50$ mJy & $\sim$4,924 deg$^2$   & 859  &  Yes & No\\ 
\citet{Lin10}$^*$ &  Cross-matching of SDSS spectroscopic sample with NVSS using $\Delta\mathrm{r}=3\arcmin$ and then matching with FIRST & $0.02<z<0.3$; absolute optical magnitude
$<21.27$ and NVSS flux density $>3.0$ mJy & $\sim$6,008 deg$^2$   & 1,040  &  No & No\\ 
\citet{Best12} &  Cross-matching of SDSS DR\,7 spectroscopic sample with NVSS using $\Delta\mathrm{r}=3\arcmin$ and identifying with FIRST & $0.01<z<0.7$; optical magnitude $<17.77$ and NVSS flux density $>5.0$ mJy & $\sim$11,664 deg$^2$   & 18,286  &  No & No\\  
\citet{Banfield15}$^\ddag$ &  Cross-identification of FIRST and ATLAS samples with WISE and SWIRE samples & FIRST flux density $>1$ mJy and ATLAS $>15$ $\mu$Jy &    &  $\sim$30,000 &  No & No\\  
\citet{Williams19}$^\ddag$ &  Cross-identification of LoTSS DR\,1 sample with Pan-STARRS and WISE samples & g-band optical magnitude $<23.3$ and {\it W1} IR-magnitude $<19.0$ and 150 MHz flux density $>0.639$ mJy & $\sim$424 deg$^2$   & 231,716  &  No & No\\  
This work &  Cross-matching of SDSS DR\,7 spectroscopic sample with FIRST using $\Delta\mathrm{r}=3\arcsec$ and then cross-identifying with NVSS sample & $z<0.7$; optical magnitude $<17.77$ and FIRST flux density $>0.6$ mJy & $\sim$11,664 deg$^2$   & 32,616  &  Yes & Yes\\  \hline
\end{tabular}   
\end{center}
\begin{minipage}{1.0\textwidth}

NOTE--Columns: (1) reference for the sample; (2) key points on the selection procedure; (3) $z$ and flux density limits imposed on the data; (4) total sky coverage; (5) total number of derived radio sources; (6) visual classification of radio morphology; (7) visual classification of optical morphology. \\
$^\dag$ Appendix B for other AGN samples\\
$^*$ Visual identification of extended sources.\\ 
$^\ddag$ On-going surveys/projects.\\ 
\end{minipage}
\end{table*}

Several attempts have been made to obtain such samples by utilizing multiwavelength datasets from surveys, which normally cover large portions of the sky. In particular, Sloan Digital Sky Survey \citep[SDSS;][]{York00}, Palomar Observatory Sky Survey \citep[POSS~\textrm{I};][]{McMahon92}, Las Campanas Redshift Survey \citep[LCRS;][]{Shectman96} at optical frequencies, NRAO VLA Sky Survey\footnote{https://www.cv.nrao.edu/nvss/} \citep[NVSS;][]{Condon98}, First Images of Radio Sky at Twenty Centimetre survey\footnote{http://sundog.stsci.edu/} \citep[FIRST;][]{Becker95, White97}, and Sydney University Molonglo Sky Survey \citep[SUMSS;][]{Mauch03} at radio frequencies, infrared from the Wide Field Survey Explorer \citep[WISE;][]{Wright10} and the Infrared Astronomical Satellite \citep[IRAS;][]{Moshir92}, and X-rays from the X-ray Multi Mirror (XMM)--{\it Newton} \citep{Rosen16}, {\it Chandra} \citep{Evans10}, and {\it Swift}-Burst Alert Telescope \citep[BAT;][]{Baumgartner13} databases have been vastly used to explore the AGN phenomena \citep[e.g.,][]{Machalski99, McMahon02, Best05, Gendre08, Best12, Mingo14, Mingo16, Gupta18, Sabater19}. 

Due to the large number of sources detected by all-sky surveys, most of these studies rely on deriving AGN samples using ``automated--(not by eye)'' selection methods. For example, at radio frequencies, using the data from the FIRST survey, \citet{Proctor11} derives a sample of radio galaxies (including quasars) by counting the number of radio components within $\sim$0\farcm96\, radius from the source. A different approach was applied by \citet{vanVelzen15}, who selected double-lobed radio sources by counting all the radio components separated by angular distance up to 1\arcmin\, and flux density limit 12 mJy. In this manner, \citeauthor{vanVelzen15} could select only FR~II sources after applying several cuts: minimum angular separation (18\arcsec), ratio of the integrated flux of the lobes ($f_{l/l}<10$), and integrated--to--core flux ratio ($F_i/F_p<5$). Radio AGN samples derived from cross-matching of sources in different wavebands are subject to contaminations due to source confusion (depending on the angular resolution) and large uncertainty in reliable association due to the complex nature of spatially extended radio sources. 

The only exceptions to this rule are the on-going Low Frequency Array (LOFAR) Two-metre Sky Survey \citep[LoTSS;][]{Shimwell17} and the Radio Galaxy Zoo \citep[][]{Banfield15} projects. LoTSS uses a combination of automated algorithms as well as visual identifications of the host galaxies of radio sources. The current LoTSS DR\,1 release covers 2\% of the sky and provides optical and/or IR identifications from the Panoramic-Survey Telescope and Rapid Response System (Pan-STARRS; \citealt{Chambers16}) and WISE surveys \citep{Shimwell19, Williams19}. The Radio Galaxy Zoo selects radio sources from FIRST and from the Australia Telescope Large Area Survey DR~3 (ATLAS; \citealt{Franzen15}) and provides IR identifications from the WISE and Spitzer Wide-Area Infrared Extragalactic Survey (SWIRE; \citealt{Lonsdale03}) samples. However, as can be seen from Table~\ref{AGNsamples}, which summarizes the main features of the previous and current efforts to obtain samples of radio sources, most of the catalogs do not provide detailed radio morphological classification, and practically none of them give the morphological classification of the host galaxy. However, we note that the main focus of majority of these studies is to provide samples of radio AGNs with optical counterparts and not necessarily the morphological classification which is the main focus of the present study. Very recently, the radio AGN sample from \citet{Best12} has been used as a parent sample to extract lists of AGNs with specific radio morphological classifications, see for example, \citet[][]{Capetti17a} for FR~I sources, \citet[][]{Capetti17b} for FR~II sources, \citet[][]{Baldi18} for FR sources with linear sizes $<$5 kpc, \citet[][]{Missaglia19} for wide--angle tail (WAT) sources, and \citet[][]{Jimenez-Gallardo19} for compact sources with linear sizes $<$60 kpc.

The present study provides a catalog of radio sources associated with optical galaxies, having their central radio component within 3\arcsec\ from the position of the optical galaxy, and comprising of {\it unresolved} or {\it extended} radio morphology. The catalog contains sources with: (1) spectroscopic redshift ($z$); (2) good quality optical spectrum from SDSS to study host galaxy and emission line properties; (3) measured radio flux densities of radio structures; (4) low flux density limit corresponding to the flux density limit of the FIRST radio survey; (5) for each source it provides a morphological classification of the radio structure and of the host galaxy. The present catalog is \textit{handmade} and the radio and the optical morphological classifications are performed \textit{visually}. It provides the {\it largest} sample of spectroscopically selected radio galaxies to date, covering $\sim$30\% of the entire sky \citep[see, in this context,][]{Ching17}. We emphasize that in contrast to previous catalogs based on galaxies from the SDSS DR\,7 release \citep{Abazajian09}, we do not impose any additional radio flux density detection limit (see Table~\ref{AGNsamples}). Therefore, our catalog of the Radio sources associated with Optical Galaxies and having Unresolved or Extended morphologies I (ROGUE~I) contains sources with flux densities reaching down to $\sim$sub--mJy levels, corresponding to the 3$\sigma$ radio source detection provided by the FIRST survey. As a consequence, the ROGUE I catalog has the same limitations as the radio and optical catalogs used for selection. However, based on the ROGUE I catalog statistically complete samples can be selected, allowing detailed investigation of e.g., the AGN phenomena as a function of BH mass, host galaxy mass, stellar population, morphological type (both radio and optical), which will be the subject of forthcoming papers. 

In Section~\ref{sec:sample} we describe in more detail all the data assembled while cross-matching. The identification procedures are described in Section~\ref{sec:methodology}, where we also describe our schemes for morphological classification and our flux density estimation methods. The results are outlined in Section~\ref{sec:result}. Comments, reclassifications, and new discoveries are presented in Section \ref{sec:resultComments}. Section~\ref{sec:discussion} gives the summary. 

\section{Sample selection}
\label{sec:sample}

The first step in the sample selection process was the identification of a parent sample of galaxies with an optical spectrum which will allow the study of stellar population and emission line properties. Our sample, consisting of 673,807 galaxies, is drawn from the SDSS Main Galaxy Sample \citep{Strauss02} and the Red Galaxy Sample \citep{Eisenstein01} based on the spectrum quality \citep[signal-to-ratio in the continuum at 4020 \AA\  $\geq$ 10;][]{Koziel-Wierzbowska17}. We note that some parts covered by SDSS DR\,7 have been observed repeatedly, therefore, in order to avoid duplication in the parent sample, we matched the galaxies in right ascension ($<$0\farcs5), declination ($<$0\farcs5), and $z$ ($\pm 0.005$). This gave us a total of 662,531 unique SDSS galaxy candidates in which we searched for the presence of a radio AGN counterpart.

In order to identify the radio sources associated with the selected optical galaxies, we used the FIRST and NVSS radio surveys conducted using the Very Large Array (VLA). These two surveys, although conducted at identical frequency (1.4 GHz) are widely different in terms of the angular resolution of the radio images and of the sensitivity for the point-like and extended/diffuse emission. The FIRST survey provides 5\farcs4 synthesized beam size images and is complete down to a flux density limit of 1 mJy for point-like sources. On the other hand, the NVSS survey provides 45\arcsec\, synthesized beam-size images and is complete down to a 2.5 mJy flux density limit. The FIRST survey is $\sim$2 times more sensitive in detecting compact sources with angular sizes $\lesssim$6\arcsec\, as compared to the NVSS survey, while the NVSS survey  is infinitely more sensitive than the FIRST survey for detecting extended/diffuse emission  due to its compact array configuration. The sky coverages of the FIRST and NVSS surveys overlap (declination $\geq -10$\dg\, and $-$40\dg, respectively), which makes them complementary to each other in order to perform a blind search for radio galaxies which might contain a point-like radio-core/hot spots and extended/diffuse emission. In addition, the part of the sky covered by the FIRST survey is almost identical to the portion of the sky covered by SDSS DR\,7 spectroscopic observations. Therefore, by combining the SDSS optical survey and the FIRST and the NVSS radio surveys we are able to identify the radio counterparts of the selected SDSS DR\,7 galaxies.

\section{Methodology}
\label{sec:methodology}

We have performed the search in two steps: (1) the optical position of a galaxy from the SDSS catalog was cross-matched with the radio position from the FIRST catalog allowing for an error of 3\arcsec; (2) once the match was found in the FIRST catalog, we made optical/radio overlay maps with angular sizes corresponding to 1 Mpc linear size at the source distance, centered at the host galaxy position, to visualise the morphologies of the radio sources. For this, we used the optical images from the Digitized Sky Survey (DSS)\footnote{http://archive.eso.org/dss/dss} and the radio images from the FIRST and NVSS surveys.  We note that optical galaxies for which we did not find a radio match in step (1) are not included in the present catalog. These 629,815 remaining galaxies from our parent SDSS sample might still host extended radio emission, but without a FIRST detection at the position of the optical host galaxy, which will be searched for in future work and lead to the publication of our second catalog, ROGUE~II. Below we describe our selection procedure in detail.

\subsection{Optical galaxies with radio cores: cross-matching of source positions}
\label{sec:crossmatch}

We searched for a radio counterpart by cross-matching the optical positions of the SDSS galaxies with the positions of the radio sources listed in the FIRST catalog. Since the radio and optical surveys have different resolutions, we chose an error circle within which a radio source can be assumed to be coincident with the optical galaxy. The search radius, $\Delta$r, is defined as 
\begin{eqnarray}
\Delta \mathrm{r} &=& [((\alpha_\mathrm{SDSS} - \alpha_\mathrm{FIRST}) \times \cos\delta_\mathrm{SDSS})^2 \nonumber \\
&& + (\delta_\mathrm{SDSS} - \delta_\mathrm{FIRST})^2]^{1/2},
\label{match_coor}
\end{eqnarray}

where $\alpha_\mathrm{SDSS}$, $\delta_\mathrm{SDSS}$ are the right ascension and declination of the optical positions of the sources from the SDSS DR\,7 list and $\alpha_\mathrm{FIRST}$, $\delta_\mathrm{FIRST}$ correspond to the right ascension and declination from the FIRST list, respectively. If the computed $\Delta$r was found to be $\leq$3\arcsec\, \citep[adopted after][see the discussion therein]{Singh15a}, the radio counterpart was considered to be coincident with the optical source. In this manner, we identified 32,616 optical galaxies with a FIRST counterpart initially treated to be a radio core.

\subsection{Source morphology: visual identification and classification}

Once a radio source was found to coincide with the optical galaxy position (Eq.~\ref{match_coor}), we derived the projected angular size corresponding to 1 Mpc linear size, according to the redshift of the optical galaxy. The luminosity distance to the source, ${\mathrm{d_L}}$, has been computed using the concordant cosmology with the Hubble constant  $ H_0= $ 69.6 km/s/Mpc, $\Omega_{M} = 0.286$, and  $\Omega_{\Lambda} = 0.714$ \citep{Spergel07}. The luminosity distance has been calculated following \citet{Hogg99}:

{\small
\begin{eqnarray}
&{\mathrm{d_L}}(z; H_0,\Omega_{M},\Omega_\Lambda)  = \frac{c(1+z)}{H_0} \times \\
&\int_{0}^z [(1+z')^2 (1+\Omega_{M} z')-z'(2+z')\Omega_\Lambda]^{-1/2} dz'] \, [Mpc],\nonumber 
\label{dl}
\end{eqnarray}
}
where $c$ is the speed of light. 

The luminosity distance has then been converted to the angular diameter distance, $\mathrm{d_m}$, 
\begin{equation}
{\mathrm{d_m}} = \frac{\mathrm{d_L}}{(1+z)^2} \mathrm{[\,Mpc]} .
\label{dm}
\end{equation}
Finally, we applied the small angle formula to convert a linear size into an angular size:
\begin{equation}
\label{smf}
\mathrm{angular \, \, size} = 206,265\, \frac{\mathrm{linear \, size}}{\mathrm{d_m}} \, [\arcsec].
\end{equation}

Subsequently, we searched for associated radio source within 1 Mpc linear size by making radio (1.4 GHz contours from FIRST and NVSS) -- optical (grayscale from DSS) overlay in the Astronomical Image Processing System (\textsc{AIPS}), using the \textsc{kntr} task. This image size was chosen to ensure that we do not miss out on a large number of galaxies with extended radio emission as the fraction of large (linear size $>$1 Mpc) double-lobed quasars is negligible \citep[see][]{deVries06}. For the optical/radio overlay, we selected contour levels starting from $\sim$3 times of the typical rms provided by the surveys (i.e., 0.6 mJy beam$^{-1}$ and 1.35 mJy beam$^{-1}$ for the FIRST and the NVSS maps, respectively). 

 \subsubsection{Radio morphological classification}
\label{galradmorph}

The FIRST and the NVSS radio maps of our entire sample of 32,616 sources were visually inspected.  All three authors participated in the classification. At first, to train ourselves and to adjust our classification scheme, we jointly classified a set of 1,000 sources. Then, we assigned a set of $\sim$11,000 sources to each of us and we carried on the classification separately.  After this round of morphological assignments, two authors together verified all the sources in terms of compatibility with previously adopted classification scheme. 
The radio morphological classification was made separately for the FIRST and the NVSS maps and the \textit{individual} FIRST and NVSS classifications were assigned to the source using the codes described in Table~\ref{tab:RadioMorph}. Figure~\ref{fig:radioMorph} gives example maps for each of our radio morphological classes. 

We classified single-component sources as compact (unresolved) or elongated according to their sizes given by FIRST (or NVSS): sources with both axes of deconvolved Gaussian equal to zero were classified as compact, while sources with at least one deconvolved dimension larger than zero were classified as elongated. Single-component NVSS sources were classified as compact when only upper limits of both Gaussian axes were available in the NVSS catalog. The classification of sources with at least two components, which we refer to as {\it extended}, in the FIRST or the NVSS catalogs used the classical \cite{Fanaroff74} classification scheme, where sources with collinear lobes brighter at the edges are classified as FR~II radio galaxies, while sources in which the brightest part of a lobe is close to the center are classified as FR~I radio galaxies. We separated also hybrid sources \citep{Gopal-Krishna00, Kapinska17} with one lobe of FR~I and another of FR~II type morphology, and one-sided FR~I or FR~II sources. Sources classified as one-sided have symmetric NVSS radio emission suggesting that the second lobe is not detected in FIRST. In our classification scheme we distinguished sources with more complex morphology, such as: Z--shaped sources (lobes forming Z or S shape structure), X--shaped (with two inclined pairs of lobes), double--double radio sources (two collinear pairs of lobes), wide-angle tail (WAT; lobes forming an obtuse angle) and narrow-angle tail (NAT; lobes forming an acute angle) sources, head-tail (HT; bright radio core connected to one-sided tail-like emission) sources, or halo (diffuse radio emission around a core) sources. Our classification of bent sources is based only on the morphological features. We note that from the data used in our search we are not able to verify if the source is a cluster member which is often used as the feature of WAT/NAT sources \citep[e.g.,][but, see also, \citealt{Missaglia19}; \citealt{Mingo19}]{Leahy93}. 
In some galaxies the location and morphology of the radio emission allowed us to classify it as coming from star-forming regions (SFR). Sources with radio emission too complex to be classified into one of the above classes were marked as sources with not clear morphology. In the cases where several sources of radio emission were measured jointly in the FIRST or NVSS catalog as one detection we classified this detection as blended. Furthermore, we found several cases in which the radio emission cannot be physically linked to the optical galaxy under consideration, these are: (i) radio emission is below the NVSS detection threshold or (ii) the radio emission is produced by a nearby galaxy or by just a part of the same galaxy, and we classified these galaxies as not detected in radio. 

The {\it final} radio morphological classification is based on the combination of the FIRST and the NVSS classifications, because of sheer differences in resolutions and the sensitivity to the extended emission of the two surveys. Due to higher angular-scale resolution of the FIRST survey,  compact and elongated types are assigned based solely on the deconvolved angular sizes of the radio components provided by the FIRST survey.

In the case of sources showing extended structures, more weight is given to the map with detailed morphological features seen in the source. For example, a small angular size source could be classified as a double-double radio galaxy (DDRG; two separate lobes on either side of the core) in FIRST, while it could be classified only as FR~II in NVSS due to its larger beam size. In such a case, the final classification is DDRG for the source. Similarly, when only a bright core could be seen in FIRST, while the lobe and the counter lobe could be detected in NVSS due to its higher sensitivity for the extended emission, then the  assigned classification is FR~II. In the case of sources for which we were not able to assign a reliable classification, we use the prefix p meaning possible.

\subsubsection{Optical morphological classification}
\label{galoptmorph}

The optical morphological classification of the galaxies associated with the radio emission was done using the 120\arcsec\, image snapshots from SDSS. The assigned morphologies were chosen from the classes given in Table~\ref{OpticalMorph}. Those use the standard Hubble classification scheme with some additional classes like distorted galaxies, ring galaxies, and galaxy mergers. We also distinguished objects where the SDSS spectrum concerns a star forming region or a part outside the nucleus of a galaxy. Figure~\ref{fig:optMorph} gives examples of our optical morphological classification scheme. In some cases we used additional codes to describe details of the morphological type like the presence of a bar or signs of interactions and used the prefix p (for possible) when we considered that the classification was uncertain.

\begin{figure*}[!htb]
\scriptsize
\hspace{0cm}{\includegraphics[width=0.3\textwidth]{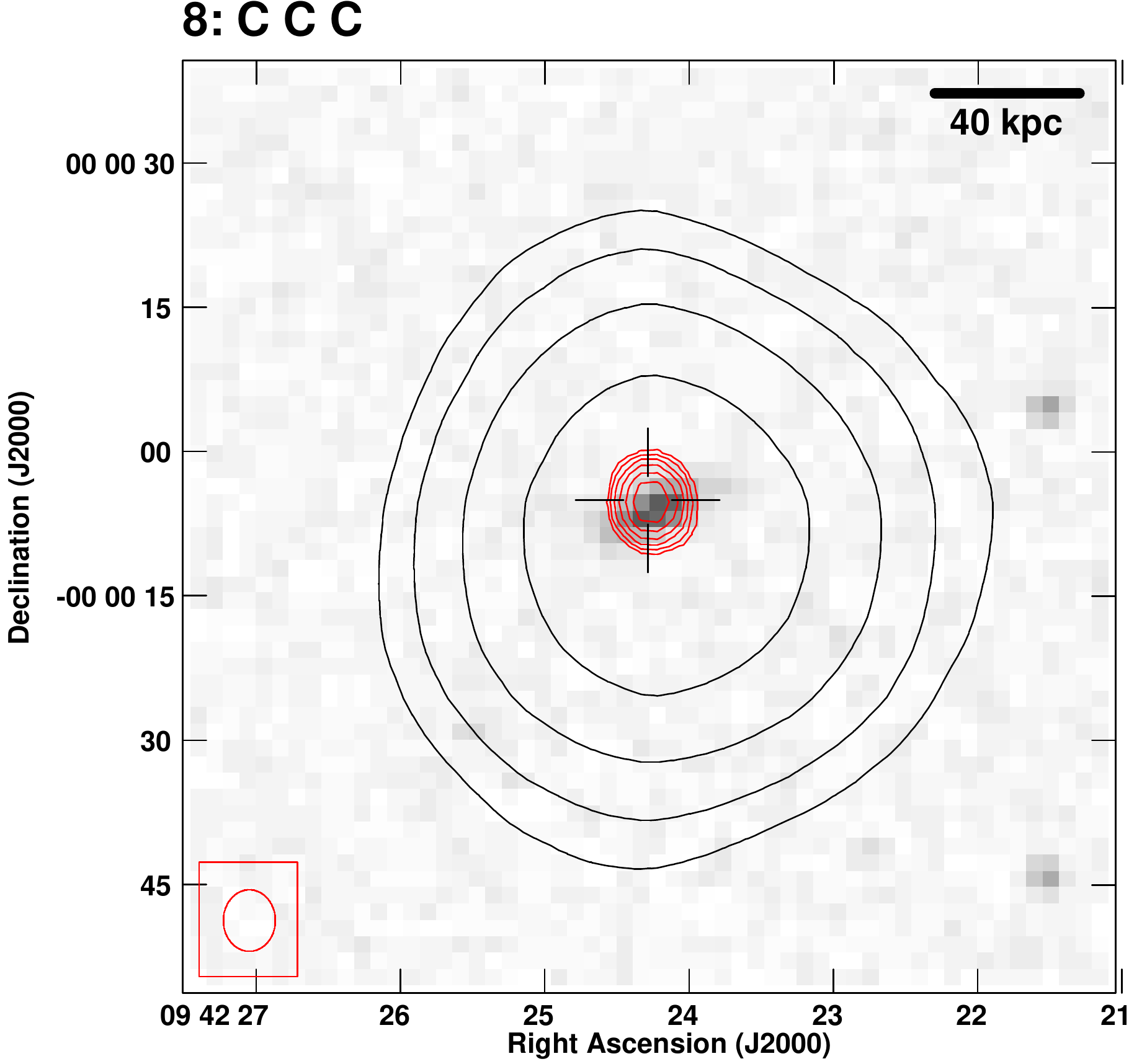}}
\hspace{0.2cm}{\includegraphics[width=0.3\textwidth]{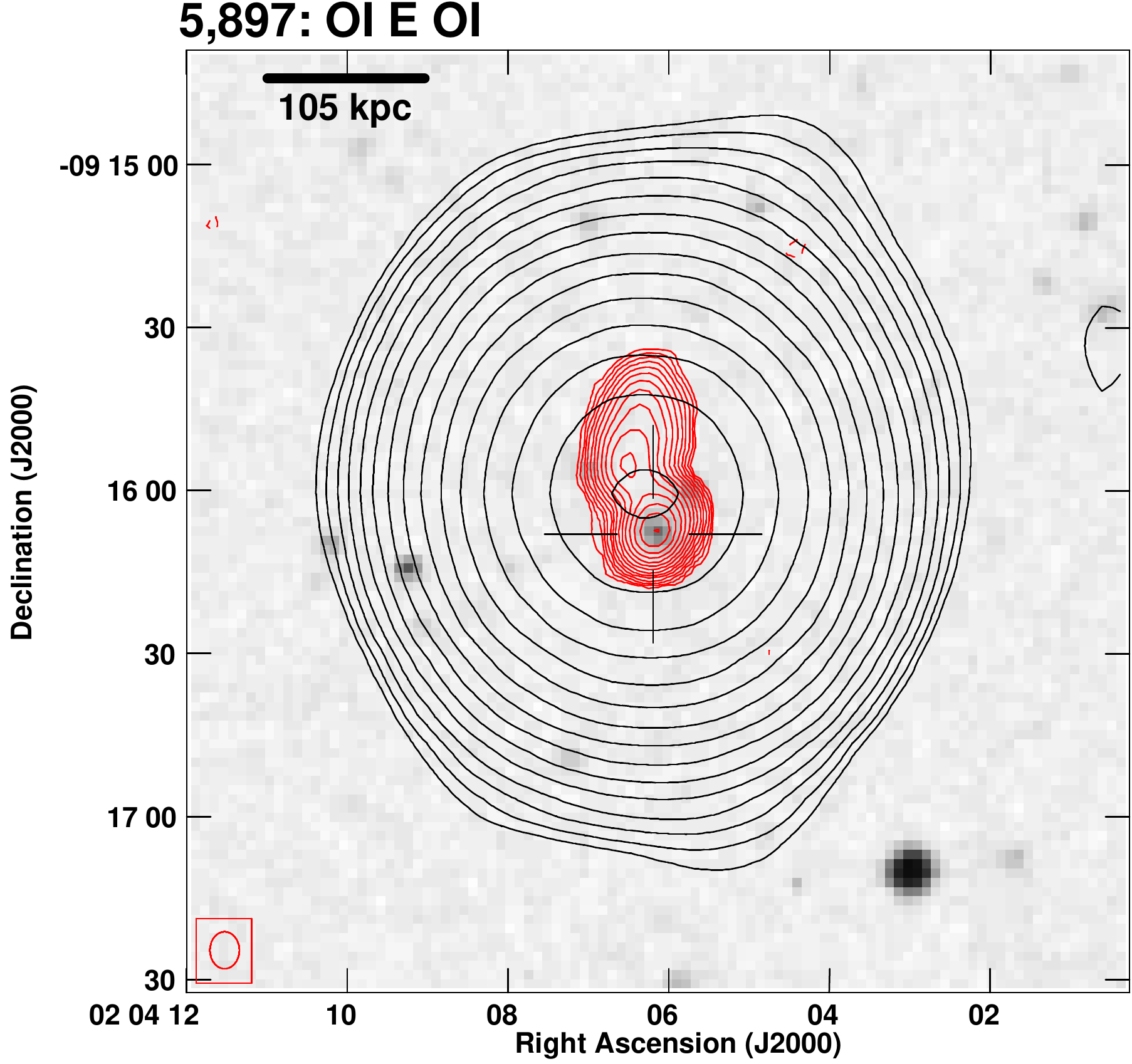}}
\hspace{0.2cm}{\includegraphics[width=0.3\textwidth]{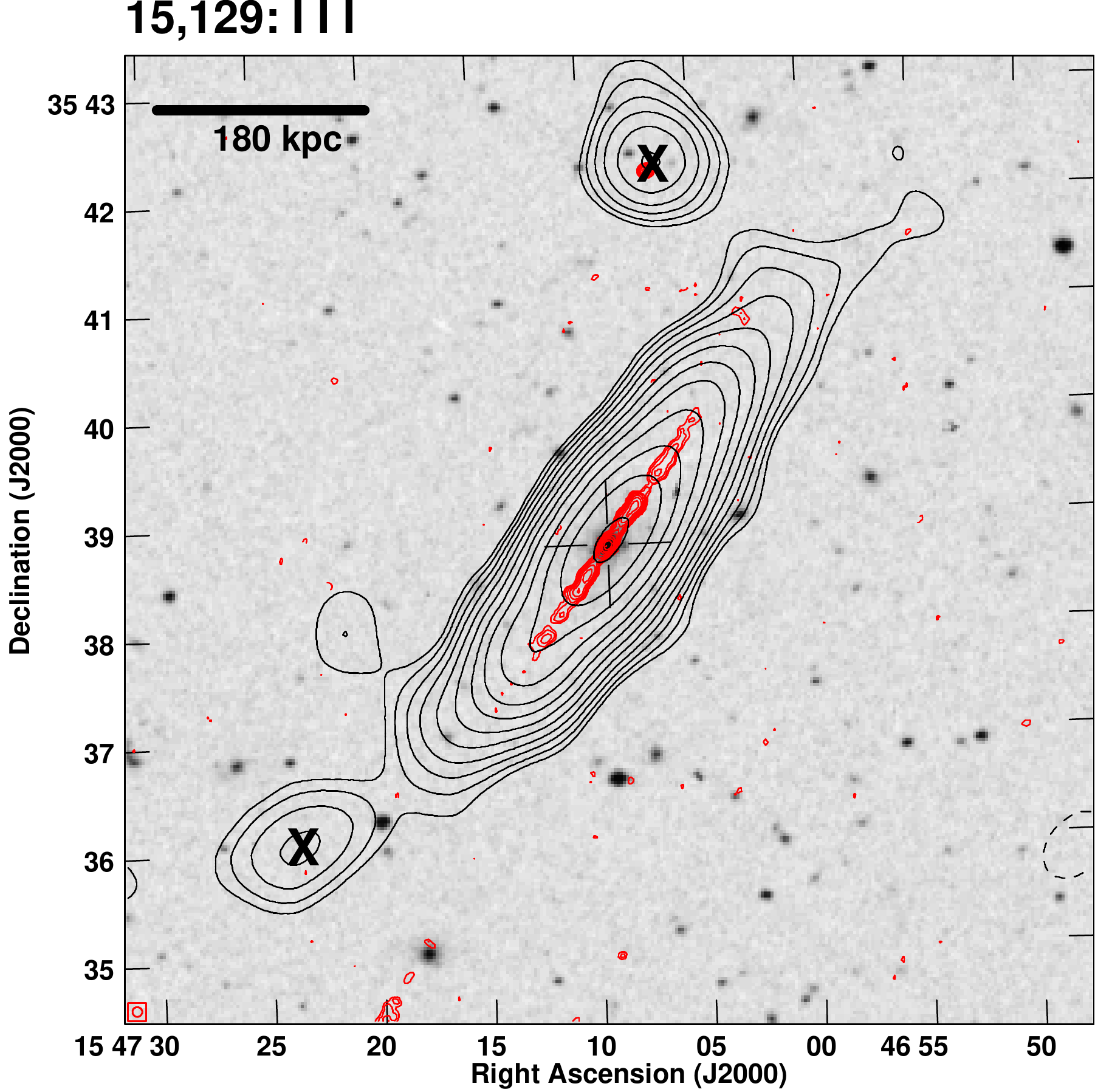}}
\hspace{0.2cm}{\includegraphics[width=0.3\textwidth]{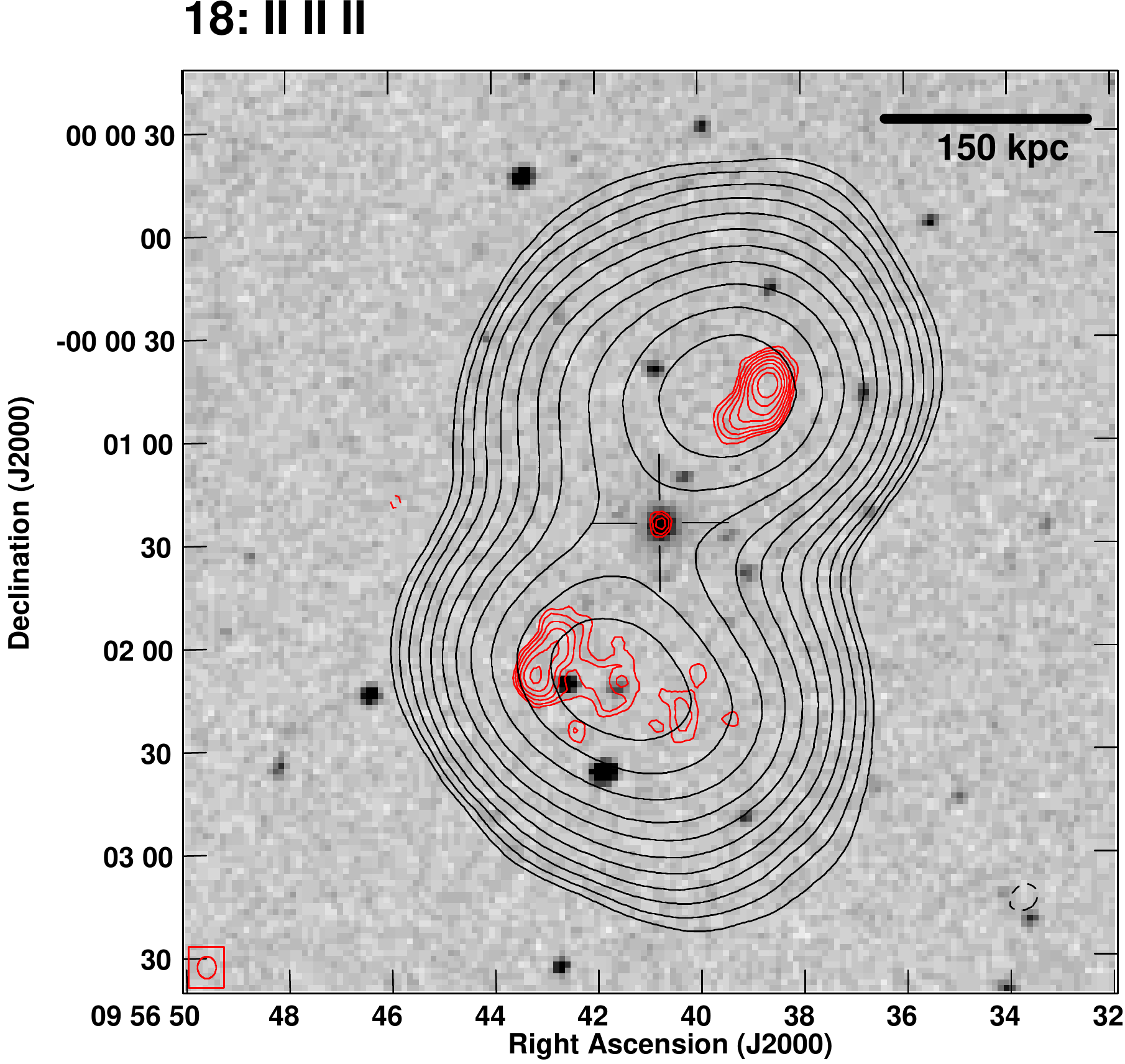}} 
\hspace{0.2cm}{\includegraphics[width=0.3\textwidth]{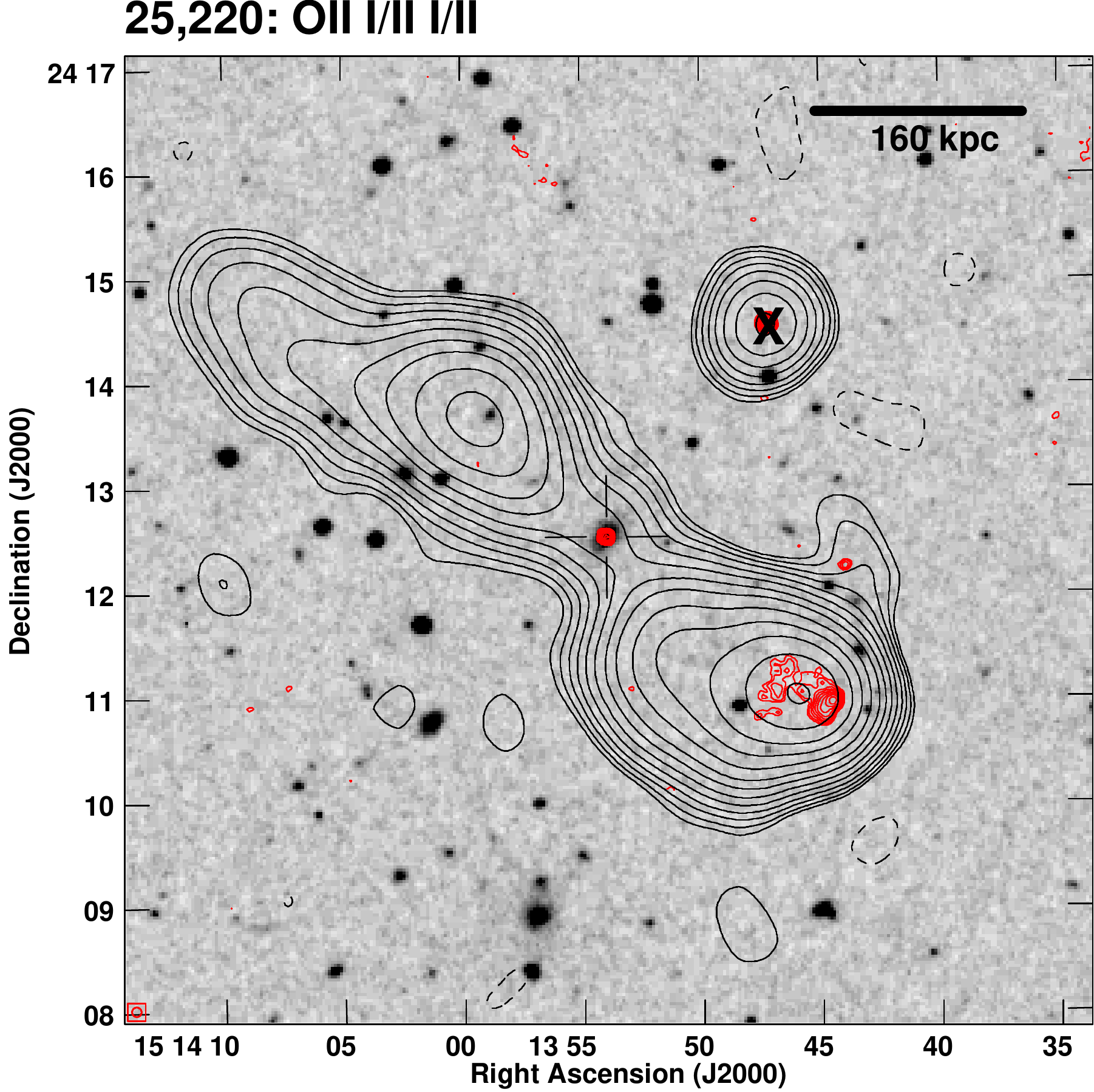}}
\hspace{0.2cm}{\includegraphics[width=0.3\textwidth]{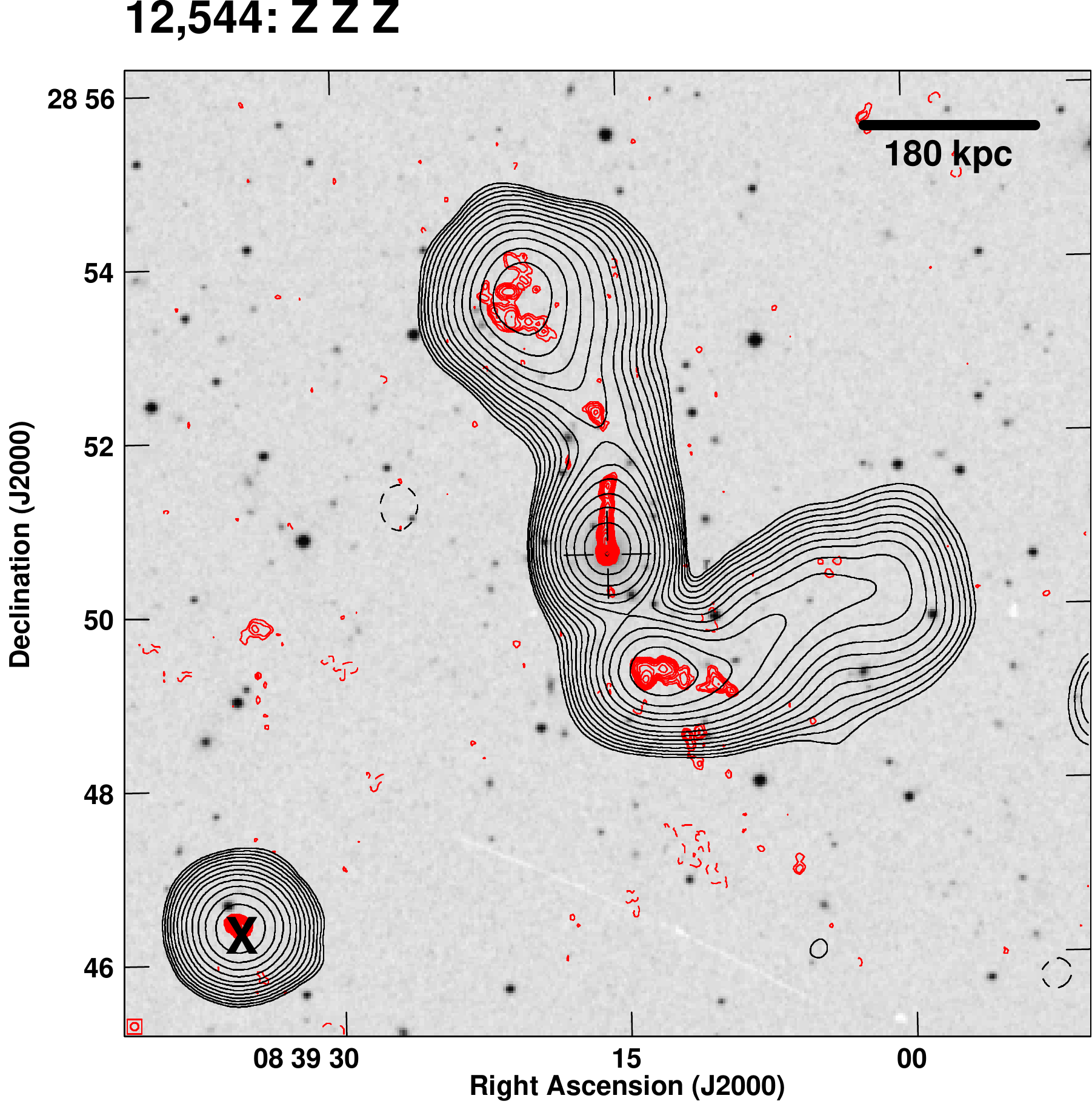}}
\hspace{0.2cm}{\includegraphics[width=0.3\textwidth]{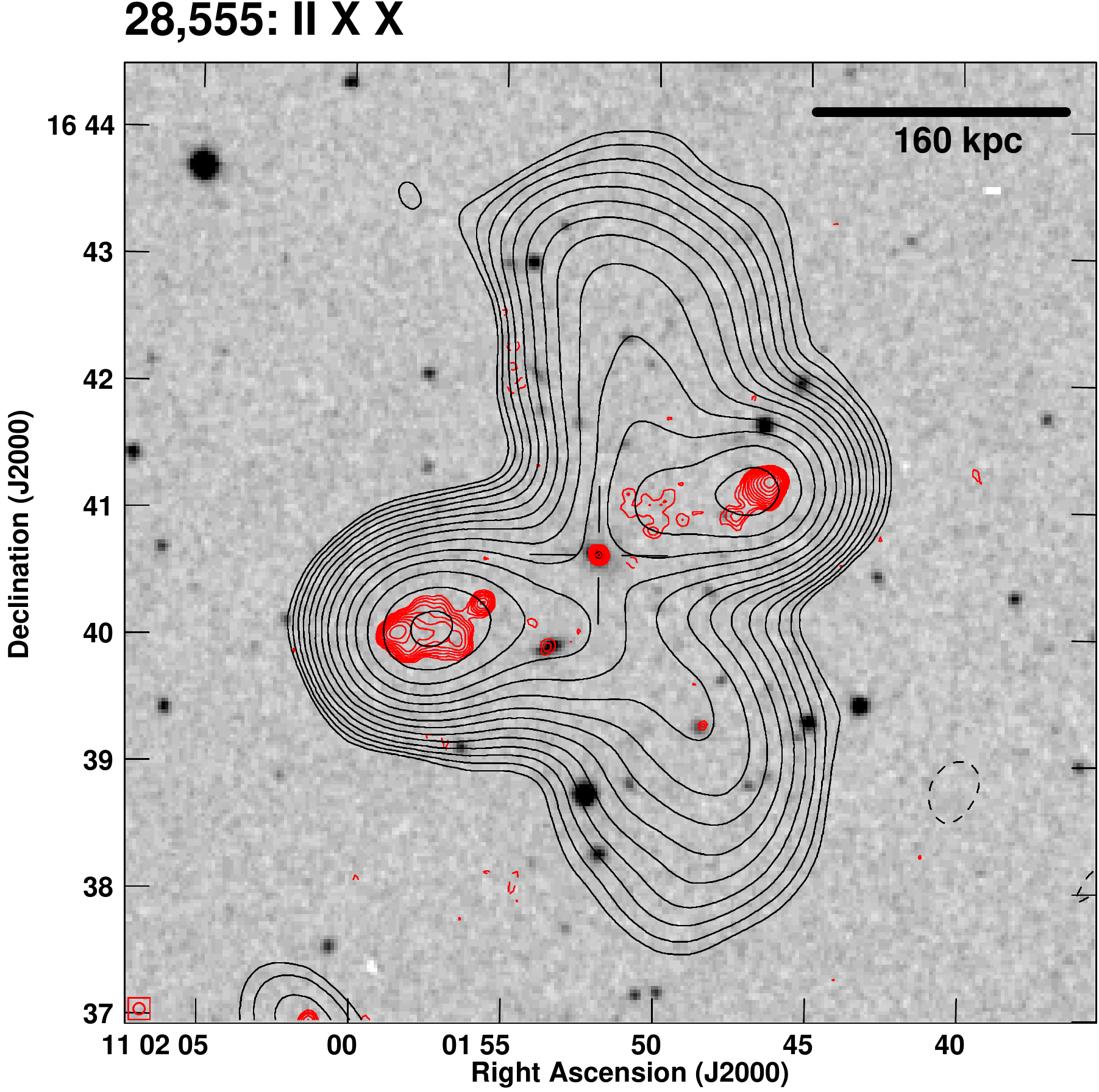}}
\hspace{0.8cm}{\includegraphics[width=0.3\textwidth]{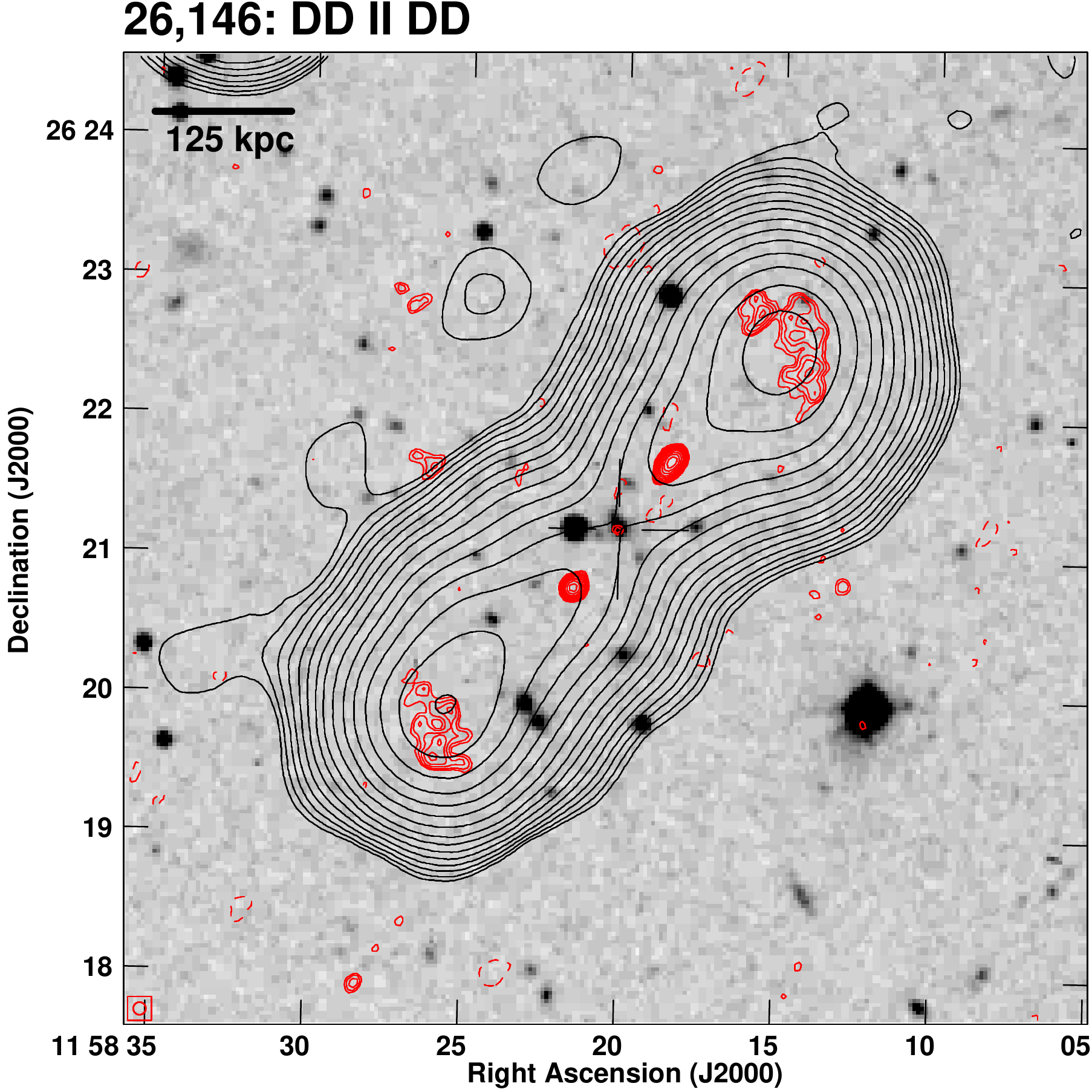}}
\hspace{0.8cm}{\includegraphics[width=0.3\textwidth]{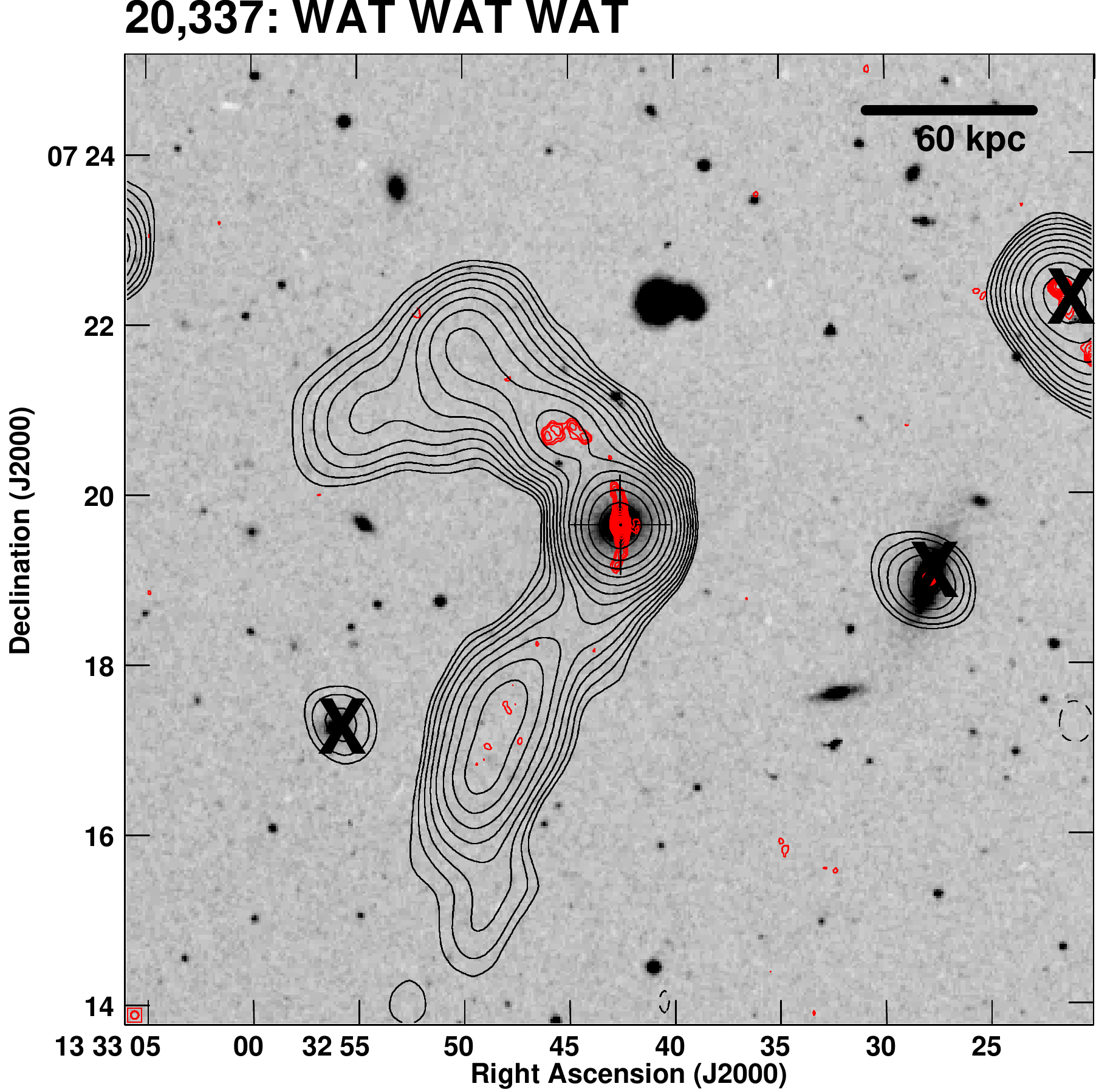}}

\caption{Examples of radio morphological classification assigned in the ROGUE~I catalog (Table~\ref{tab:RadioMorph}). The 1.4 GHz radio contours from the FIRST (red) and the NVSS (black) maps are overlaid on the optical DSS (gray scale) image, centered at the host galaxy position marked by a plus sign. Background/foreground sources are marked with ``X'' sign. The FIRST beam is placed inside the square box at the bottom left corner and the NVSS beam ($\sim$9 times of size of the FIRST beam) is not shown here for clarity. The contours levels, for the FIRST and the NVSS maps are drawn at 0.6 mJy beam$^{-1}$ and 1.35 mJy beam$^{-1}$ ($\sim3\sigma$), respectively, which increase by factors of ($\sqrt{2}$)$^{n}$ where $n$ ranges from 0,1,2,3,...20. The contours at $-$3$\sigma$ are shown by the dashed lines. In the title of each image we give the catalog number and the codes adopted for radio morphology based on FIRST, NVSS, and final classifications, respectively.}

\label{fig:radioMorph}
\end{figure*}

\renewcommand{\thefigure}{\arabic{figure} (Cont.)}

\begin{figure*}[!htb]
\scriptsize
\ContinuedFloat
\hspace{0.2cm}{\includegraphics[width=0.3\textwidth]{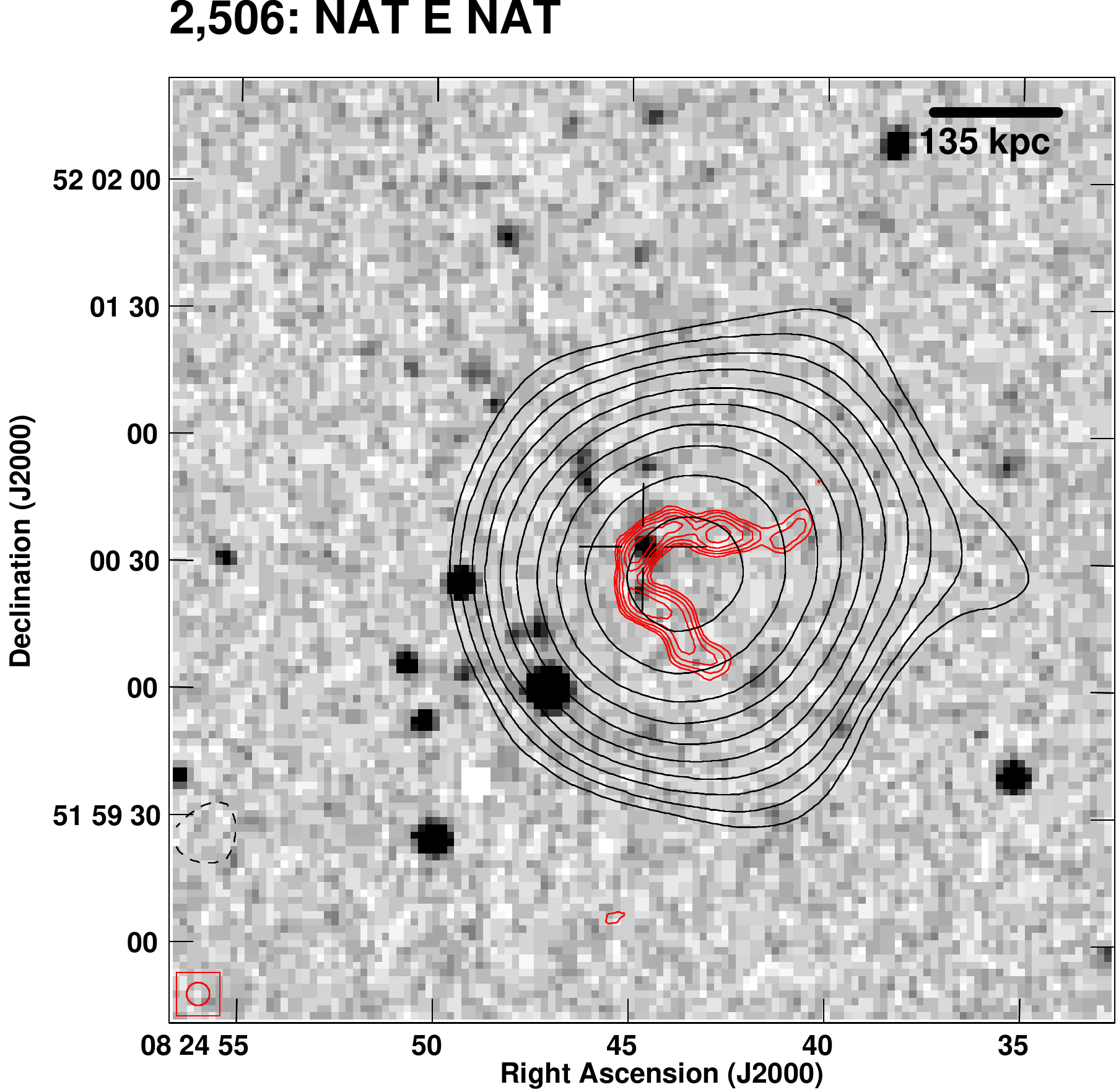}}
\hspace{0.2cm}{\includegraphics[width=0.3\textwidth]{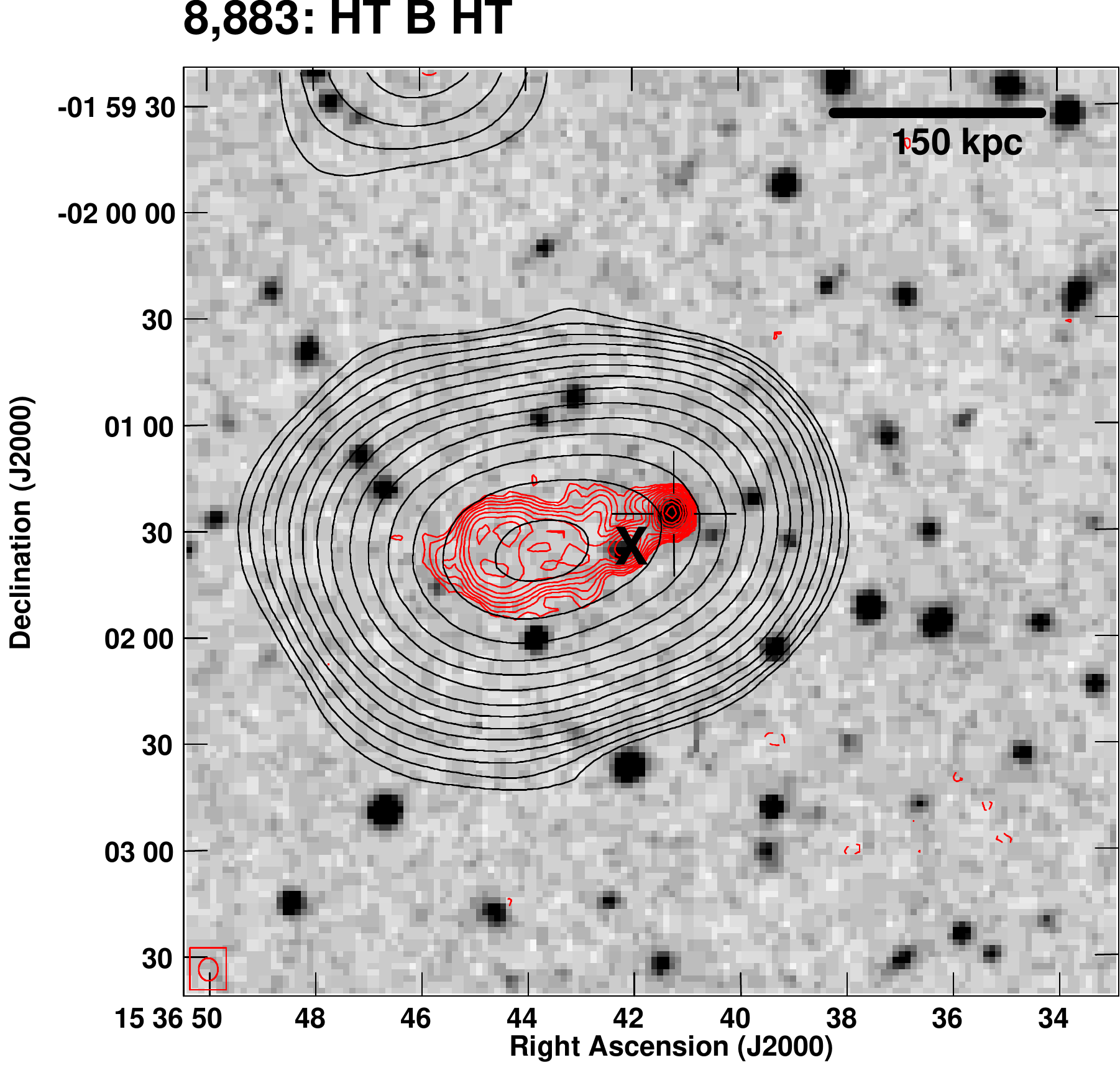}}
\hspace{0.2cm}{\includegraphics[width=0.3\textwidth]{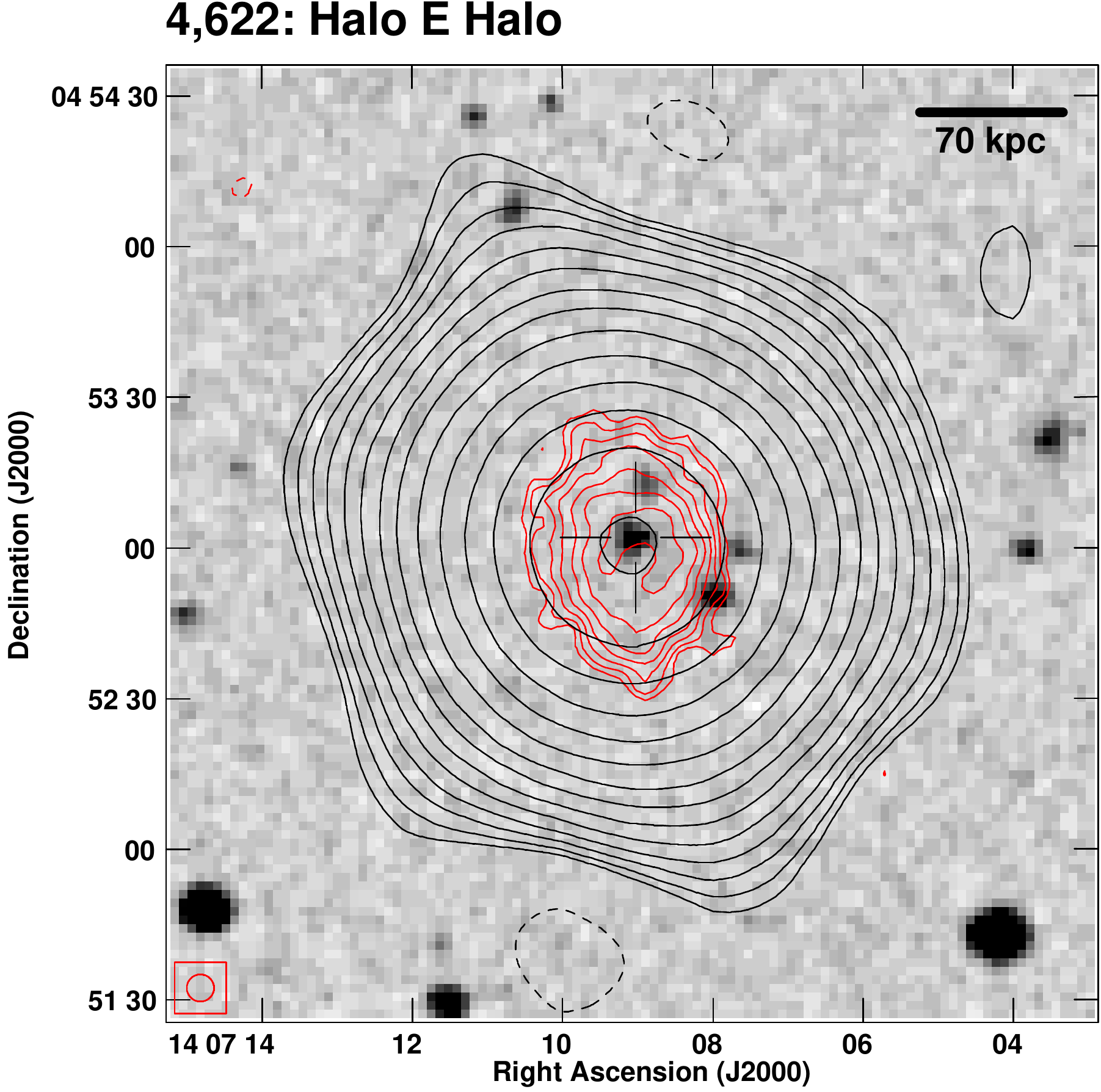}}
\hspace{0.2cm}{\includegraphics[width=0.3\textwidth]{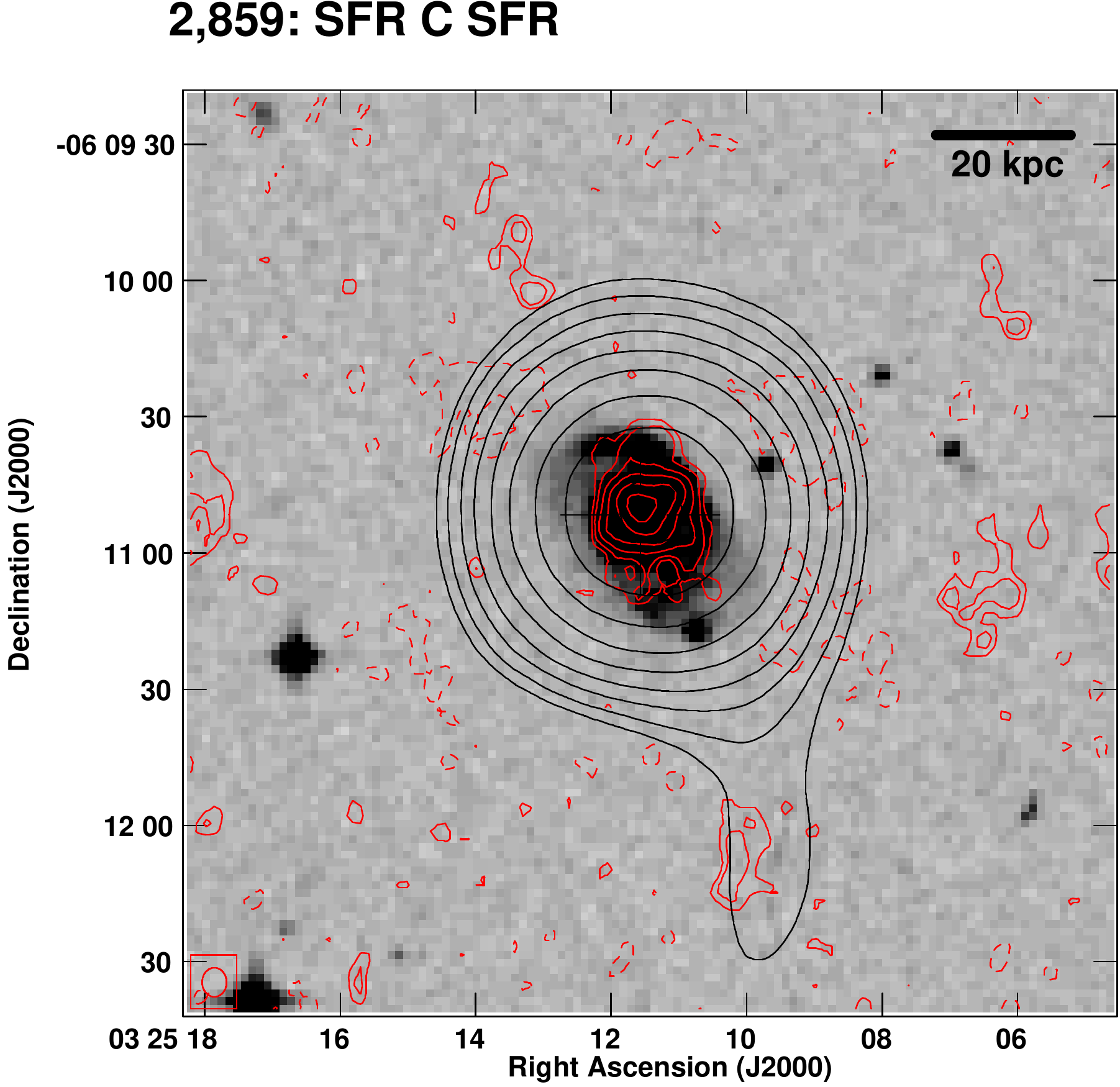}}
\hspace{0.8cm}{\includegraphics[width=0.3\textwidth]{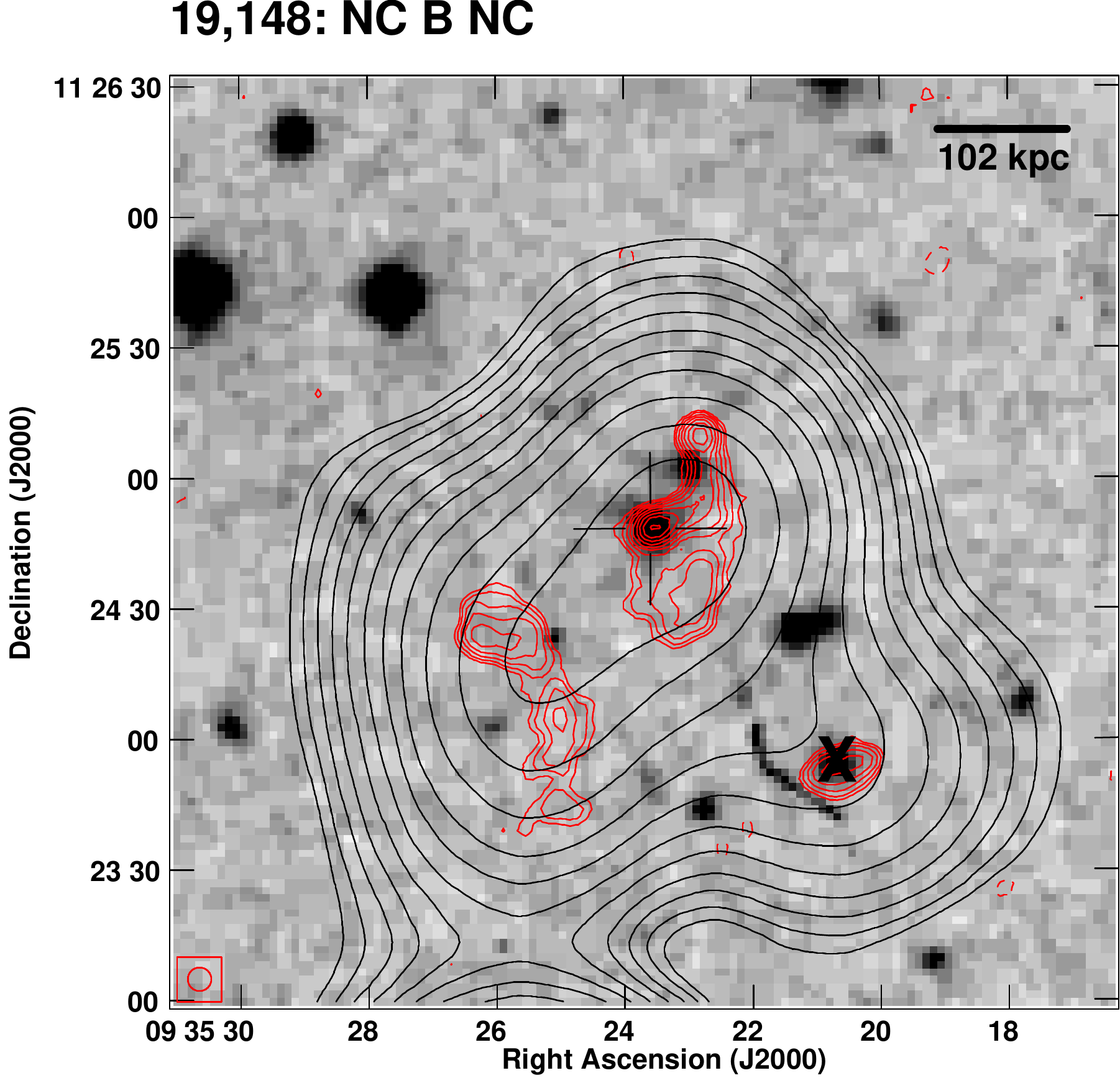}}
\hspace{0.8cm}{\includegraphics[width=0.3\textwidth]{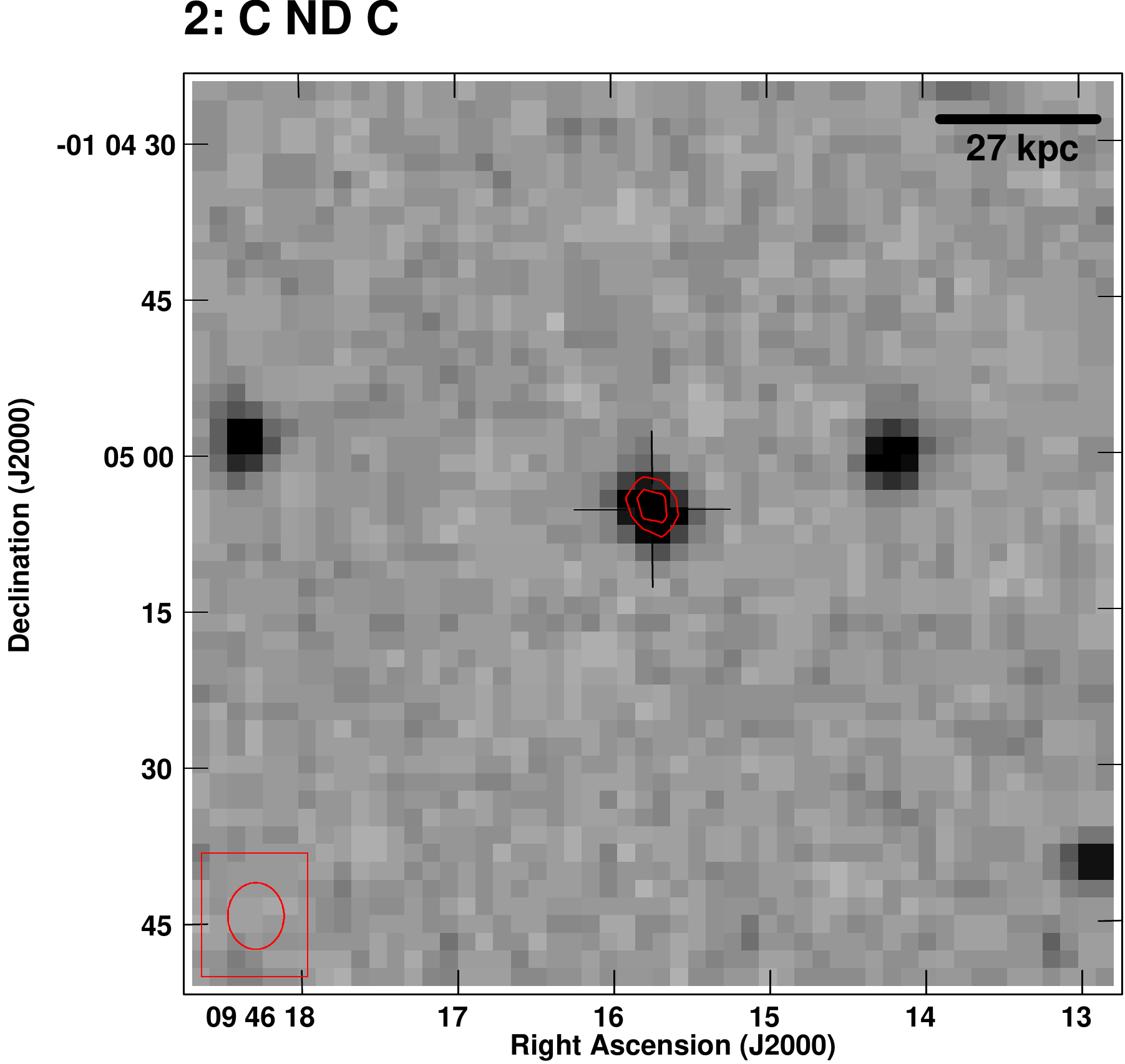}}
\caption{Examples of radio morphological classification assigned in the ROGUE~I catalog.\label{}}
\end{figure*}

\renewcommand{\thefigure}{\arabic{figure}}
\begin{figure*}[!ht]
{\includegraphics[width=0.3\textwidth]{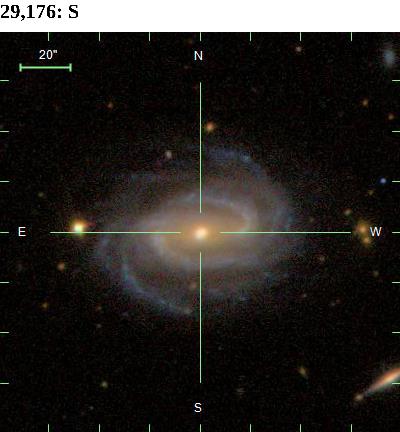}}\vspace{0.5cm}
{\includegraphics[width=0.3\textwidth]{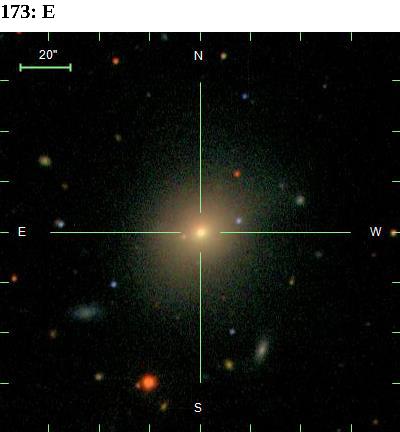}}
{\includegraphics[width=0.3\textwidth]{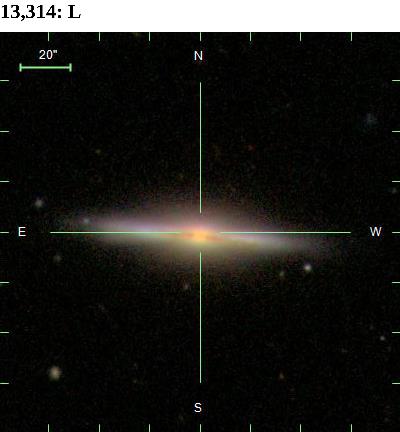}}
{\includegraphics[width=0.3\textwidth]{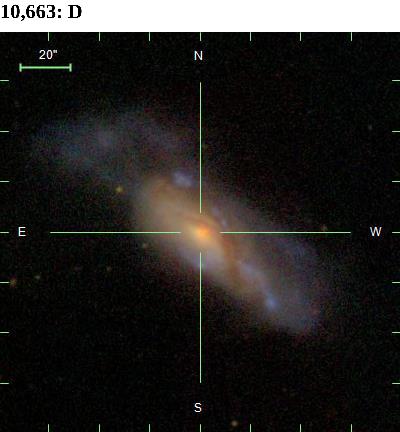}}\vspace{0.5cm}
{\includegraphics[width=0.3\textwidth]{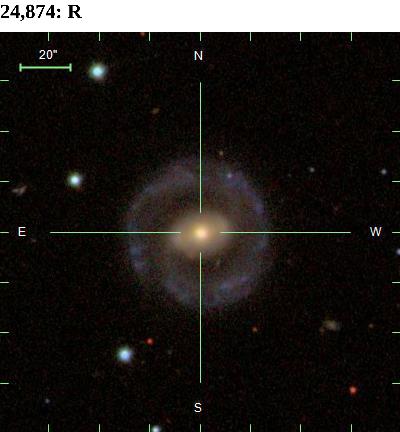}}
{\includegraphics[width=0.3\textwidth]{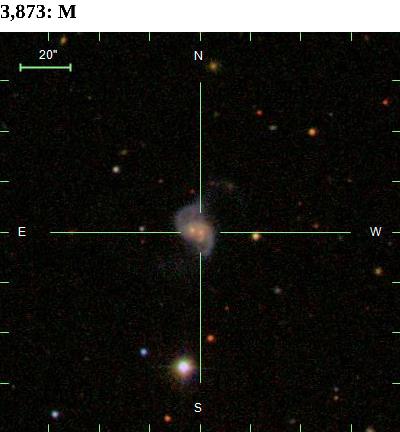}}
{\includegraphics[width=0.3\textwidth]{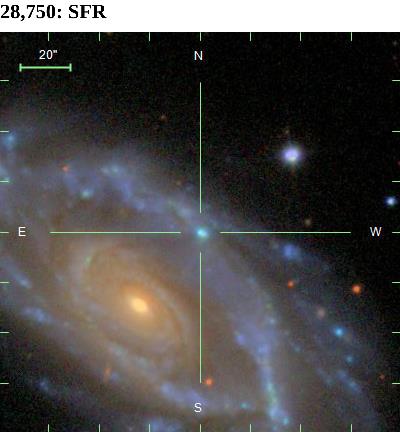}}\vspace{0.5cm}
\hspace{0.8cm}{\includegraphics[width=0.3\textwidth]{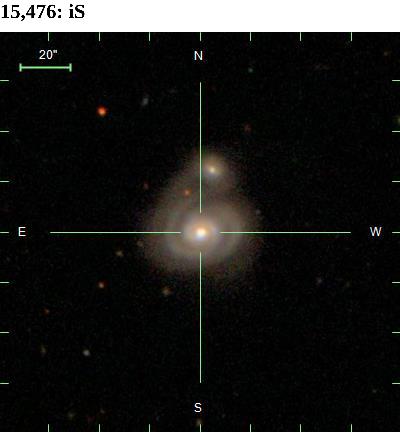}}
\hspace{0.8cm}{\includegraphics[width=0.3\textwidth]{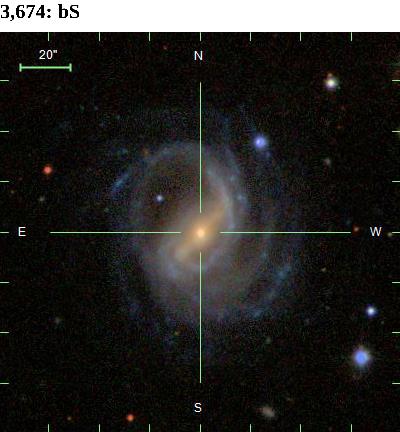}}
\caption{Examples of optical morphologies assigned in the ROGUE~I catalog (Table~\ref{OpticalMorph}) through {\it visual} inspection of the 120\arcsec\, SDSS image snapshots. Crosses indicate the position of the SDSS aperture used to measure the spectra. In the header of each image we give the catalog number and the code adopted for optical morphology based on the SDSS images.}
\label{fig:optMorph}
\end{figure*}

\renewcommand{\thefigure}{\arabic{figure} (Cont.)}

\begin{figure*}[htb!]
\ContinuedFloat
\centering{\includegraphics[width=0.3\textwidth]{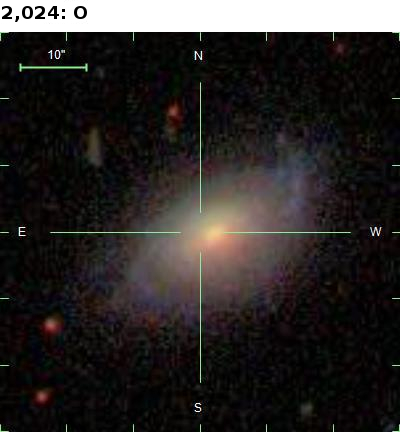}}
\caption{Examples of optical morphologies assigned in the ROGUE~I catalog.}
\end{figure*}

\renewcommand{\thefigure}{\arabic{figure}}

\begin{deluxetable*}{llp{13cm}}
\tabletypesize{\footnotesize}
\tablecolumns{3}
\tablewidth{0pt}
\tablecaption{Radio morphologies of the sources listed in the ROGUE~I catalog with adopted codes and descriptions. 
\label{tab:RadioMorph}}
\tablehead{
\multicolumn{1}{c}{Radio morphology} &  Code &  Description \\ 
\multicolumn{1}{c}{(1)} &  (2) &  (3) 
}
\startdata 
Compact & C & point-like single-component \\
Elongated & E & elliptical profile single-component \\ 
FR~I & I & linear structure brighter near core \citep{Fanaroff74}\\
FR~II & II & linear structure brighter near edges \citep{Fanaroff74}\\
hybrid & I/II & hybrid morphology with one lobe of FR~I and another of FR~II morphology \citep{Gopal-Krishna00}\\
One-sided FR~I & O I & one-sided source with FR~I lobe \\ 
One-sided FR~II & O II & one-sided source with FR~II lobe\\
Z--shaped & Z & Z-- or S--shaped radio morphology \\
X--shaped & X & X--shaped radio morphology \citep{Cheung07}\\
Double-double RG & DD & two pairs of collinear lobes \citep{Lara99}\\ 
Wide-angle tail & WAT & bent source with angle between lobes $>90^{\circ}$\\ 
Narrow-angle tail& NAT & bent source with angle between lobes $<90^{\circ}$\\ 
Head-tail & HT & bright core (head) and a tail \citep{Owen79}\\
Halo & Halo & diffuse radio emission around the core\\
Star-forming region & SFR & emission from the host galaxy 
\\
Not clear & NC & radio source with unclear morphology \\
Blended & B & radio emission blended with other source\\
Not detected & ND & optical galaxy is not the host of the radio emission\\
\hline
Possible & p & uncertain attribution of the above types\\\hline
\enddata
\end{deluxetable*}

\begin{deluxetable*}{llp{12cm}}
\tabletypesize{\footnotesize}
\tablecolumns{3}
\tablewidth{0pt}
\tablecaption{Optical morphologies of the host galaxies of sources in the ROGUE~I catalog with adopted codes and descriptions. \label{OpticalMorph}}
\tablehead{
\multicolumn{1}{c}{Optical morphology} &  Code &  Description \\ 
\multicolumn{1}{c}{(1)} &  (2) &  (3)  
}
\startdata 
Spiral galaxy & S & disc galaxy with visible spiral arms, face-on or edge-on\\
Elliptical galaxy & E & elliptical galaxy\\
Lenticular galaxy & L & disc galaxy without spiral arms \\
Distorted & D & galaxy with distorted, perturbed morphology\\
Ring galaxy & R & galaxy with ring-like shape\\
Galaxy merger & M & Merging galaxies, mainly major merger \\
Star-forming region & SFR & SDSS spectrum of star-forming region, and not the galaxy center \\
Off-center & O & off-center spectrum, not corresponding to star-forming region\\
\hline
Interacting galaxy & i & if the signs of interaction are visible (iS, iL, iE)\\
Barred galaxy & b & spiral or lenticular galaxies with prominent bars (bS, bL) \\\hline
Possible & p & uncertain attribution of the above types 
\enddata
\end{deluxetable*}

We need to point out that in this paper we do not study the origin of radio emission of sources in the ROGUE~I catalog. Therefore, in this catalog star-forming (SF) galaxies can be present as well as radio AGNs. The separation into AGN and SF will be discussed elsewhere (Koziel-Wierzbowska, in prep.). Details of the radio and optical morphological classifications for the galaxies presented in the ROGUE~I catalog are given in Section~\ref{sec:result}.  
 
\subsection{Estimation of radio flux density, radio luminosity, and absolute optical magnitude}

In the ROGUE~I catalog, we also present the \textit{core} and \textit{total} radio flux densities of all sources for which the radio emission can be safely separated from the emission of other nearby sources. \textit{Core} flux densities are the flux densities of the compact central components taken from the FIRST catalog (i.e., the radio sources resulting from Section ~\ref{sec:crossmatch}). In order to estimate the \textit{total} radio flux densities of the ROGUE~I sources, we employed a number of procedures depending on the radio morphology and proximity of neighbouring sources. Below, we outline in detail our methodology for estimating the total radio flux densities. 

1) In the case of compact or elongated radio emitters, i.e. the majority of sources in the ROGUE~I catalog, the total flux densities were obtained directly from the NVSS catalog. 

2) In the case of sources with extended radio morphology for which the radio emission consists of many components, their total flux density was estimated as a sum of the flux densities of separate components listed in the NVSS catalog. 

3) For sources blended with foreground or background point-like sources, the total flux densities were estimated as the difference between the NVSS flux density of the source components and the FIRST flux density of the blended source.

4) The total flux densities of a few sources blended with elongated sources for which we were able to separate individual components were measured manually from the NVSS intensity maps. The manual flux measurements were done with the \textsc{AIPS} software using the \textsc{JMFIT} task developed to fit Gaussian components to a defined part of an image. This method was used in order to be consistent with the NVSS catalog flux measurement method \citep{Condon98}.

5) Moreover, during our analysis, we noticed that in the case of a few sources, the FIRST flux density is higher than given in the NVSS catalog. This can happen in case of variable sources. For such sources we provide values from the FIRST catalog as the total flux density.

6) The FIRST flux densities are given as a measure of total flux density also for sources not detected in the NVSS survey.

The observed monochromatic radio luminosity was computed from the flux density, S, as follows:
\begin{equation}
\mathrm{L_{obs} = 4\,\pi\,d_L^2\,S\, [W\,Hz^{-1}]}\,  
\label{Lobs}
\end{equation}

Next, we also compute for the extended sources, the rest frame monochromatic total luminosity using a spectral index $\alpha=-0.75$ \citep{Yuan18}.

\begin{equation}
\mathrm{L_{rest} = \frac{4\,\pi\,d_L^2\,}{(1+z)^{\alpha+1}} S\, [{W\, Hz^{-1}]}}\,  
\label{Lrest}
\end{equation}

The optical absolute magnitude, M$_r$, was calculated from the SDSS apparent magnitude m$_r$ using the Eq.~\ref{absmag}. 
\begin{equation}
\mathrm{M_r = m_r - A_r + 5 - 5 \log(d_L)}
\label{absmag}
\end{equation}
We applied a correction for Galactic reddening using the values of galactic extinction, A$_r$ \citep{Schlegel98}, however, we did not apply any K-correction. 

\section{Catalog}
\label{sec:result}

Table~\ref{Catalog} presents the radio and optical morphological classifications of the first 20 optical galaxies of the ROGUE~I catalog. 
The catalog of our entire sample of 32,616 sources is published in the machine-readable format. The catalog and the analysed radio-optical overlays are also available at \url{http://rogue.oa.uj.edu.pl/}. The columns are as follows: \\
Column 1: catalog number of the source. \\
Column 2: plate number in the SDSS database. \\
Column 3: MJD in the SDSS database. \\
Column 4: fiber number in the SDSS database. \\
Column 5: Right ascension from the SDSS database. \\
Column 6: Declination from the SDSS database. \\
Column 7: redshift, $z$, from the SDSS database. \\
Column 8: right ascension from the FIRST database. \\
Column 9: declination from the FIRST database. \\
Column 10: optical morphological classification from inspection of SDSS images.\\
Column 11: radio morphological classification from inspection of FIRST contour images. \\
Column 12: radio morphological classification from inspection of NVSS contour images. \\
Column 13: final radio morphological classification. \\
Column 14: computed luminosity distance. \\
Column 15: observed flux density of the core at 1.4 GHz. \\
Column 16: error on the flux density of the core at 1.4 GHz. \\
Column 17: observed flux density of the total emission at 1.4 GHz. \\
Column 18: error on the flux density of the total emission at 1.4 GHz. \\
Column 19: reference for the estimation of total flux density, i.e., if directly obtained or, in some rare cases,  measured manually. \\
Column 20: apparent optical magnitude of galaxy from the SDSS database. \\ 

\clearpage
\begin{minipage}[t][25cm][t]{1.2\textwidth}
\begin{rotatetable*}
\vspace{-5cm}
\begin{deluxetable*}{lccccccccccccccccccc}
\small
\tabletypesize{\tiny}
\tablecolumns{20}
\tablewidth{0pt}
\tablecaption{First 20 galaxies from the ROGUE~I catalog.\label{Catalog}}
\tablehead{
\multirow{2}{*}{No.} & \multicolumn{6}{c}{SDSS}  & \multicolumn{2}{c}{FIRST} & \multicolumn{4}{c}{Classification} & \colhead{$d_{L}$} & \colhead{S$_{\mathrm{core}}$} & \colhead{ eS$_{\mathrm{core}}$} &  \colhead{S$_{\mathrm{total}}$}&
\colhead{eS$_{\mathrm{total}}$}& \multirow{2}{*}{Flag} & \colhead{m$_r$}  \\
 & Plate & MJD & Fiber & $\alpha_{\mathrm{opt}}$ & $\delta_{\mathrm{opt}}$ & z & $\alpha_{\mathrm{rad}}$ & $\delta_{\mathrm{rad}}$ & SDSS & FIRST & NVSS & Final & (Mpc) & (mJy) & (mJy)& (mJy) & (mJy) &  & (mag)  \\
(1) & (2) & (3) & (4) & (5) & (6) & (7) & (8) & (9) & (10) & (11) & (12) & (13) & (14) & (15) & (16) & (17) & (18) & (19) & (20)
}
\startdata
1 & 266 & 51630 & 25 & 146.95607 & -0.342297 & 0.134663 & 09 47 49.453 & -00 20 33.55 & E & E & C & E & 639.1 & 100.2 & 0.14 & 100.2 & 0.14 & F & 17.54\\
2 & 266 & 51630 & 42 & 146.565613 & -1.084756 & 0.09758 &  09 46 15.738 &  -01 05 04.99 &      L &      E &     ND &      E & 451.8 & 1.3 & 0.14 & 1.3 & 0.14 &  F & 17.15\\
3 & 266 & 51630 & 77 & 146.809128 & 0.02636 & 0.126075 &  09 47 14.183 &  +00 01 35.25 &      E &      C &      C &      C & 595 & 2.75 & 0.19 & 2.8 & 0.4 &  N & 16.65\\
4 & 266 & 51630 & 90 & 146.14357 & -0.741639 & 0.203829 &  09 44 34.458 &  -00 44 29.44 &      D &      E &      B &      E & 1009.6 & 2.57 & 0.15 & 2.57 & 0.15 &  F & 16.6\\
5 & 266 & 51630 & 100 & 146.007797 & -0.642273 & 0.005024 &  09 44 01.896 &  -00 38 32.19 &      D &      E &      C &      E & 21.7 & 1.3 & 0.14 & 3.3 & 0.6 &  N & 16.16\\
6 & 266 & 51630 & 119 & 146.737137 & -0.252201 & 0.13054 &  09 46 56.879 &  -00 15 08.11 &      E &      E &      C &      E & 617.8 & 4.98 & 0.14 & 7.2 & 0.5 &  N & 16.76\\
7 & 266 & 51630 & 141 & 146.373795 & -0.36845 & 0.053307 &  09 45 29.731 &  -00 22 04.32 &     iS &      E &      B &      E & 239.2 & 3.8 & 0.14 & 3.8 & 0.14 &  F & 16.03\\
8 & 266 & 51630 & 223 & 145.601166 & -0.001393 & 0.14577 &  09 42 24.263 &  -00 00 05.23 &      D &      C &      C &      C & 696.8 & 4.87 & 0.14 & 5.5 & 0.4 &  N & 17.16\\
9 & 266 & 51630 & 255 & 145.52623 & -0.747411 & 0.218403 &  09 42 06.297 &  -00 44 51.13 &      E &      E &      C &      E & 1091.1 & 3.58 & 0.15 & 4.1 & 0.4 &  N & 17.83\\
10 & 266 & 51630 & 506 & 146.462982 & 0.63869 & 0.030345 &  09 45 51.057 &  +00 38 21.23 &      D &      E &      B &      E & 133.9 & 2.81 & 0.15 & 2.81 & 0.15 &  F & 15.71\\
11 & 266 & 51630 & 543 & 146.806839 & 0.665554 & 0.02008 &  09 47 13.587 &  +00 39 55.85 &      S &      E &      C &      E & 87.9 & 13.09 & 0.15 & 17.7 & 0.7 &  N & 17.16\\
12 & 266 & 51630 & 545 & 146.799088 & 0.702682 & 0.030555 &  09 47 11.672 &  +00 42 07.69 &      D &      E &      C &      E & 134.8 & 4.53 & 0.15 & 5.4 & 0.5 &  N & 15.45\\
13 & 266 & 51630 & 572 & 146.781509 & 0.737954 & 0.261903 &  09 47 07.515 &  +00 44 17.15 &      E &    WAT &      C &    WAT & 1340.9 & 9.37 & 0.15 & 49.3 & 1.9 &  N & 17.35\\
14 & 266 & 51630 & 613 & 147.080475 & 0.788018 & 0.211183 &  09 48 19.281 &  +00 47 16.60 &      E &      C &      C &      C & 1050.6 & 7.84 & 0.14 & 8 & 0.5 &  N & 17.29\\
15 & 267 & 51608 & 9 & 148.829819 & -0.740928 & 0.292166 &  09 55 19.154 &  -00 44 27.02 &      E &      E &      C &      E & 1520.4 & 2.6 & 0.14 & 3.9 & 0.5 &  N & 18.36\\
16 & 267 & 51608 & 19 & 148.606583 & -0.92869 & 0.358335 &  09 54 25.603 &  -00 55 43.81 &      E &      E &      C &      E & 1927.9 & 185.64 & 0.13 & 185.64 & 0.13 &  F & 17.9\\
17 & 267 & 51608 & 27 & 149.112656 & -0.47563 & 0.086629 &  09 56 27.061 &  -00 28 32.13 &      L &      C &     ND &      C & 398 & 1.27 & 0.15 & 1.27 & 0.15 &  F & 16.33\\
18 & 267 & 51608 & 34 & 149.169876 & -0.023346 & 0.139254 &  09 56 40.762 &  -00 01 24.26 &      E &     II &     II &     II & 662.9 & 2.15 & 0.14 & 199.8 & 5.23 &  N & 16.26\\
19 & 267 & 51608 & 47 & 148.43251 & -1.026422 & 0.110105 &  09 53 43.793 &  -01 01 35.04 &     pM &      E &      E &      E & 514.1 & 11.11 & 0.14 & 11.11 & 0.14 &  F & 16.32\\
20 & 267 & 51608 & 97 & 148.237686 & -0.791982 & 0.089783 &  09 52 57.012 &  -00 47 31.18 &      E &      E &      E &      E & 413.5 & 17.26 & 0.15 & 23.6 & 1.5 &  N & 15.69\\
\enddata
\vspace{0.4cm}
\tablecomments{Table is published in its entirety in the machine-readable format.
      A portion is shown here for guidance regarding its form and content. The columns are: (1) catalog number; (2) plate number from SDSS; (3) MJD from SDSS; (4) fiber number from SDSS; (5) Right ascension from SDSS; (6) Declination from SDSS; (7) redshift; (8) Right ascension from FIRST; (9) Declination from FIRST; (10) result of optical morphological classification (Table\ref{OpticalMorph}); (11) - (12) results of individual radio morphological classification (Table~\ref{tab:RadioMorph}); (13) final radio morphological classification; (14) computed luminosity distance of the source (Eq.~\ref{dl}); (15) flux density of radio core from FIRST; (16) uncertainty of the flux density of radio core from FIRST; (17) total radio flux density; (18) uncertainty of the total radio flux density; (19) reference for the total radio flux density of the source (N--NVSS catalog; F--FIRST catalog; S--NVSS corrected for background source; M--manually obtained); (20) apparent optical magnitude from SDSS.}      
\end{deluxetable*}
\end{rotatetable*}
\end{minipage}
\clearpage

\section{Comments on the catalog} \label{sec:resultComments}
\subsection{Number of sources with given radio and optical  morphologies}

The vast majority of sources in the ROGUE~I catalog possess single-component compact or elongated radio morphologies, forming together a sample of 29,237 ($\sim$90\%) radio sources, 876 sources are classified as SFR, blended, or not detected, while the remaining 2,503 galaxies ($\sim$8\%) are extended radio sources with complex radio structures. In the group of extended radio sources (including I, II, Hybrid, OI, OII, DD, X, Z, WAT, NAT, HT, Halo, and NC classes), 1,519 ($\sim$61\% of extended sources) are considered as Fanaroff--Riley type I, II, hybrid, or one-sided FR~I and FR~II,  while 436 ($\sim$17\% of extended sources) are possible classifications of the above types.
Bent sources securely classified as wide--angle tail, narrow--angle tail, or head--tail radio sources form a large group of 390 ($\sim$16\%) objects,  and 73 ($\sim$3\%) bent sources having possible classifications. Double--double, Z--shaped, X--shaped, and halo radio sources (secure and possible) form a small group of 67 objects ($\sim$3\%). 
Table~\ref{RadioMorphNumbers} gives a summary of the radio morphologies in the ROGUE~I catalog. 

\begin{deluxetable}{lll}
\tablecolumns{3}
\tablewidth{1.0\textwidth} 
\tablecaption{Summary of the radio morphologies of galaxies in the ROGUE~I catalog.
\label{RadioMorphNumbers}}
\tablehead{
\multicolumn{1}{c}{Radio morphology} &  \colhead{Code} &  Number 
}
\startdata 
Compact & C & 4,785 \\
Elongated & E & 24,452\\
Fanaroff-Riley I & I (pI) & 269 (147)\\
Fanaroff-Riley II & II (pII) & 730 (141)\\
Hybrid & I/II (pI/II) & 115 (101)\\
One-sided FR~I & OI (pOI) & 191 (33)\\
One-sided FR~II & OII (pOII) & 214 (14)\\
Z--shaped & Z (pZ) & 18 (7)\\
X--shaped & X (pX) & 7 (7)\\
Double-double & DD (pDD) & 8 (12)\\
Wide-angle bent & WAT (pWAT) & 273 (36)\\
Narrow-angle bent & NAT (pNAT) & 101 (25)\\
Head-tail & HT (pHT) & 16 (12)\\
Halo & Halo (pHalo) & 3 (5)\\
Star-forming region & SFR & 423 \\
Not clear & NC & 18 \\
Blended & B & 414\\ 
Not detected & ND & 39\\\hline
\enddata
\tablecomments{Values in brackets correspond to the numbers of objects with possible classification.}
\end{deluxetable}

Out of 32,616 galaxies listed in the ROGUE~I catalog, we classified 19,535 objects as elliptical and possible elliptical galaxies, comprising together the most numerous group, i.e. $\sim$60\%. Other large groups of galaxies consist of: spiral and possible spiral --- 5,174 ($\sim$16\%), distorted --- 3,946 ($\sim$12\%), lenticular and possible lenticular --- 2,367 ($\sim$7\%). Secure and possible merger, ring, interacting, and barred galaxies, as well as star-forming regions constitute a group of 1570 objects ($\sim$5\%). The numbers of galaxies with different optical morphologies are listed in Table~\ref{OpticalMorphNumbers}.

\begin{deluxetable}{lll}
\tablecolumns{3}
\tablewidth{1pt}
\tablecaption{Summary of the optical morphologies of galaxies in the ROGUE~I catalog.\label{OpticalMorphNumbers}}
\tablehead{
\multicolumn{1}{c}{Optical morphology} &  \colhead{Code} &  Number 
}
\startdata 
Elliptical & E (pE) & 18,416 (1,119)\\
Interacting elliptical & iE (piE) & 795 (6)\\
Distorted & D & 3,946\\
Spiral & S (pS) & 2,927 (2,247)\\
Interacting Spiral & iS (piS) & 115 (14) \\
Barred spiral & bS (pbS) & 142 (3)\\
Lenticular & L (pL) & 1,580 (787)\\ 
Interacting lenticular & iL & 16\\
Barred lenticular & bL & 39\\
Merger galaxy & M (pM) & 235 (93)\\
Ring galaxy & R (pR) & 88 (12)\\
Star-forming region & SFR & 12\\
Off-center & O & 24\\\hline
\enddata
\tablecomments{Values in brackets correspond to the numbers of objects with possible classification.
}
\end{deluxetable}

Table~\ref{ExtendedRadioOpticalNumbers} shows that most of the sources with extended radio morphology are hosted by elliptical galaxies, i.e. 2,445 ($\sim$98\%). The rest of them ($\sim$2\%) are distorted, spiral, lenticular (also barred) galaxies and galaxy mergers. 

Table~\ref{CompactRadioOpticalNumbers} presents the number of unresolved and elongated radio sources corresponding to different optical morphological classes. Again also here the majority of the host galaxies are elliptical; this is a selection effect arising from our sampling of radio flux densities down to FIRST detection threshold ($\sim$0.6 mJy beam$^{-1}$ corresponding to L$_\mathrm{obs,total} \sim$10$^{22}$ W Hz$^{-1}$ for z$\sim$0.1) where the radio-active galaxy population still dominates over the star-forming galaxy population. However, we note that in unresolved and elongated sources, the variety of optical morphological types is much larger, suggesting that these classes are a mixture of objects in which the radio emission is connected to different phenomena (AGN vs. SF). 

We also examined the morphologies of the galaxies where the radio emission is considered to originate from a star-forming region. None of the radio emission associated with these objects is identified with elliptical galaxies.

\begin{deluxetable}{lccccc}
\tablecolumns{6}
\tablewidth{0pt}
\tablecaption{Number of extended radio sources corresponding to different optical morphological classes. \label{ExtendedRadioOpticalNumbers}}
\tablehead{
\multirow{2}{*}{\diagbox[innerwidth=1.3cm, height=8.5ex]{Radio}{Optical}} & \colhead{E + iE}  & D & S + bS + iS  & L+bL + iL & M \\ 
\multicolumn{1}{c}{} & (pE+piE) & & (pS + pbS + piS) & (pL+pbL) & (pM)}
\startdata 
I & 265 (1) & 3 & - & - & -\\
pI & 144 & 1 & 1 & (1) & - \\
II & 710 (4) & 10 & 2 (2) & 1 & (1) \\
pII & 134 (3) & 2 & - & 1 (1) & - \\
I/II & 112 (1) & 2 & - & - & - \\	
pI/II & 98 (1) & 1 & - & (1) & - \\
OI & 183 & 7 & - & (1) & - \\
pOI & 32 & 1 & - & - & - \\
OII & 208 (1) & 4 & - & 1 & - \\
pOII & 12 (1) & 1 & - & - & - \\
DD & 7 & 1 & - & -  & -\\	
pDD & 11 (1) & - & - & - & - \\	
WAT & 264 (2) & 6 &	- & - & (1)\\
pWAT & 36 & - &	- & - & -\\
NAT & 101 & - & - & - & -\\
pNAT & 25 & - & - & - & -\\
HT & 16 & - & - & - & - \\				
pHT & 11 & 1 & - & - & - \\		
X & 6 & 1 & - & - & - \\			
pX & 6 & 1 & - & - & - \\			
Z & 18 & - & - & - & - \\	
pZ & 7 &  - & - & - & -\\	
Halo & 3 & - & - & - & - \\			
pHalo & 4 & 1 & - & - & - \\			
NC & 17 & - & - & (1) & -\\			
\hline
\enddata
\tablecomments{Values in brackets correspond to sources with possible classification.}
\end{deluxetable}

\begin{deluxetable*}{cccccccccccccc} 
\tablecolumns{14}
\tablewidth{0pt}
\tablecaption{Number of unresolved and elongated radio sources corresponding to different optical morphological classes. \label{CompactRadioOpticalNumbers}}
\tablehead{
\multirow{2}{*}{\diagbox[innerwidth=1.3cm, height=8.5ex]{Radio}{Optical}} & \colhead{E} & iE & D & S & iS  & bS & L & iL  & bL &  R & M & SFR & O\\ 
\multicolumn{1}{c}{} & (pE) & (piE) &  & (pS) & (piS) & (pbS) & (pL) & & &  (pR) &(pM) & & 
}
\startdata 
\multicolumn{1}{l}{C} & 2,609 (249) & 88 (1) & 662 & 281 (371) & 9 (2) & 15 & 284 (169) & 0 & 7 & 16 (2) & 15 (4) & 1 & 0 \\
\multicolumn{1}{l}{E} & 13,326 (848) & 512 (5) & 3,141 & 2,377 (1,828) & 91 (7) & 117 (3) & 1,264 (607) & 14 & 32 & 71 (9) & 152 (40) & 5 & 3\\
\enddata
\tablecomments{Values in brackets correspond to sources with possible classification.}
\end{deluxetable*}

\subsection{Redshift and luminosity distributions of the ROGUE~I sources}

\begin{figure}[!htb]
\begin{center}
\includegraphics[angle=0,scale=0.35]{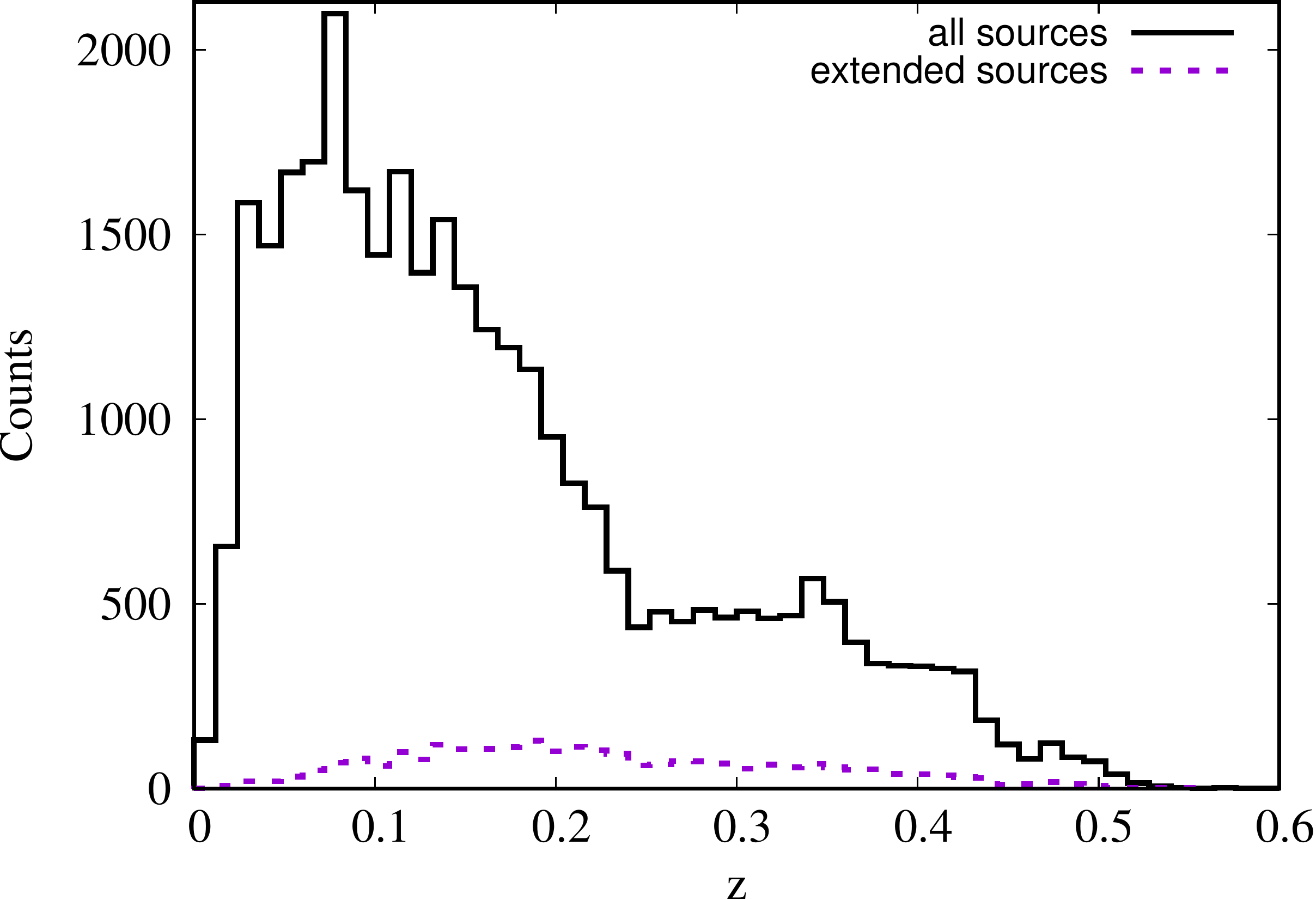}
\includegraphics[angle=0,scale=0.35]{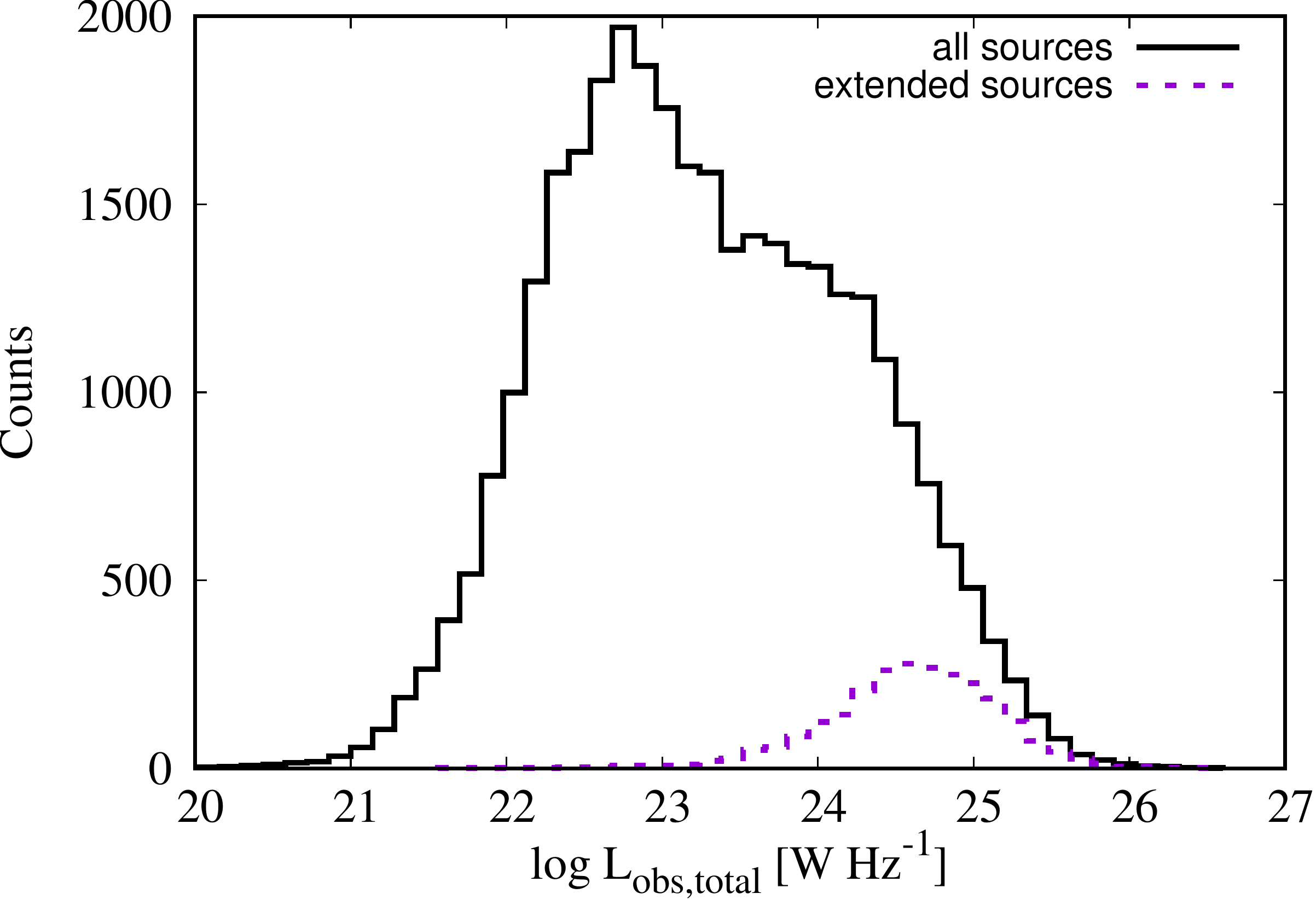}
\caption{Histograms of $z$ (top panel) and L$_\mathrm{obs,total}$ (bottom panel) at 1.4 GHz of all (black solid lines) and secure as well as possible extended (violet dashed lines) radio sources listed in the ROGUE~I catalog. \label{redshiftDist}} 
\end{center}
\end{figure}

The sources of the ROGUE~I catalog cover a wide range of redshifts, $z = 0.0021 - 0.636$, and a wide range of total radio luminosities at 1.4 GHz, L$_\mathrm{obs,total} = 10^{18.86-26.59}$ W Hz$^{-1}$, and core luminosities L$_\mathrm{obs,core} = 10^{18.86-26.26}$ W Hz$^{-1}$. Figure~\ref{redshiftDist} shows the distributions of the radio sources as a function of $z$ and L$_\mathrm{obs,total}$. The peak of the redshift distribution is at about 0.1 and there is a long tail towards higher redshifts. The redshift range of the extended sources is between 0.0162 and 0.5443 with L$_\mathrm{obs,total} = 10^{22.25-26.50}$ W Hz$^{-1}$. As can be inferred from the local radio luminosity function \citep[e.g.][]{Best12}, SF galaxies dominate at low luminosities in the distribution of L$_\mathrm{total}$. At higher luminosities, where also extended sources are found, the majority of the sources are probably AGNs.

\begin{figure}[!htb]
\begin{center}
\includegraphics[angle=270,scale=0.28]{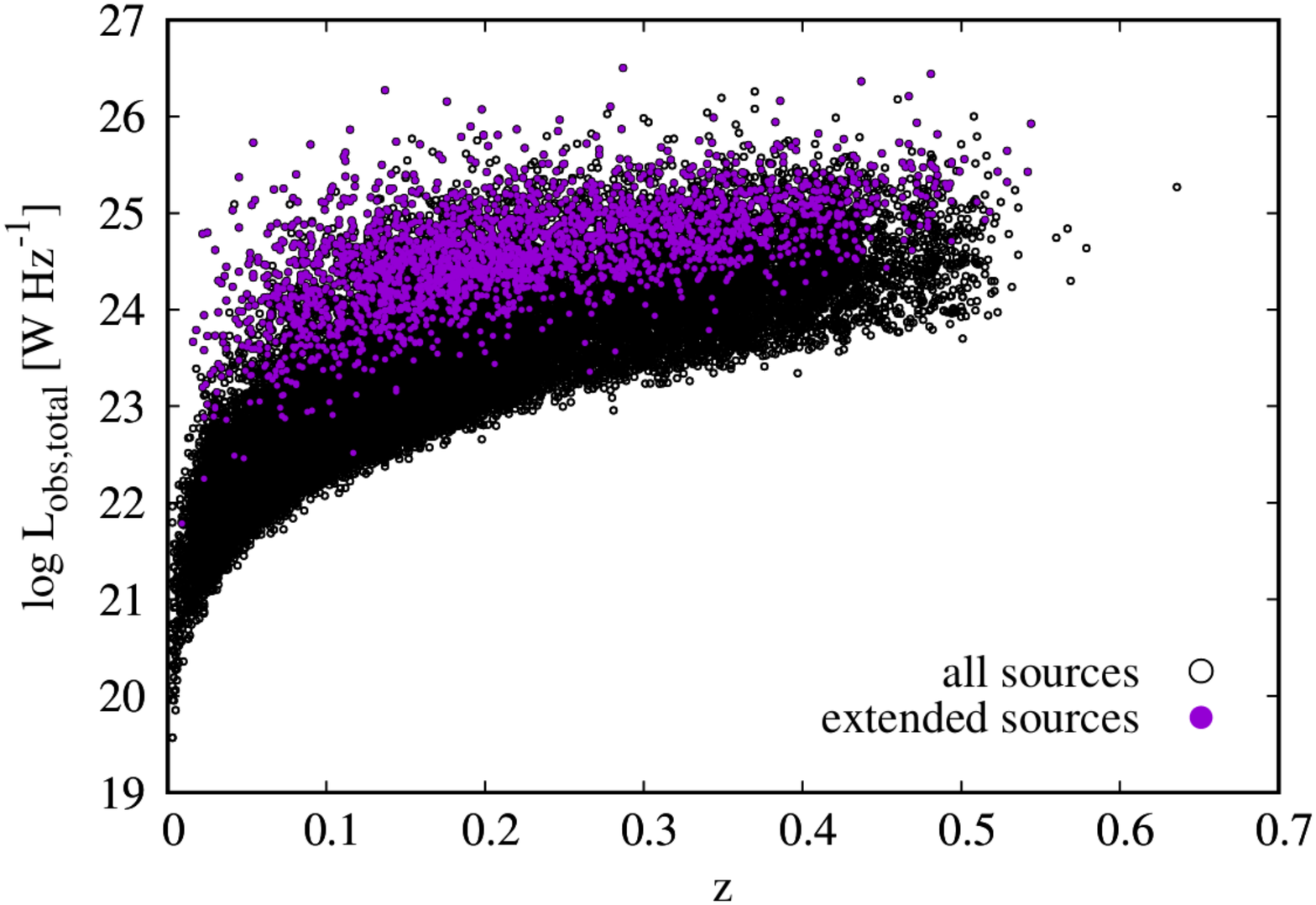}
\includegraphics[angle=270,scale=0.28]{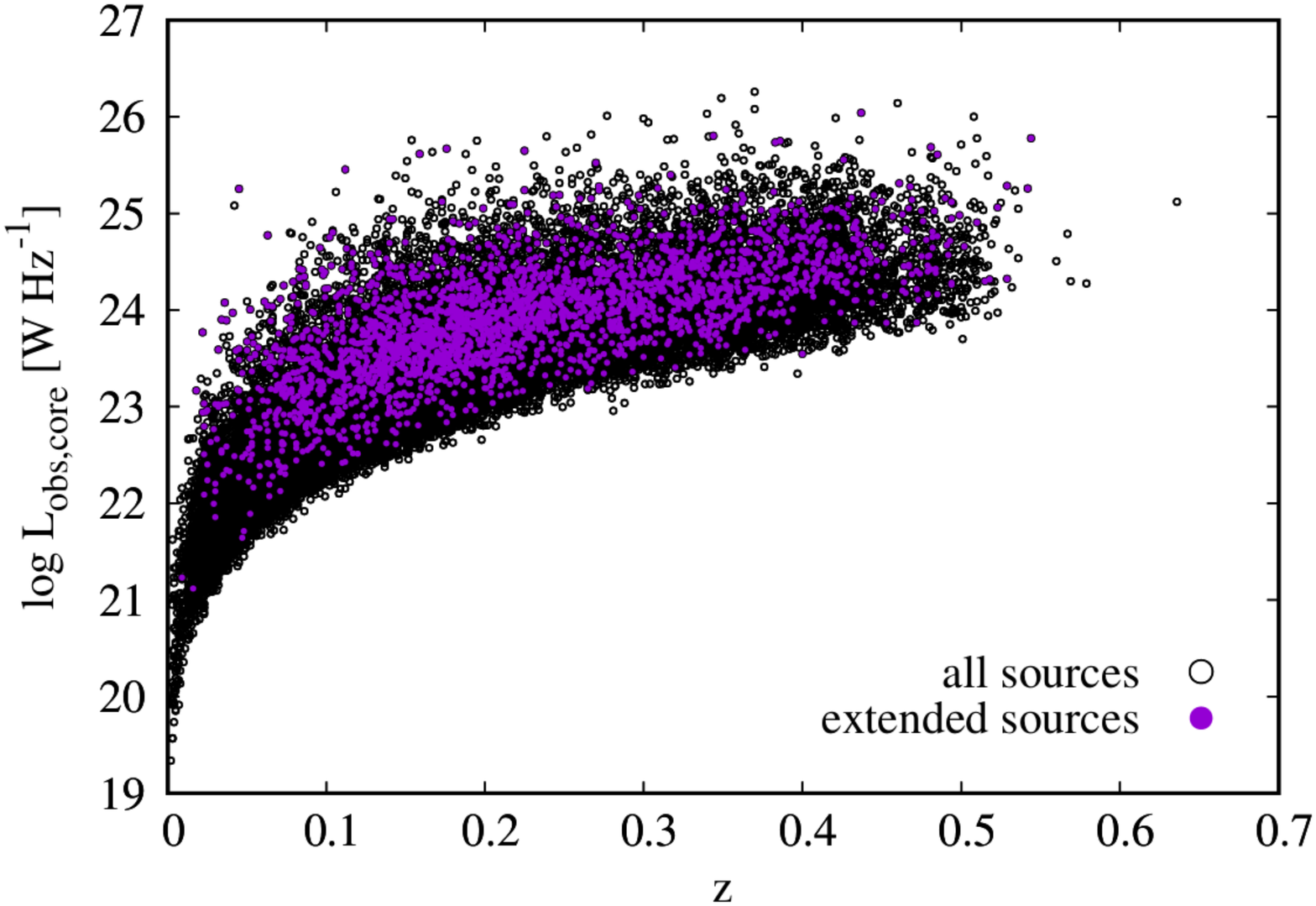}
\caption{Distributions of L$_\mathrm{obs,total}$ vs. $z$ (top panel) and L$_\mathrm{obs,core}$ vs. $z$ (bottom panel) of all (black open circles) and extended (violet filled circles) sources listed in the ROGUE~I catalog. \label{Mr_Lrad_All}} 
\end{center}
\end{figure}

The evolution of the total and core luminosities with $z$ are shown in the top and bottom panels of Figure~\ref{Mr_Lrad_All}, respectively. The distribution of L$_\mathrm{obs,total}$ shows the detection threshold of the FIRST survey. Extended radio sources are evidently shifted with respect to the whole population of the ROGUE~I sources. This is a result of the selection method in which sources are classified as extended only when there is more than one component in the FIRST or the NVSS maps that can be identified as belonging to one source. It means that a source has to have at least two components to be identified as extended, thereby increasing the total radio flux density threshold as compared to the unresolved or elongated radio sources presented here.

We notice an offset of the extended sources in relation to all sources in the core radio luminosity distribution. This is due to the fact that at low luminosities, the lobes of extended radio structures cannot be detected by the FIRST survey as the low-luminosity (if any) extended emission is resolved out and only the core is detected. In such cases, the radio source is classified as compact or elongated (see Section~\ref{sec:methodology}). However, the hint that some of these objects can have extended emission comes from a comparison of flux densities from the FIRST and the NVSS catalogs: in the presence of extended structure  NVSS should have an excess of radio flux density compared to the FIRST flux density. We find that $\sim$27\% of the compact and elongated sources have the NVSS flux measurements larger than the FIRST flux measurements by 20\% or more \citep[see also the discussion in ][]{vanVelzen15}. Therefore, these could be potential candidates for low-luminosity (mostly FR~I) extended radio sources.

\begin{figure}[!htb]
\begin{center}
\includegraphics[angle=0,scale=0.35]{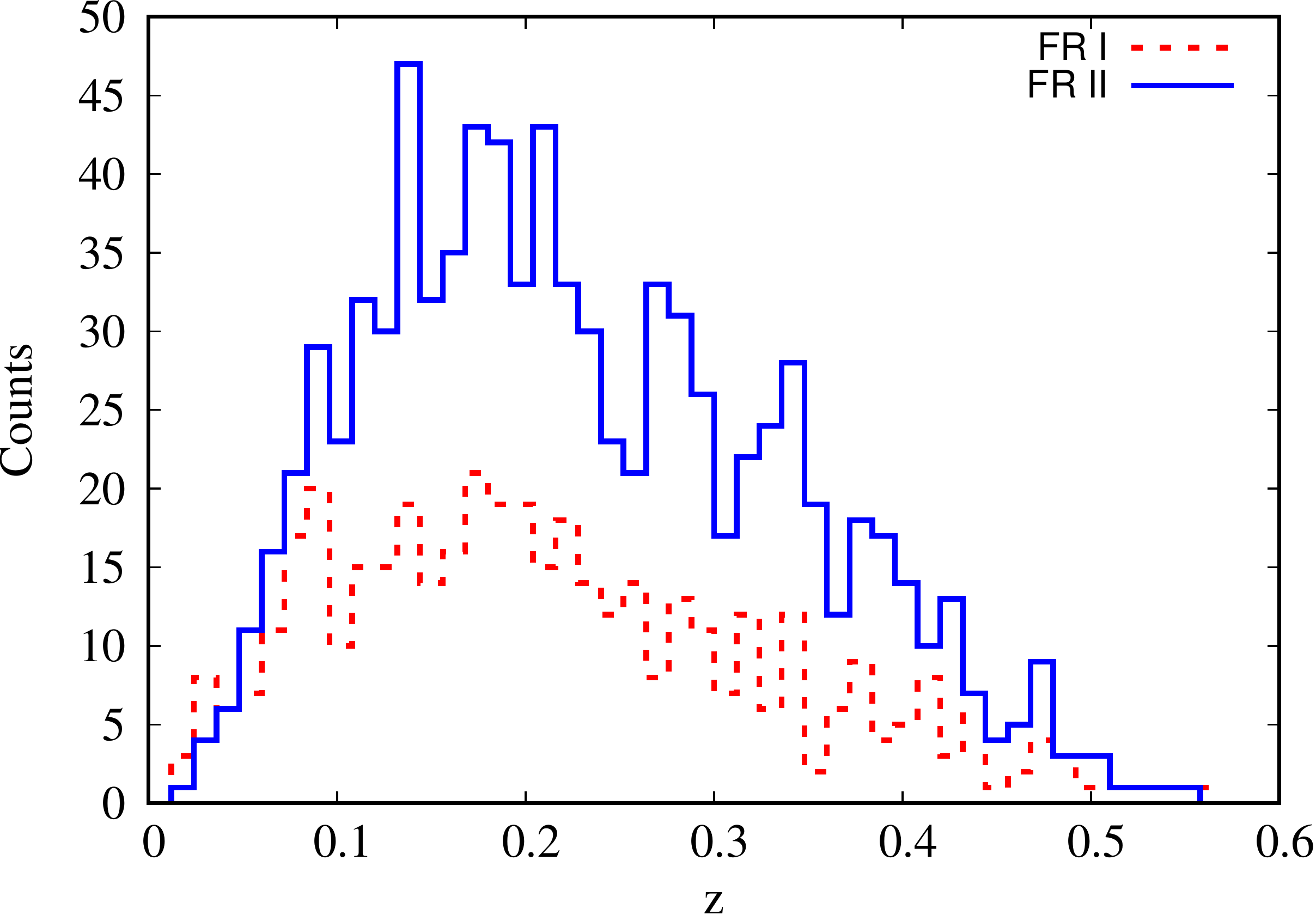}
\includegraphics[angle=0,scale=0.35]{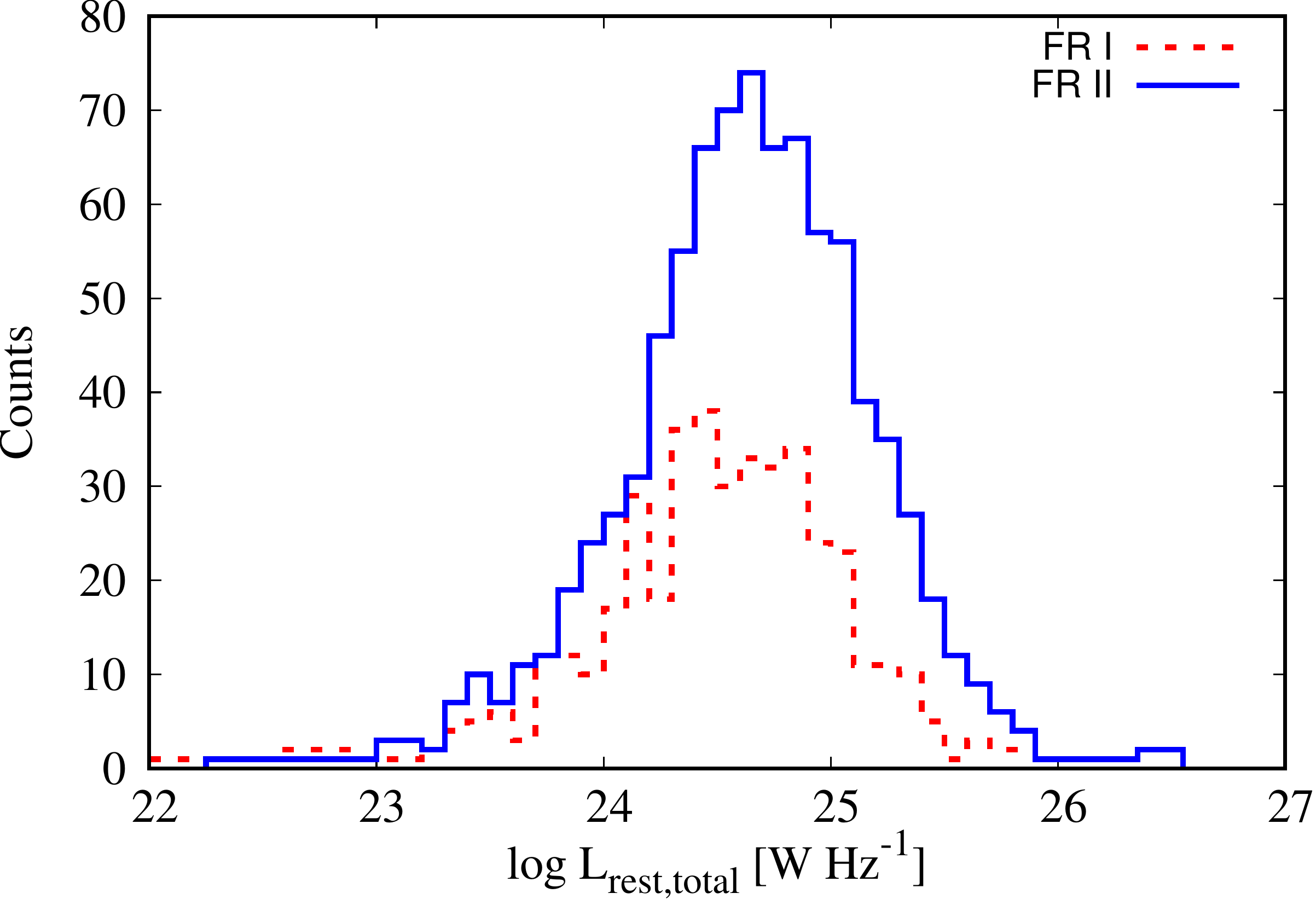}
\caption{Histograms of $z$ (top panel) and $L_\mathrm{rest,total}$ (bottom panel) of secure and possible FR~I (416 objects; red dashed lines) and FR~II (871 objects; blue solid lines) radio sources listed in the ROGUE~I catalog.}
\label{morphDist} 
\end{center}
\end{figure}

Figure~\ref{morphDist} shows the distribution of secure and possible FR~I and FR~II galaxies as a function of $z$ (top panel) and L$_\mathrm{rest,total}$ (bottom panel). The redshift range of both FR~Is and FR~IIs is similar. This is unexpected since the lower luminosity lobes in FR~I sources should be much harder to detect at larger redshifts.

Both the position of the maximum and the width of the distributions of the FR~Is and FR~IIs total radio luminosity are very similar. This is surprising since FR~IIs are considered to have larger radio luminosities. In the next section we discuss in detail the distribution of our sources in the radio--optical luminosity plane, the so-called Ledlow--Owen diagram \citep[]{Ledlow96} and compare it to previous studies.

\subsection{The radio-optical luminosity plane for FR~I and ~II sources}

\begin{figure}[!htb]
\begin{center}
\includegraphics[angle=0,scale=0.35]{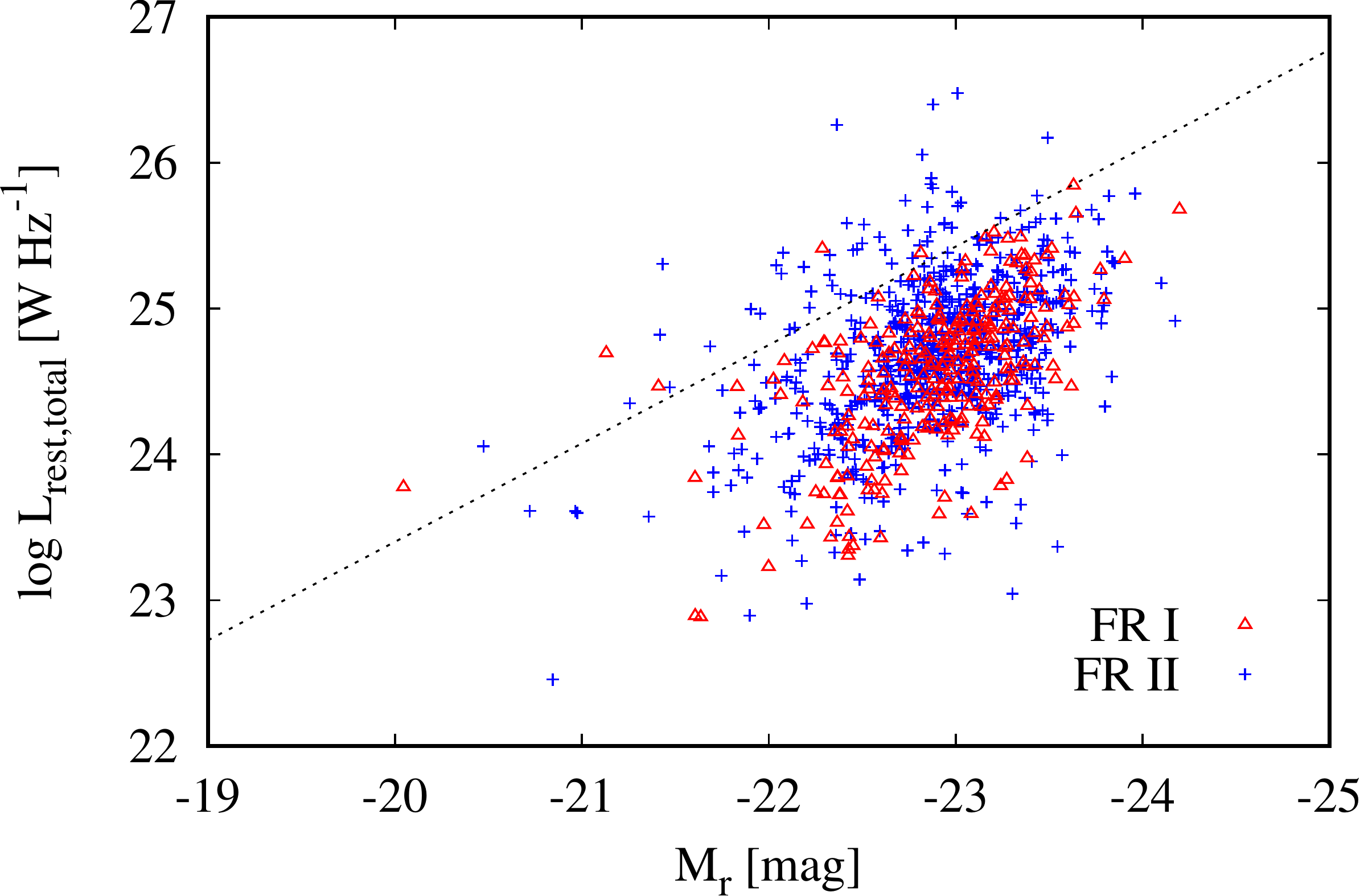}
\caption{Distribution of $L_\mathrm{rest,total}$ at 1.4 GHz vs. absolute magnitude, M$_{r}$ of the FR~I (269 sources; red triangles) and FR~II (730 sources; blue crosses), using only secure classifications in the ROGUE~I catalog. The dotted line marks the division of the FR~I and FR~II type of radio sources \citep{Ledlow96}. Here we reproduce the division line with a correction factor of 0.34 mag resulting from the conversion of color from Cousin to the SDSS filter systems \citep[see also,][]{Capetti17b}}. 
\label{LOdiag} 
\end{center}
\end{figure}

Studies based on the most luminous (with a radio flux density larger than 8 Jy) radio sources detected in the 3C catalogs \citep{Edge59,Bennett62, Laing83} found that FR~I and FR~II radio sources can be separated by their radio luminosity: FR~II radio sources are more luminous and the separating luminosity is about $2\times10^{25}$ W Hz$^{-1}$ at 0.178 GHz \citep{Fanaroff74}. Later studies \citep[e.g.][]{Owen94, Ledlow96} found that the luminosity at which these two types of radio sources are separated depends on the luminosity of the optical host galaxy: it is larger for more luminous hosts. This dependence of the separation of FR~Is and FR~IIs on the host luminosity suggested that, beside the jet power, also the environment plays a crucial role in shaping radio sources. However, lower flux density limit surveys \citep[eg.,][]{Gendre13, Miraghaei17, Capetti17b} show that the separation between FR~Is and FR~IIs is no longer discernible when extended down to lower radio luminosities. The area in the Ledlow-Owen diagram reserved before just for FR~Is now is populated also by radio-faint FR~II type sources. Recently, \citet{Capetti17a, Capetti17b} have published a list of FR~I (FRICAT) and FR~II (FRIICAT) radio sources obtained by {\it visual} inspection of the FIRST and the NVSS maps with optical galaxies up to $z < 0.15$ from the SDSS DR\,7 release and the radio AGN sample of \citet{Best12}. Since they also use {\it visual} identification for radio morphological classification, it is useful to compare our results with theirs. In Figure~\ref{LOdiag}, we show the Ledlow--Owen diagram for our FR~I and FR~II sources with secure classification only (Table~\ref{Catalog}). Consistently with \citet{Capetti17b}, we find that most of FR~IIs are found below the original division line in the Ledlow--Owen diagram. We note that the total number of FR~II sources in ROGUE~I is larger than given by FRIICAT for the same redshift limit (202 from ROGUE~I vs. 122 from FRIICAT). This is a result of the lower flux density limit adopted in ROGUE~I (0.6 mJy vs. 5 mJy in \citealt{Capetti17b}). The number of FR~II sources above the \citet{Ledlow96} division line is lower for the same $z$ limit (13 from ROGUE~I vs. 33 from FRIICAT). This is mainly due to the fact that \citet{Capetti17b} includes also radio sources without radio core, which are absent in the ROGUE~I catalog. We also note that the number of FR~I sources detected by us is lower for the same $z$ limit (99 from ROGUE~I vs. 209 from FRICAT). This results from the inclusion of radio sources with only one detection in the FIRST catalog as FR~Is in their classification. In the ROGUE~I catalog these sources are classified as elongated. We note that \citet{Mingo19} also find about a factor of three more FR~I sources than FR~II sources in the LoTSS data release due to significantly lower surface brightness limit of the LoTSS survey as compared to the FIRST survey.

\subsection{Sources with uncommon radio structures}\label{uncommonSources}

Comparing maps at the same frequencies but with different angular-scale sensitivities, we were able to classify some of the radio sources more precisely, henceforth our morphological classifications differ from those in the literature. The following tables list all the objects for which we propose new morphological designations, such as giant radio sources (GRSs; Table~\ref{GNewObjects}), possible GRSs (Table~\ref{PGNewObjects}), double-double radio sources (Table~\ref{DDNewObjects}), X--shaped radio sources (Table~\ref{XNewObjects}), and Z--shaped radio sources (Table~\ref{ZshapedObjects}). 

We classify GRSs following the definition given by \cite{Kuzmicz.etal.2018a}, i.e. sources with linear size $>$700 kpc. We measured the sizes of all the extended radio sources for which the radio structure is larger than 2/3 of the analyzed maps (i.e., $\sim$660 kpc). In the case of sources possessing prominent hot spots, i.e. FR~IIs, the size was estimated as the sum of the lengths of both lobes, being distances from the core to the most distant FIRST/NVSS components. Regarding the gradually darkening structures, i.e. FR~Is, we measured the size manually taking into account deconvolution with the synthesized beam. 

In a few cases, for example when (1) the lobes are faint and not listed in the FIRST/NVSS catalogs (i.e., in practice only one contour is present in the map), (2) the sizes are slightly smaller than 700 kpc ($\ga$690 kpc), or (3) the assigned final morphology is  possible, the sources are considered as a GRS candidate.

We also cross-matched our list of double-double and X--shaped radio sources with \cite{Lal07, Cheung07, Kuzmicz17}. We found that 16 (including 12 possible) 
X--shaped radio sources from ROGUE~I catalog do not appear in their lists, therefore they are newly discovered sources.

\begin{deluxetable*}{lccccccclcc}
\small
\tabletypesize{\scriptsize}
\tablecolumns{11}
\tablewidth{0pt}
\tablecaption{New giant radio sources listed in ROGUE~I. \label{GNewObjects}}
\tablehead{\colhead{} & \multicolumn{3}{c}{SDSS} &  & \multicolumn{2}{c}{FIRST} & \multicolumn{2}{c}{Classification} & AS  & LS \\
No. & Plate & MJD & Fiber & z & $\alpha$ & $\delta$ & Optical & Radio &  ($^{\prime\prime}$)  & (kpc) \\
(1) & (2) & (3) & (4) & (5) & (6) & (7) & (8) & (9)  & (10) & (11) 
}
\startdata
402$^{\dagger}$ & 290 & 51941 & 44 & 0.1353 & 12 37 45.9 & $-$01 14 16.2 & E & pDD  & 693 & 1673 \\
2,635$^{\star\dagger}$ & 447 & 51877 & 421 & 0.4029 & 08 45 25.5 & $+$52 29 15.8 & E & DD & 212 & 1155 \\
3,684 & 528 & 52022 & 454 & 0.3212 & 13 36 03.5 & $+$03 07 45.4 & E  & II & 159 & 749 \\
10,340 & 1007 & 52706 & 415 & 0.2413 & 10 10 38.0 &  $+$51 11 19.9 & E & II  &  187 & 718 \\
10,864 & 1048 & 52736 & 495 & 0.4340 & 14 53 31.1 & $+$48 26 35.4 & E & II & 164 & 934 \\
11,817 & 1205 & 52670 & 248 & 0.1842 & 08 00 46.3 & $+$24 43 17.1 & E & II & 312 & 1003 \\
11,827 & 1205 & 52670 & 491 & 0.1370 & 08 05 31.3 & $+$25 48 11.4 & E & II & 332 & 810 \\
12,749 & 1281 & 52753 & 178 & 0.3645 & 13 12 16.3 & $+$48 47 45.4 & E &  pWAT &  150 & 768 \\
13,085 & 1301 & 52976 & 265 & 0.2426 & 09 11 54.7 & $+$08 12 31.0 & E & II & 208 & 802 \\
14,373 & 1373 & 53063 & 554  & 0.0777 & 12 53 03.2 & $+$45 00 44.8 & E & I & 486 & 719 \\
15,090 & 1415 & 52885 & 307  &  0.2821 & 16 50 25.3 & $+$21 44 57.8  & E & II & 168 & 723 \\
15,353 & 1430 & 53002 & 5 & 0.1449 & 10 36 36.3 & $+$38 35 07.5 & E & I/II & 301 & 770 \\
16,242 & 1573 & 53226 & 357 & 0.1481 & 16 22 06.0 & $+$24 49 16.6 & E & II  & 463 & 1207 \\
16,309 & 1576 & 53496 & 575 & 0.2659 & 16 20 31.1 & $+$27 17 37.5 & E & II & 209 & 862 \\
17,467 & 1648 & 53171 & 370 & 0.2821 & 15 02 08.9 & $+$33 31 14.3 & E & II  & 192 & 826 \\
18,519$^\ddag$ & 1709 & 53533 & 491 & 0.1665 & 14 31 51.1 & $+$10 29 59.4 & E & pX & 271 & 778 \\
19,682 & 1767 & 53436 & 67 & 0.3876 & 12 27 53.4 & $+$14 16 45.6 & E & II & 166 & 883 \\
21,094 & 1841 & 53491 & 418 & 0.2384 & 14 31 03.4 & $+$33 45 41.6 & E & II & 192 & 731\\
21,583 & 1920 & 53314 & 285 & 0.1883 & 07 46 33.7 & $+$17 08 09.6 & E & II & 438 & 1389 \\
21,944 & 1939 & 53389 & 320 & 0.3597 & 09 23 16.2 & $+$28 54 57.9 & E & II  & 162 & 822 \\
23,741 & 2087 & 53415 & 557 & 0.2150 & 09 19 42.2 & $+$26 09 24.1 & E & II & 213 & 749 \\
24,327 & 2116 & 53854 & 319 & 0.2404 & 13 50 00.7 & $+$29 47 21.4 & E & II & 183 & 701 \\
25,187 & 2154 & 54539 & 376 & 0.3584 & 15 08 58.5 & $+$28 26 28.2 & E & II & 182 & 921 \\
25,302 & 2159 & 54328 & 102 & 0.3358 & 15 24 44.6 & $+$19 59 57.1 & E & II  & 248 & 1203 \\
25,511 & 2169 & 53556 & 29 & 0.1154 & 15 52 06.7 & $+$22 47 39.2 & D & I/II & 668 & 1407  \\
25,565$^{\S}$ & 2171 & 53557 & 389 & 0.0683 & 15 52 22.4 & $+$22 33 11.8 & E & II  & 578 & 760\\
26,948 &  2284 & 53708 & 269 & 0.4100 & 09 01 36.7 & $+$21 46 33.8 &  E  & I/II & 130 & 716 \\
27,059 & 2291 & 53714 & 114 & 0.0345 & 09 23 31.5 & $+$24 26 46.7 & D &  I/II & 1080 & 746 \\
28,725 &  2494 & 54174 & 488 & 0.1781 & 11 21 45.0 & $+$17 24 25.3 & E & I/II & 255 & 773 \\
28,749 & 2495 & 54175 & 564 & 0.1665 & 11 23 32.3  & $+$20 04 17.6 & E & II & 263 & 755 \\
29,643 & 2577 & 54086 & 54 & 0.1602 & 09 21 01.5 & $+$11 29 44.6 & E &  II & 259 & 720\\
29,804 & 2585 & 54097 & 327 & 0.1235 & 09 59 40.4 & $+$17 25 28.2 & E &  II & 530 & 1183 \\
29,989 & 2593 & 54175 & 397 & 0.1368 & 10 34 03.9 & $+$18 40 49.0 & E & II & 540 & 1316 \\
\hline
\enddata
\tablecomments{Columns: (1) catalog number; (2) plate number from SDSS; (3) MJD from SDSS; (4) Fiber number from SDSS; (5) redshift; (6) Right ascension from FIRST; (7) Declination from FIRST; (8) Optical morphological classification (Table~\ref{OpticalMorph}); (9) final radio morphological classification; (10) angular size; (11) projected linear size. \\ $^{\dagger}$classified also as a new double-double or possible double-double radio source.\\
$^\ddag$classified also as a new possible X--shaped radio source. \\
$^\S$formerly classified as FR~I radio galaxy \citep{Capetti17a}.\\$^\star$Object was included in the FR~II radio galaxy sample of \cite{KozielWierzbowska11}.}
\end{deluxetable*}

\begin{deluxetable*}{lccccccclcc}
\small
\tabletypesize{\scriptsize}
\tablecolumns{11}
\tablewidth{0pt}
\tablecaption{Candidates for new giant radio sources listed in ROGUE~I.\label{PGNewObjects}}
\tablehead{\colhead{} & \multicolumn{3}{c}{SDSS} &  & \multicolumn{2}{c}{FIRST} & \multicolumn{2}{c}{Classification} & AS & LS \\
No. & Plate & MJD & Fiber & z & $\alpha$ & $\delta$ & Optical & Radio &  ($^{\prime\prime}$)  & (kpc) \\
(1) & (2) & (3) & (4) & (5) & (6) & (7) & (8) & (9)  & (10) & (11) 
}
\startdata
370 & 287 & 52023 & 573 & 0.2510 & 12 14 34.6 & $+$00 47 28.3 & E & II & 204 & 807\\
1,415 & 375 & 52140 & 399 & 0.2102 & 22 20 55.9 & $+$00 18 20.0 & E & II & 202 & 698\\
2,500 & 442 & 51882 & 259 & 0.1603 & 08 17 36.1 & $+$49 59 31.6 & E & I/II & 251 & 699 \\
7,161 & 817 & 52381 & 623 & 0.2707 & 16 38 09.9 & $+$40 58 39.9 & E & II & 167 & 698 \\
7,346 & 832 & 52312 & 555  & 0.3242 & 09 17 17.9 & $+$44 34 26.1 & E & II & 146 & 692 \\
11,034 & 1058 & 52520 & 395 &  0.4046 & 16 25 13.3  & $+$33 41 51.4 & E & I & 127 & 694 \\
11,417 & 1175 & 52791 & 74 & 0.1994 & 16 52 47.4 & $+$32 34 59.4 & E & II & 244 & 810 \\
12,374 & 1237 & 52762 & 105 & 0.1766 & 10 16 14.4 & $+$08 15 13.8 & E & pI/II  &  278 & 837\\
12,766 & 1282 & 52759 & 26 & 0.3318 & 13 30 41.8 & $+$48 27 54.8 & E & II & 255 & 1227 \\
13,360 & 1318 & 52781 & 221 & 0.1071 & 12 57 17.6 & $+$56 39 12.1 & E & pI/II & 582 & 1148 \\
14,074$^{\ddagger}$ & 1360 & 53033 & 175 & 0.0921 & 10 30 53.6 & $+$41 13 15.8 & E & pX & 530 & 915  \\
14,686 & 1388 & 53119 & 40 & 0.1991 & 15 36 59.2 & $+$31 05 38.8  & E &  I & 270 & 895 \\
14,841 & 1396 & 53112 & 120 & 0.3350 & 14 41 35.0 & $+$41 56 32.7 &  E &  pI/II & 295  &  1429 \\
16,484 & 1587 & 52964 & 238 & 0.0878 & 08 36 07.8 & $+$26 48 43.7 & E &  pII & 552 & 912 \\
19,048 & 1737 & 53055 & 197 & 0.1851 & 07 48 18.8 & $+$45 44 46.3 & E & pI/II & 228 & 713 \\
20,418$^{\dagger}$ & 1805 & 53875 & 413 & 0.1501 & 13 51 10.8 & $+$07 28 46.2 & pE & pDD & 338 & 891 \\
25,432 & 2165 & 53917 & 363 & 0.2052 & 15 37 21.2 & $+$24 55 58.7 & E & pII & 217 & 736 \\
25,587 & 2172 & 54230 & 332 & 0.0897 & 15 52 09.1 & $+$20 05 48.3 & E & pII & 1186 & 1998\\
26,971 & 2285 & 53700 & 401 & 0.3289 & 09 02 33.7 & $+$20 23 43.9 & E & II & 145 & 694 \\
29,884 & 2588 & 54174 & 389 & 0.2593 & 10 12 07.2 & $+$16 19 26.2 & E & pII & 206 & 834 \\
29,900 & 2589 & 54174 & 382 & 0.4518 & 10 18 06.1 & $+$17 48 09.1 & E & I & 131 & 764 \\
29,948 & 2591 & 54140 & 268 & 0.2982 & 10 24 24.4 & $+$17 09 17.2 & E & pII & 188 & 841 \\
\hline
\enddata
\tablecomments{Columns: (1) catalog number; (2) plate number from SDSS; (3) MJD from SDSS; (4) Fiber number from SDSS; (5) redshift; (6) Right ascension from FIRST; (7) Declination from FIRST; (8) Optical morphological classification (Table~\ref{OpticalMorph}); (9) final radio morphological classification; (10) angular size; (11) projected linear size.\\
$^{\dagger}$classified also as a new double-double or possible double-double radio source.\\
$^\ddag$classified also as a new possible X--shaped radio source. \\} 
\end{deluxetable*}

\begin{deluxetable*}{lcccccccl}
\small
\tabletypesize{\scriptsize}
\tablecolumns{9}
\tablewidth{0pt}
\tablecaption{New double-double radio sources listed in ROGUE~I.\label{DDNewObjects}}
\tablehead{\colhead{} & \multicolumn{3}{c}{SDSS} &  & \multicolumn{2}{c}{FIRST} & \multicolumn{2}{c}{Classification}\\
No. & Plate & MJD & Fiber & z & $\alpha$ & $\delta$ & Optical & Radio  \\
(1) & (2) & (3) & (4) & (5) & (6) & (7) & (8) & (9) 
}
\startdata
2,635$^{\star\dagger}$ & 447 & 51877 & 421 & 0.4029 & 08 45 25.5 & $+$52 29 15.8 & E & DD \\
8,602$^\star$ & 906 & 52368 & 169 & 0.1446 & 10 46 32.2 & $+$54 35 59.7 & E & DD \\
9,180$^\star$ & 941 & 52709 & 201 & 0.0721 & 09 47 08.8 & $+$42 11 25.6 & E & DD\\
19,429 & 1753 & 53383 & 486 & 0.1720 & 11 24 22.8 & $+$15 09 58.0 & E & DD \\\hline
402$^{\star\dagger}$ & 290 & 51941 & 44 & 0.1353 & 12 37 45.9 & $-$01 14 16.2 & E & pDD \\
1,232$^\star$ & 358 & 51818 & 161 & 0.333 & 17 32 50.2 & $+$56 34 27.0 & E & pDD \\
4,035 & 548 & 51986 & 18 & 0.1800 & 08 33 46.8 & $+$45 15 18.6 & E & pDD \\
4,581 & 580 & 52368 & 461 & 0.0354 & 10 59 14.6 & $+$05 17 31.3 & E & pDD \\
4,861 & 599 & 52317 & 129 & 0.1045 & 12 13 26.0 & $+$63 59 09.1 & E & pDD \\
6,099 & 725 & 52258 & 79 & 0.1593 & 23 06 32.1 & -09 30 18.0 & E & pDD \\
15,170 & 1419 & 53144 & 481 & 0.3038 & 16 08 10.8 & $+$32 54 18.9 & E & pDD \\
16,107$^\star$ & 1465 & 53082 & 522 & 0.2249 & 13 43 00.4 & $+$46 27 19.9 & E & pDD \\
20,418$^{\dagger}$ & 1805 & 53875 & 413 & 0.1501 & 13 51 10.8 & $+$07 28 46.2 & pE & pDD \\ 
24,209$^\S$ & 2110 & 53467 & 344 & 0.0162 & 13 23 45.0 & $+$31 33 56.7 & E & pDD \\
25,983 & 2218 & 53816 & 458 & 0.0732 & 11 29 12.2 & $+$27 33 14.1 & E & pDD \\
30,866$^\S$ & 2656 & 54484 & 499 & 0.0226 & 12 08 05.6 & $+$25 14 14.1 & E & pDD \\\hline
\enddata
\tablecomments{Columns: (1) catalog number; (2) plate number from SDSS; (3) MJD from SDSS; (4) Fiber number from SDSS; (5) redshift; (6) Right ascension from FIRST; (7) Declination from FIRST; (8) Optical morphological classification (Table~\ref{OpticalMorph}); (9) final radio morphological classification. \\  
$^\S$formerly classified as FR~I radio galaxy \citep{Kharb12}.\\
$^\star$ included in the FR~II radio galaxy sample of \cite{KozielWierzbowska11}.\\
$^{\dagger}$ classified also as a new giant or a candidate for a new giant radio source.\\}
\end{deluxetable*}

\begin{deluxetable*}{lcccccccl}
\small
\tabletypesize{\scriptsize}
\tablecolumns{9}
\tablewidth{0pt}
\tablecaption{New X--shaped radio sources listed in ROGUE~I.\label{XNewObjects}}
\tablehead{\colhead{} & \multicolumn{3}{c}{SDSS} &  & \multicolumn{2}{c}{FIRST} & \multicolumn{2}{c}{Classification}\\
No. & Plate & MJD & Fiber & z & $\alpha$ & $\delta$ & Optical & Radio  \\
(1) & (2) & (3) & (4) & (5) & (6) & (7) & (8) & (9) 
}
\startdata
18,205 & 1695 & 53473 & 492 & 0.2077 & 12 57 21.7 & $+$12 28 19.3 & E & X \\
25,798$^\star$ & 2209 & 53907 & 286 & 0.2791 & 16 30 16.6 & $+$14 35 11.4 & E & X \\
27,729 & 2368 & 53758 & 58 & 0.191 & 09 32 38.3 & $+$16 11 58.0 & E & X \\ \hline
3,155 & 484 & 51907 & 497 & 0.2698 & 09 09 51.0 & $+$58 47 07.0 & E & pX \\
4,287 & 561 & 52295 & 303 & 0.3626 & 10 40 21.9 & $+$59 58 41.3 & E & pX \\
6,713 & 776 & 52319 & 99 &  0.1112 & 11 37 21.4 & $+$61 20 00.9 & E & pX \\
14,074$^{\dagger}$ & 1360 & 53033 & 175 & 0.0921 & 10 30 53.6 & $+$41 13 15.8 & E & pX \\
18,519$^\dagger$ & 1709 & 53533 & 491 & 0.1665 & 14 31 51.1 & $+$10 29 59.4 & E & pX \\
23,094 & 2012 & 53493 & 629 & 0.1148 & 11 44 27.2 & $+$37 08 32.4 & E & pX \\ \hline
\enddata
\tablecomments{Columns: (1) catalog number; (2) plate number from SDSS; (3) MJD from SDSS; (4) Fiber number from SDSS; (5) redshift; (6) Right ascension from FIRST; (7) Declination from FIRST; (8) Optical morphological classification (Table~\ref{OpticalMorph}); (9) final radio morphological classification. \\
$^\star$ included in the FR~II radio galaxy sample of \cite{KozielWierzbowska11}.\\
$^{\dagger}$ classified also as a new giant or a candidate for a new giant radio source.\\}
\end{deluxetable*}

Although sources with Z--shaped symmetry are traditionally included in the group of X--shaped sources \citep{Gopal-Krishna03, Cheung07}, the radio morphologies are significantly different for these two classes. The former class has a single pair of jets gradually (S--shaped) or abruptly (Z--shaped) changing direction of propagation (Figure~\ref{fig:radioMorph}; Z--shaped source), while the latter possesses two pairs of lobes, where the axis of the second pair of lobes also crosses the core at an angle to the first pair (Figure~\ref{fig:radioMorph}; X--shaped source). This is in the agreement with \citet{Roberts15,Roberts18}, who divide the two classes where the former class is identified as a separate category and the latter class is termed as \textit{true} X-shaped sources.

In Appendix~\ref{appendix}, we present the radio-optical overlays for the newly discovered sources and give notes on selected sources. In particular, see Section~\ref{Giants} and Figure~\ref{fig:NewGiantsMaps} for GRSs,  Section~\ref{PGiants} and Figure~\ref{NewPGiantsMaps} for possible GRSs, Section~\ref{DD} and Figure~\ref{NewDDMaps} for DDs, Section~\ref{X} and Figure~\ref{NewXMaps} for X--shaped sources, and Section~\ref{Z} and Figure~\ref{Z} for Z--shaped sources.

\begin{deluxetable*}{lcccccccc}
\small
\tabletypesize{\scriptsize}
\tablecolumns{9}
\tablewidth{0pt}
\tablecaption{Z--shaped radio sources listed in ROGUE~I.\label{ZshapedObjects}}
\tablehead{
\colhead{} & \multicolumn{3}{c}{SDSS} &  & \multicolumn{2}{c}{FIRST} & \multicolumn{2}{c}{Classification}\\
No. & Plate & MJD & Fiber & z & $\alpha$ & $\delta$ & Optical & Radio  \\
(1) & (2) & (3) & (4) & (5) & (6) & (7) & (8) & (9) 
}
\startdata
\hline
1,606 & 385 & 51877 & 375 & 0.1835 & 23 39 00.3 & $+$00 42 58.2 & E & Z \\
2,798 & 456 & 51910 & 365 & 0.3485 & 02 43 20.6 & $-$07 14 45.9 & E & Z \\
2,809 & 457 & 51901 & 193 & 0.0782 & 02 52 27.6 & $-$07 56 04.8 & E & Z \\
4,142 & 552 & 51992 & 471 & 0.0992 & 09 02 36.8 & $+$52 03 48.0 & E & Z\\
7,225 & 826 & 52295 & 491 & 0.0618 & 08 21 10.4 & $+$35 47 35.9 & E & Z \\ 
7,502 & 842 & 52376 & 209 & 0.1489 & 12 04 25.3 & $+$03 45 09.3 & E & Z \\
7,977$^{\dag}$ & 875 & 52354 & 521 & 0.1539 & 10 40 22.5 & $+$50 56 23.0 & E & Z \\
11,229 & 1164 & 52674 & 103 & 0.1331 & 15 02 29.0 & $+$52 44 02.2 & E & Z \\
12,408 & 1238 & 52761 & 550 & 0.0626 & 10 23 22.6 & $+$08 52 01.6 & E & Z\\
12,544$^{\ddag}$ & 1269 & 52937 & 243 & 0.0788 & 08 39 15.8 & $+$28 50 39.1 & E & Z \\
13,184 & 1307 & 52999 & 67 & 0.1615 & 10 04 56.7 & $+$09 47 04.7 & E & Z \\	
18,459 & 1707 & 53885 & 313 & 0.3442 & 14 18 13.3 & $+$09 52 37.0 & E & Z \\
21,690 & 1926 & 53317 & 156 & 0.0959 & 08 18 54.1 & $+$22 47 46.1 & E & Z\\
22,328 & 1971 & 53472 & 124 & 0.3522 & 12 32 11.6 & $+$31 30 56.6 & E & Z \\
22,897$^\star$ & 2002 & 53471 & 462 & 0.0728 & 13 19 04.2 & $+$29 38 35.4 & E & Z \\
25,124$^\S$ & 2151 & 54523 & 113 & 0.0541 & 15 04 57.1 & $+$26 00 58.3 & E & Z \\
28,729 & 2494 & 54174 & 626 & 0.1422 & 11 24 57.4 & $+$17 17 43.2 & E & Z\\
32,462 & 2954 & 54561 & 299 & 0.1161 & 15 26 42.0 & $+$00 53 30.1 & E & Z\\ \hline
6,195 & 733 & 52207 & 202 & 0.2641 & 21 49 39.7 & $+$10 57 27.4 & E & pZ \\ 
10,652 & 1038 & 52673 & 475 & 0.0851 & 12 46 47.5 & $+$54 53 15.2 & E & pZ\\
10,919 & 1051 & 52468 & 483 & 0.2531 & 15 23 33.5 & $+$45 03 36.6 & E & pZ\\
13,950 & 1352 & 52819 & 491 & 0.3860& 15 05 57.1 & $+$37 02 07.2 & E & pZ \\
18,391 & 1704 & 53178 & 2 & 0.0914 & 14 8 33.0 & $+$12 24 25.5 & E & pZ\\
24,512 & 2123 & 53793 & 634 & 0.2939 & 14 12 24.4 & $+$27 17 59.7 & E & pZ\\
28,012 & 2424 & 54448 & 561 & 0.1647 & 08 30 59.5 & $+$12 52 53.2 & E & pZ \\
\hline
\enddata
\tablecomments{Columns: (1) Catalog number; (2) Plate number from SDSS; (3) MJD from SDSS; (4) Fiber number from SDSS; (5) Redshift; (6) Right ascension from FIRST; (7) Declination from FIRST; (8) Optical morphology classification (Table~\ref{OpticalMorph}); (9) Final radio morphology classification. \\Objects earlier classified as: $^\dag$X-shaped \citep{Cheung07}, $^\ddag$WAT \citep{Donoghue93}, $^\star$FR~II \citep{KozielWierzbowska11}, and $^\S$FR~I with lobes \citep{Croston18}.}
\end{deluxetable*}

\subsection{Comparison with other catalogs}

The catalog of \citet[][BH12 hereafter]{Best12} is the recent and widely explored catalog of radio sources with optical counterparts from SDSS\,DR\,7 \citep[see,][]{Capetti17a, Capetti17b, Baldi18}. As stated in Table~\ref{AGNsamples}, the BH12 catalog was made by cross-matching SDSS\,DR\,7 with NVSS and FIRST using automatic methods described in \citet{Best05} with modifications introduced in \citet{Donoso2009}. The BH12 catalog contains 18,286 radio objects selected with a flux density limit of 5 mJy. All the objects in this catalog were classified as AGNs (about 15,000) or SF galaxies (about 3,000) using three different classification schemes \citep[see Appendix A in][]{Best12}. Compared to BH12, the ROGUE~I catalog was selected using the same optical and radio surveys, but without applying any additional radio flux density limit. 

As we already mentioned, in the SDSS catalogs there are galaxies with repeated observations, therefore we performed our comparison based on the SDSS coordinates and not plate, MJD, and fiber numbers. Out of the 14,383 BH12 sources that fulfill our spectrum and photometry quality criteria 
(see Section~\ref{sec:sample}), 11,882 sources are also in ROGUE~I. However, two of them are sources with off-centered spectra, therefore, the ROGUE~I and the BH12 catalogs contain 11,880 unique sources in common. 

BH12 contains 2,500 sources that are not listed in ROGUE~I, however, all these sources do not have a FIRST radio detection within 3\arcsec\, from the optical galaxy. 
On the other hand, in the ROGUE~I catalog there are 20,728 sources not included in BH12, 846 of which having total radio flux densities higher than 5 mJy, including 169 \textit{extended}, 53 compact, 578 elongated radio sources, and 46 SFRs. This comparison shows that in automatic search even bright sources with radio cores can be missed, therefore, in such projects selection criteria should be chosen with caution.

Since ROGUE~I contains extended sources with assigned morphological types, our comparison should be also made with catalogs of extended sources. In the \citet{Lin10} catalog,  SDSS DR\,6 was cross-matched with NVSS and FIRST. The \citeauthor{Lin10} sample is limited in redshift (0.02 $<$ z $<$ 0.3) and contains galaxies that are more luminous than the characteristic magnitude in the galaxy luminosity function \citep{Blanton03}. They applied a radio flux density limit of 3 mJy and a search radius of 3\arcmin\, between the optical galaxy and the NVSS radio source. The \citet{Lin10} catalog contains about 10,500 objects of which 1,040 have extended morphologies. \citeauthor{Lin10} proposed a classification scheme similar to the standard \citet{Fanaroff74} one, which was based on the ratio, r$_{S}$, of the separation between the brightest regions on either sides of the host galaxy and the total size of the radio source.

The comparison of ROGUE~I and extended sources from  \citet{Lin10} gives 505 common objects, 154 sources which are not in our optical galaxy sample, and 381 sources which have no FIRST detection within 3\arcsec. Among common objects, the majority of sources with the highest values of r$_{S}$ are ROGUE~I FR~II radio sources. Going to lower values of r$_{S}$, more sources with more complex structures are found, which is in agreement with our and \citeauthor{Lin10} classification schemes. However, we note that the values of this ratio for some sources do not match their ROGUE~I morphology. It is a result of incompatible identification of radio components as a part of the radio source. This shows that the proper identification of all parts of radio sources will be a challenging problem in future automatic searches. It also shows that measuring sizes and classification similar to the one proposed by \citeauthor{Lin10} can be inaccurate in the case of bent sources, or sources with more than one pair of lobes \citep[see also comments in][]{Mingo19}.

\section{Summary}
\label{sec:discussion}

We have presented ROGUE~I, the \emph{largest handmade} catalog of radio sources associated with optical galaxies. It has been constructed using the SDSS DR\,7 spectroscopic catalog of galaxies and the FIRST and the NVSS radio catalogs. All ROGUE~I objects have spectroscopic redshifts and good quality optical spectra that can be used to derive basic host galaxy properties, as well as  stellar velocity dispersions from which BH masses can be estimated. ROGUE~I consists only of sources with a central FIRST component which, in the case of AGNs, can be interpreted as a radio core. A second catalog, ROGUE~II, which will deal with radio galaxies with SDSS counterparts but \textit{without} a FIRST core, is in preparation. ROGUE~I provides the morphological classification of the host galaxies as well as of the associated radio sources, and a careful estimation of the total radio flux densities.

The main results of our visual classifications are as follows: 
\begin{enumerate}
\item Unresolved (compact) and elongated 
radio sources dominate in the ROGUE~I catalog. They constitute $\sim$90\% of the total number. About 8\% of the sources in the sample exhibit extended  morphology. 
\item Radio sources (secure and possible classifications) of \citeauthor{Fanaroff74} I, II, hybrid, and one-sided types constitute $\sim$78\% of the extended sources, bent (wide-angle, narrow-angle, head-tail) sources $\sim$18\%, while sources with intermittent or reoriented jet activity (double-double, X-shape, Z-shape sources) $\sim$3\%. 
\item Although the FIRST and the NVSS catalogs together with SDSS DR\,7 have been extensively explored in the past, our selection procedure allowed us to discover or reclassify a number of objects as giant, double--double, X--shaped, and Z--shaped radio sources. Moreover, we have classified much bigger samples of \citeauthor{Fanaroff74} I and II types (416 and 871, respectively, including both secure and possible assignments), than presented in \citet{Capetti17a,Capetti17b} due to higher $z$ range and lower radio detection thresholds in our study. We note that the ROGUE~II catalog, which will comprise of radio sources without cores, will further increase the numbers of extended radio sources.
\item We identify a total of 81 GRSs (55 new and 26 from the sample of \citealt{Kuzmicz.etal.2018a}) among the group of 2,503 extended radio sources in ROGUE~I. This corresponds to $\sim$3\% of the extended radio source population, in agreement with the fraction of GRSs in the local Universe  \citep[i.e., $z<$1;][]{Saripalli12}. 
\item The optical morphological classification of the host galaxies revealed that $\sim$62\% of radio sources detected at 1.4\,GHz have elliptical, $\sim$17\% spiral, and $\sim$7\% lenticular hosts. A significant number of sources ($\sim$12\%) have host galaxies with distorted morphologies.
\item In accord with earlier studies, most of the FR~II radio sources in ROGUE~I have low radio luminosities, comparable to the luminosities of the FR~I sources.
\end{enumerate}
Comparisons with automatically selected catalogs show that visual analysis, although more time-consuming, still gives better results, and that the selection and classification schemes used in such procedure can be more complex than in automatic searches. Although our method would be very difficult to apply to the catalogues based on the large radio surveys like LOFAR or EMU \citep{Norris11}, our sample can serve as a database for training automatic methods of radio source identification and classification \citep[as][]{Alger18, Ma18}.

\acknowledgements
The authors thank the anonymous referee for the useful comments, Gra\.zyna Stasi{\'n}ska, Natalia Vale Asari, Marek Sikora, and Marek Jamrozy for discussions, and Marian Soida for his help in setting up the computing facility. We also thank Richard L. White for queries related to the FIRST database, and Benjamin Alan Weaver and Aniruddha R. Thakar for help with the SDSS data. DKW acknowledges the support of Polish National Science Centre (NCN) grant via 2016/21/B/ST9/01620. AG acknowledges the full support of NCN via 2018/29/B/ST9/02298. N\.Z work is supported by the NCN through the grant DEC-2014/15/N/ST9/05171.
The Digitized Sky Survey was produced at the Space Telescope Science Institute under US government grant NAGW-2166. 
Funding for the SDSS and SDSS-II has been provided by the Alfred P. Sloan Foundation, the Participating Institutions, the National Science Foundation, the U.S. Department of Energy, the National Aeronautics and Space Administration, the Japanese Monbukagakusho, the Max Planck Society, and the Higher Education Funding Council for England. The SDSS Web Site is http://www.sdss.org/. The SDSS is managed by the Astrophysical Research Consortium for the Participating Institutions. The Participating Institutions are the American Museum of Natural History, Astrophysical Institute Potsdam, University of Basel, University of Cambridge, Case Western Reserve University, University of Chicago, Drexel University, Fermilab, the Institute for Advanced Study, the Japan Participation Group, Johns Hopkins University, the Joint Institute for Nuclear Astrophysics, the Kavli Institute for Particle Astrophysics and Cosmology, the Korean Scientist Group, the Chinese Academy of Sciences (LAMOST), Los Alamos National Laboratory, the Max-Planck-Institute for Astronomy (MPIA), the Max-Planck-Institute for Astrophysics (MPA), New Mexico State University, Ohio State University, University of Pittsburgh, University of Portsmouth, Princeton University, the United States Naval Observatory, and the University of Washington.

We thank the staff of the GMRT that made these observations possible. GMRT is run by the National Centre for Radio Astrophysics of the Tata Institute of Fundamental Research.

\appendix
\section{Notes for selected sources}
\label{appendix}

\setcounter{figure}{0}
\renewcommand{\thefigure}{A.\arabic{figure}}

Figures~\ref{fig:NewGiantsMaps}, \ref{NewPGiantsMaps},  \ref{NewDDMaps}, \ref{NewXMaps}, and \ref{NewZMaps} present radio--optical overlays of the newly discovered and/or reclassified sources listed in Section~\ref{uncommonSources}. The header of each plot gives the number of the object in the ROGUE~I catalog. The red radio contours are generated based on the FIRST 1.4 GHz maps, while the black contours on the NVSS maps. The gray scale is the optical DSS image. The map is centered at the optical host galaxy position marked by a plus sign. For a guidance, a bar with the linear scale is located at the upper right corner, while the FIRST beam is displayed inside a square box at the bottom left corner of each contour map. Below, we briefly describe the selected sources. It is worth to note that the optical hosts of all the sources presented here are elliptical galaxies with two exceptions, namely 25,511 and 27,059 giant sources hosted by distorted galaxies. All the redshifts are taken from the SDSS.

\subsection{Giant radio sources}\label{Giants}

\begin{itemize}
  \item[]2,635: the radio source is located at a distance of $z=0.40$. An angular size of the radio emission, measured between FIRST hot spots,
  is 212\arcsec\, and a projected linear size reaches 1.2 Mpc. The source is included in the FR~II radio galaxy sample in \citet{KozielWierzbowska11}. The FIRST map reveals an elongated core with a pair of bright inner lobes and faint outer lobes. The internal FIRST structure matches with the central elongated NVSS core, while the external pair of lobes also agree well with the lobes visible on the NVSS map. Comparing the radio structures from both maps, we additionally classify the source as a possible restarting double--double radio galaxy.
  
  \item[]12,749: we classify this giant, located at $z=0.36$, as a possible WAT source, reaching an angular size of 150\arcsec\, and a projected linear size of 0.8 Mpc, measured on the NVSS map. The FIRST map shows a compact core and a faint emission from one lobe. The elongated radio emission from the core as well as both lobes are visible on the NVSS map.
  
  \item[]14,373: this giant located at $z=0.08$, reaching an angular size of 486\arcsec\, and a projected linear size of 0.7 Mpc, is classified as FR~I radio galaxy in the ROGUE~I catalog. The FIRST map reveals a complex extended structure, containing a clear core and several components. The NVSS map displays an elongated morphology with the brightness maximum close to the core and smooth dimness in the lobes.

  \item[]15,353: the source located at $z=0.14$ reaches an angular size of 301\arcsec\, and a projected linear size of 0.8 Mpc. An FR~II morphology is revealed on both maps, where the FIRST map uncovers a core and a two-sided jet-like structure as well as clear outer lobes, while in the NVSS data an elongated radio emission with backflows is also visible. The radio size was measured manually taking into account the backflows.

  \item[]16,242: the source is located at $z=0.15$, reaches an angular size of  463\arcsec\, and a projected linear size of 1.2 Mpc and has an FR II morphology. On the FIRST map, a compact core and a faint emission from the south-east lobe are visible, while the NVSS map exhibits an elongated emission from the central parts and from a pair of symmetric lobes.

  \item[]25,511: in \citet{Proctor16} the source is classified as a giant (or possible giant) with an asymmetric morphology. The coordinates of the system given by \cite{Proctor16}, $\alpha$ = 15 52 10.97 and $\delta$ = 22 45 08.0, were chosen based on a fit of the NVSS catalog entry, pointing to a faint elliptical galaxy with photometric $z=0.78$ as a possible optical host. The estimated angular size of the radio structure reaches 668\arcsec and, at such a redshift, it corresponds to a projected linear size of 5 Mpc, making this source  the largest one known to date. However, we argue that the actual host is a distorted galaxy located at $z=0.12$. This redshift gives for the radio source a projected linear size  of 1.5 Mpc. We classify the source as an hybrid FR:  the FIRST map reveals only a compact core while on the NVSS map an extended hybrid FR structure is clearly visible. 
  
 \item[]25,565: \cite{Proctor16} classified this source, based on the NVSS data, as a giant or possible giant with a possible double--double morphology. The coordinates of the host galaxy proposed by them, $\alpha$ = 15 52 35.17 and $\delta$ = 22 34 18.8, were estimated visually with the symmetry assumption. Taking these coordinates, we found two possible optical hosts. Both of them are faint elliptical galaxies with no optical spectra and only photometric redshifts in the SDSS database. Using the given redshifts, the radio structure with an angular size 578\arcsec reaches a huge size, i.e. 3.9 Mpc for $z=0.39$ and 4.9 Mpc for $z=0.59$. Based on FIRST, NVSS, and SDSS data, we also suggest that the source is of FR~II type, but the FIRST data reveals an elongated jet-like structure associated with an elliptical galaxy ($\alpha$ = 15 52 22.4, $\delta$ = 22 33 11.8) at $z=0.07$, which we interpret as the host. This host with only small one-sided jet was classified as an FR~I radio galaxy by \cite{Capetti17a}. If we take into account the whole radio structure, the angular size corresponds to a projected linear size of 0.9 Mpc, and therefore we classify it as a giant. Due to the uncertainty in the identification of the host galaxy, we also investigated the TIFR GMRT Sky Survey \citep[TGSS;][]{Intema17} high resolution 150 MHz maps to verify the radio morphology of possible optical hosts. The maps reveal a significant emission at the location of the putative host pointed by us. It is also worth mentioning that the host galaxy is a member of a galaxy cluster located at z=0.07. It is, therefore, possible that the large asymmetry of the radio structure is the result of interactions between the lobe and the intracluster ambient medium. 
  
  \item[]29,884: we classify this giant radio source ($z=0.26$, angular size of 206\arcsec\, and projected linear size 0.8 Mpc) as an FR~II radio galaxy. A strong compact core and a faint emission in one lobe is visible on the FIRST map. The NVSS map shows a strong slightly elongated core and two much weaker but still clear lobes.
\end{itemize}

\begin{figure*}[t!]
\hspace{0cm}{\includegraphics[width=0.3\textwidth]{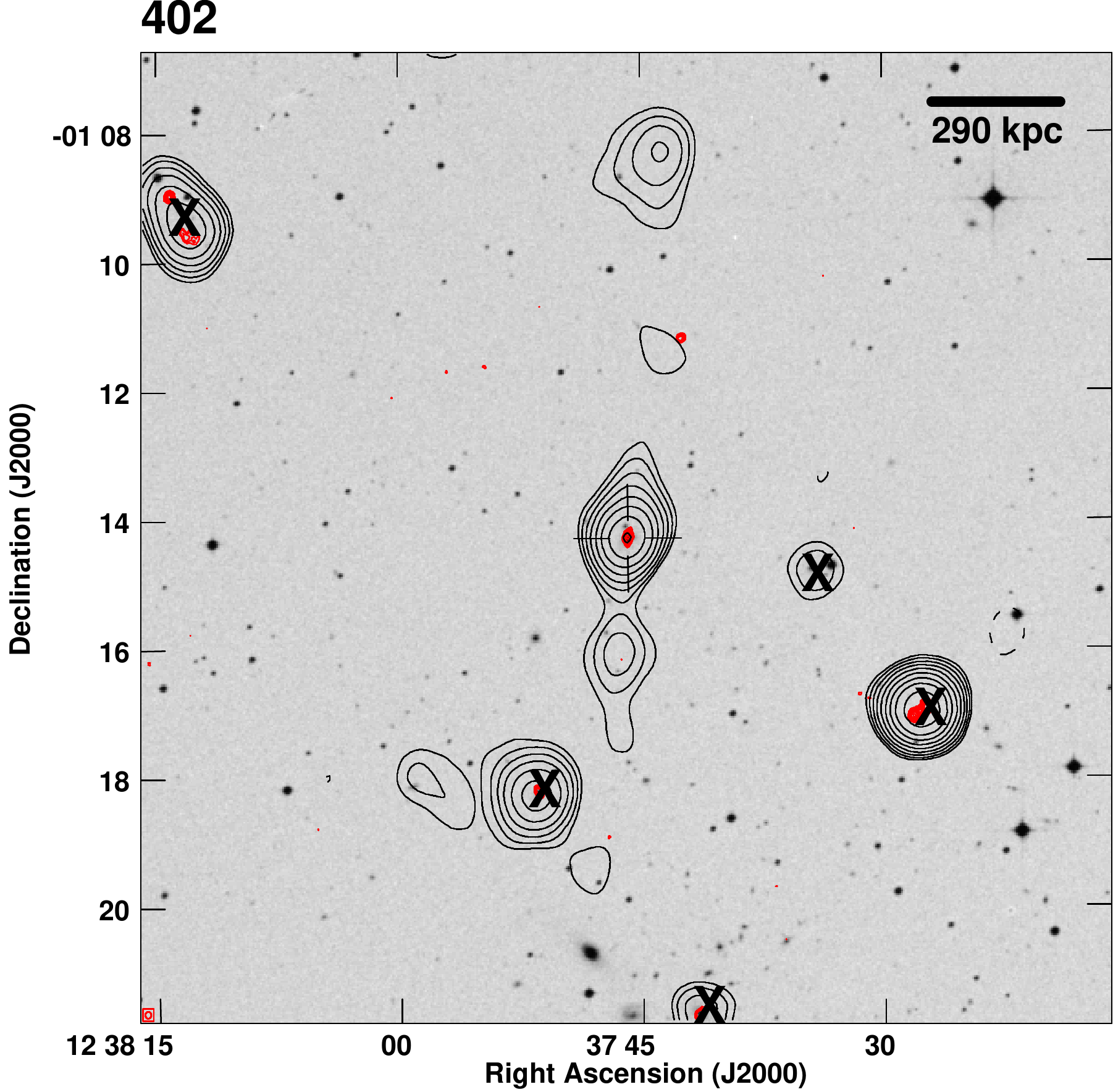}}
\hspace{0.2cm}{\includegraphics[width=0.3\textwidth]{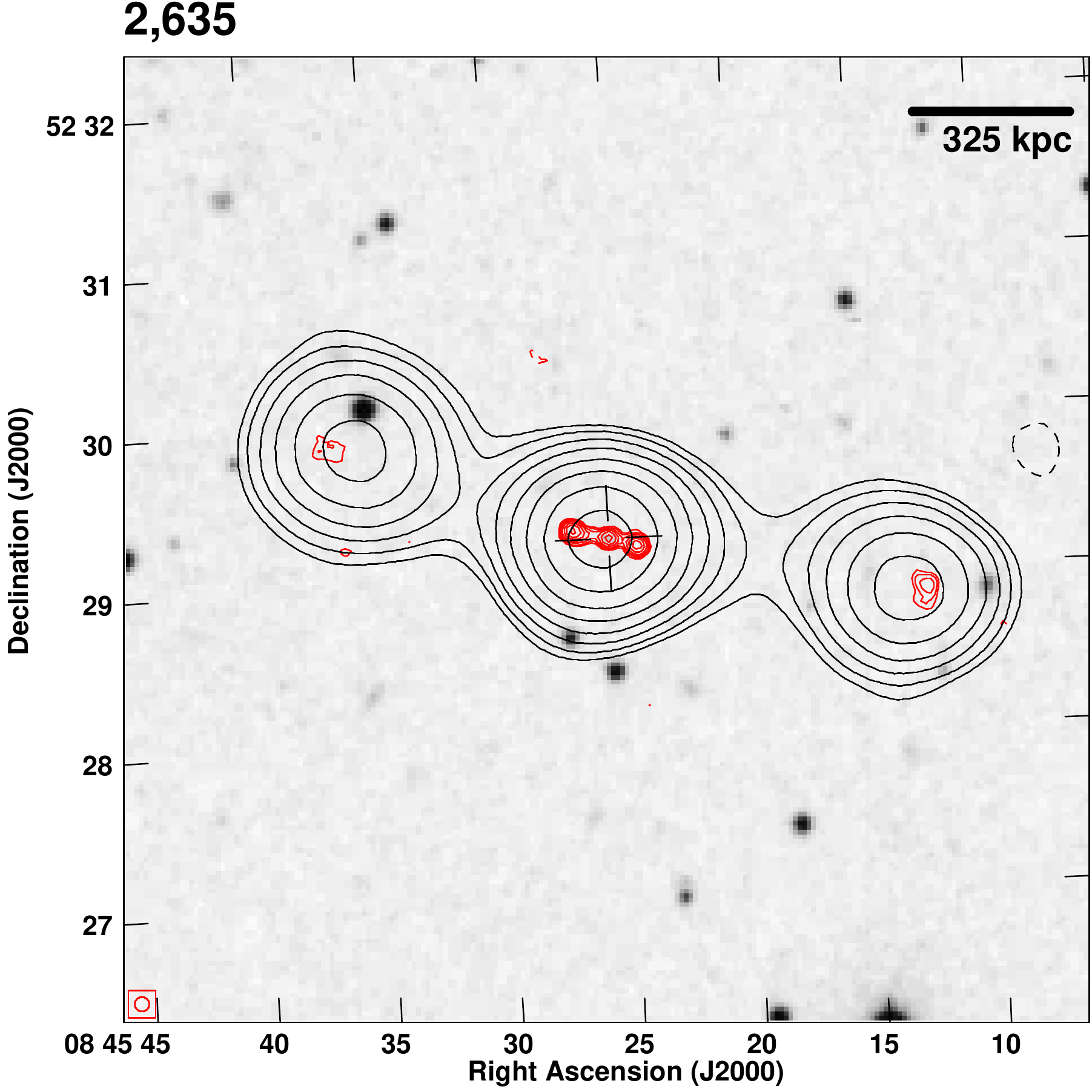}}
\hspace{0.2cm}{\includegraphics[width=0.3\textwidth]{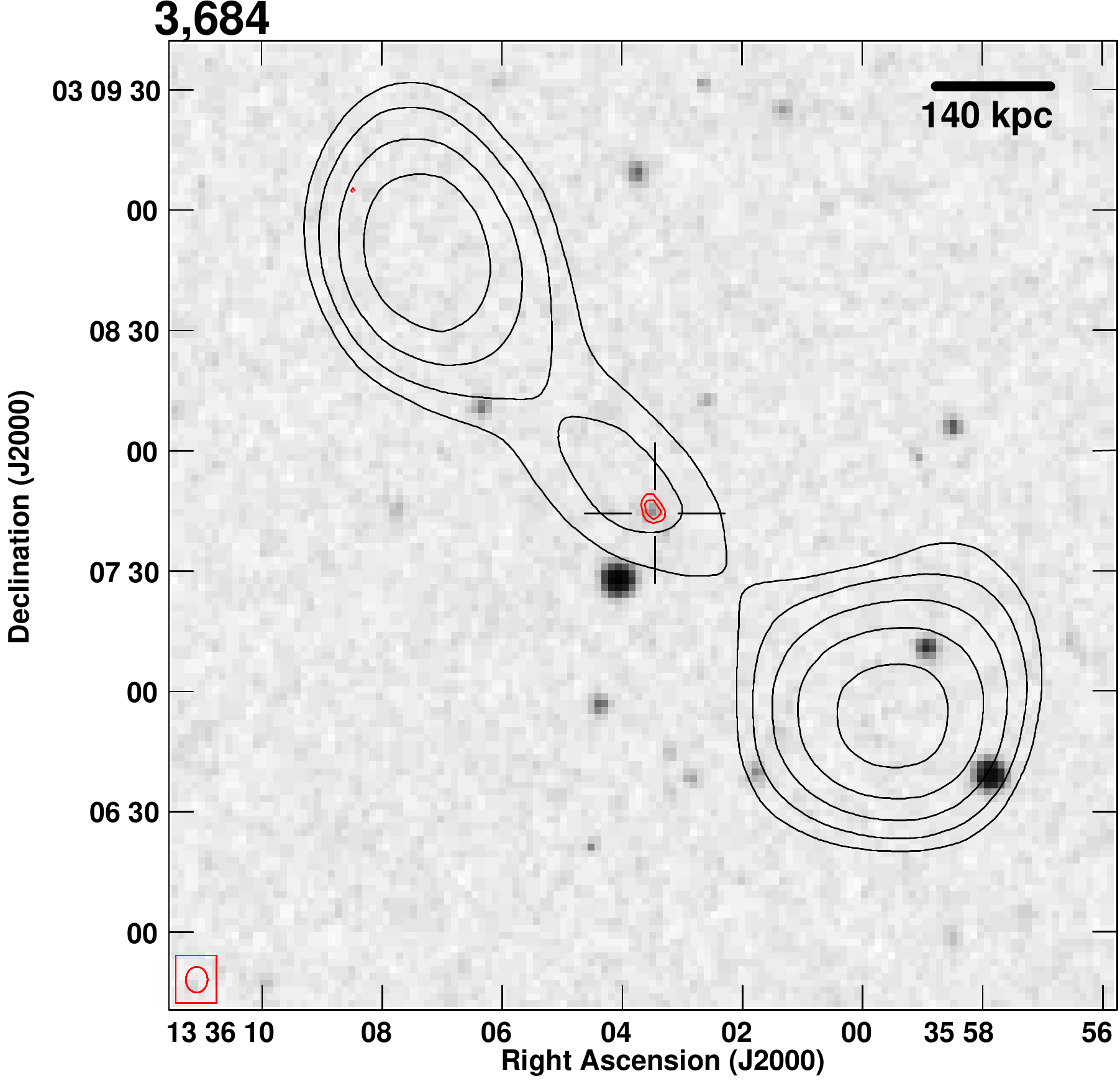}}
\hspace{0.2cm}{\includegraphics[width=0.3\textwidth]{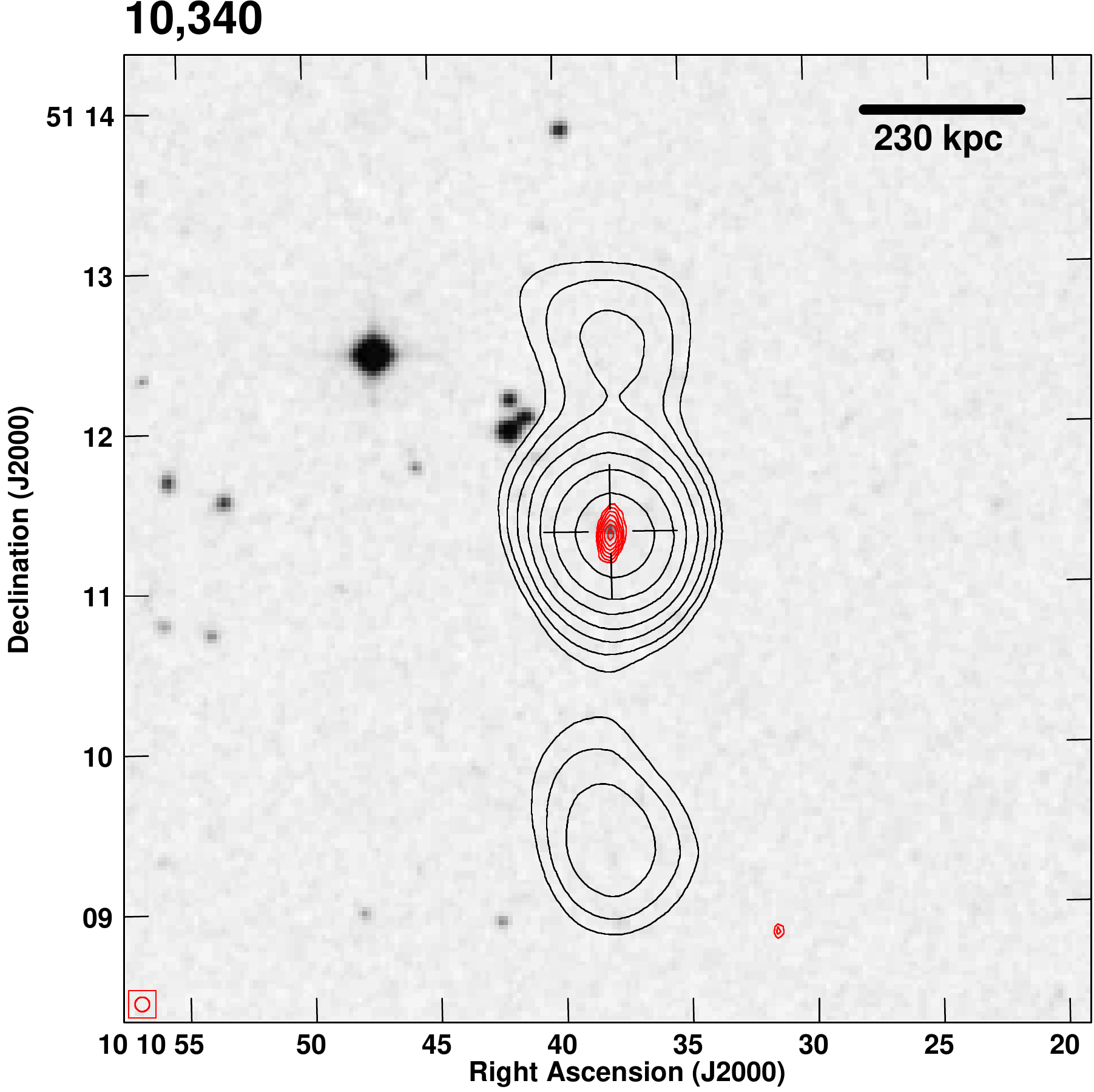}}
\hspace{0.2cm}{\includegraphics[width=0.3\textwidth]{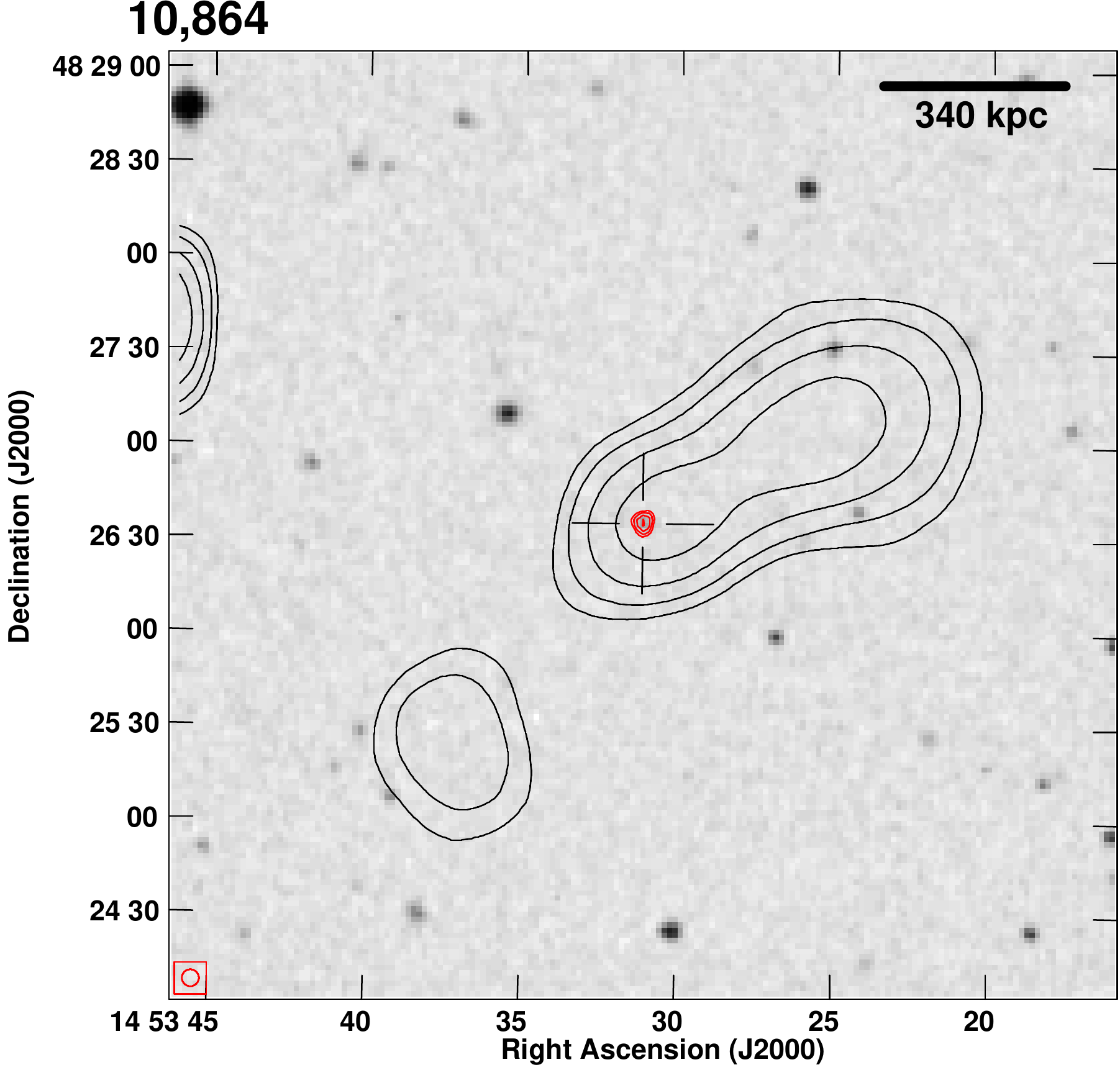}}
\hspace{0.2cm}{\includegraphics[width=0.3\textwidth]{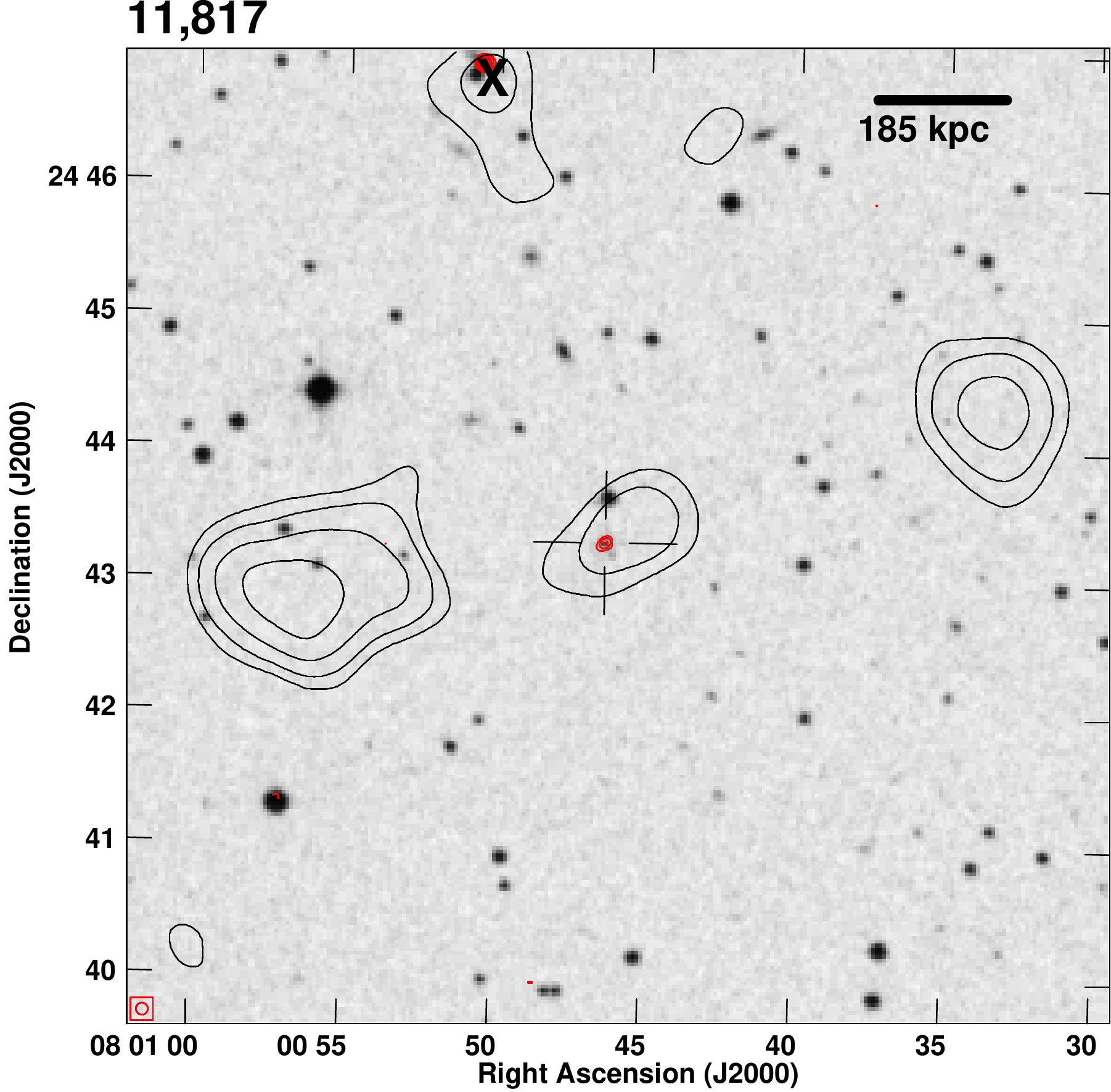}}
\hspace{0.2cm}{\includegraphics[width=0.3\textwidth]{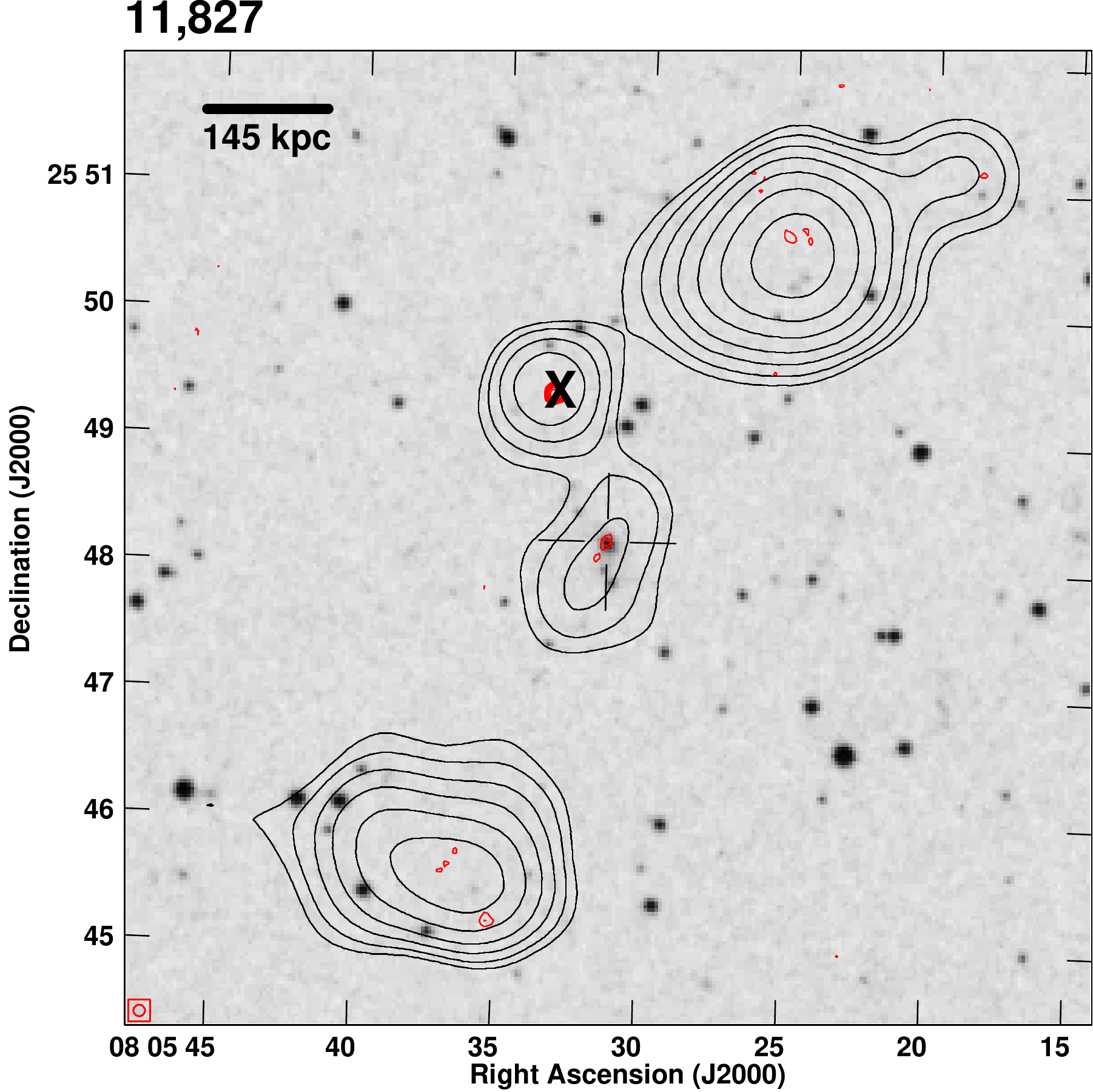}}
\hspace{0.2cm}{\includegraphics[width=0.3\textwidth]{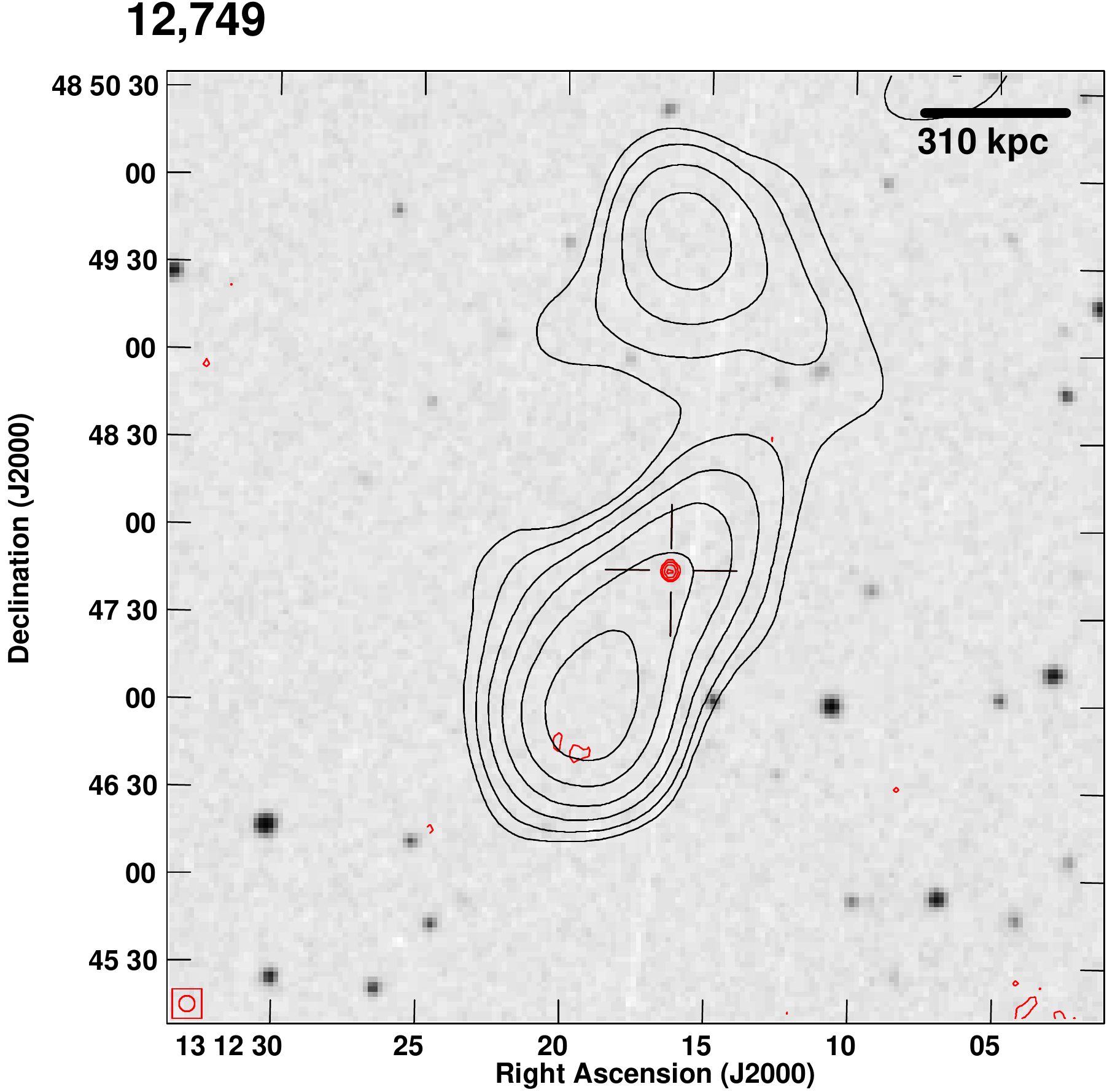}}
\hspace{0.2cm}{\includegraphics[width=0.3\textwidth]{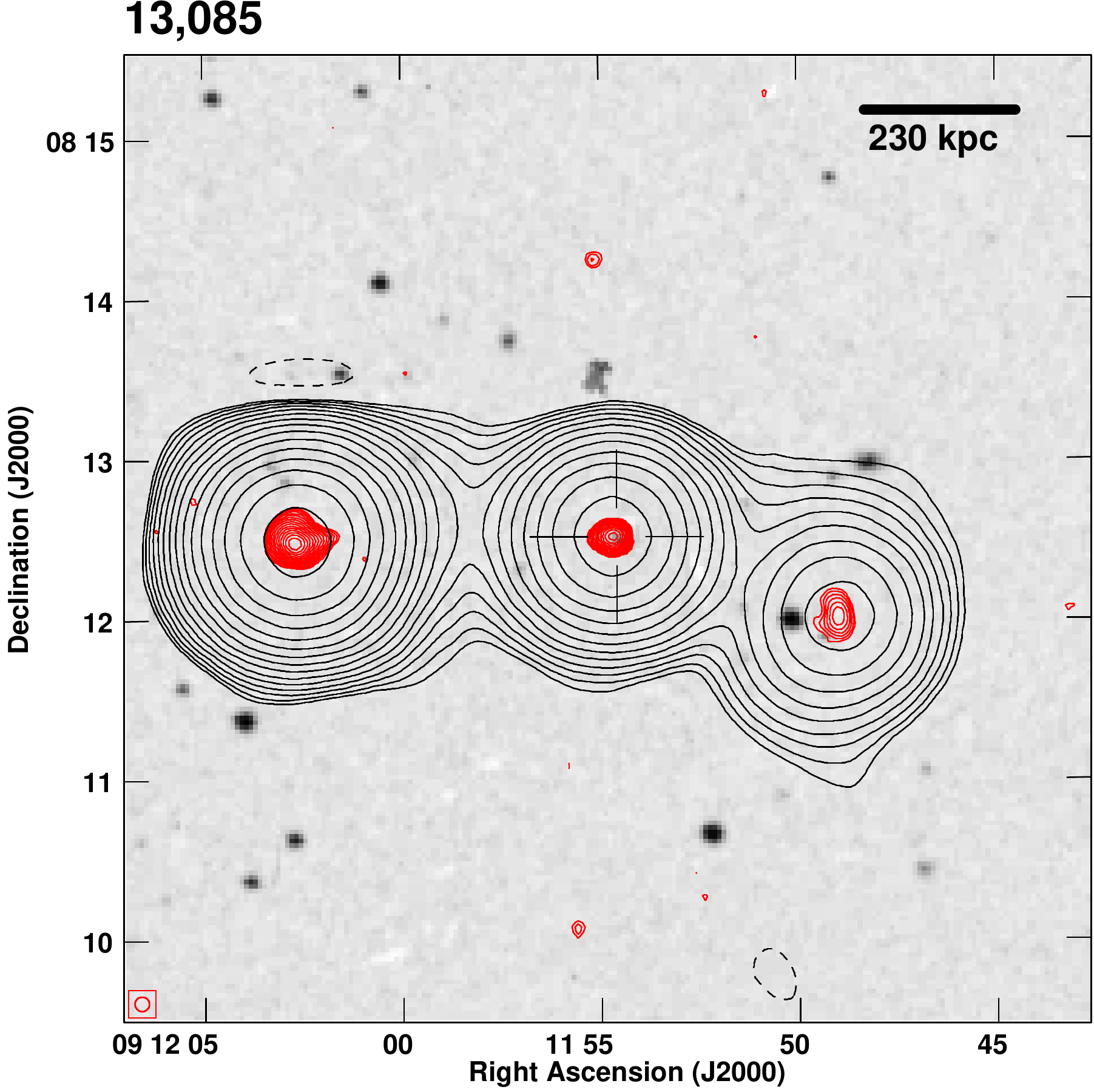}}
\hspace{0.2cm}{\includegraphics[width=0.3\textwidth]{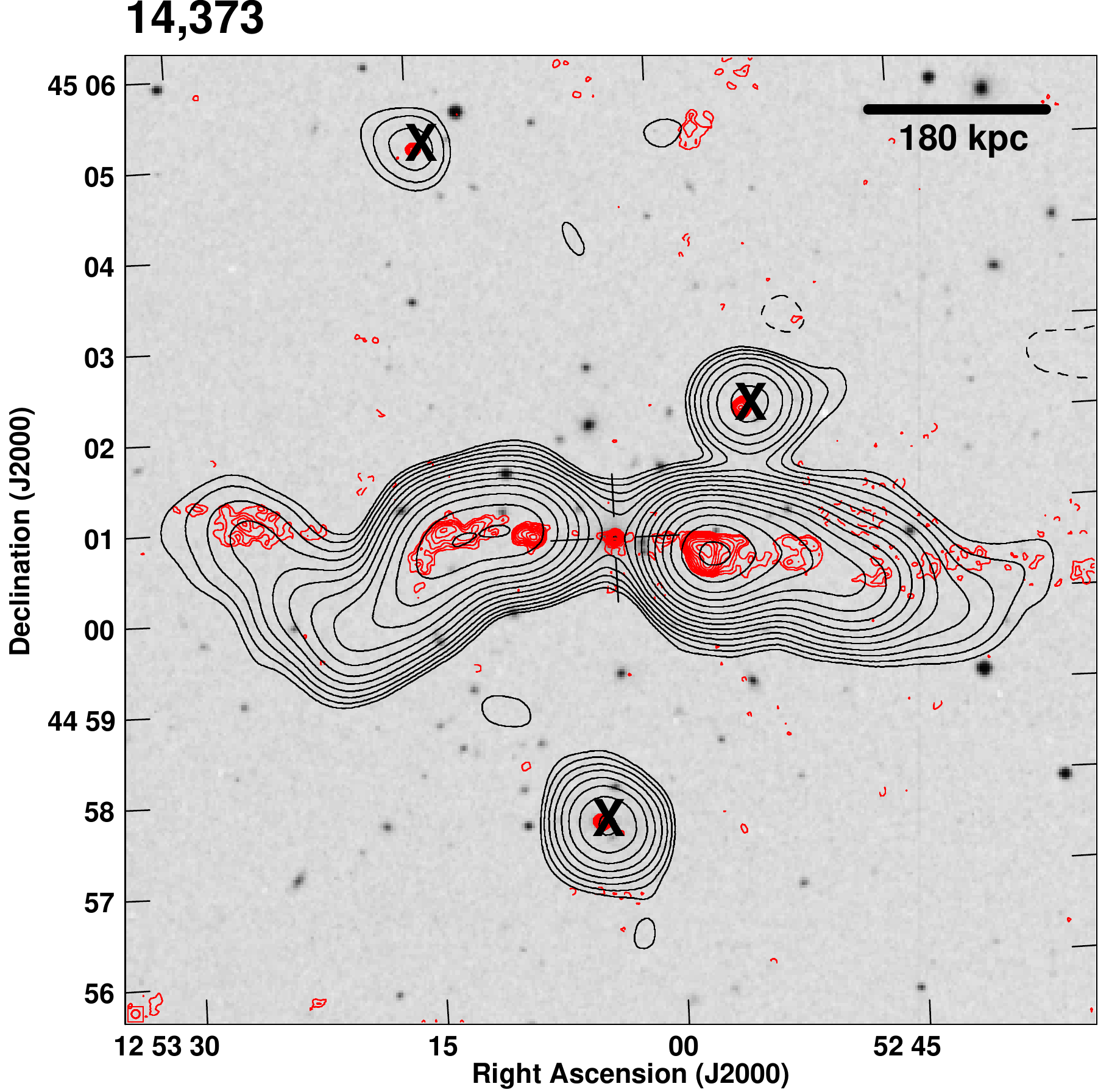}}
\hspace{0.8cm}{\includegraphics[width=0.3\textwidth]{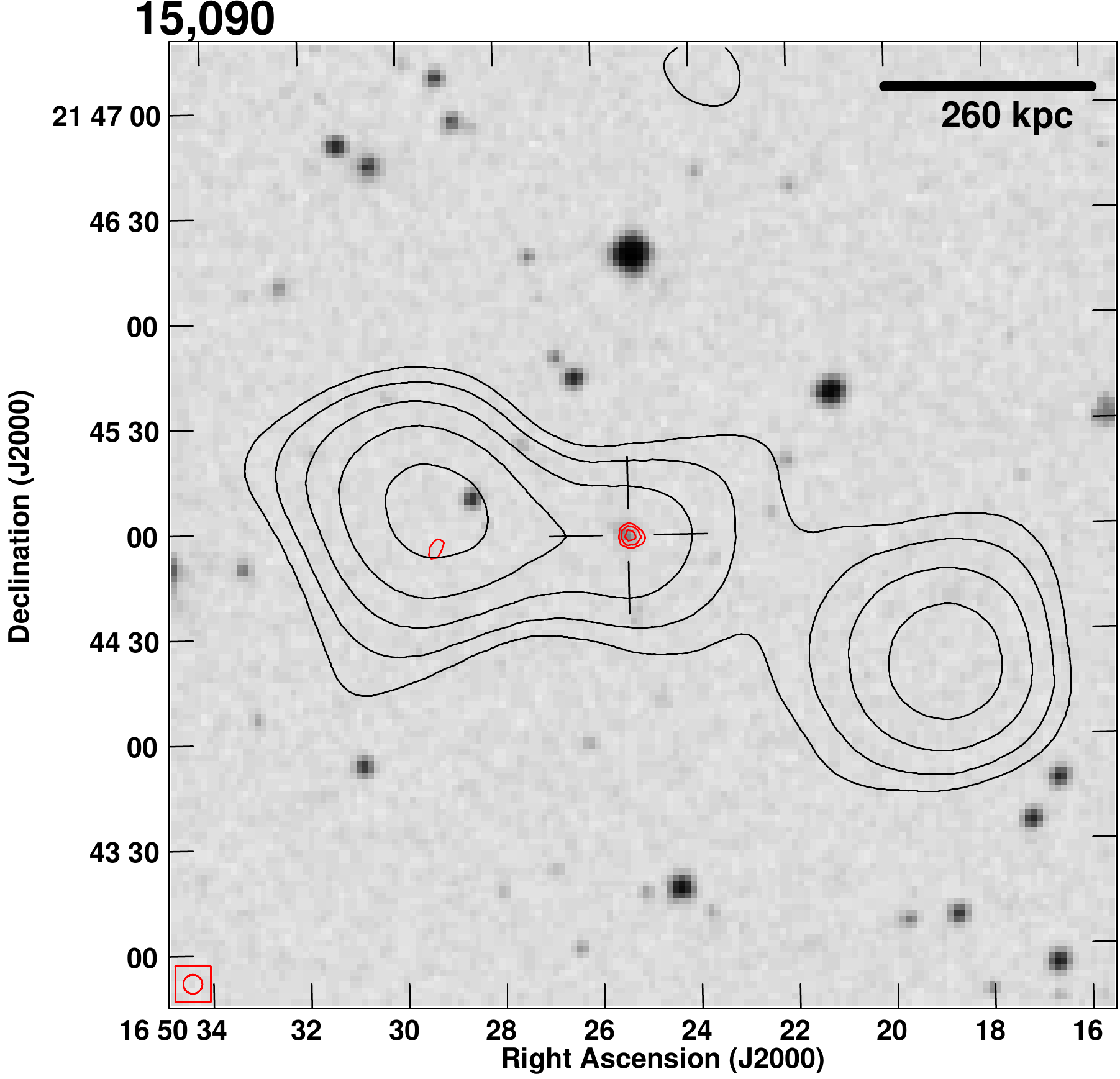}}
\hspace{0.8cm}{\includegraphics[width=0.3\textwidth]{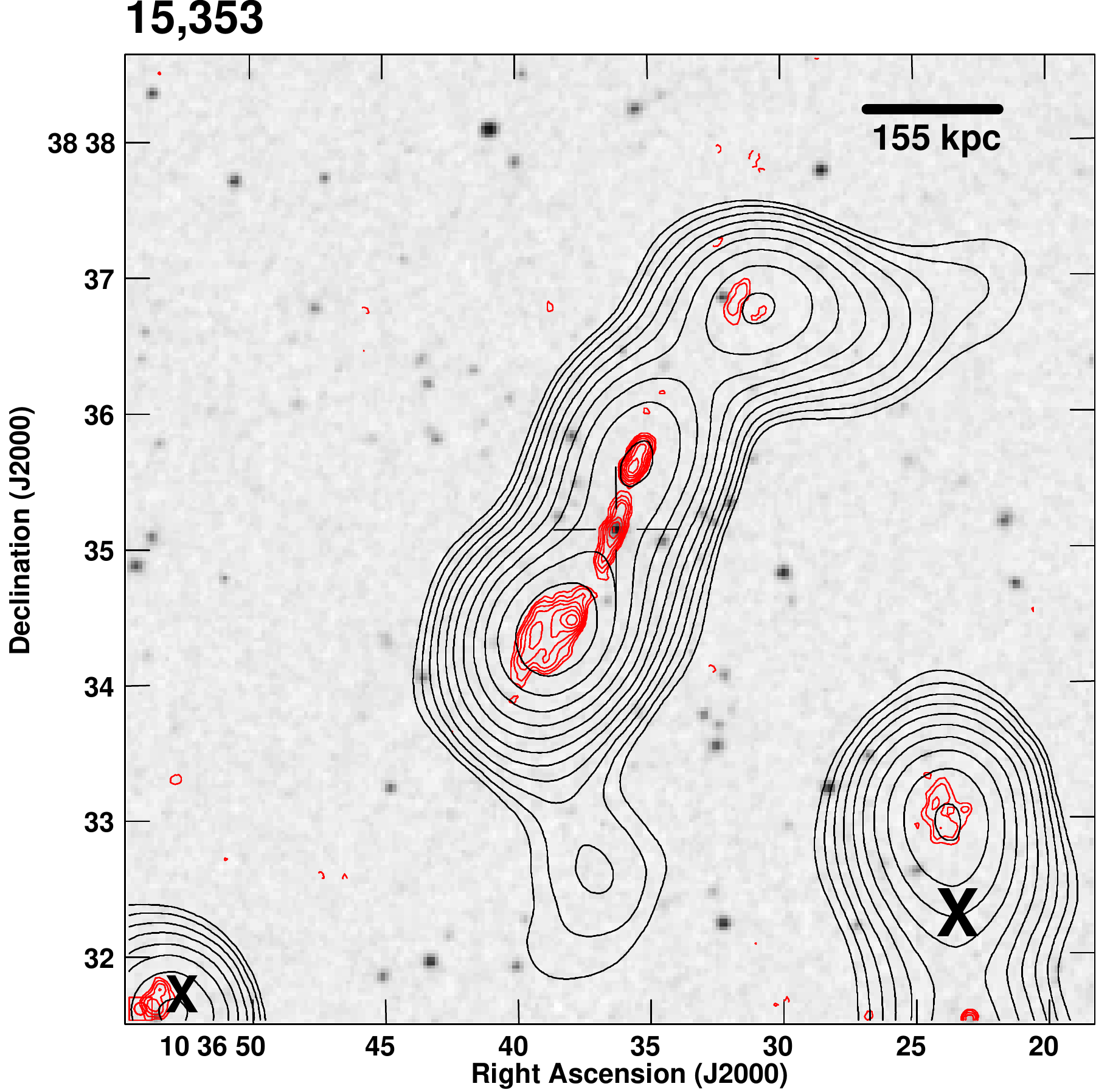}}

\caption{Newly discovered and/or classified giant radio sources in ROGUE~I. The 1.4\,GHz radio contours from the FIRST (red) and the NVSS (black) maps are overlaid on the optical DSS (gray scale) image, centered at the host galaxy position marked by a plus sign. Each panel contains catalog number of the galaxy. The FIRST beam is placed inside the square box. The contours levels are the same as in Fig.\ref{fig:radioMorph}}
\label{fig:NewGiantsMaps}
\end{figure*}

\renewcommand{\thefigure}{A.\arabic{figure} (Cont.)}

\begin{figure*}[htb!]
\ContinuedFloat
\hspace{0.0cm}{\includegraphics[width=0.3\textwidth]{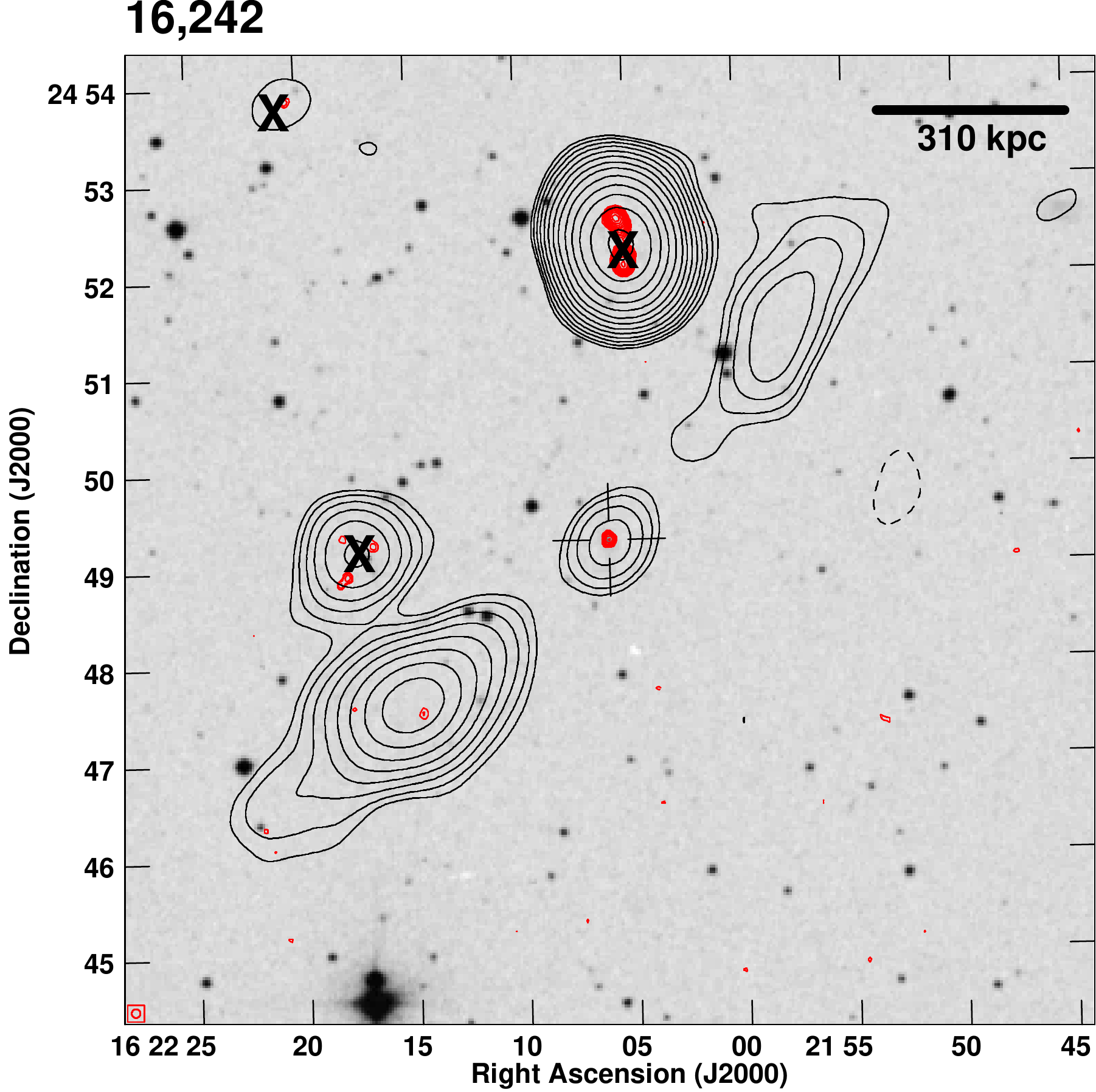}}
\hspace{0.2cm}{\includegraphics[width=0.3\textwidth]{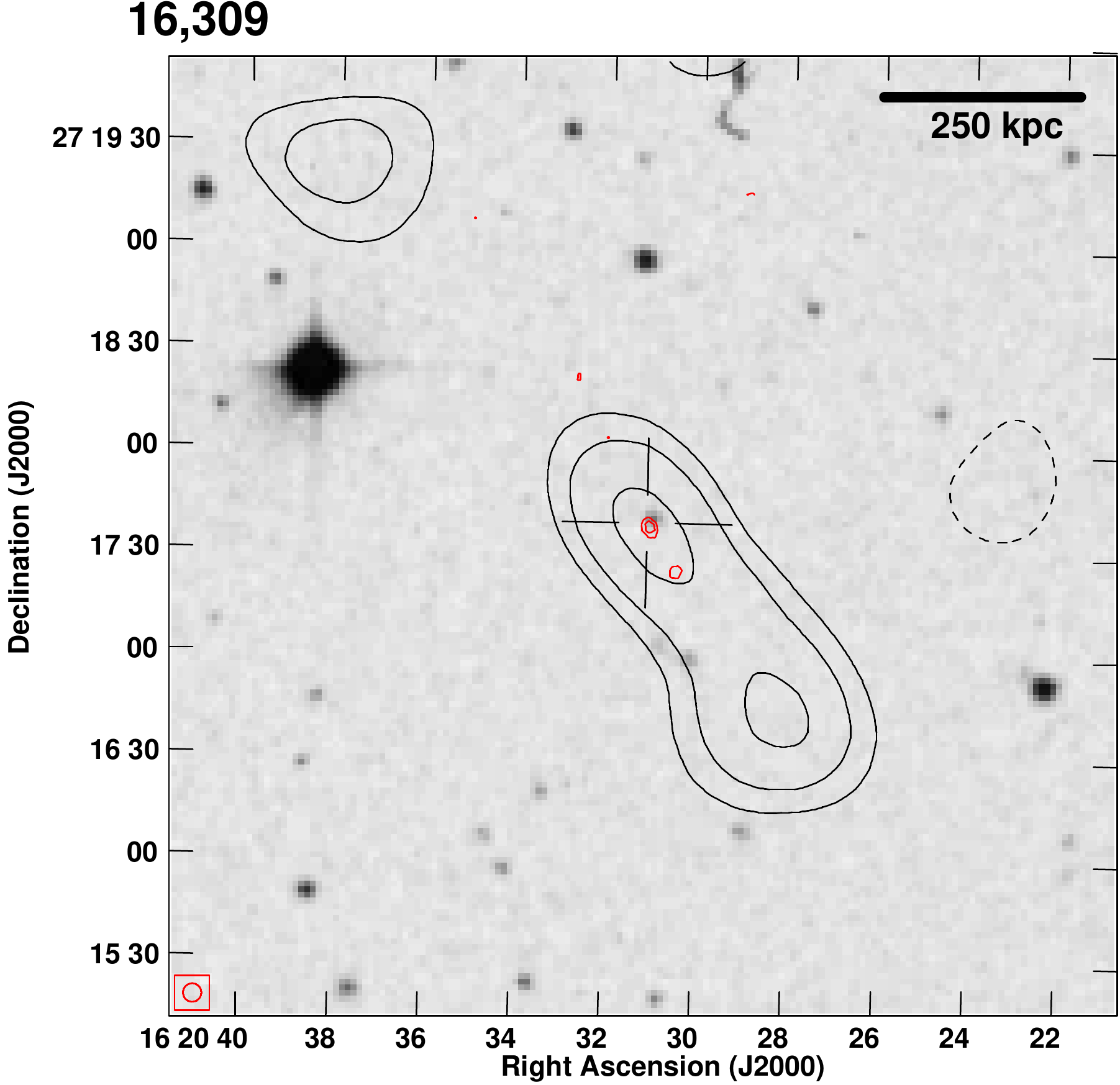}}
\hspace{0.2cm}{\includegraphics[width=0.3\textwidth]{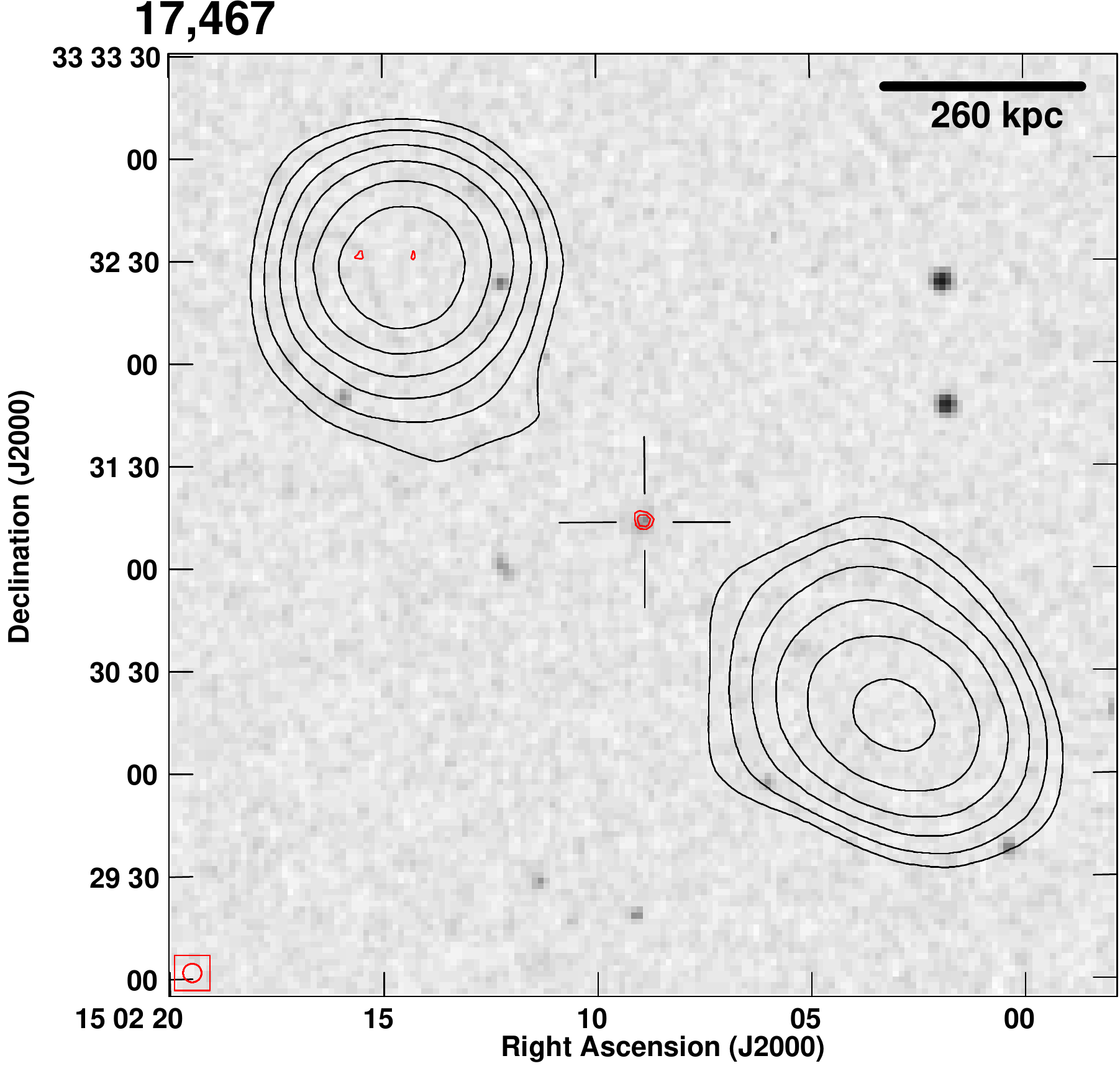}}
\hspace{0.2cm}{\includegraphics[width=0.3\textwidth]{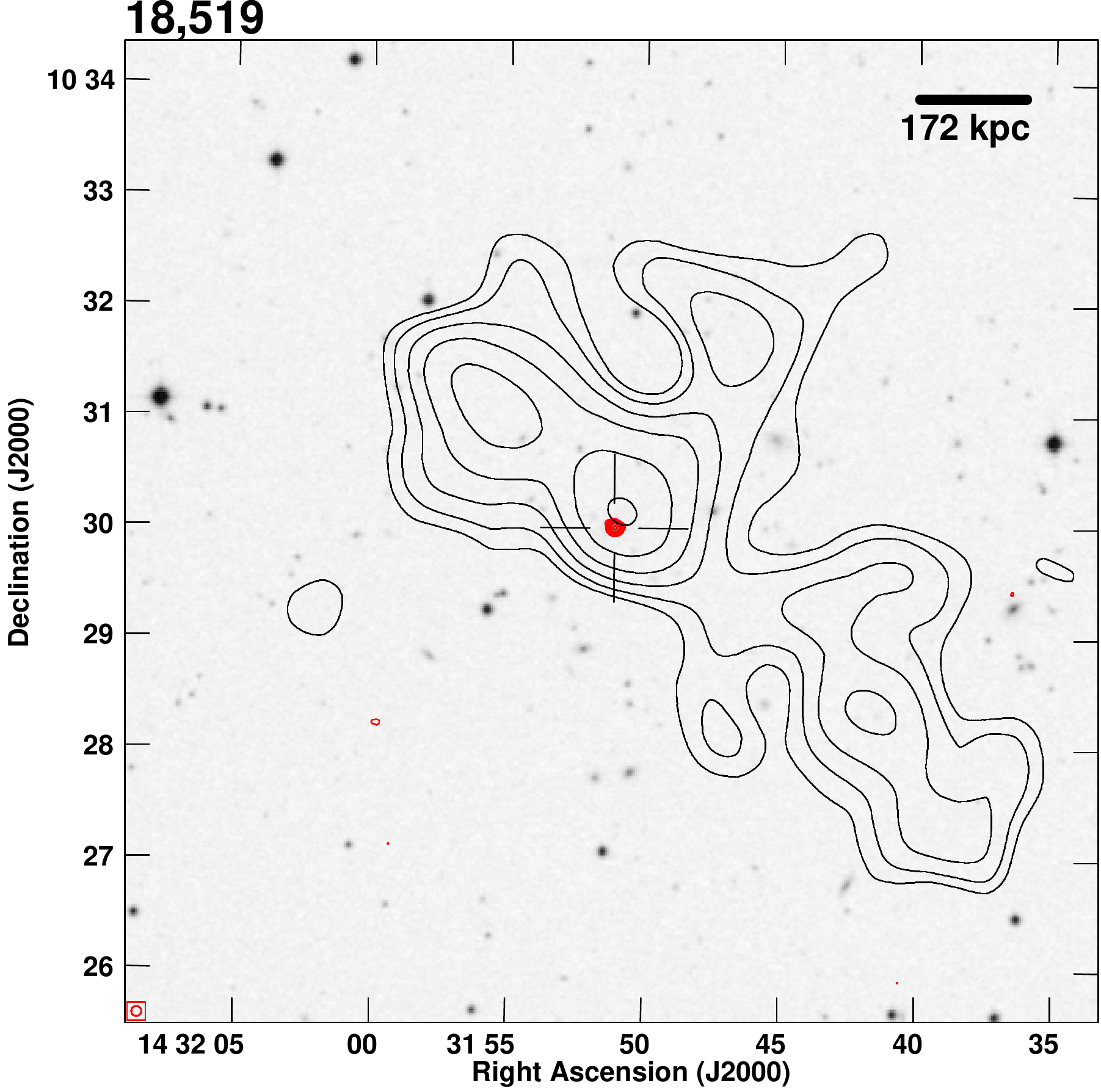}}
\hspace{0.2cm}{\includegraphics[width=0.3\textwidth]{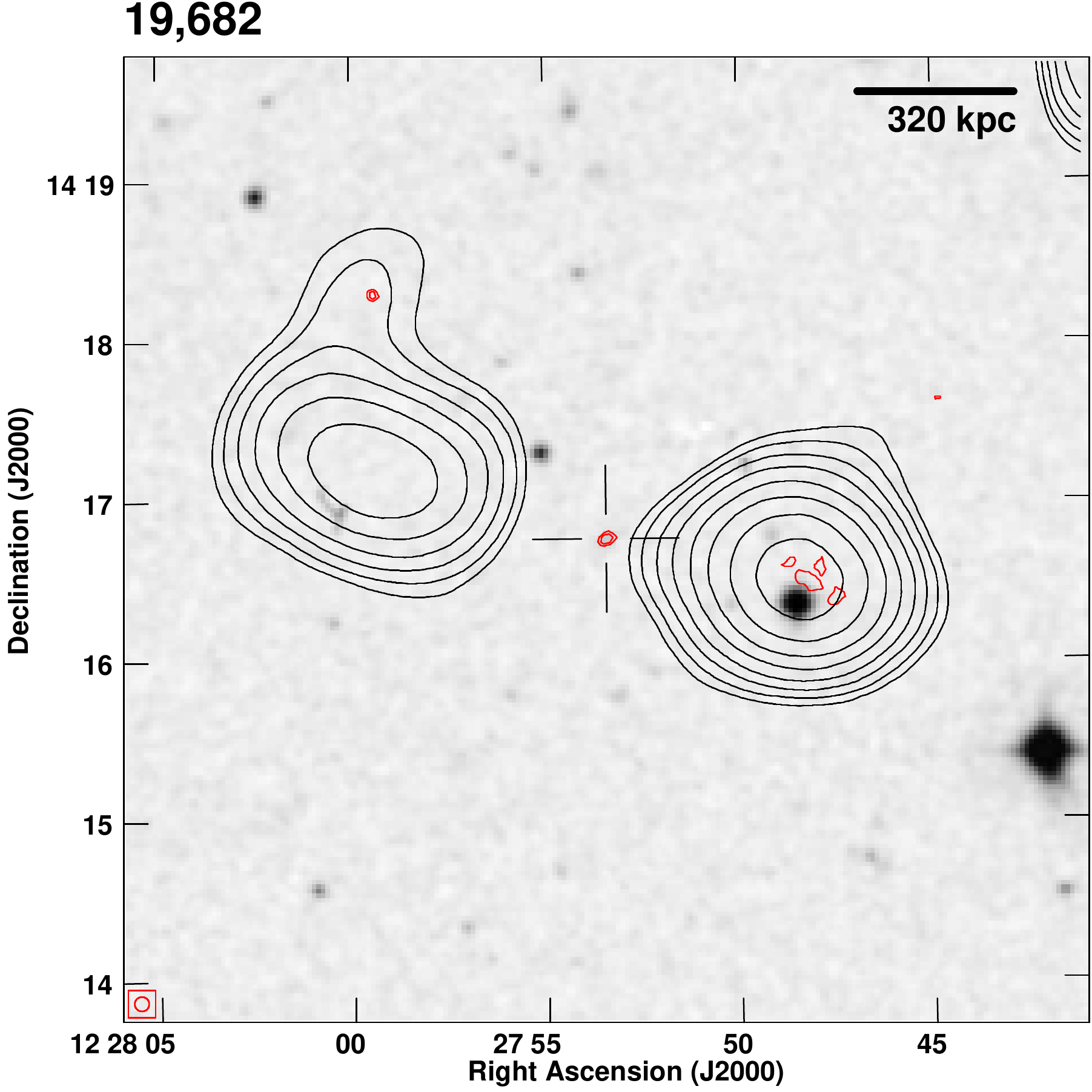}}
\hspace{0.2cm}{\includegraphics[width=0.3\textwidth]{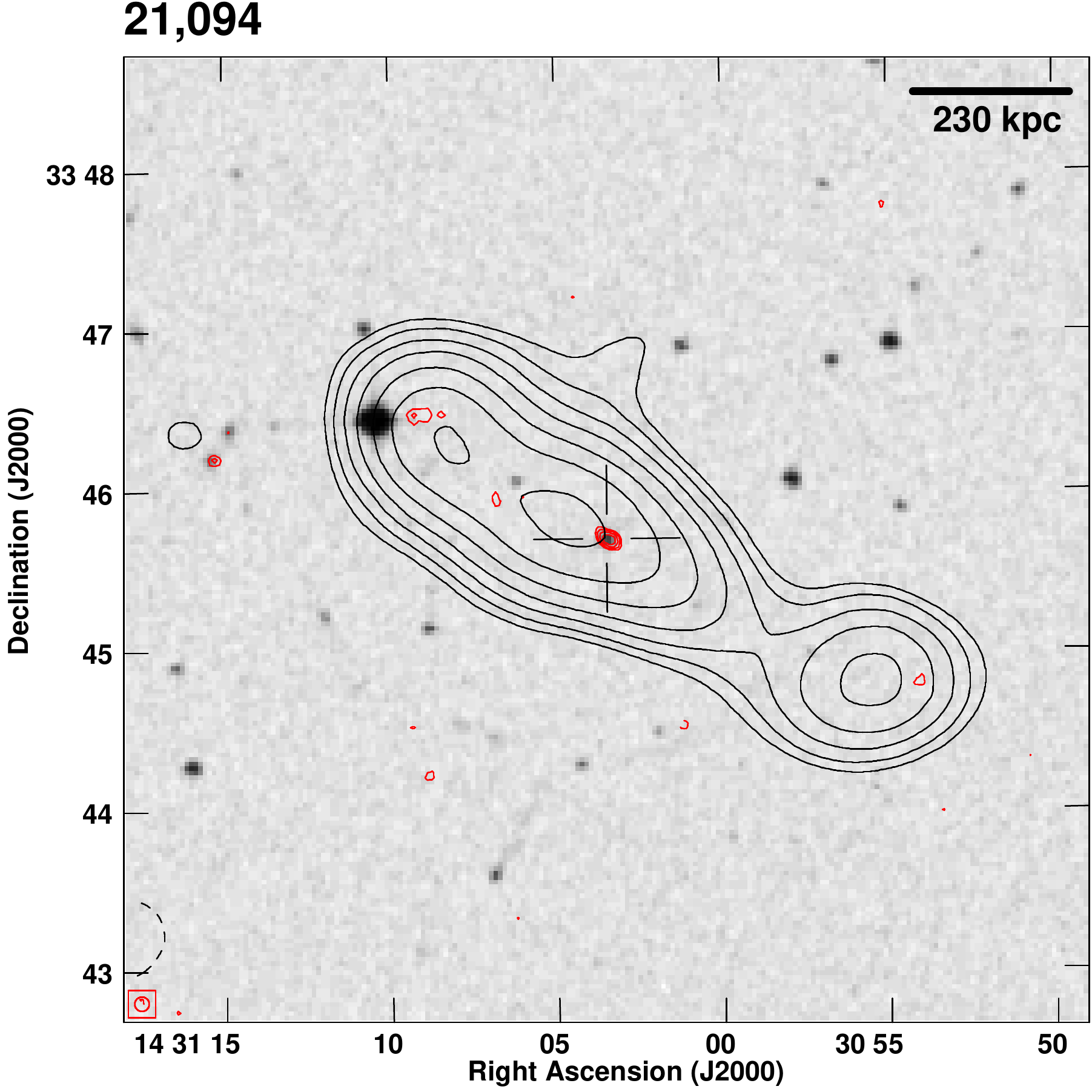}}
\hspace{0.2cm}{\includegraphics[width=0.3\textwidth]{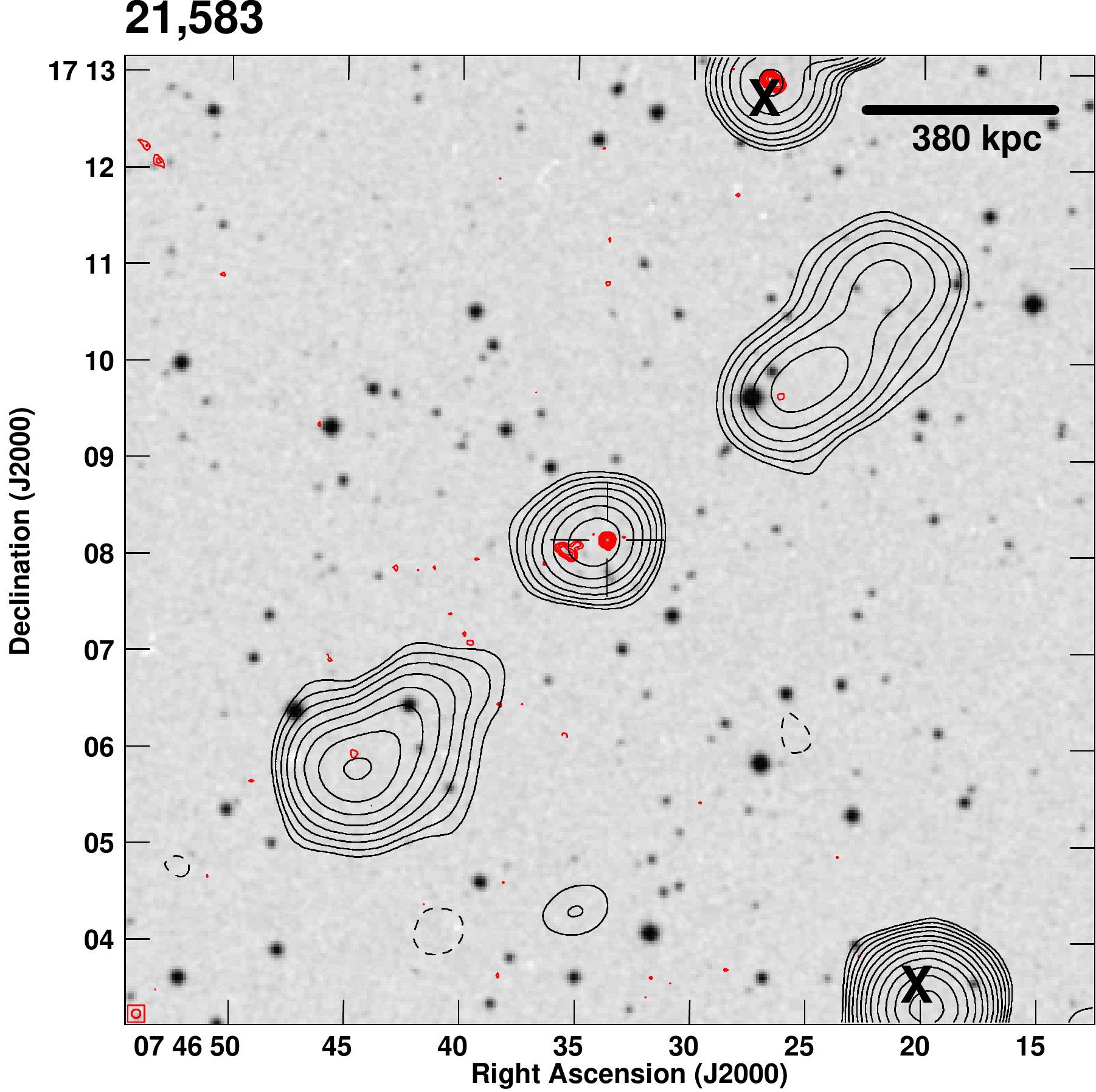}}
\hspace{0cm}{\includegraphics[width=0.3\textwidth]{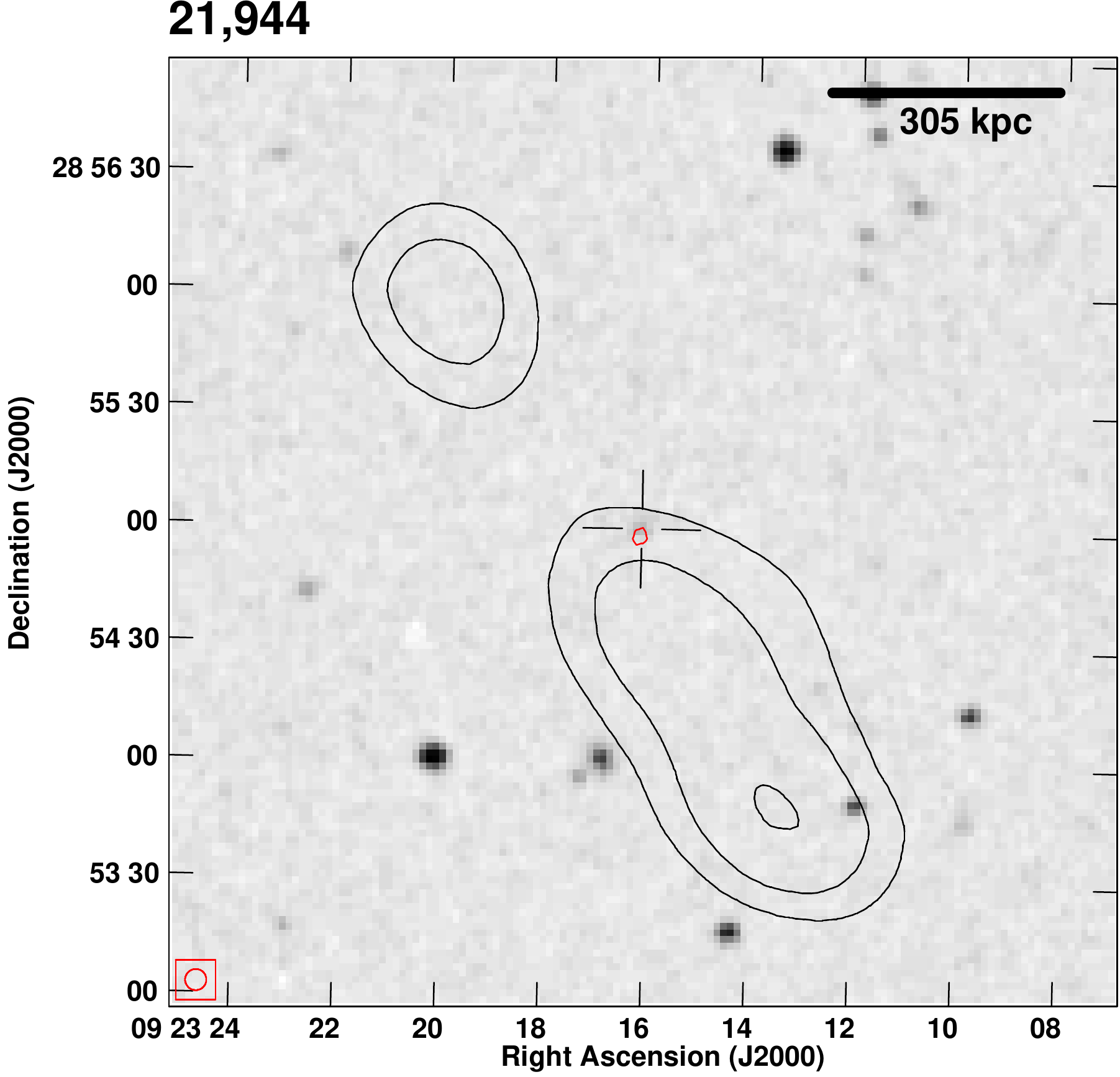}}
\hspace{0.2cm}{\includegraphics[width=0.3\textwidth]{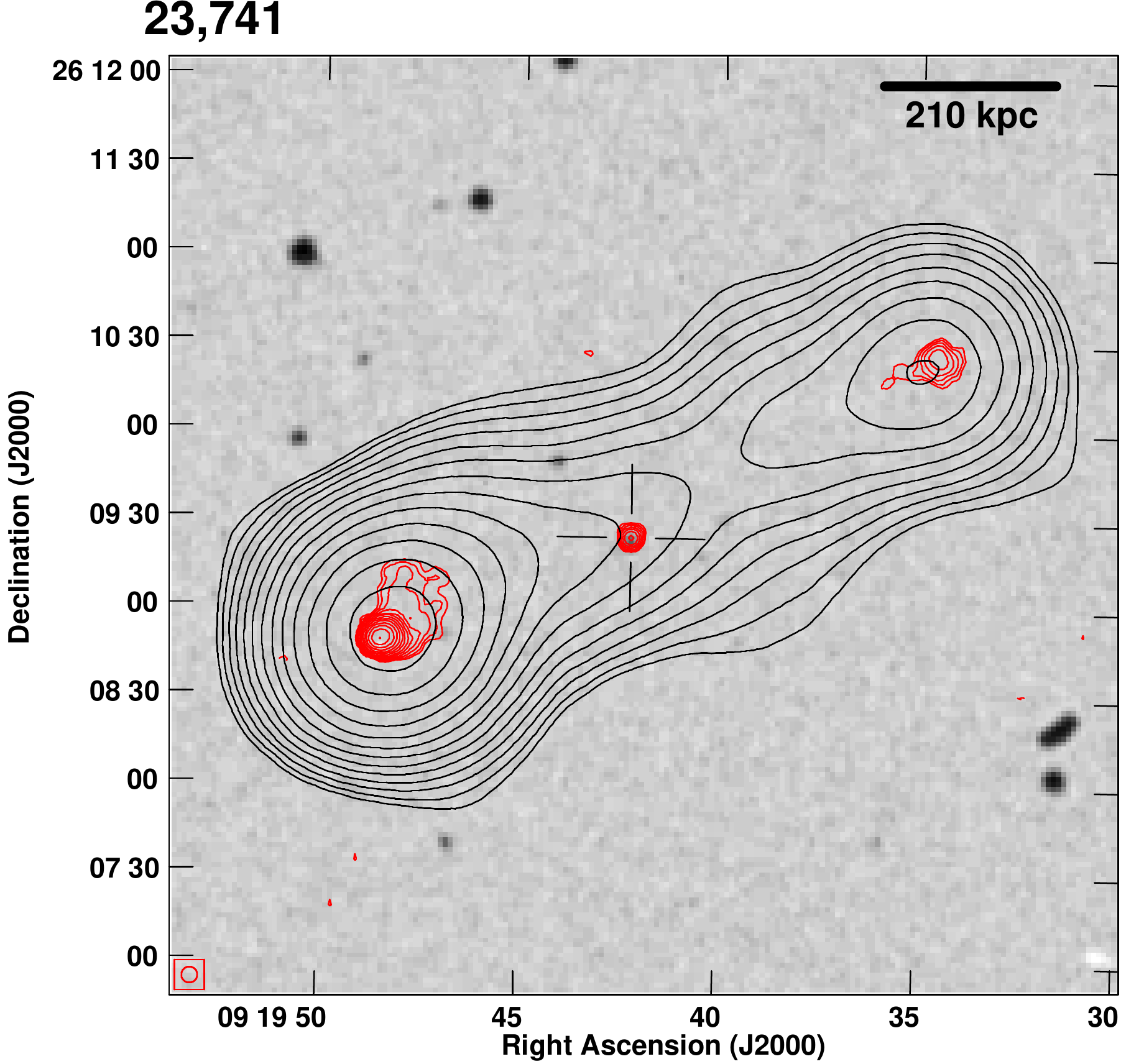}}
\hspace{0.8cm}{\includegraphics[width=0.3\textwidth]{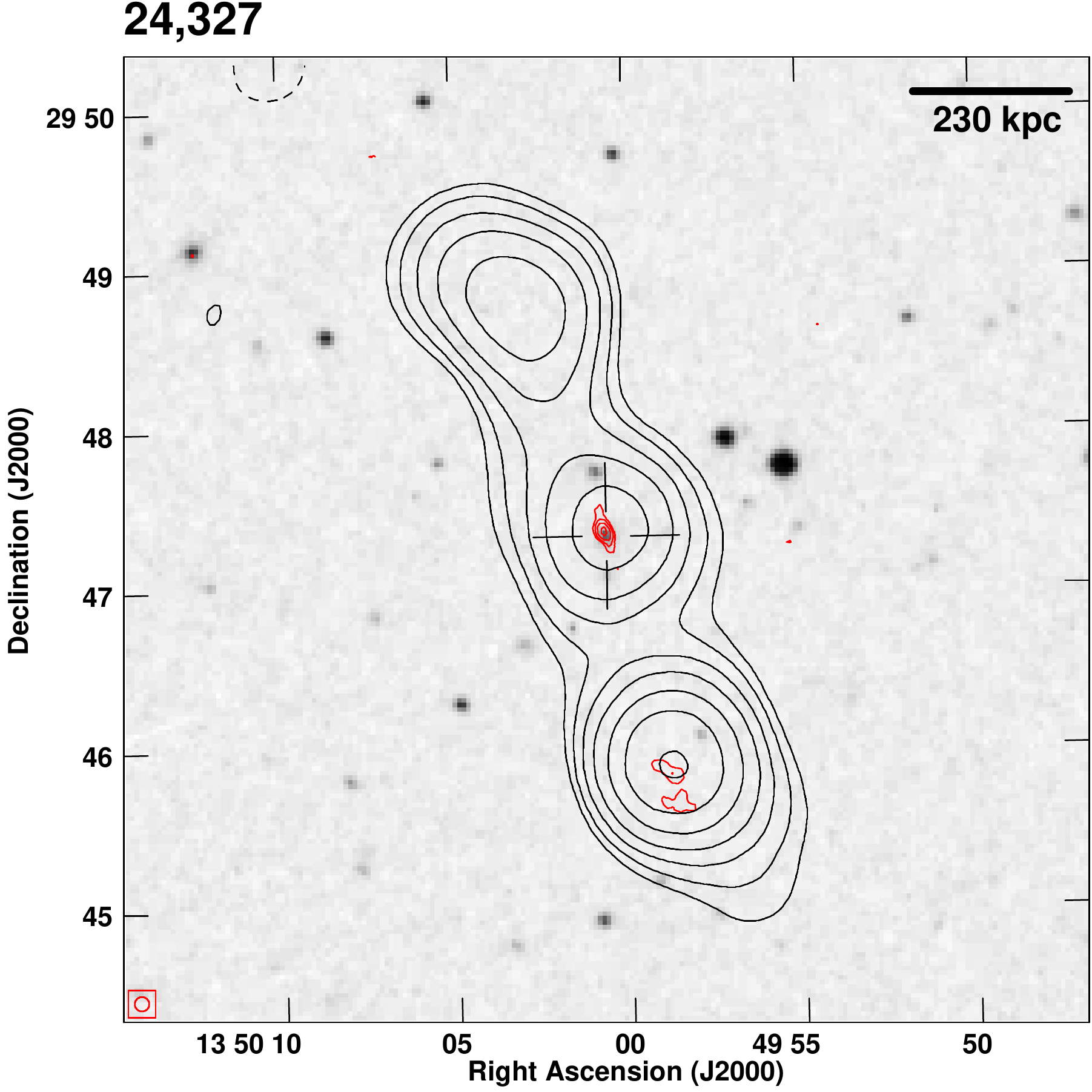}}
\hspace{0.8cm}{\includegraphics[width=0.3\textwidth]{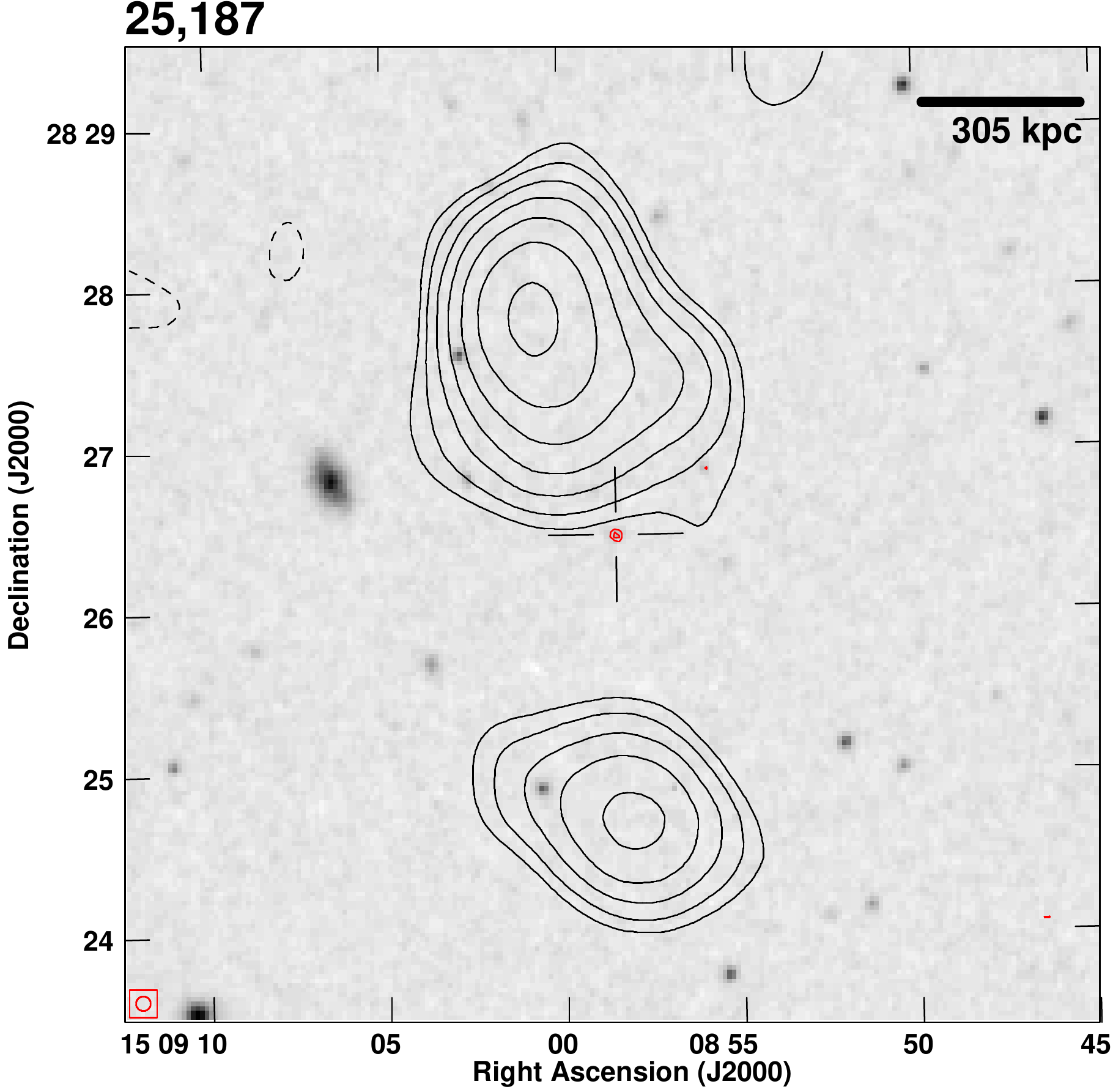}}
\hspace{0.8cm}{\includegraphics[width=0.3\textwidth]{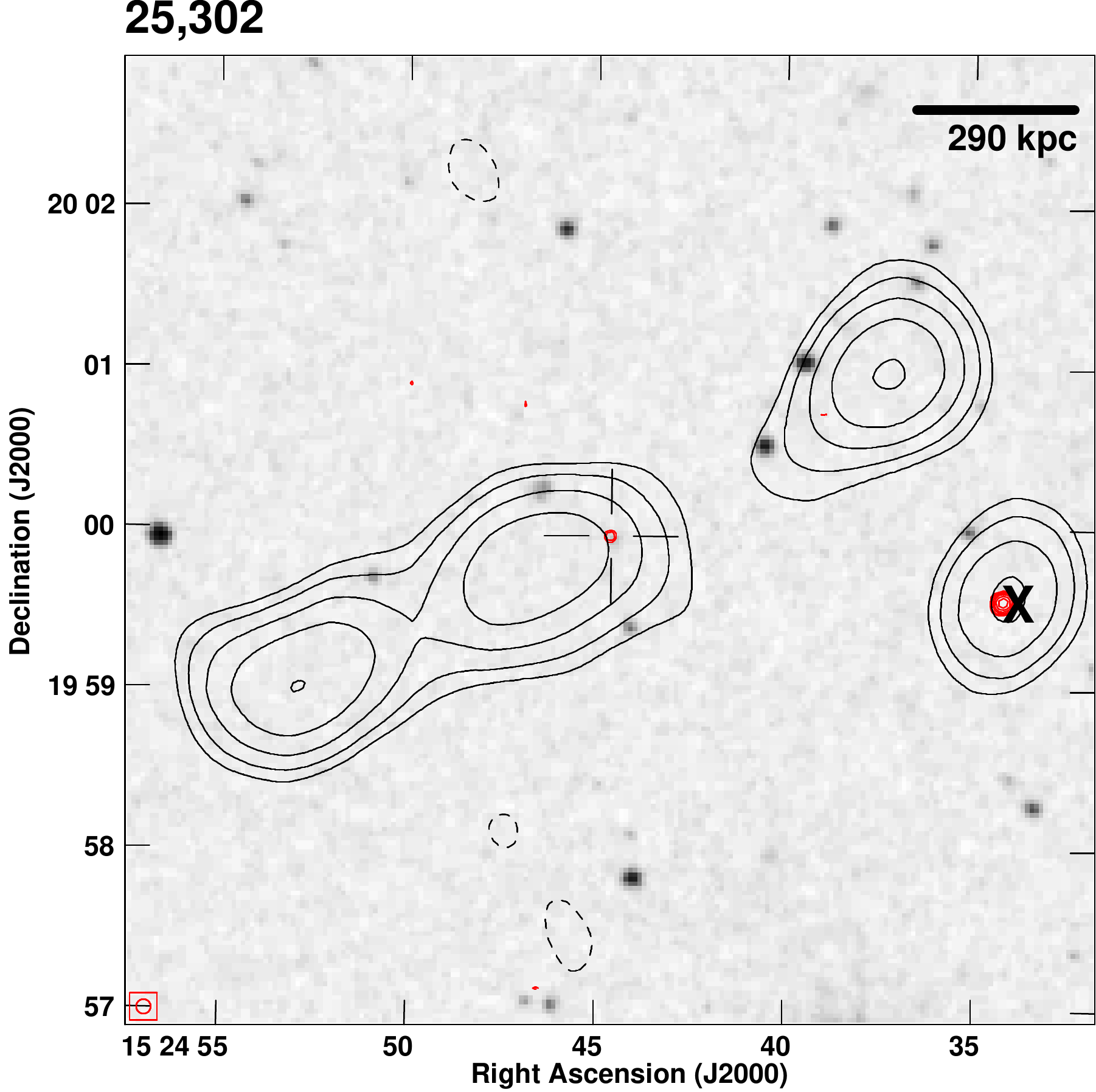}}
\caption{Newly discovered and/or classified giant radio sources in ROGUE~I.}
\end{figure*}

\renewcommand{\thefigure}{A.\arabic{figure} (Cont.)}


\begin{figure*}[htb!]
\ContinuedFloat
\hspace{0.2cm}{\includegraphics[width=\textwidth]{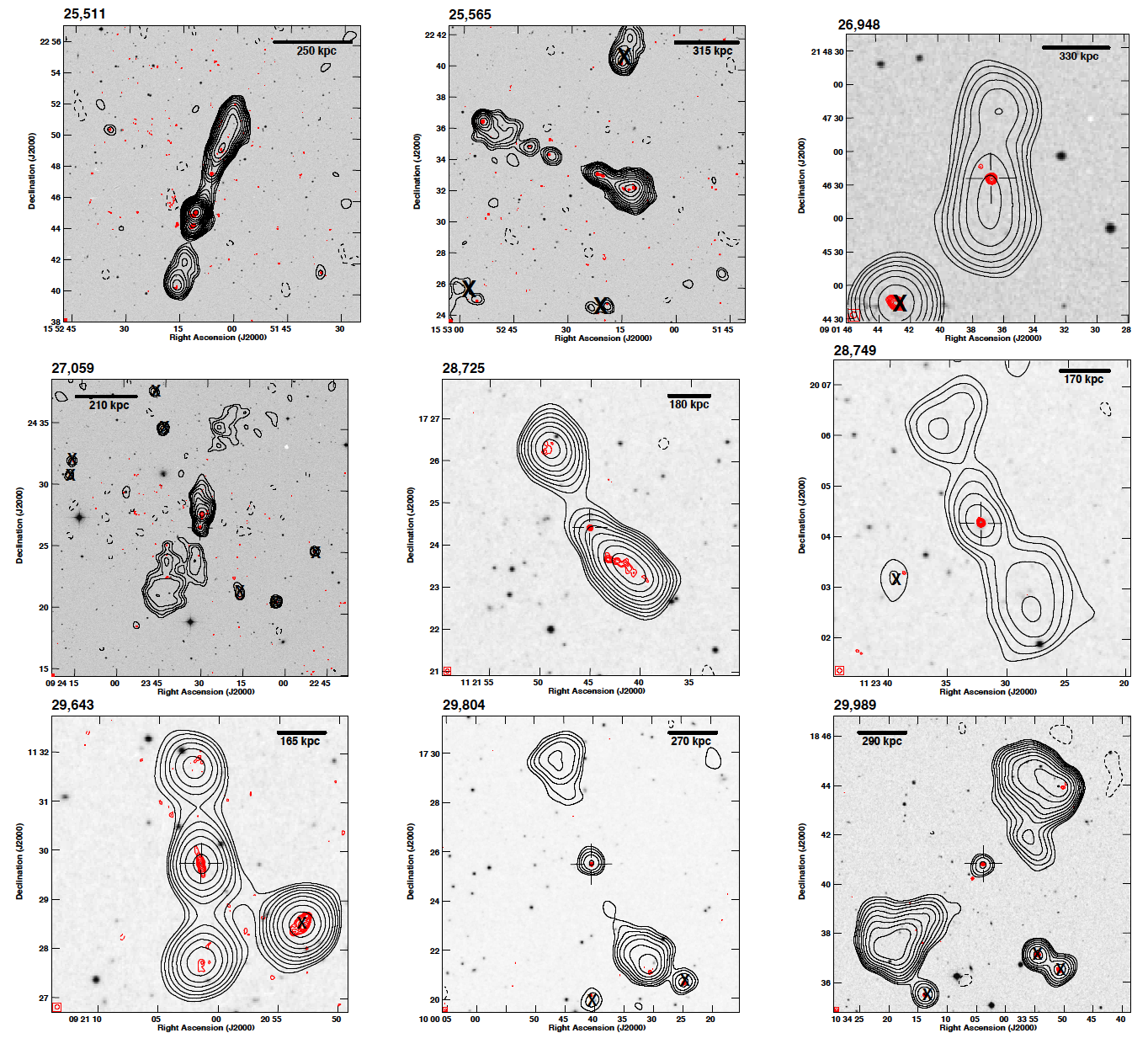}}
\caption{Newly discovered and/or classified giant radio sources in ROGUE~I.}
\end{figure*}

\subsection{Possible giant radio sources}
\label{PGiants}

\begin{itemize}
  \item[]370: based on FIRST and NVSS maps, we classify this possible giant as an FR~II radio galaxy. On the FIRST map, an elongated core and a pair of hot spots are visible. The NVSS map displays more extended structure with two clear lobes exceeding structure visible in the FIRST. The source is located at $z=0.25$, and including tails visible on NVSS maps, it reaches an angular size of 204\arcsec, which corresponds to a projected linear size of 0.8 Mpc.

  \item[]11,417: the radio structure of this possible giant located at $z=0.2$ reaches an angular size of 244\arcsec, corresponding to a projected linear size of 0.8 Mpc. The FIRST map reveals a slightly elongated core and a faint radio emission, suggesting the existence of a southern lobe, but the northern lobe is undetected. The core and the southern lobe are evident on the NVSS map, while the presence of the northern lobe is visible (indicated by one radio contour).

  \item[]12,374: this source located at $z=0.18$ has an angular size of 278\arcsec\, and a projected linear size of 0.8 Mpc. We classify it as a possible giant due to its complex morphology visible on both maps. The FIRST map shows an elongated core and three side sources, from which one can be interpreted as foreground/background emission. 
  The NVSS map, however, reveals a connection between linear lobes with possible  hybrid morphology. 

  \item[]12,766: the source, classified as FR~II and located at $z=0.33$ has an angular size of 255\arcsec, corresponding to a projected linear size of 1.2 Mpc. Only a faint slightly elongated core appears in the FIRST map, while the NVSS map reveals extended, slightly bent structure with prominent lobes and a hint of the faint extension of a southeastern lobe.
  
  \item[]13,360: this possible hybrid radio galaxy is located at z = 0.11. Its angular size is 582\arcsec, corresponding to a linear size of 1.1 Mpc. The FIRST map reveals only a compact core, while an extended FR hybrid radio structure is visible on the NVSS map. Analysing the optical as well as the radio data, we suspect that this possible giant can be a superposition of radio emission from foreground/background galaxies, and therefore we consider it as a possible giant.
  
  \item[]14,074: This source located at $z=0.09$ has an angular size of 530\arcsec corresponding to a projected linear size of 1 Mpc. The FIRST map reveals a FR~II type morphology. The NVSS map additionally uncovers a possible X--shaped radio structure with an asymmetric one-sided strong emission towards the north-south direction with hot spots and a fainter diffuse emission perpendicular to it. 
  
  \item[]16,484: the source is located at $z=0.09$, with an angular size of 552\arcsec\, and a projected linear size of 0.9 Mpc. We classify it as a possible FR~II. On the FIRST map, merely a core is visible, while the NVSS map reveals a complex morphology with a faint extended central part and two detached bright lobes. Also, a point-like, background/foreground source is visible next to the core.  
  
  \item[]19,048: this possible hybrid radio galaxy is located at $z=0.19$, with an angular size of 228\arcsec\, corresponding to a projected linear size of 0.7 Mpc. The FIRST map reveals a core and a pair of lobes, while the NVSS map shows an extended radio structure with a distinct south lobe and a diminishing north one.
   
  \item[]20,418: this double--double giant radio galaxy at $z=0.15$ has an angular size 338\arcsec\, corresponding to a projected linear size of 0.9 Mpc. The FIRST map reveals a core surrounded by a halo-like structure, a two-sided jet, and outer lobes. On the NVSS map, the elongated central emission and both outer lobes are apparent.
 
  \item[]25,432: this faint source, with redshift $z=0.21$, the angular size of 217\arcsec, and the projected linear size of 0.7 Mpc, is classified as a possible FR~II radio galaxy. The FIRST map shows only a core, whereas a core as well as both lobes are visible on the NVSS map.
  
  \item[]25,587: \citet{Kuzmicz.etal.2018a} classifies the radio structure as a giant belonging to a host galaxy J155209.19+200523.2, different from the ROGUE~I host galaxy. 
  
  \item[]29,900: the radio structure of this FR~I galaxy located at $z=0.45$, reaches an angular size of 131\arcsec, corresponding to a linear size of 0.7 Mpc. It has a faint core and a prominent south lobe. Only two contours suggesting the position of the north lobe are apparent on the NVSS map.
\end{itemize}

\renewcommand{\thefigure}{A.\arabic{figure}}

\begin{figure*}[ht!]
\hspace{0cm}{\includegraphics[width=0.3\textwidth]{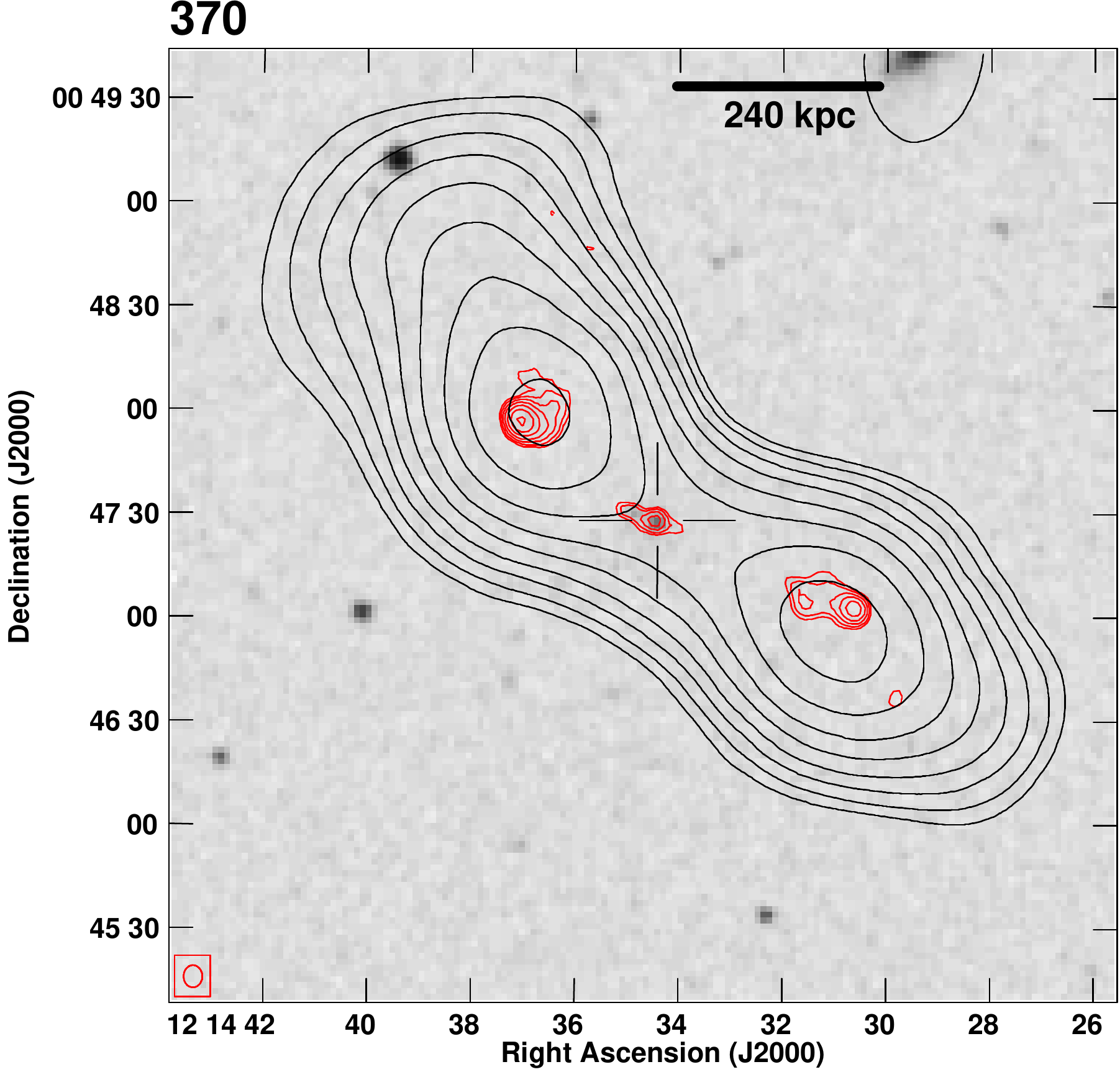}}
\hspace{0.2cm}{\includegraphics[width=0.3\textwidth]{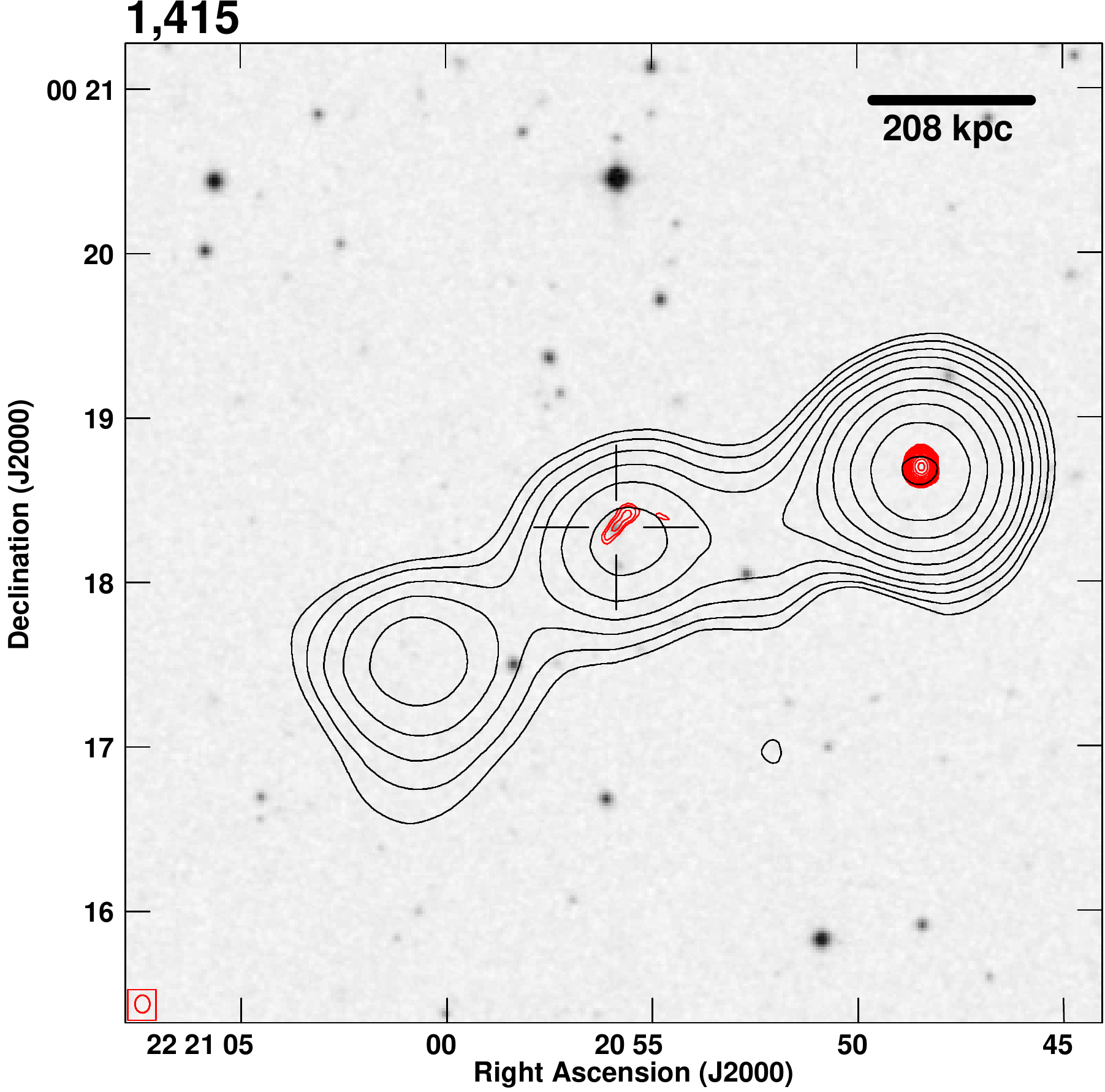}}
\hspace{0.2cm}{\includegraphics[width=0.3\textwidth]{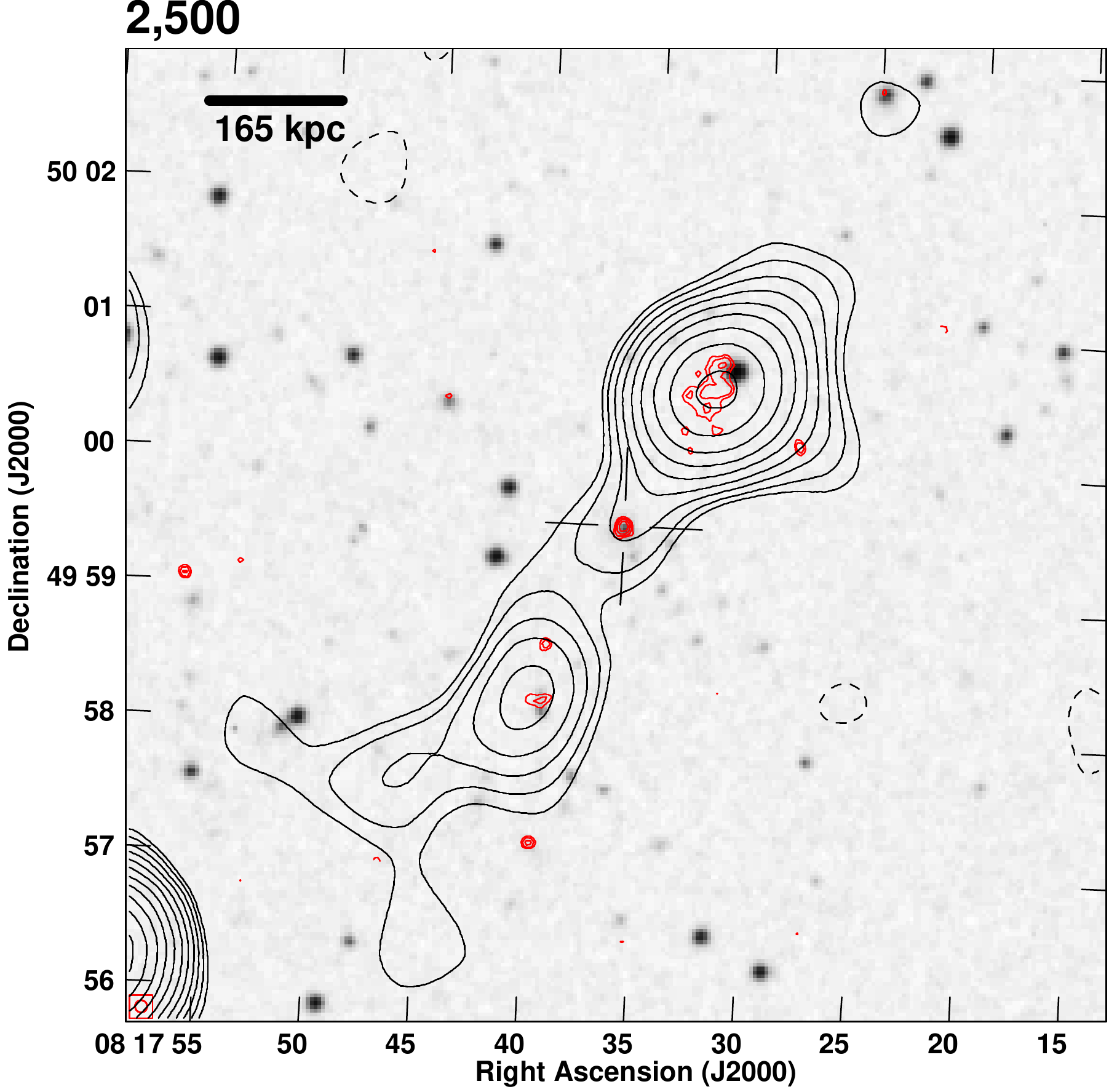}}
\hspace{0.2cm}{\includegraphics[width=0.3\textwidth]{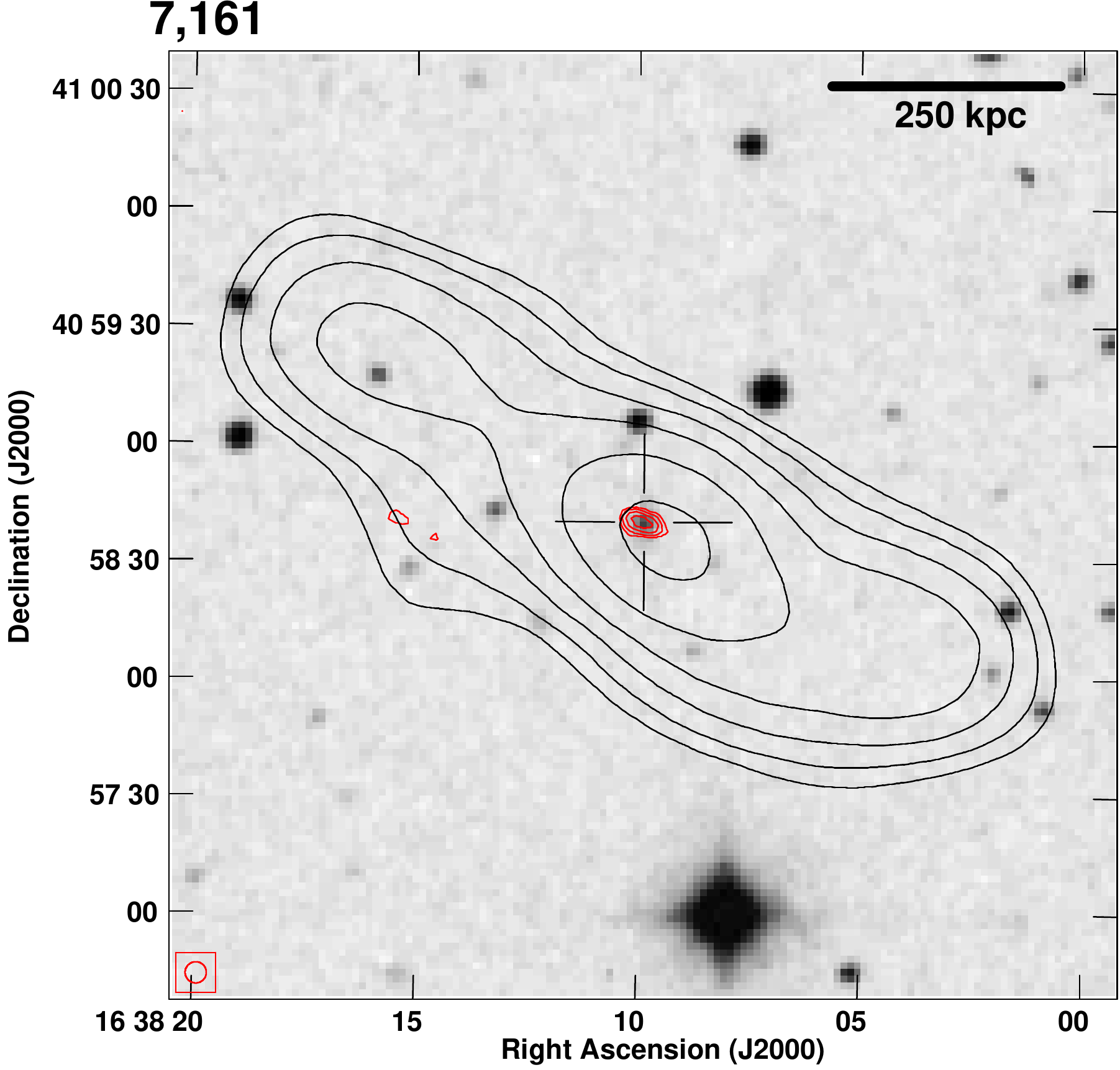}}
\hspace{0.2cm}{\includegraphics[width=0.3\textwidth]{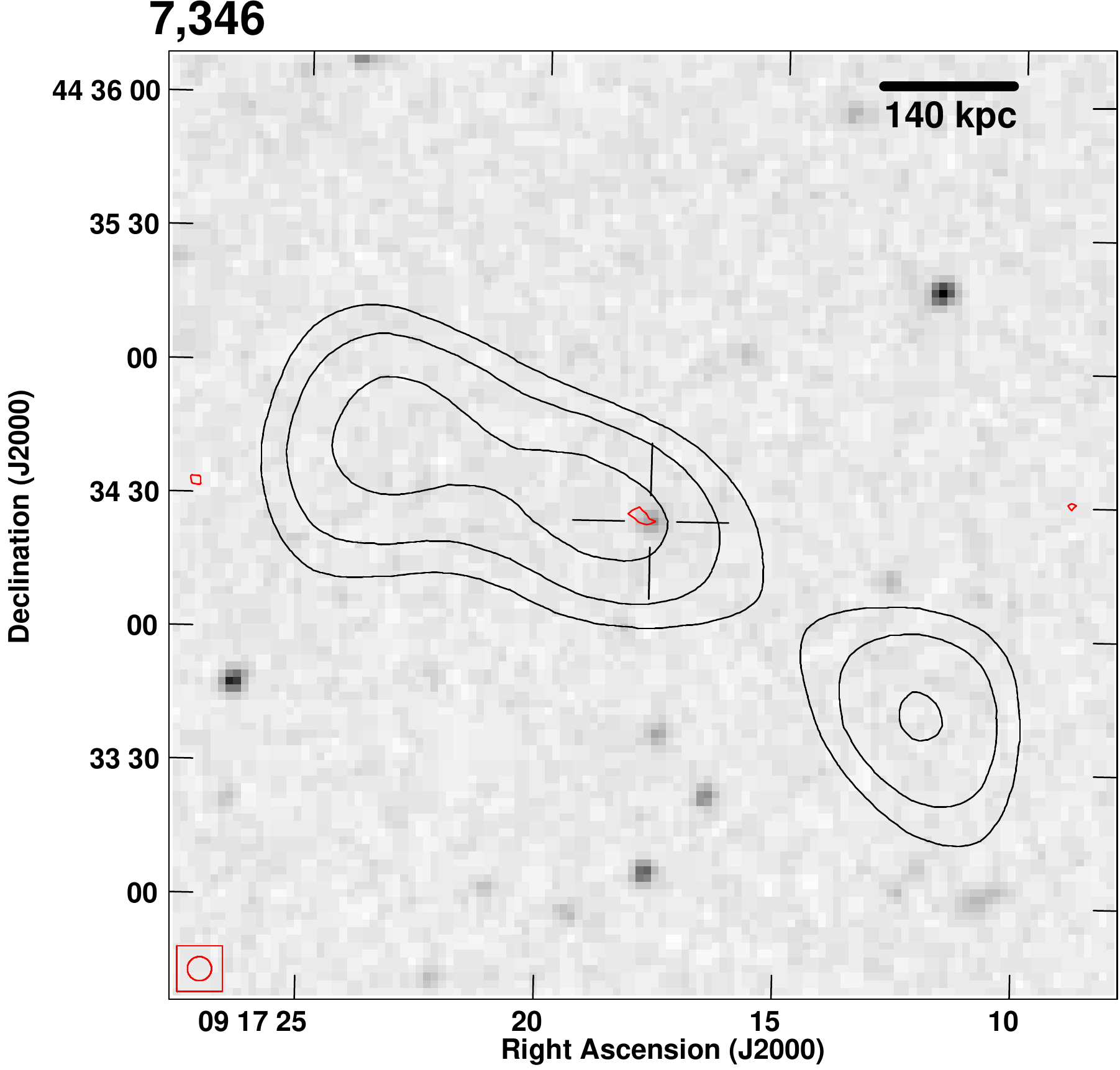}}
\hspace{0.2cm}{\includegraphics[width=0.3\textwidth]{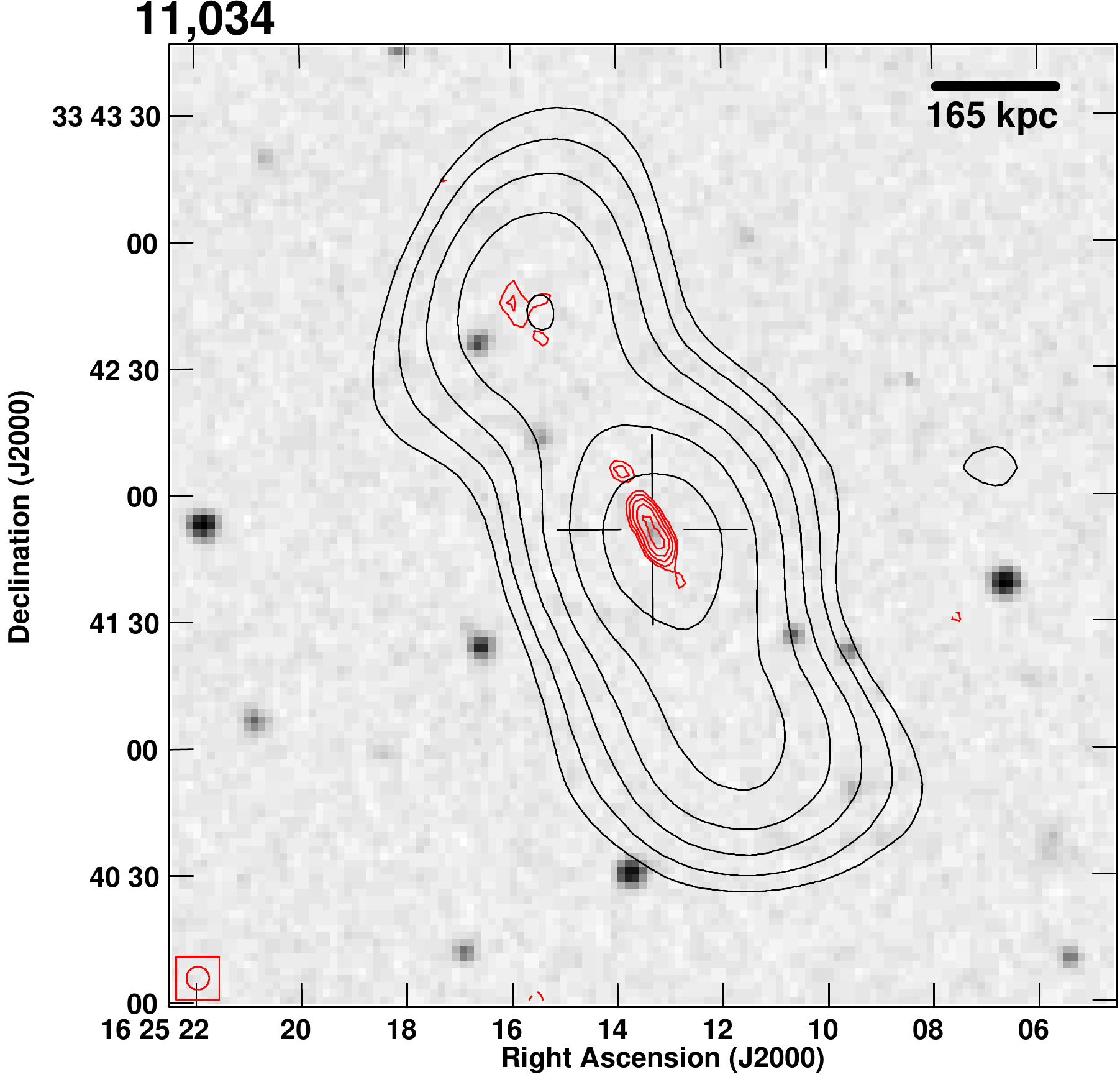}}
\hspace{0.2cm}{\includegraphics[width=0.3\textwidth]{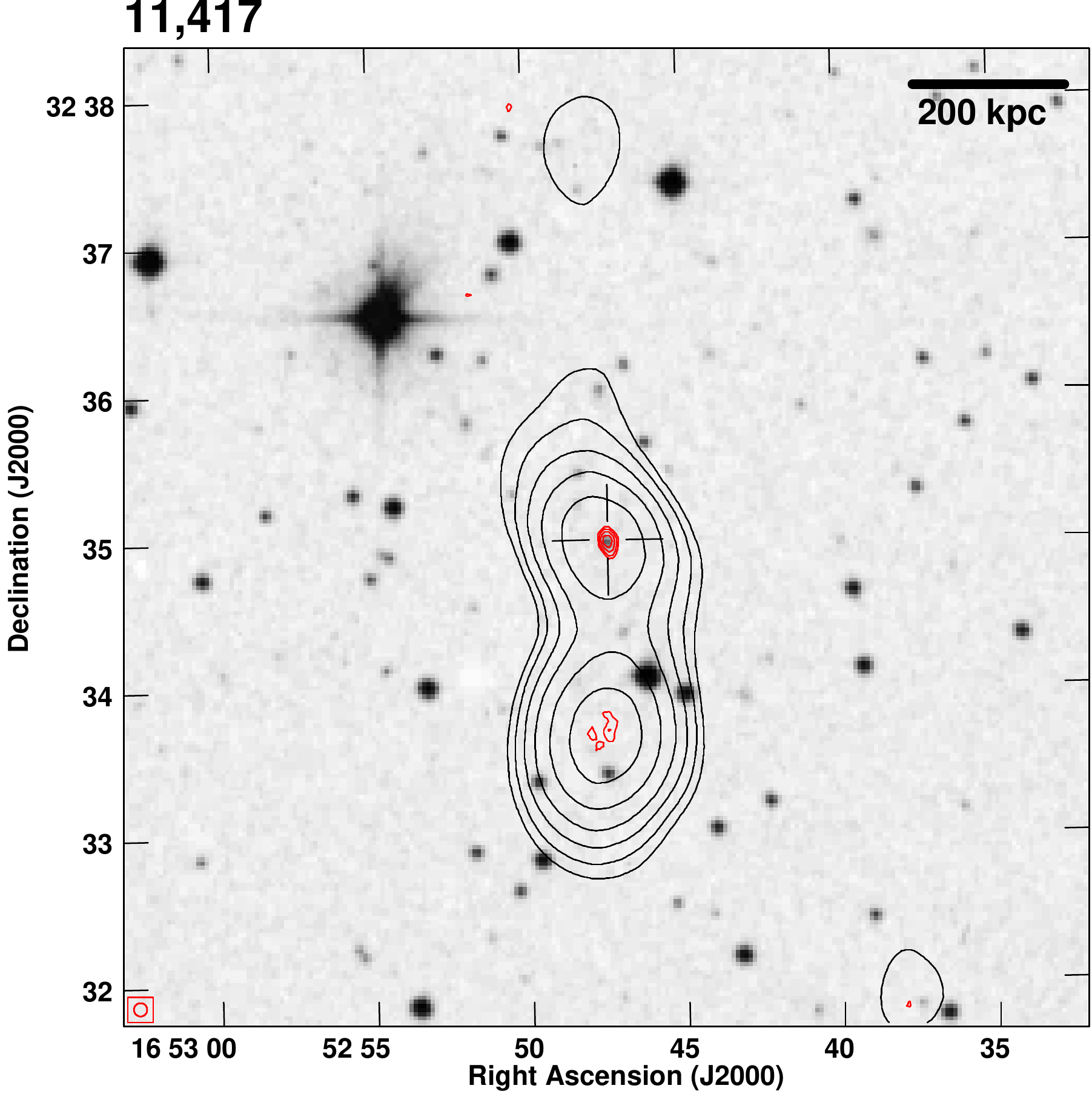}}
\hspace{0.2cm}{\includegraphics[width=0.3\textwidth]{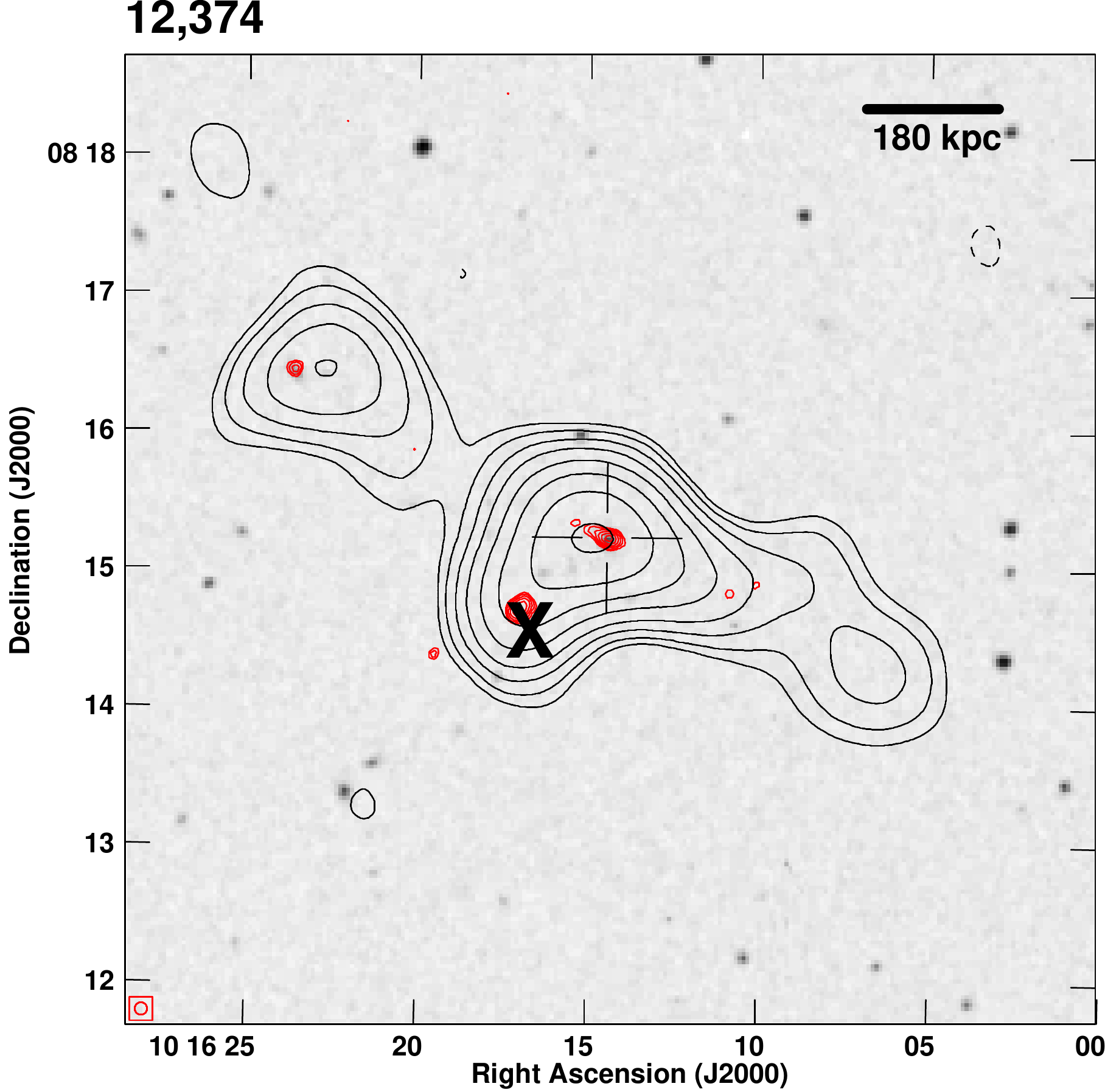}}
\hspace{0.2cm}{\includegraphics[width=0.3\textwidth]{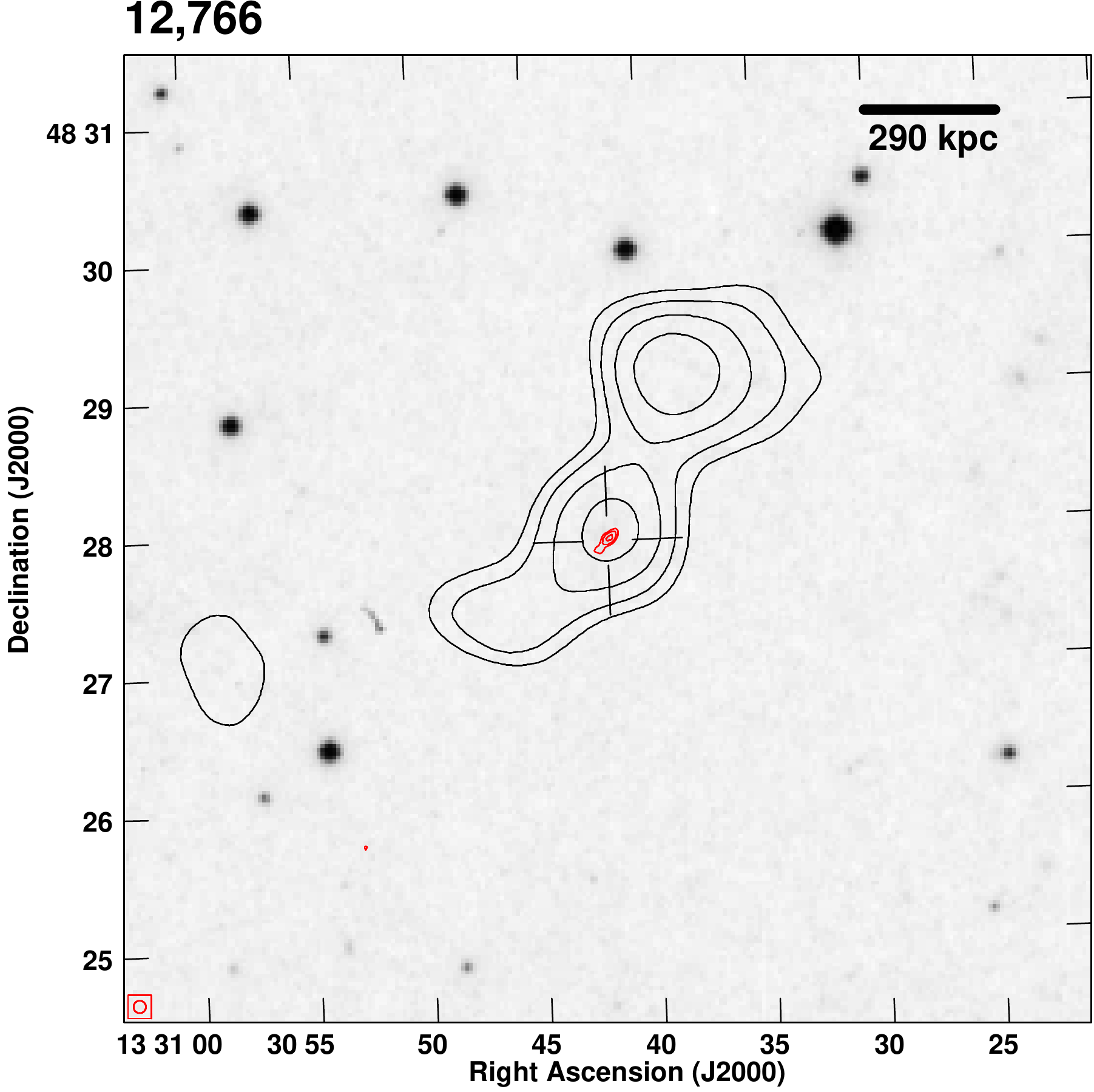}}
\hspace{0.8cm}{\includegraphics[width=0.3\textwidth]{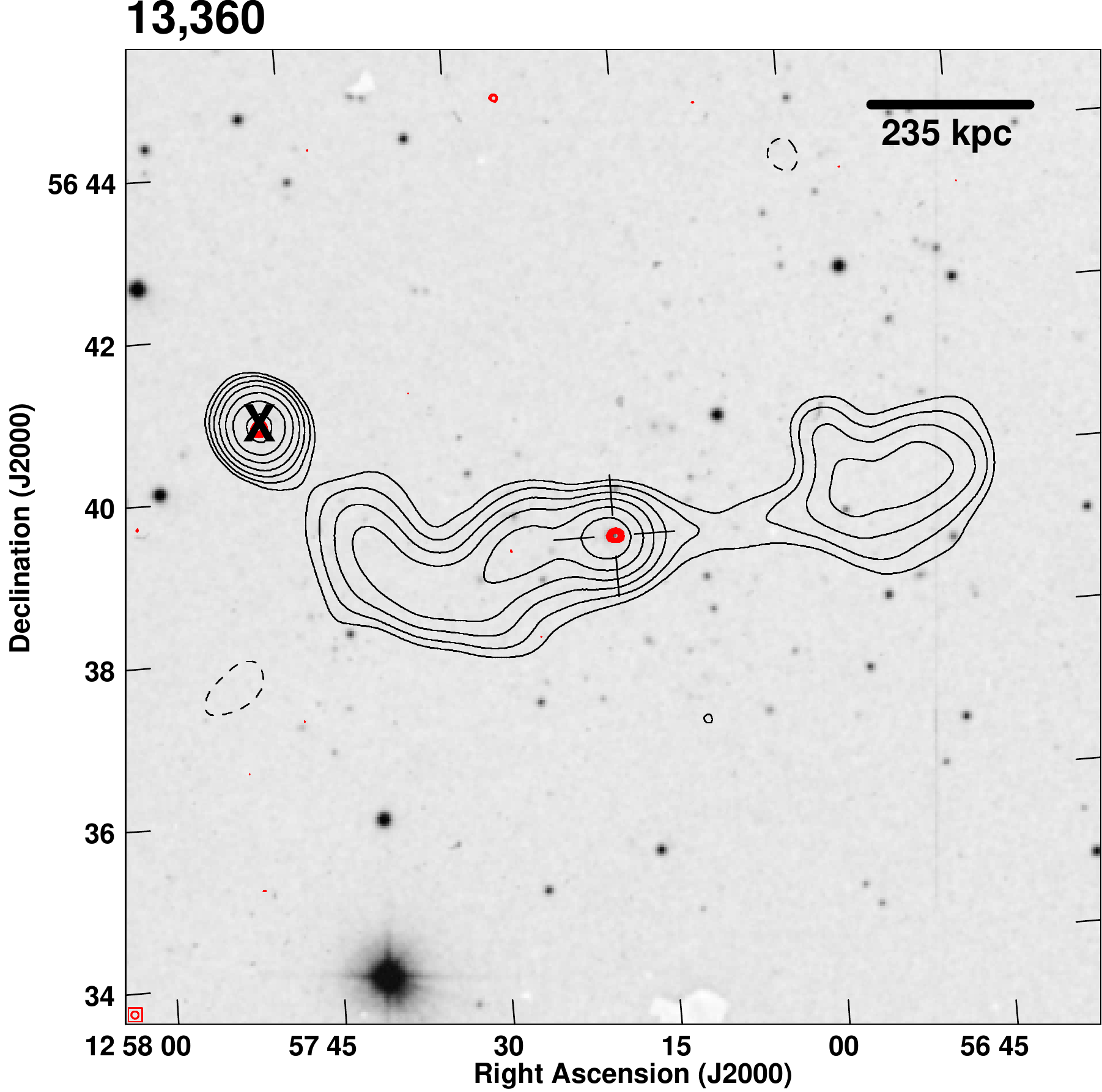}}
\hspace{0.8cm}{\includegraphics[width=0.3\textwidth]{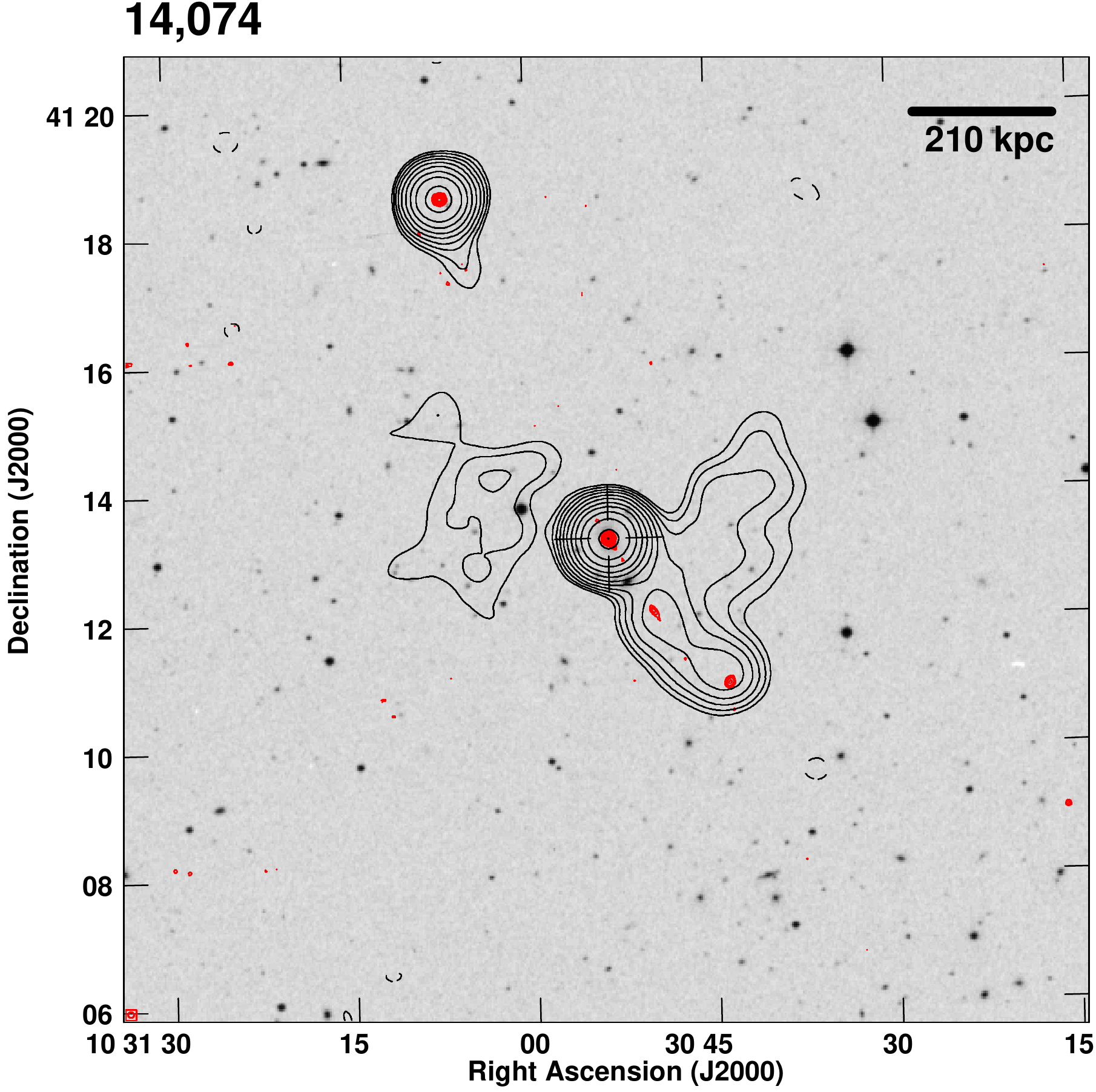}}
\hspace{0.8cm}{\includegraphics[width=0.3\textwidth]{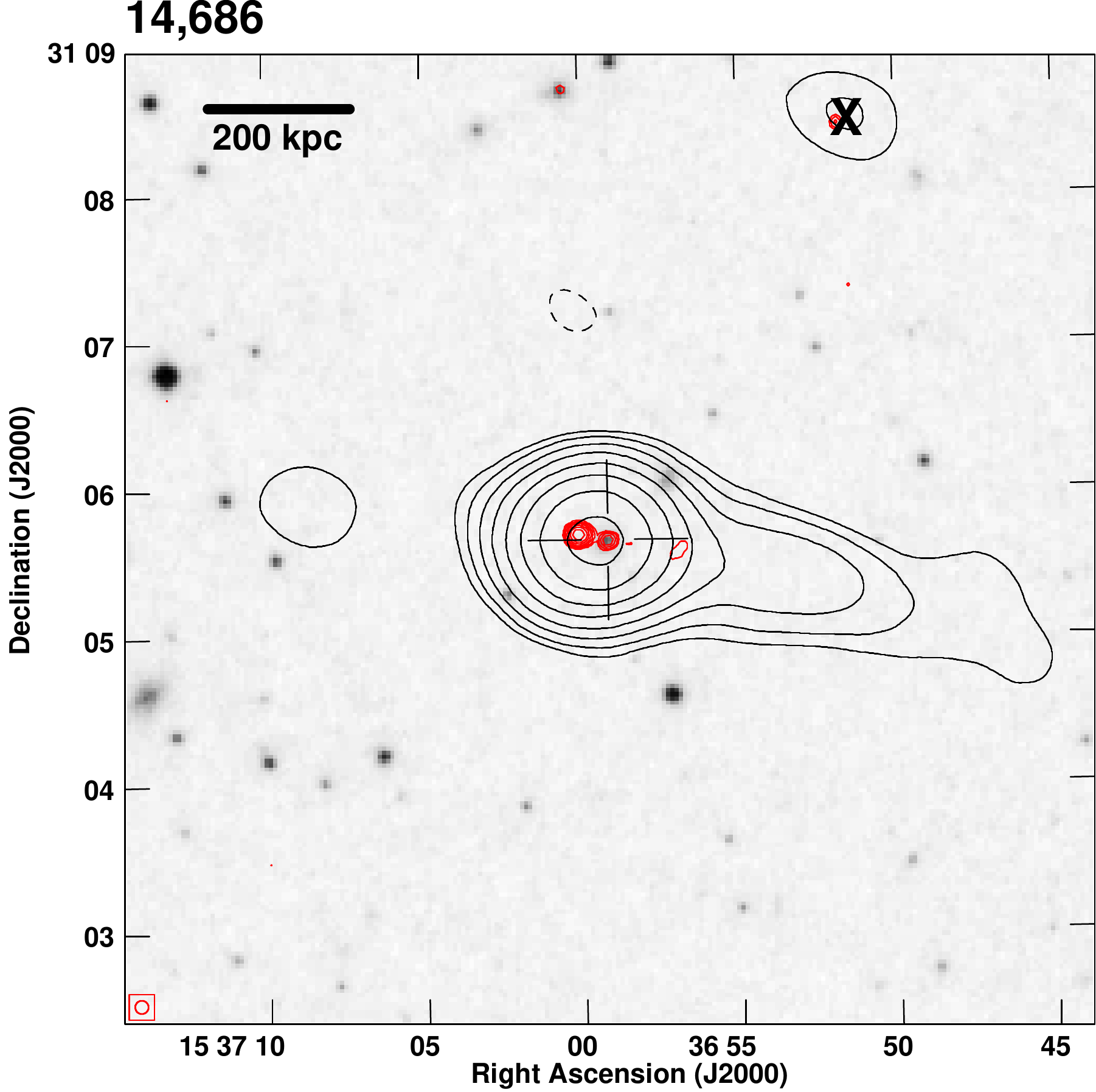}}
\caption{Newly discovered and/or classified possible giant radio sources in ROGUE~I. The layout of the maps is the same as in Figure~\ref{fig:NewGiantsMaps}}
\label{NewPGiantsMaps}
\end{figure*}

\renewcommand{\thefigure}{A.\arabic{figure} (Cont.)}

\begin{figure*}[ht!]
\ContinuedFloat
\hspace{0cm}{\includegraphics[width=0.3\textwidth]{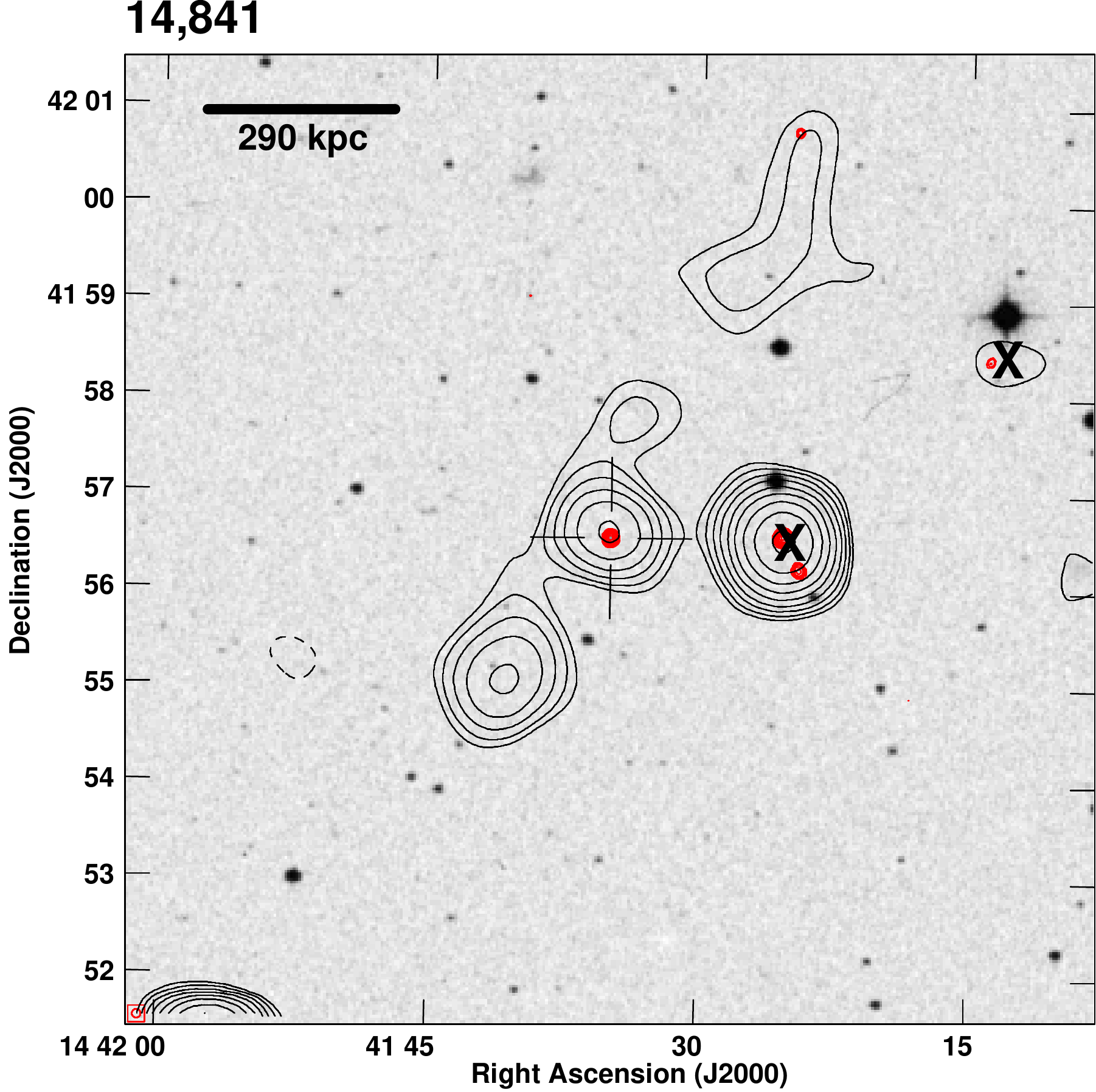}}
\hspace{0.2cm}{\includegraphics[width=0.3\textwidth]{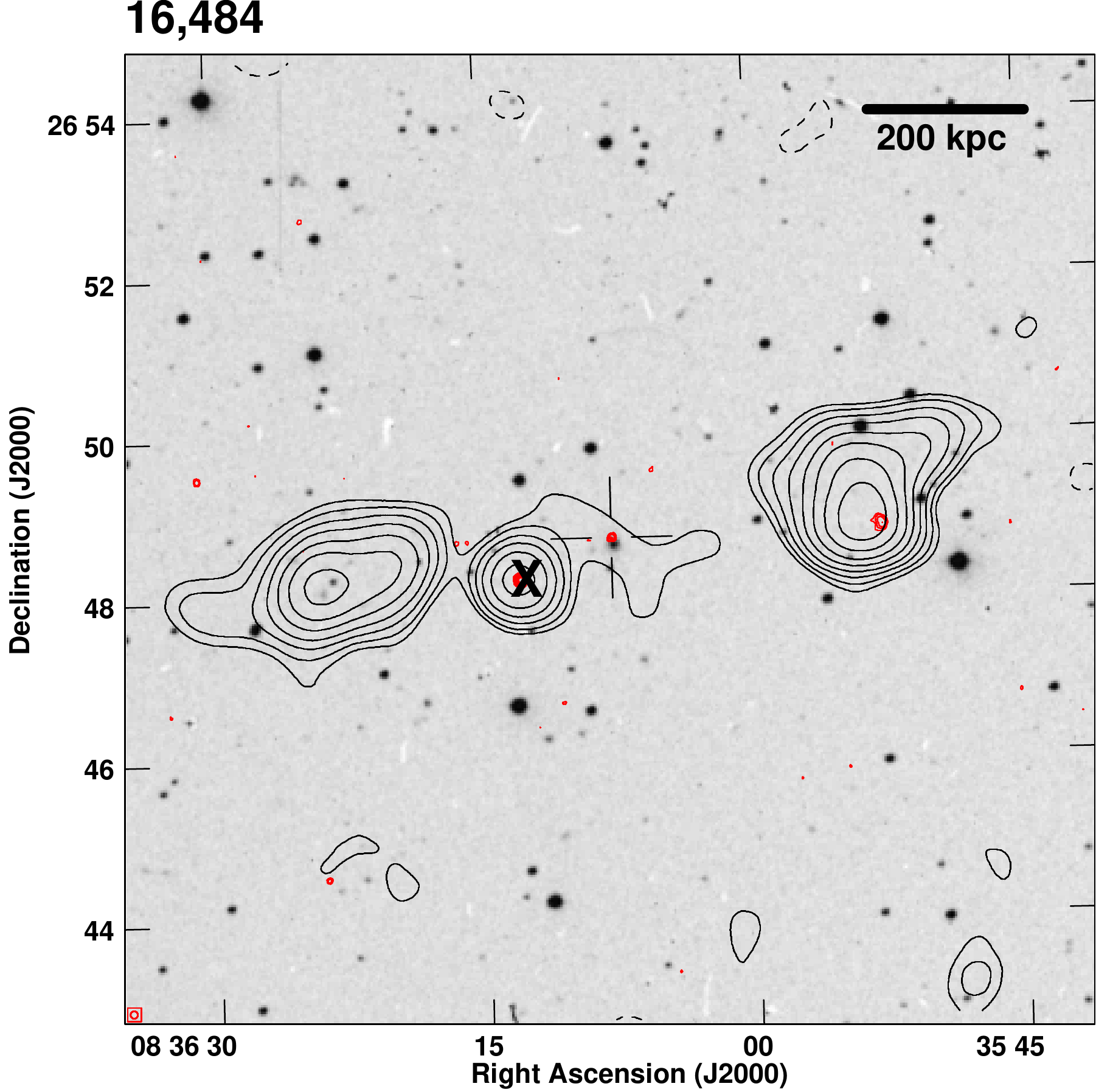}}
\hspace{0.0cm}{\includegraphics[width=0.3\textwidth]{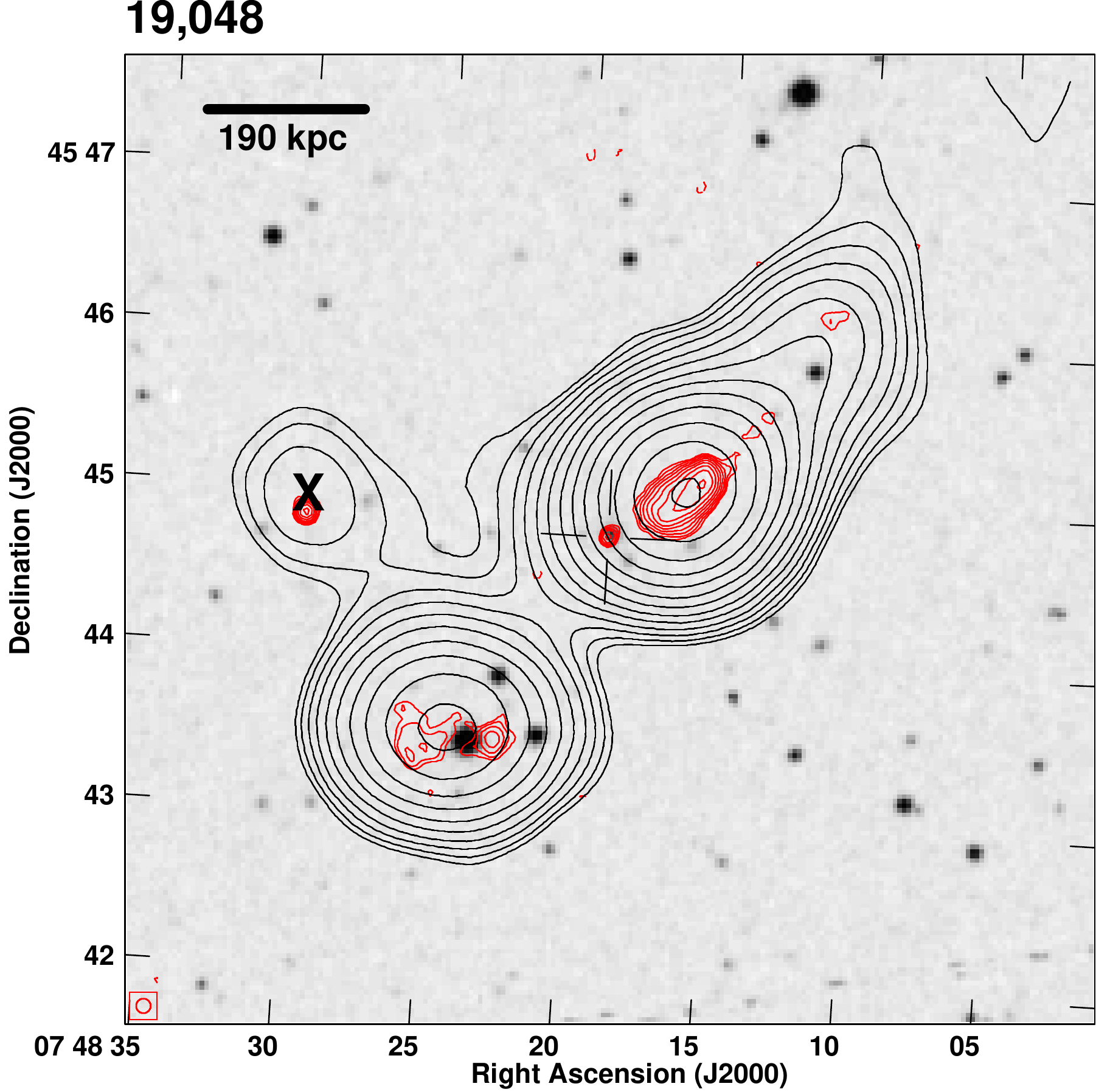}}
\hspace{0.2cm}{\includegraphics[width=0.3\textwidth]{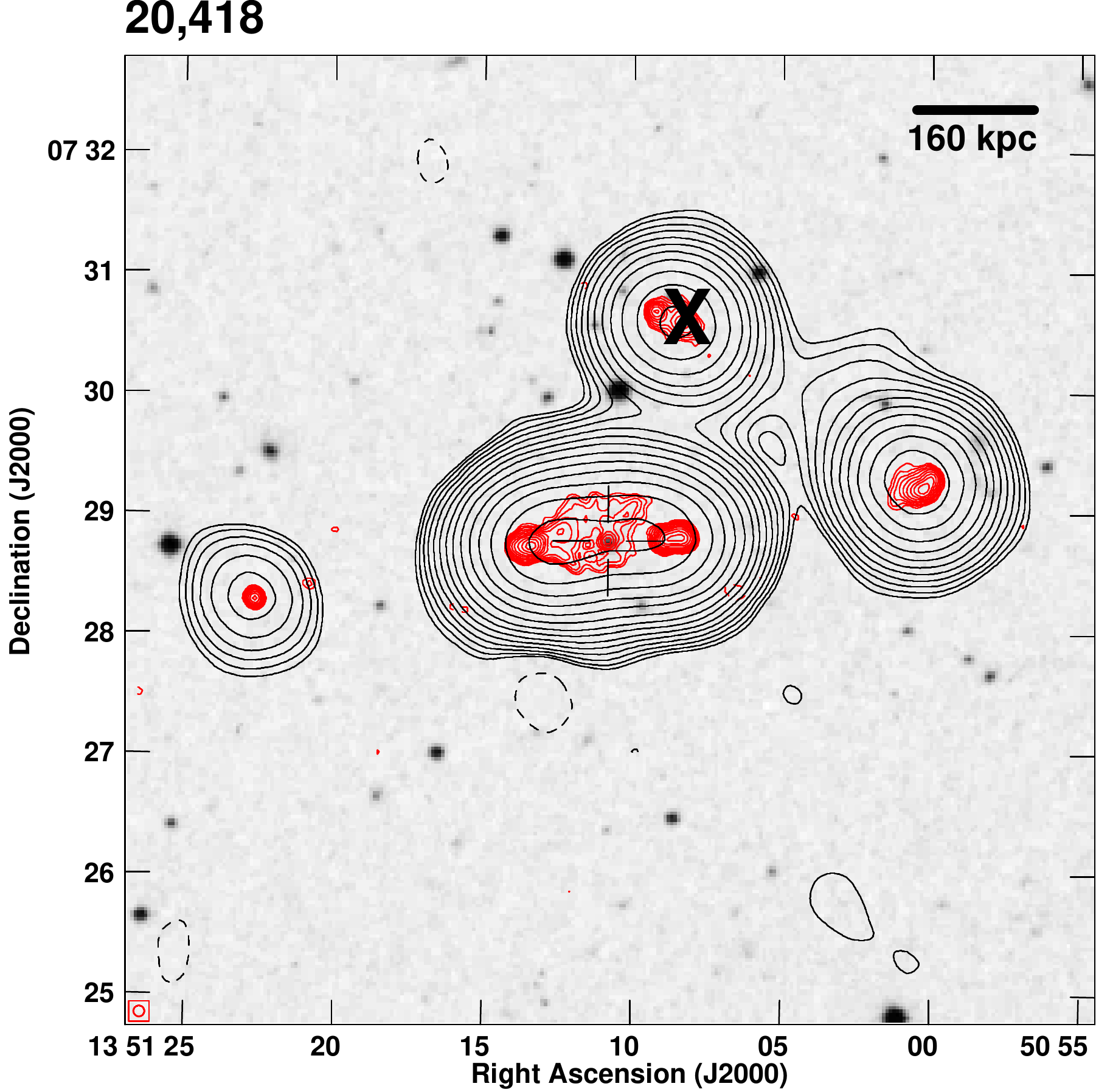}}
\hspace{0.2cm}{\includegraphics[width=0.3\textwidth]{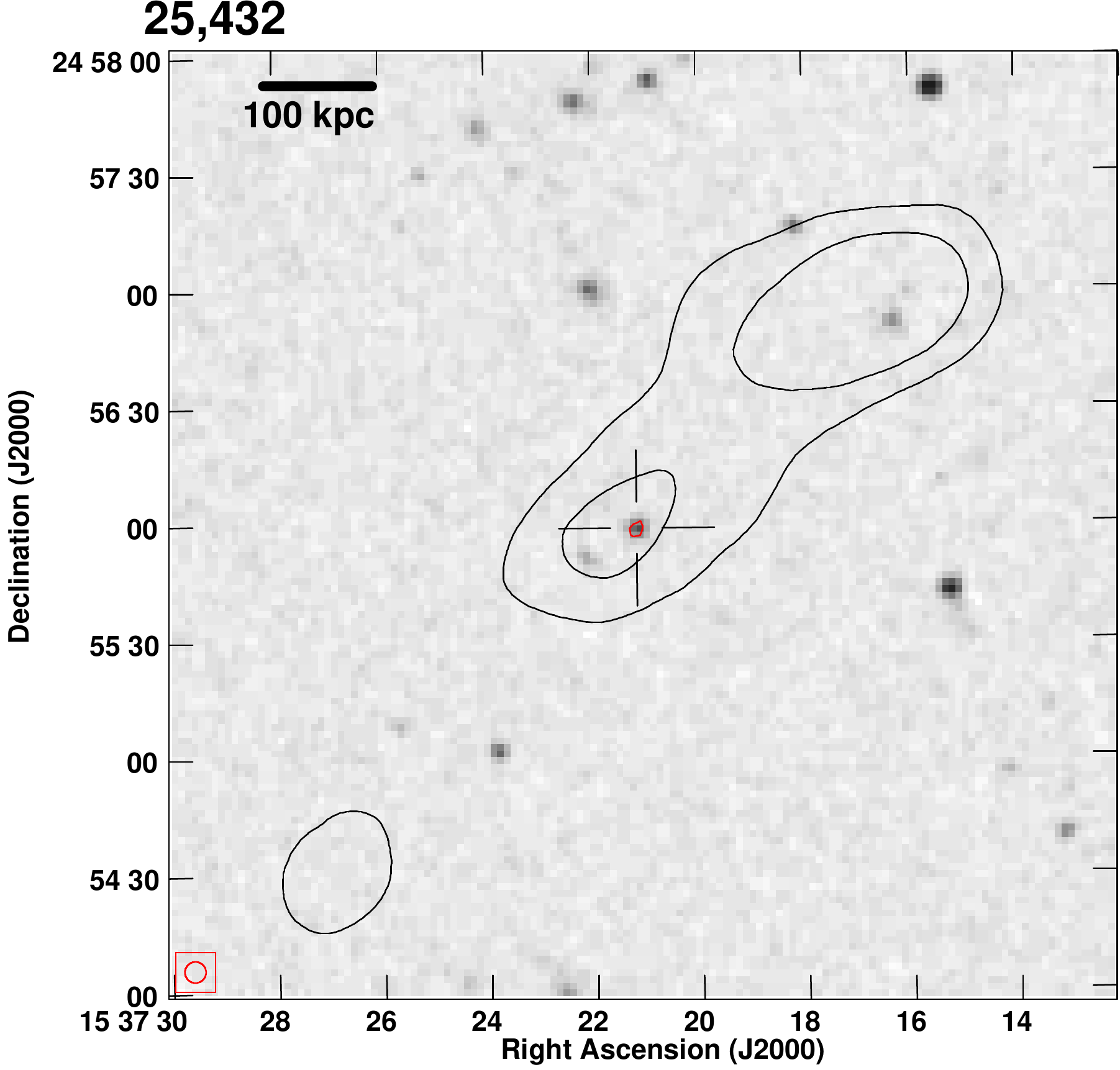}}
\hspace{0.2cm}{\includegraphics[width=0.3\textwidth]{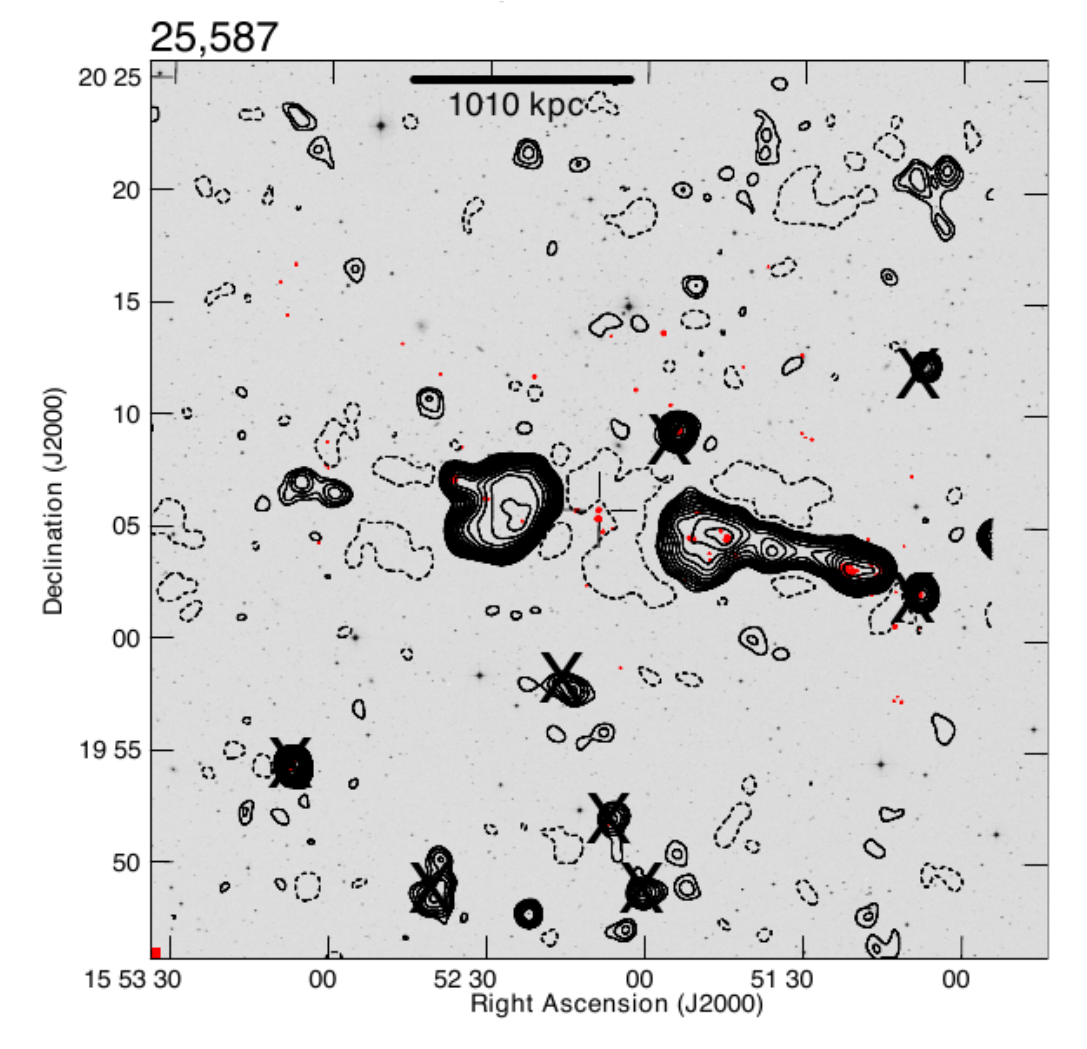}}
\hspace{0.2cm}{\includegraphics[width=0.3\textwidth]{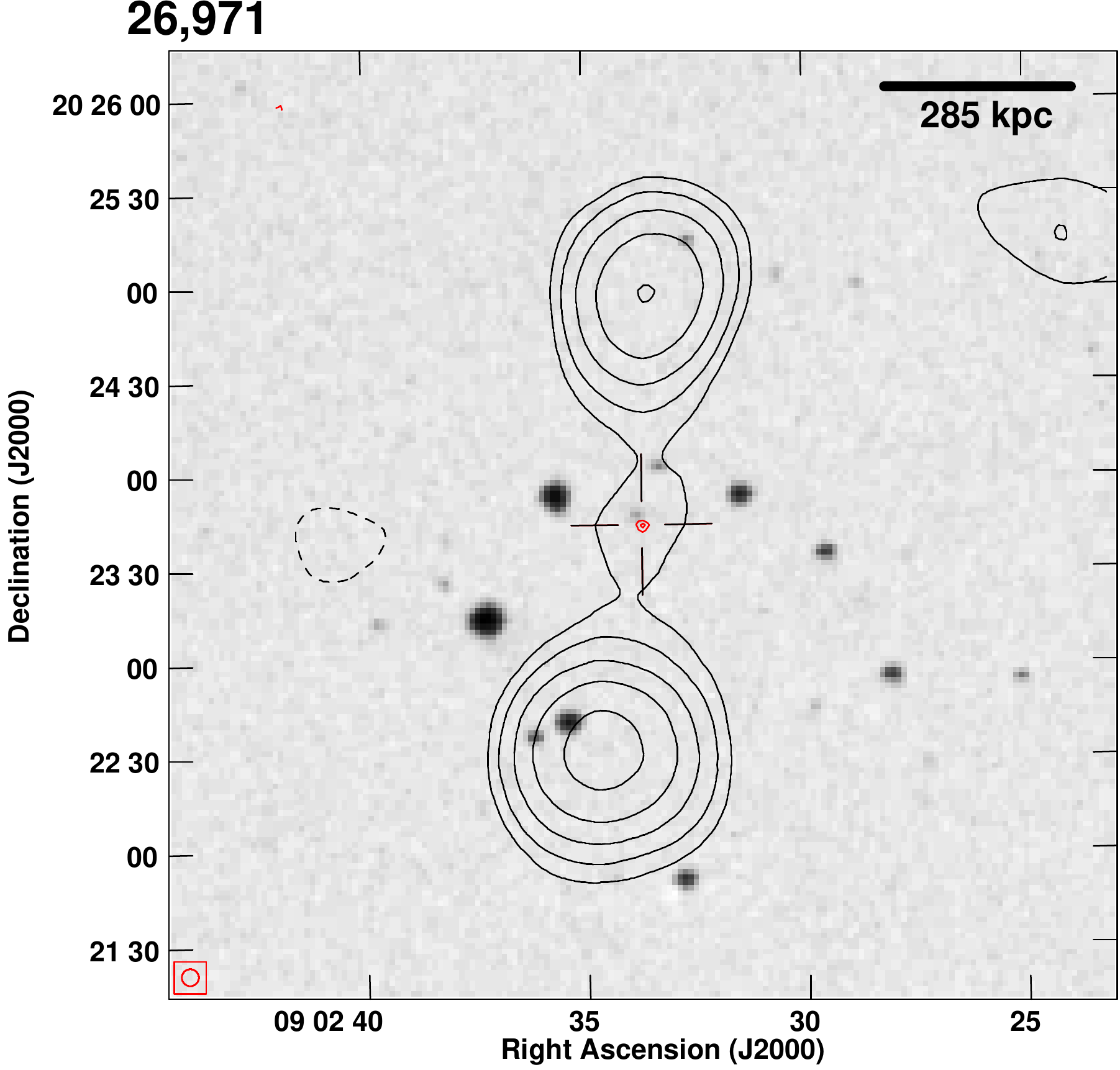}}
\hspace{0.8cm}{\includegraphics[width=0.3\textwidth]{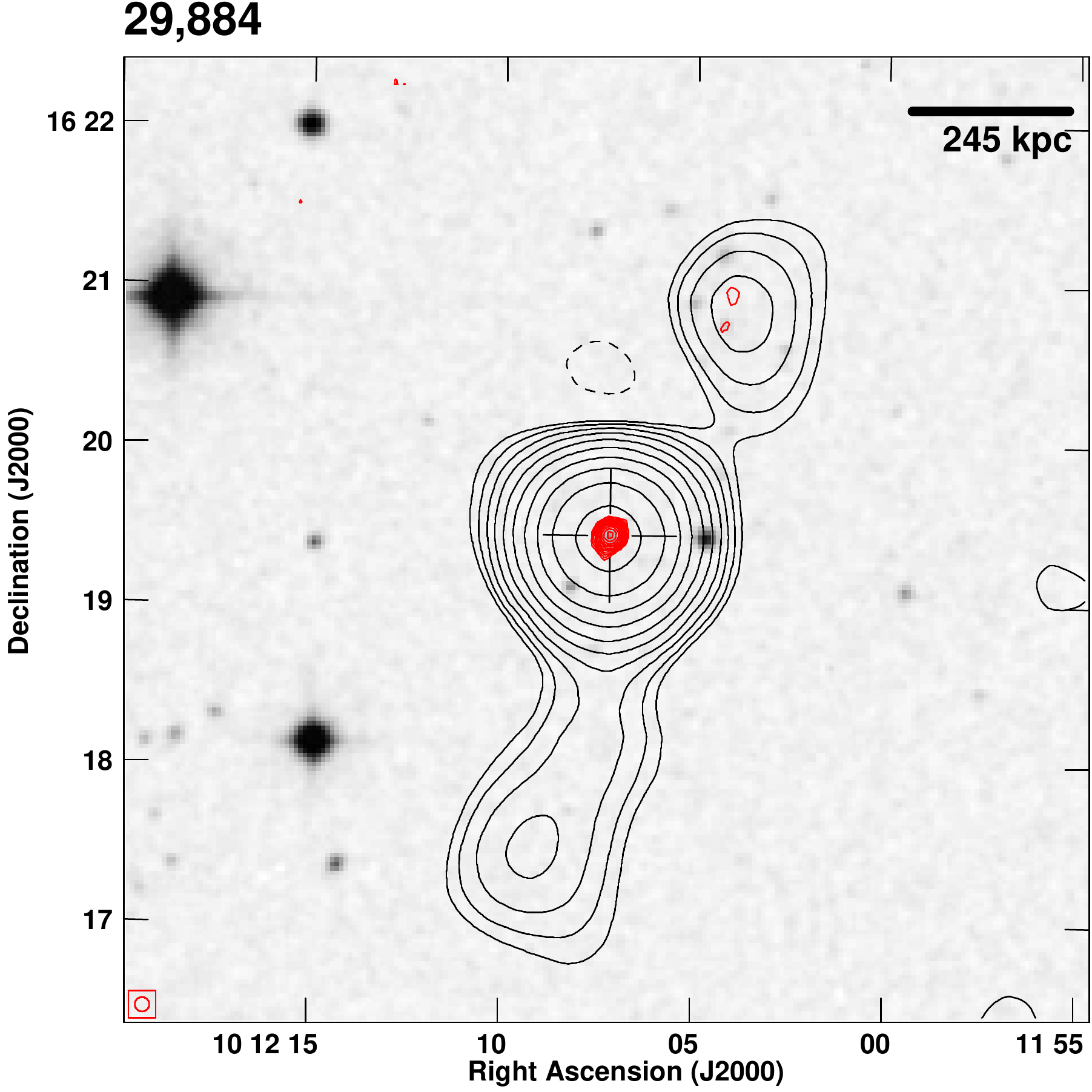}}
\hspace{0.8cm}{\includegraphics[width=0.3\textwidth]{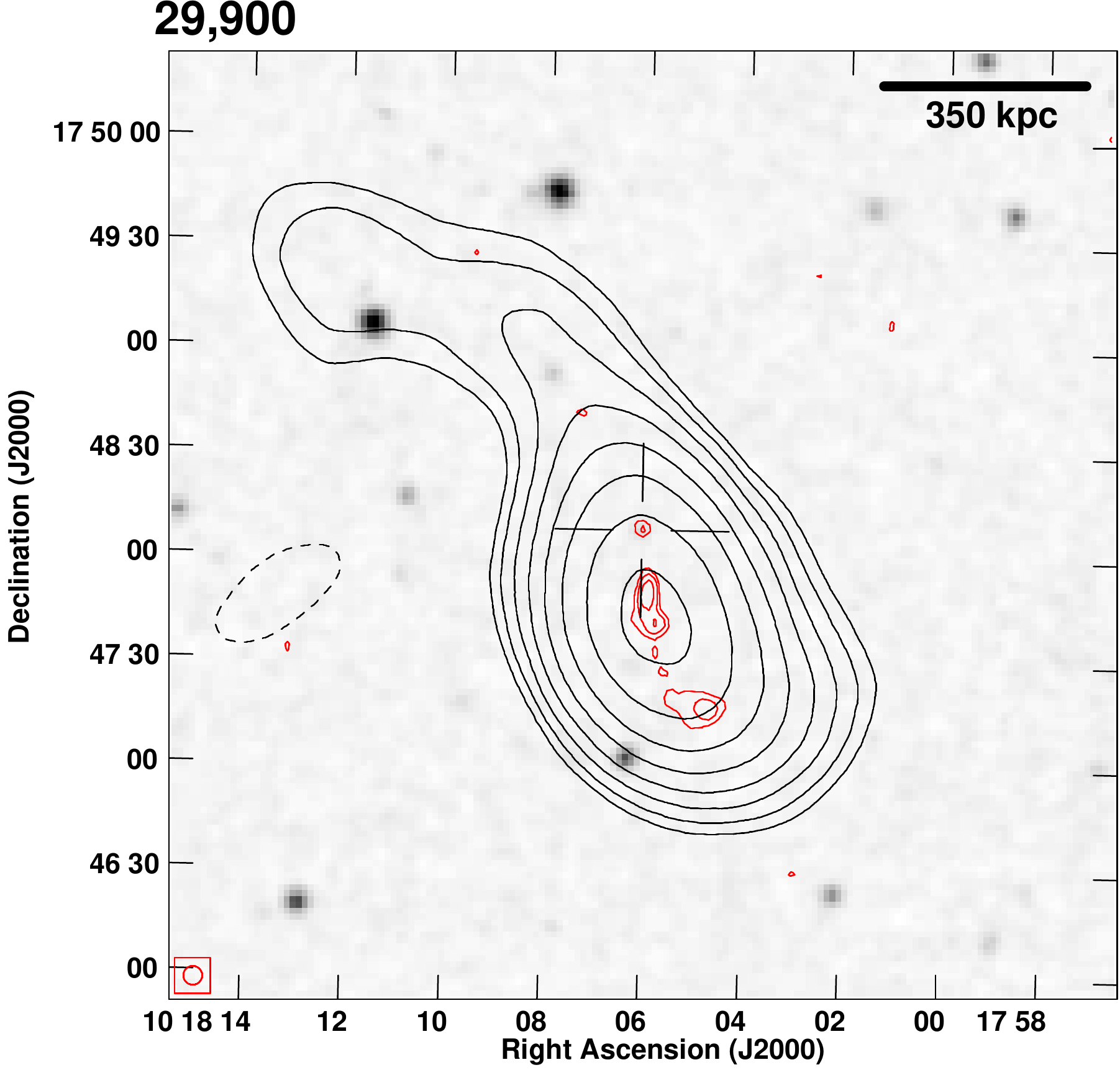}}
\hspace{0.2cm}{\includegraphics[width=0.3\textwidth]{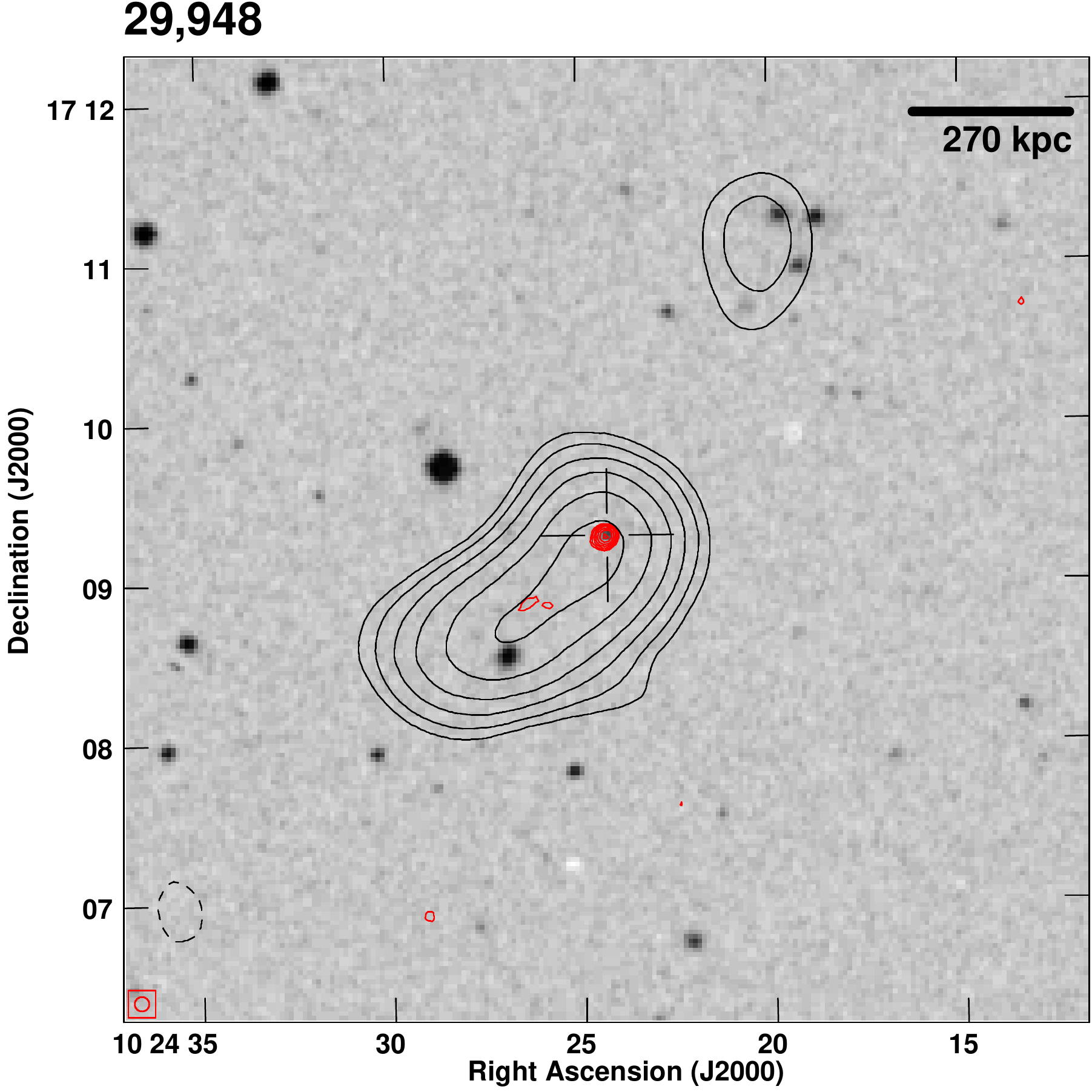}}
\caption{Newly discovered and/or classified possible giant radio sources in ROGUE~I.} 
\end{figure*}

\subsection{ Double--double radio sources}
\label{DD}

\begin{itemize}

  \item[]8,602: this source is included in the FR~II radio galaxy sample of \cite{KozielWierzbowska11}.
  Both, the FIRST and NVSS maps display an extended structure consisting of two pairs of lobes, where the outer lobes are brighter than the inner ones --- the opposite of what is usually observed in double--double radio sources. 

  \item[]9,180: this source appears in the FR~II sample of \cite{KozielWierzbowska11}. The FIRST map displays strong elongated lobes next to the core and a fainter one-sided outer lobe, whereas on the NVSS map both outer lobes are clearly seen. 

  \item[]19,429: on the FIRST map, a point-like core, symmetric lobe-like structures, as well as a fainter one-sided outer lobe are visible. The NVSS map displays a central part extending in the direction of structures visible on the FIRST map and additional slightly asymmetric outer lobes. The radio emission from the northern lobe is blended with two compact foreground/background sources.  

  \item[]1,232: this source appears in the FR~II sample of \cite{KozielWierzbowska11}. Analysing the FIRST map, we noticed that the source has more complex structure than just FR~II.
  The west side of the source consists of inner and outer lobes, while the outer lobe is missing on its east side. This may be caused by a projection or by ageing and light-time effect.
 
  \item[]4,035: the FIRST map reveals a complex morphology consisting of several radio components, while the NVSS map exhibits an elongated shape. The radio structures next to the core visible in the FIRST map could be either jets or inner lobes. Therefore we classify the source as a possible double--double radio galaxy. 

  \item[]15,170: the FIRST map displays a radio structure composed of several lobes along with radio emission from a background/foreground source. The source has an elongated and slightly bent morphology visible in the NVSS data. 

  \item[]16,107: \cite{KozielWierzbowska11} added the source to the FR~II radio galaxy sample. On the FIRST map, a compact core and an extended structure consisting of two pairs of detached lobes are visible, while the NVSS map reveals a possible FR~II morphology. 

  \item[]24,209: in the literature, the source is classified as an FR~I radio galaxy \citep{Kharb12}. Investigating the FIRST data, we noticed a small, two-sided 
 structure which can be interpreted as a pair of inner lobes. 
 The NVSS map reveals a clear large structure with FR~I morphology.

  \item[]25,983: the FIRST map reveals a radio structure consisting of inner and outer lobes, however, the emission close to the core may be jet-like. The source has an FR~I morphology in the NVSS data. 

  \item[]30,866: \cite{Kharb12} classified the source as an FR~I radio galaxy. The FIRST map displays an elongated core with a one-sided jet and a pair of lobes. On the NVSS map, in addition to the internal structure, two pairs of external lobes are visible. 
\end{itemize}
 
\renewcommand{\thefigure}{A.\arabic{figure}}

\begin{figure*}[ht!]
\hspace{0.2cm}{\includegraphics[width=0.3\textwidth]{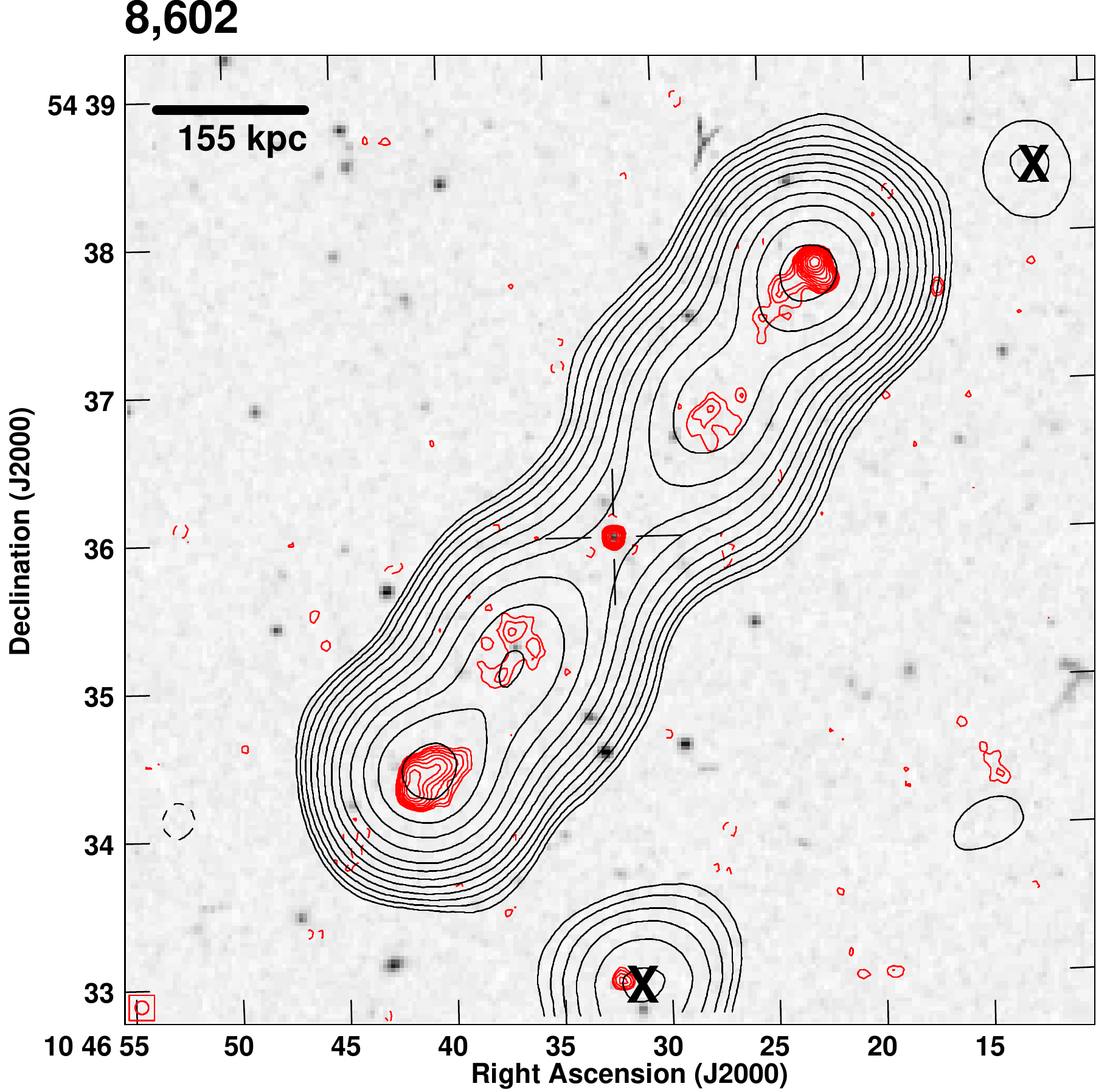}}
\hspace{0.2cm}{\includegraphics[width=0.3\textwidth]{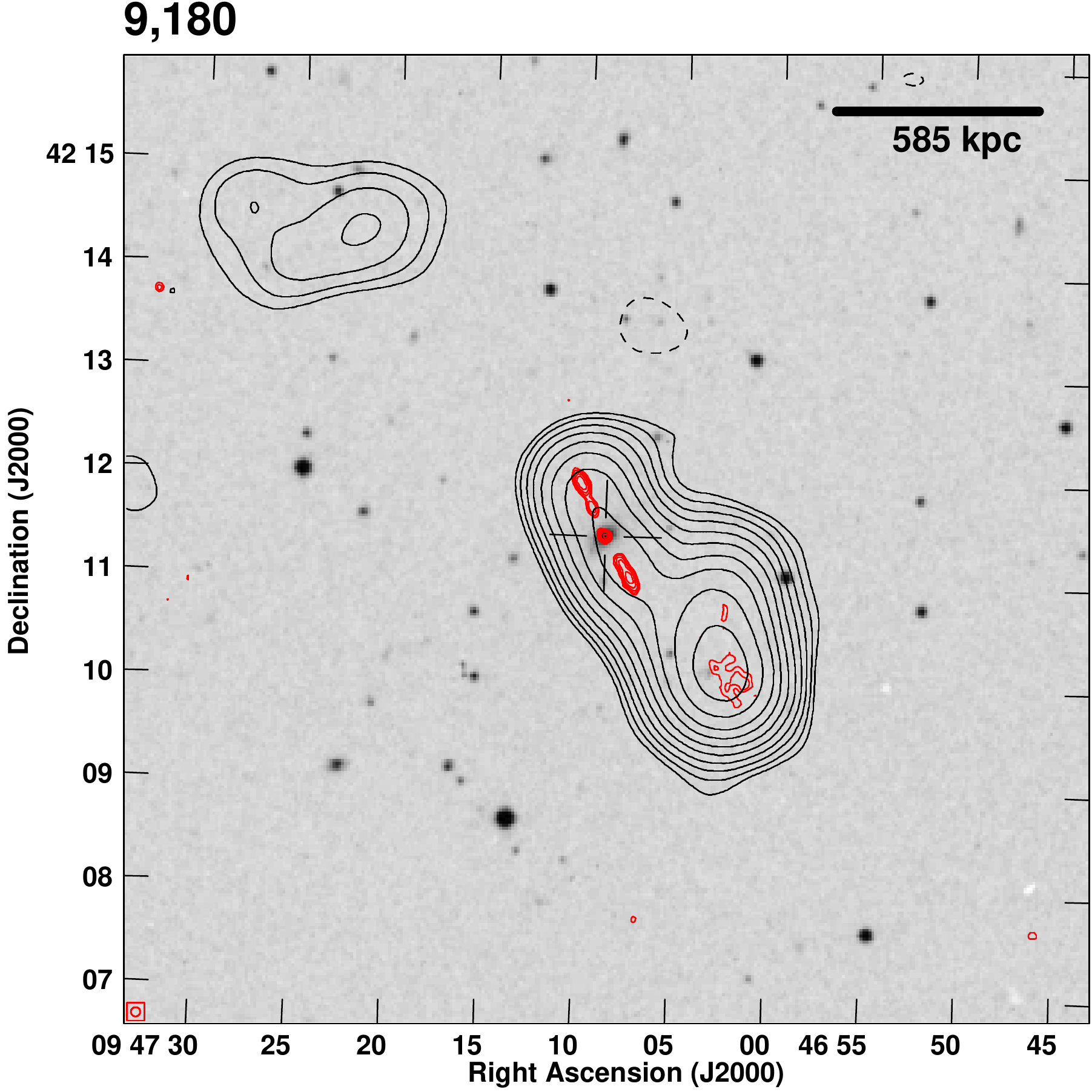}}
\hspace{0.2cm}{\includegraphics[width=0.3\textwidth]{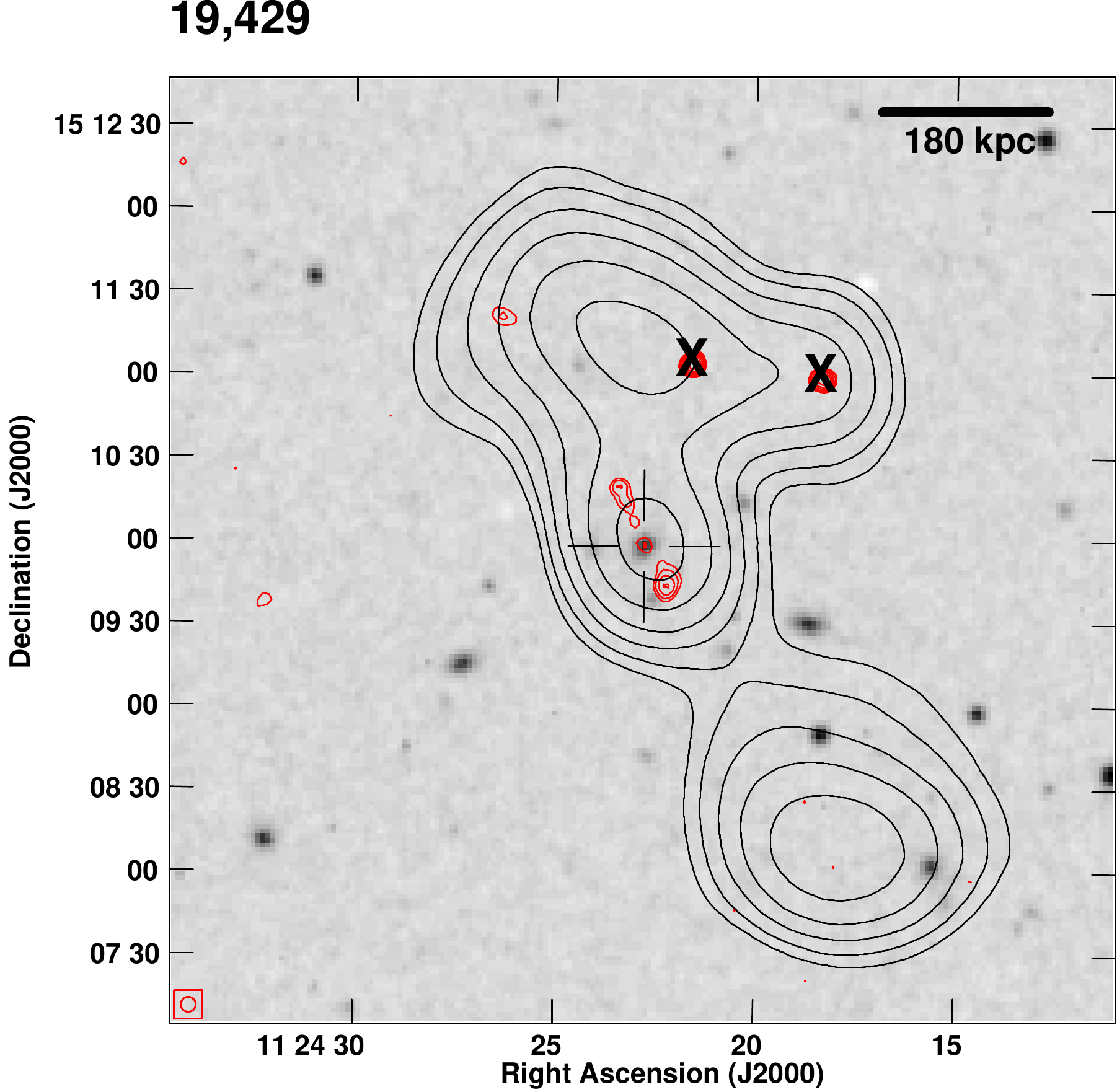}}
\hspace{0.2cm}{\includegraphics[width=0.3\textwidth]{com_dd20418_cBS.pdf}}
\hspace{0.2cm}{\includegraphics[width=0.3\textwidth]{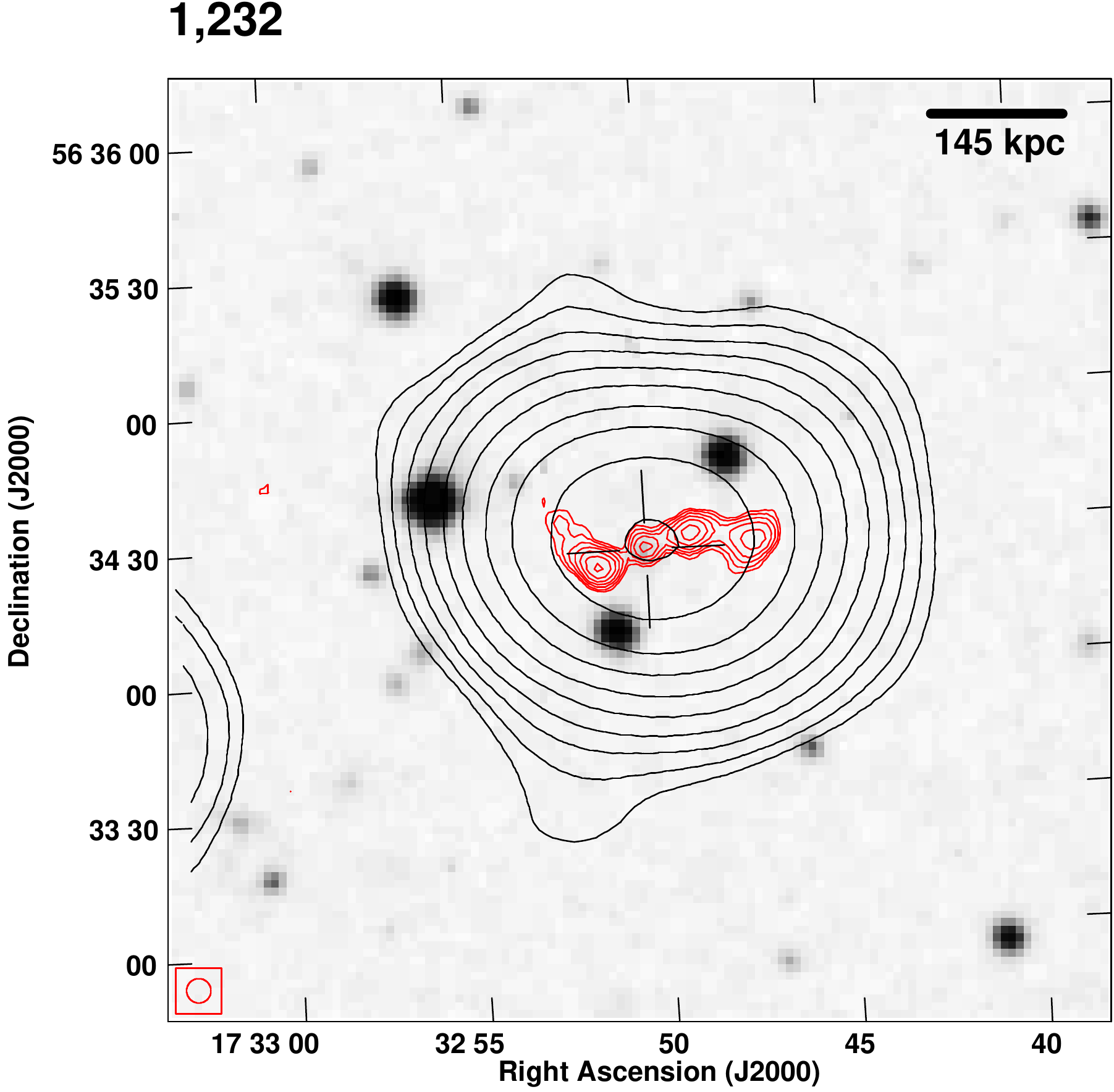}}
\hspace{0.2cm}{\includegraphics[width=0.3\textwidth]{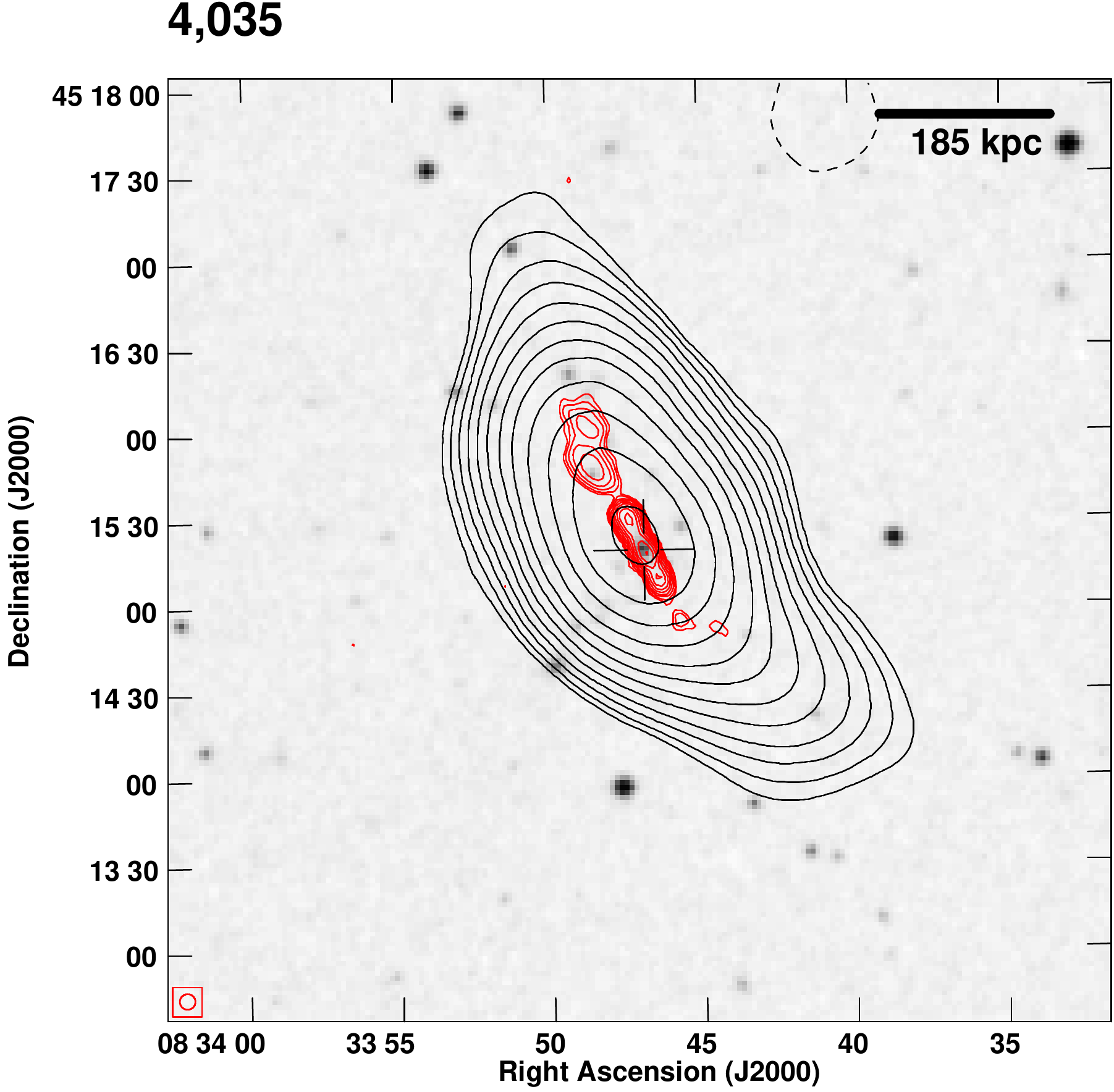}}
\hspace{0.2cm}{\includegraphics[width=0.3\textwidth]{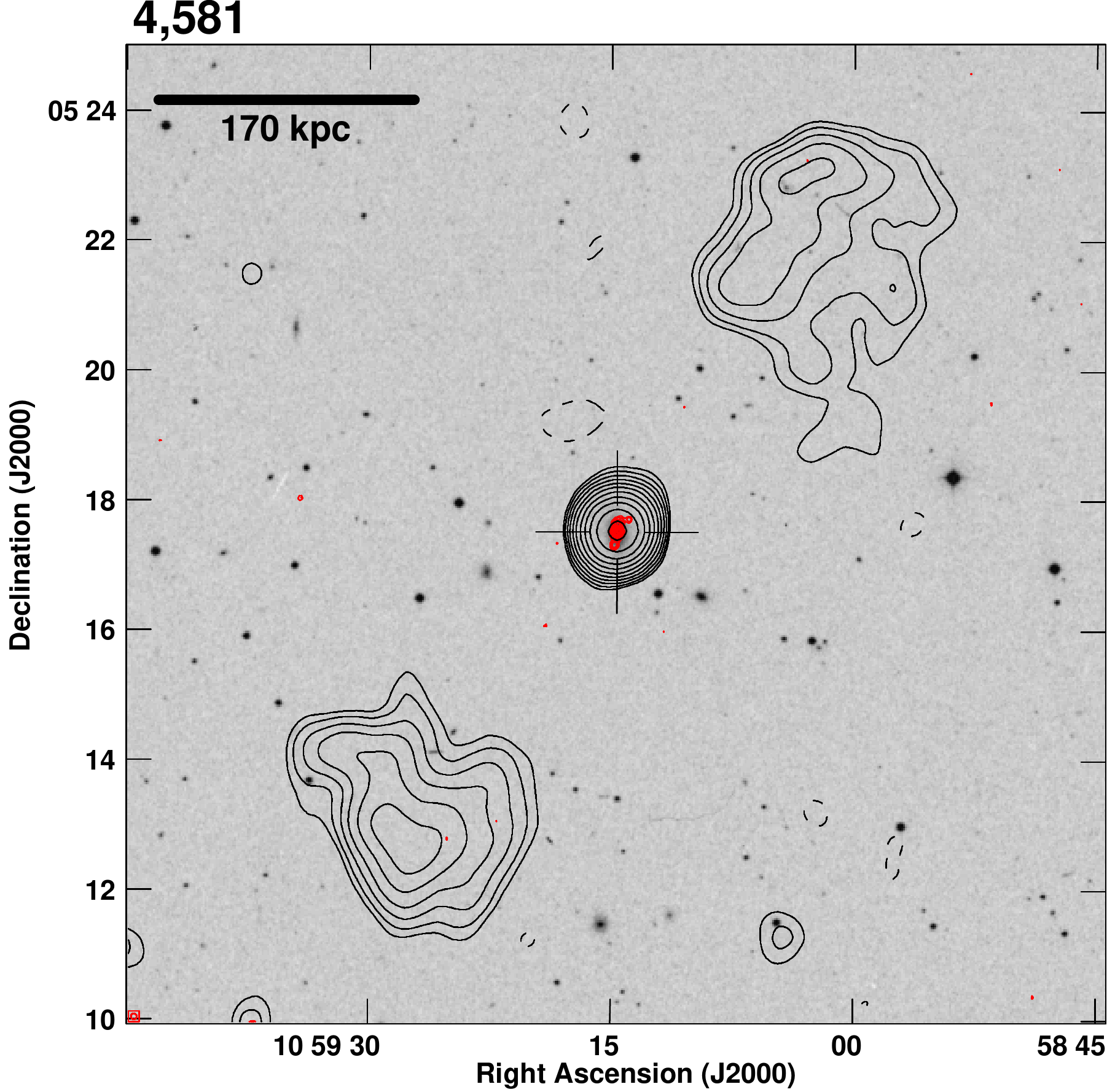}}
\hspace{0.2cm}{\includegraphics[width=0.3\textwidth]{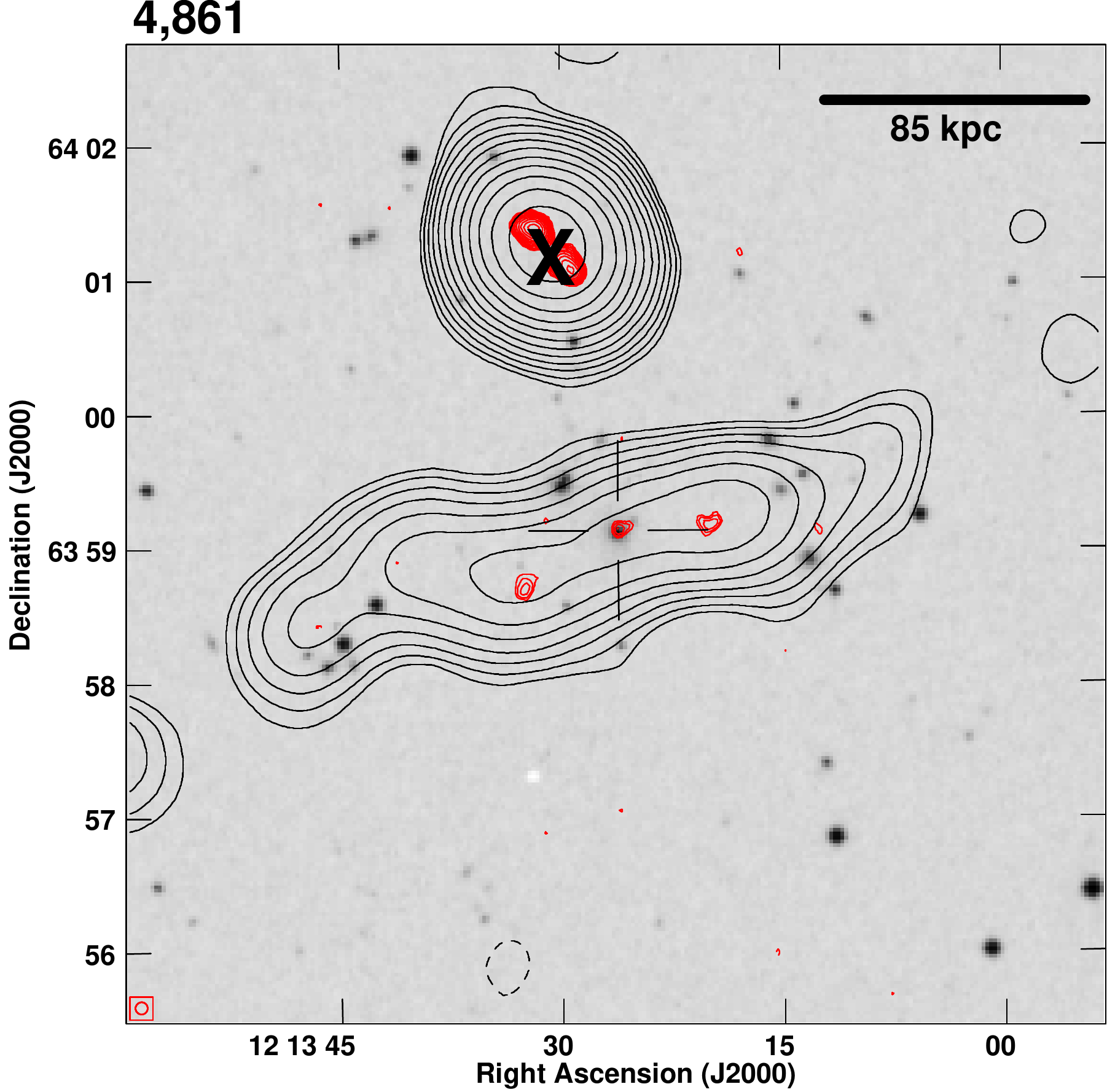}}
\hspace{0.2cm}{\includegraphics[width=0.3\textwidth]{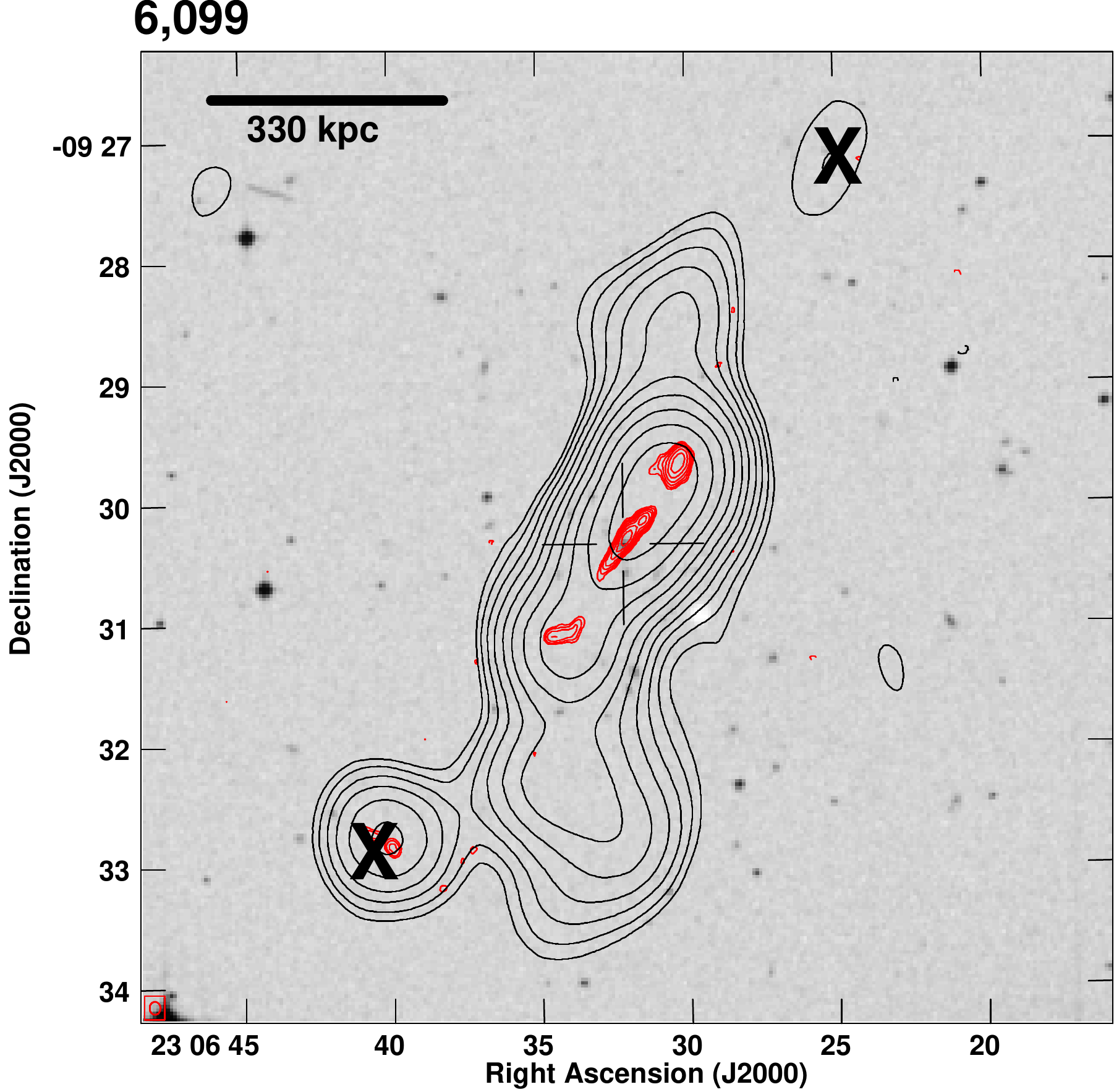}}
\hspace{0.2cm}{\includegraphics[width=0.3\textwidth]{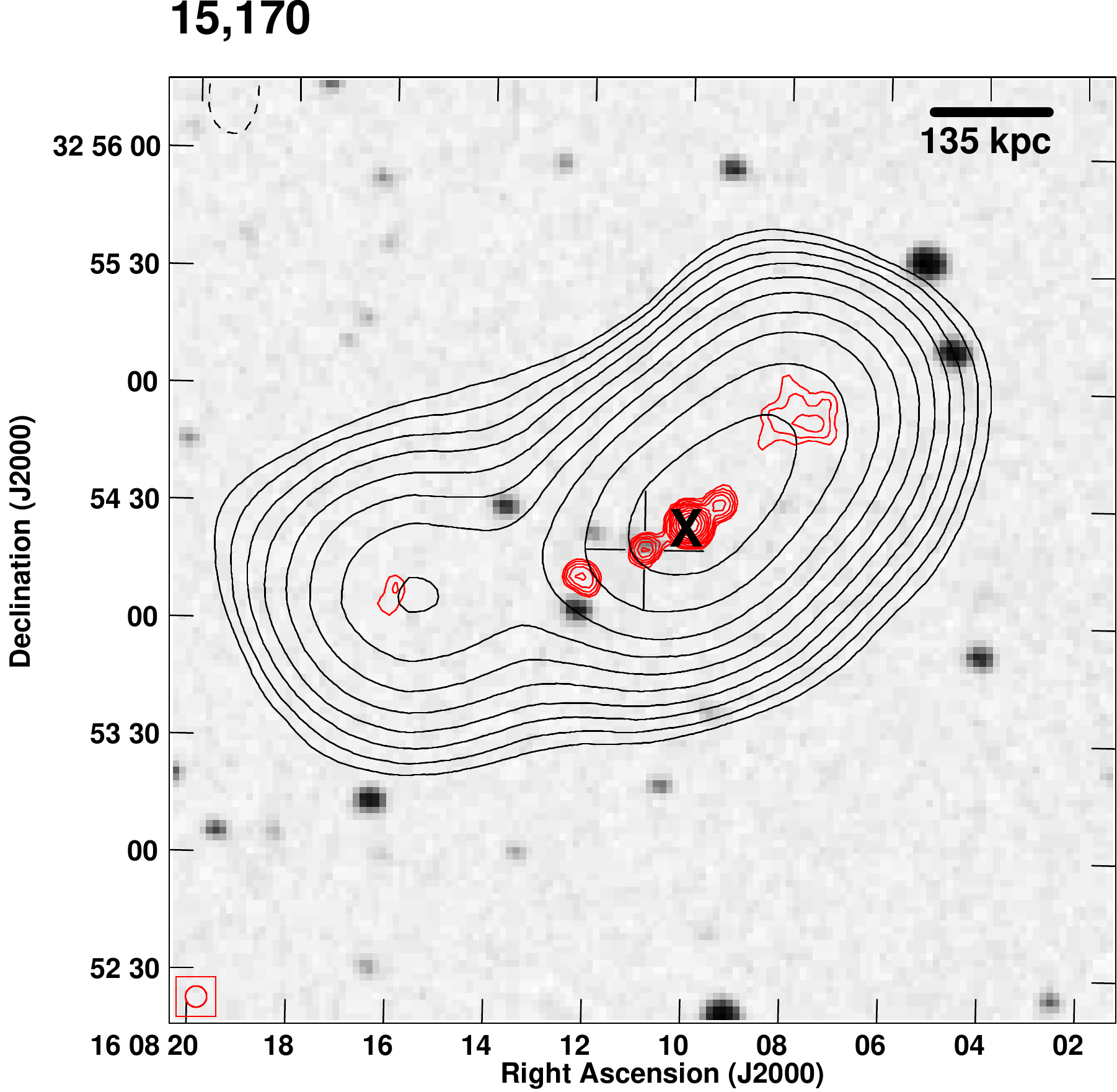}}
\hspace{0.8cm}{\includegraphics[width=0.3\textwidth]{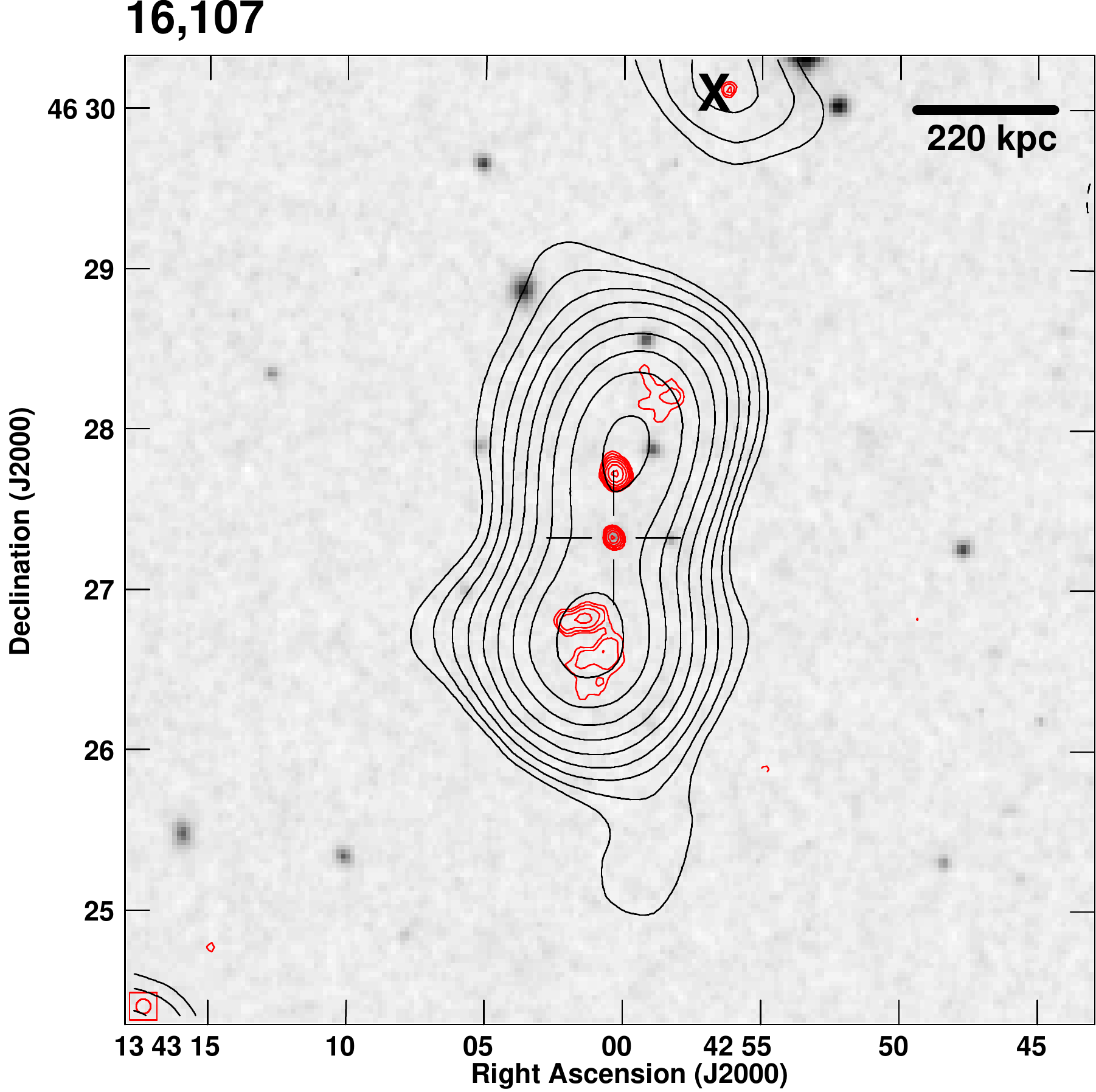}}
\hspace{0.8cm}{\includegraphics[width=0.3\textwidth]{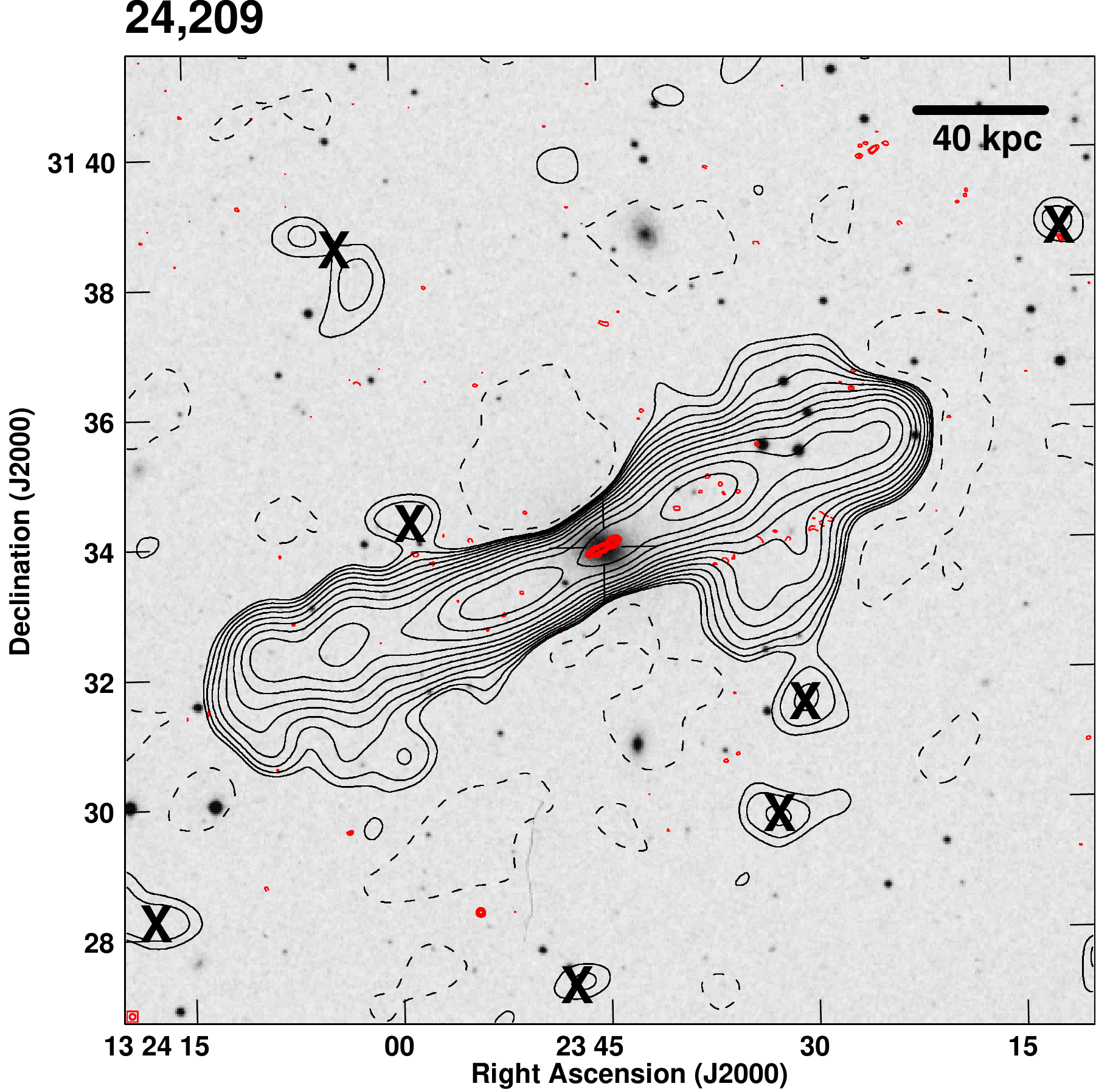}}
\caption{Newly discovered and/or classified as double--double or possible double--double radio galaxies in ROGUE~I. The layout of the maps is the same as in Figure~\ref{fig:NewGiantsMaps}.}
\label{NewDDMaps}
\end{figure*}

\renewcommand{\thefigure}{A.\arabic{figure} (Cont.)}

\begin{figure*}[ht!]
\ContinuedFloat
\hspace{0.0cm}{\includegraphics[width=0.3\textwidth]{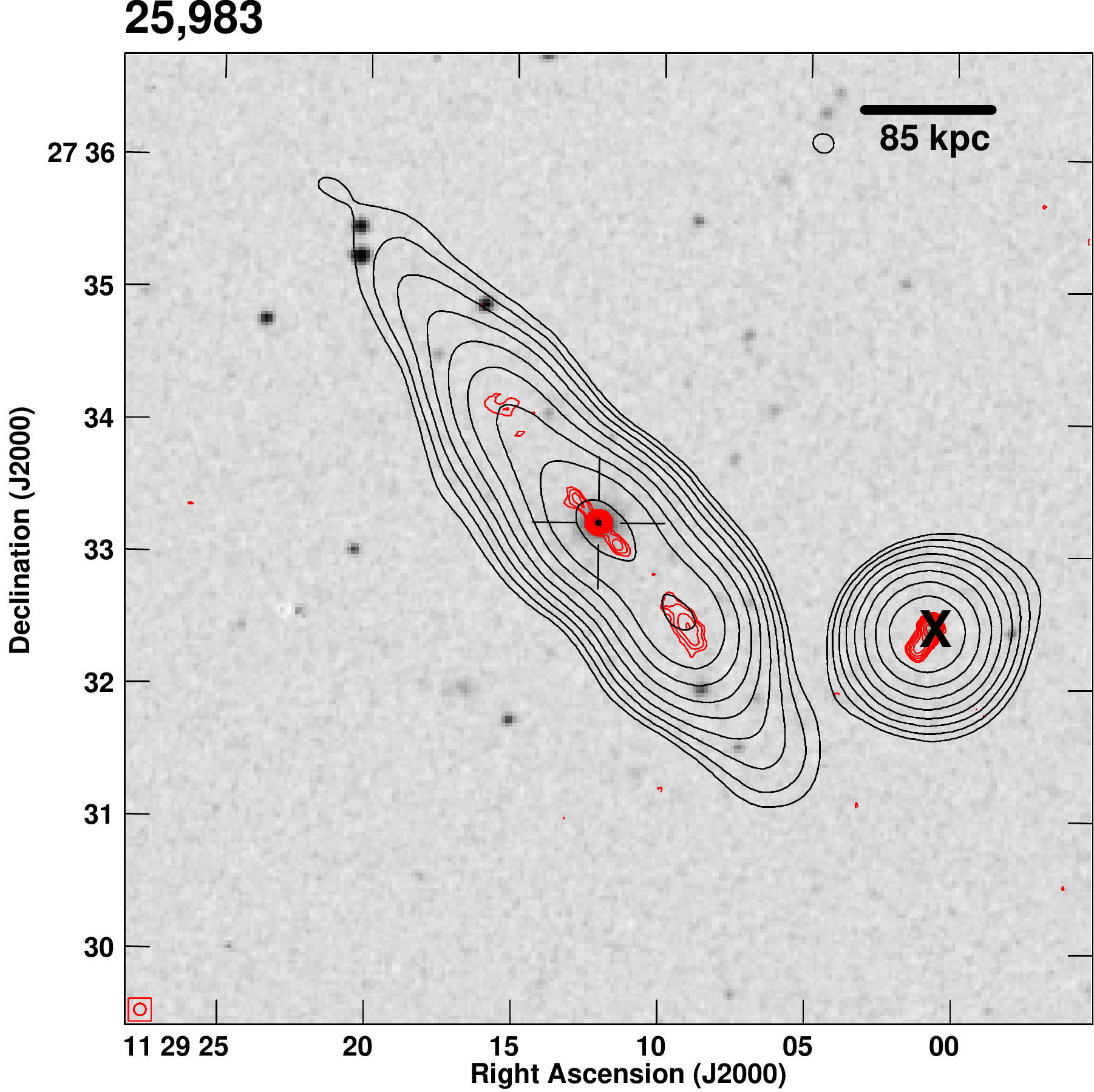}}
\hspace{0.2cm}{\includegraphics[width=0.3\textwidth]{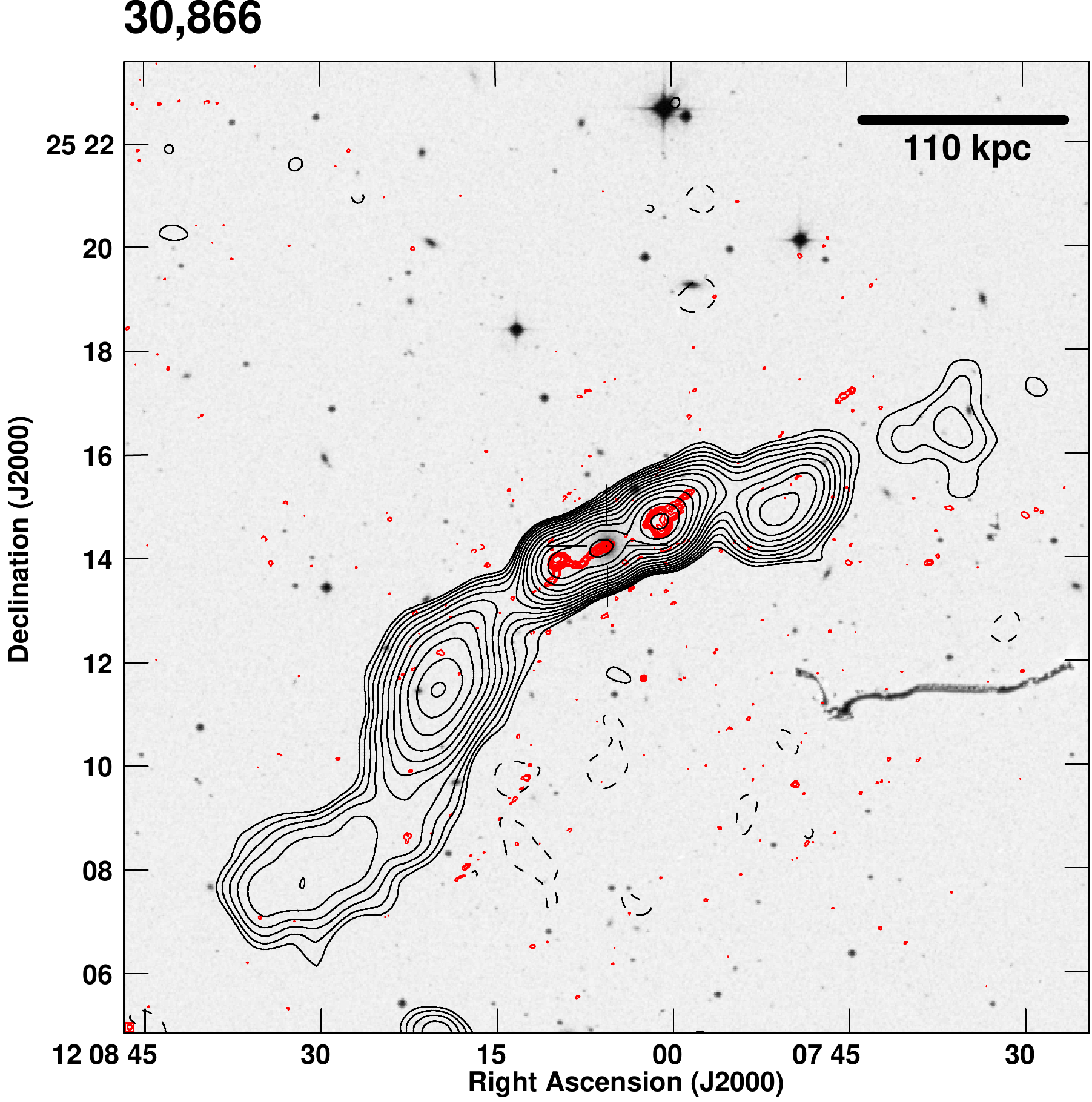}}
\caption{Newly discovered and/or classified as double--double or possible double--double radio galaxies in ROGUE~I. }
\end{figure*}

\subsection{X--shaped radio sources}
\label{X}

\begin{itemize}
  \item[]18,205: the FIRST map reveals a pair of classical strong FR~II type lobes with prominent hot spots and indication of a faint structure perpendicular to the former lobes. The NVSS map shows an extended structure with two axes of symmetry forming an X--shaped morphology, i.e. one along the FR~II lobes and another connecting faint extensions, almost perpendicular to the aforementioned structure. 
  \item[]25,798: the source  appears in the FR~II radio galaxy sample of  \cite{KozielWierzbowska11}. The FIRST map displays a complex morphology consisting of a strong core, symmetric FR~II type lobes, and an additional lobe inclined to them. On the NVSS map, an asymmetric two-axis boxy structure, blended with a foreground/background source is visible. 
  
  \item[]27,729: in the FIRST data, the source consists of two pairs of lobes rotated relative to each other by $\sim$30\degr. The shorter pair of lobes is fainter, while the the longer lobes are stronger with hot spots. On the NVSS map, an extended and slightly asymmetric FR~II type structure is visible. It is classified as a giant radio source by \cite{Kuzmicz.etal.2018a} with a linear size $\sim$0.8 Mpc.
  
  \item[]3,155: the FIRST map exhibits a possible X--shaped radio structure with two pairs of underdeveloped lobes. Only a uniform circular structure is visible on the NVSS map.
  
  \item[]4,287: the source has an X--shaped type radio structure with barely seen both pairs of lobes and circular featureless emission visible in the FIRST and the NVSS maps, respectively.
  
  \item[]6,713: the source has a clear FR~II type morphology in FIRST map and an additional emission at an inclined axis from the main jet axis in NVSS map. 
  
  \item[]23,094: the sources has clear FR\,II type morphology in FIRST map while an extended emission inclined to the main jet axis (could be a backflow or from second pair of `old' lobes) is seen in NVSS map. 
  \end{itemize} 
  
\renewcommand{\thefigure}{A.\arabic{figure}}


\begin{figure*}[htb!]
\hspace{0.0cm}{\includegraphics[width=\textwidth]{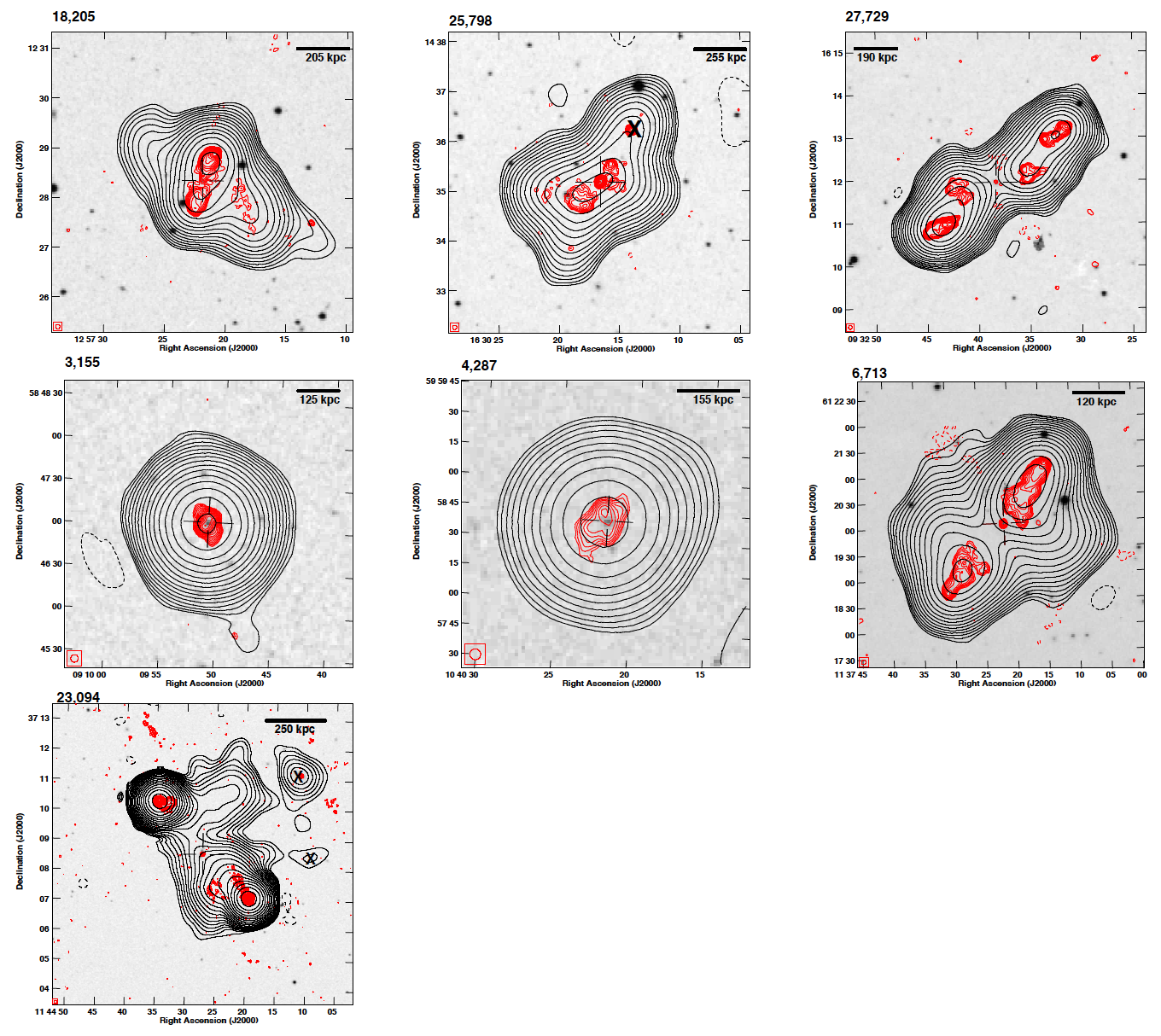}}
\caption{Newly discovered and/or classified as X--shaped or possible X--shaped radio sources in ROGUE~I. The layout of the maps is the same as in Figure~\ref{fig:NewGiantsMaps}.}
\label{NewXMaps}
\end{figure*}

\subsection{Z--shaped radio sources}
\label{Z}

\begin{itemize}
  \item[]13,184: the core, jet-like structure and southern lobe are clearly visible on the FIRST map, while the northern lobe is outlined by only one contour. The NVSS map reveals an elongated central part and two FR~II type lobes with a Z--shaped morphology. 
  \item[]25,124: the FIRST map shows a complex radio structure consisting of many components and forming an Z--shaped morphology. 
  \end{itemize} 
  
\renewcommand{\thefigure}{A.\arabic{figure}}

\begin{figure*}[htb!]
\scriptsize
\hspace{0cm}{\includegraphics[width=0.3\textwidth]{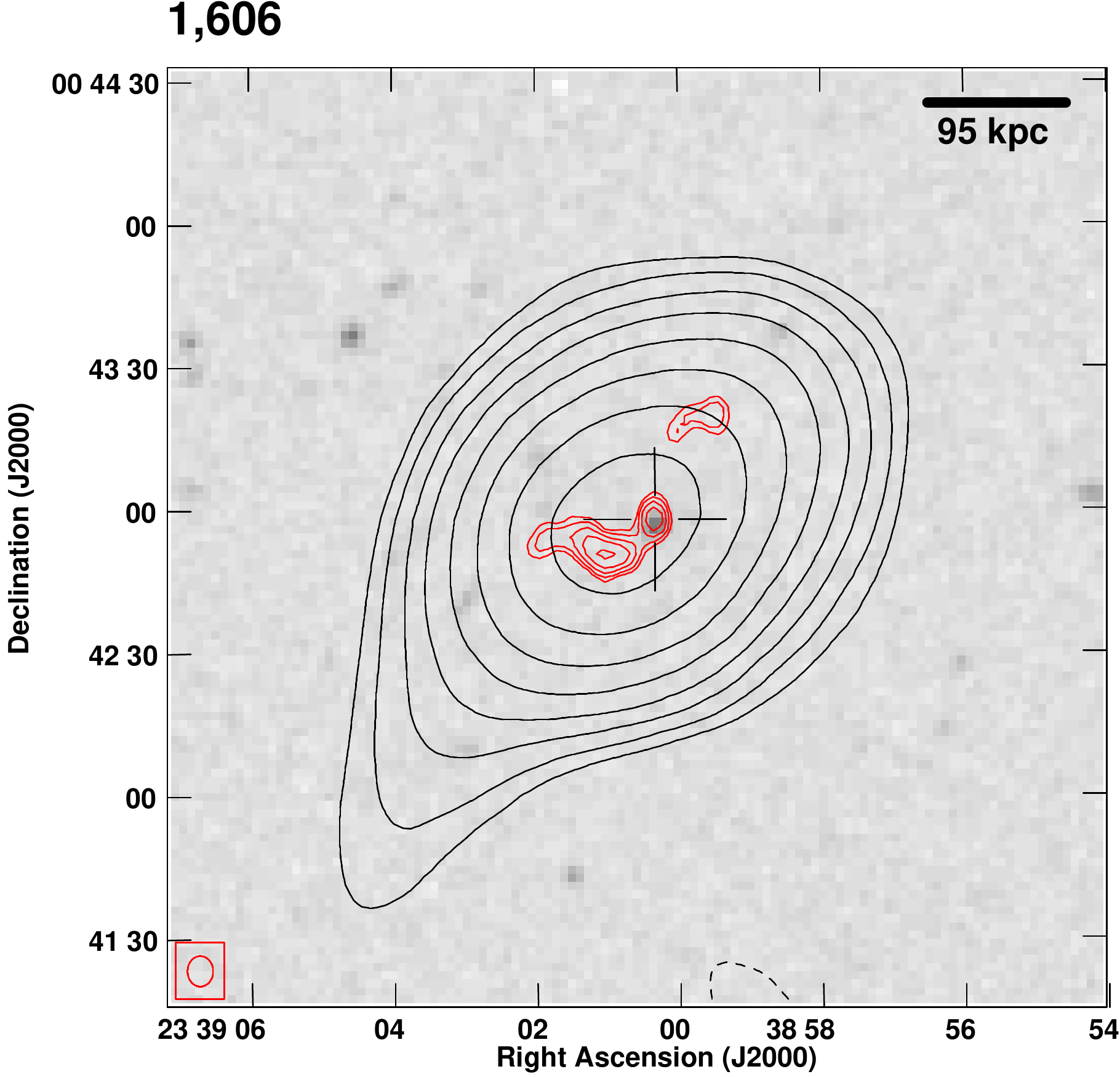}}
\hspace{0.2cm}{\includegraphics[width=0.3\textwidth]{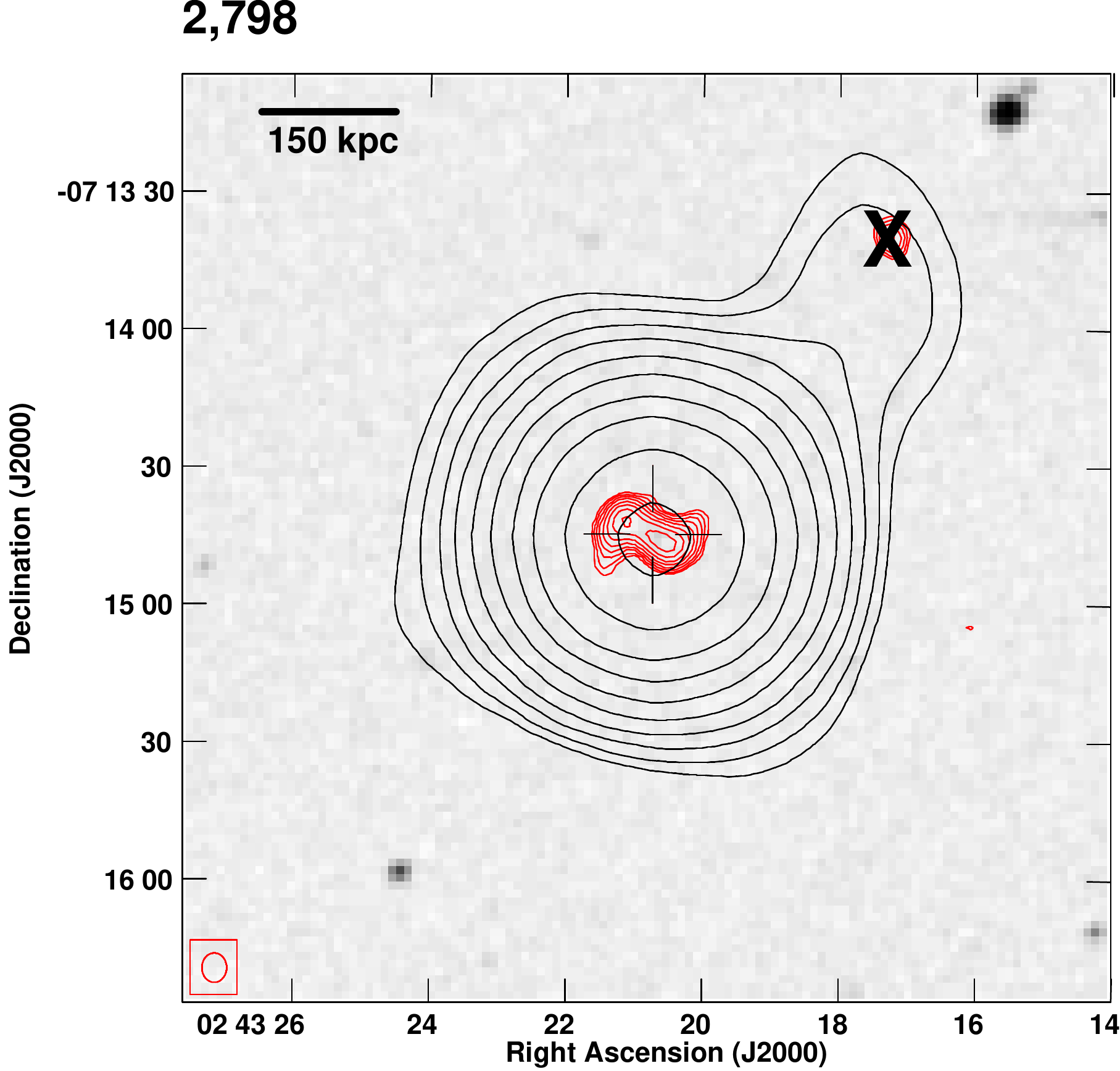}}
\hspace{0.2cm}{\includegraphics[width=0.3\textwidth]{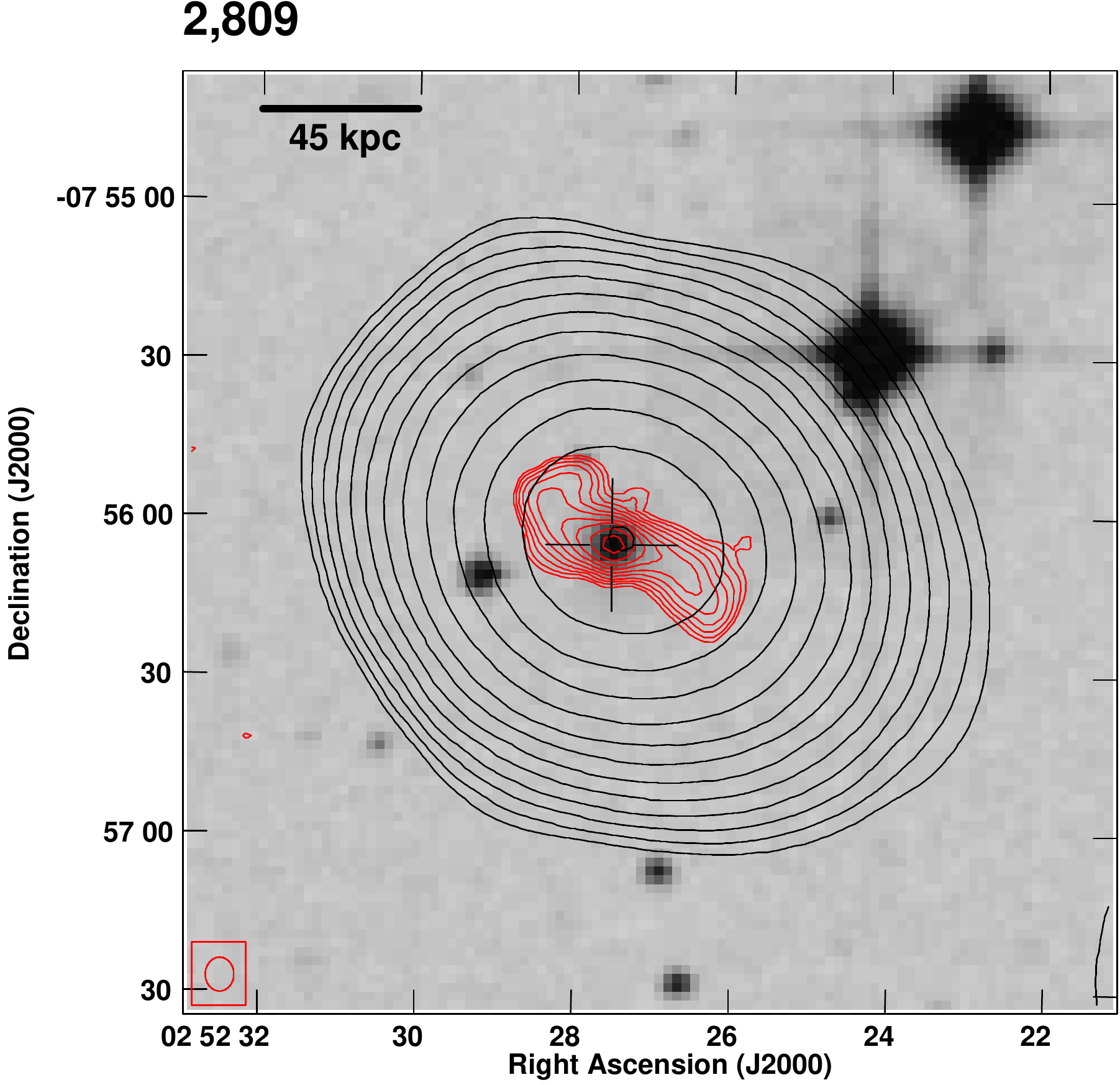}}
\hspace{0.2cm}{\includegraphics[width=0.3\textwidth]{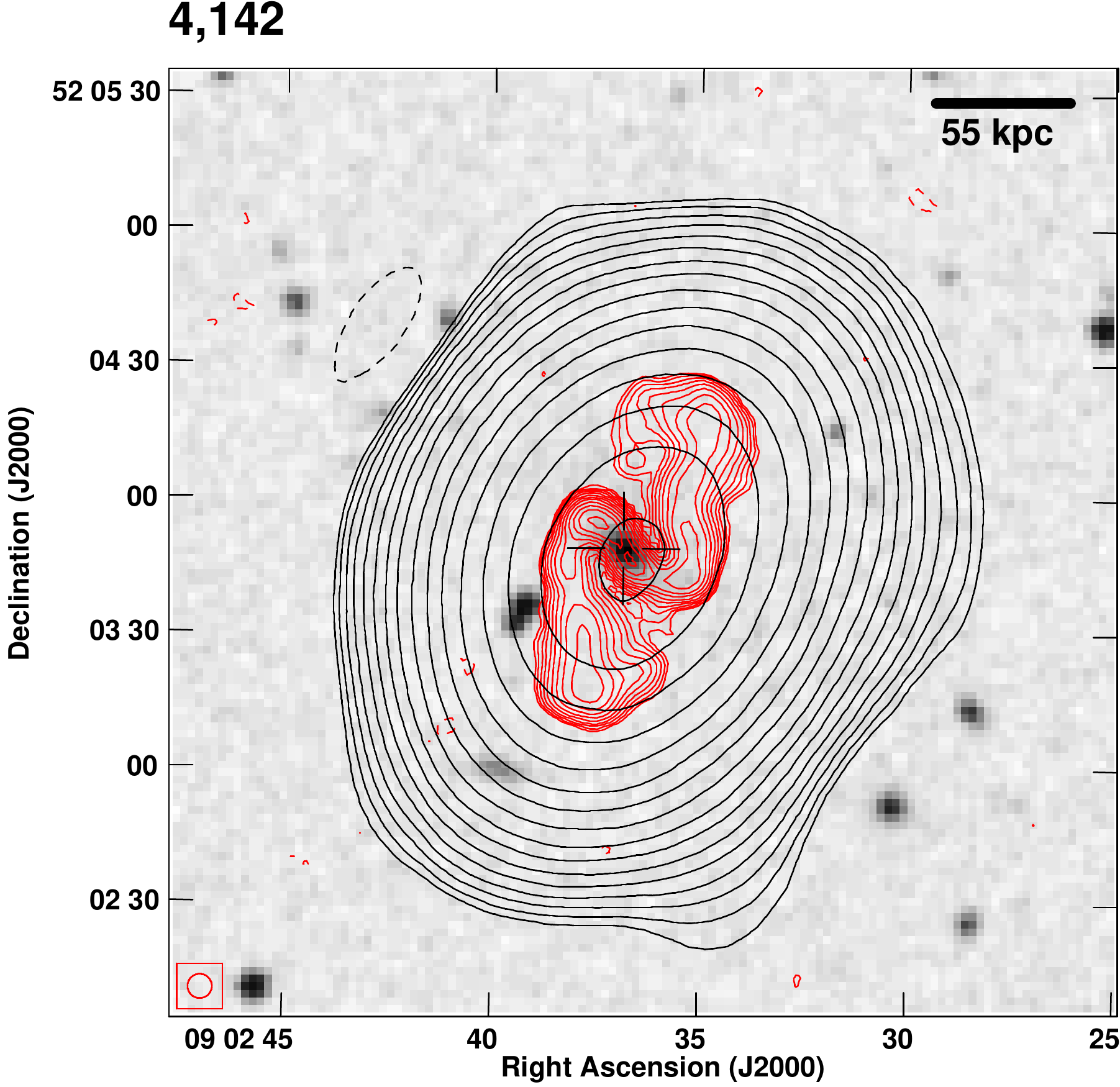}}
\hspace{0.2cm}{\includegraphics[width=0.3\textwidth]{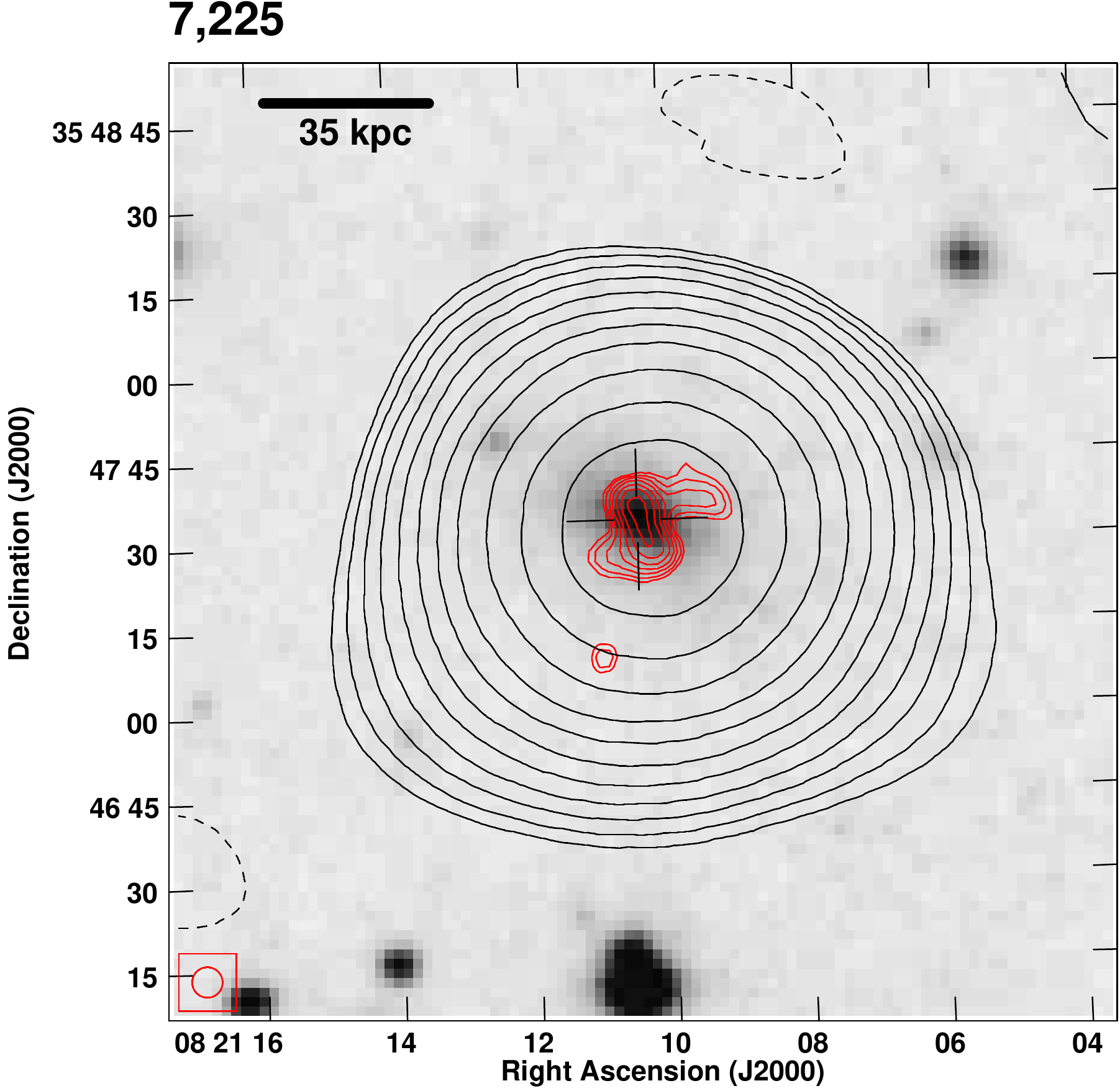}}
\hspace{0.2cm}{\includegraphics[width=0.3\textwidth]{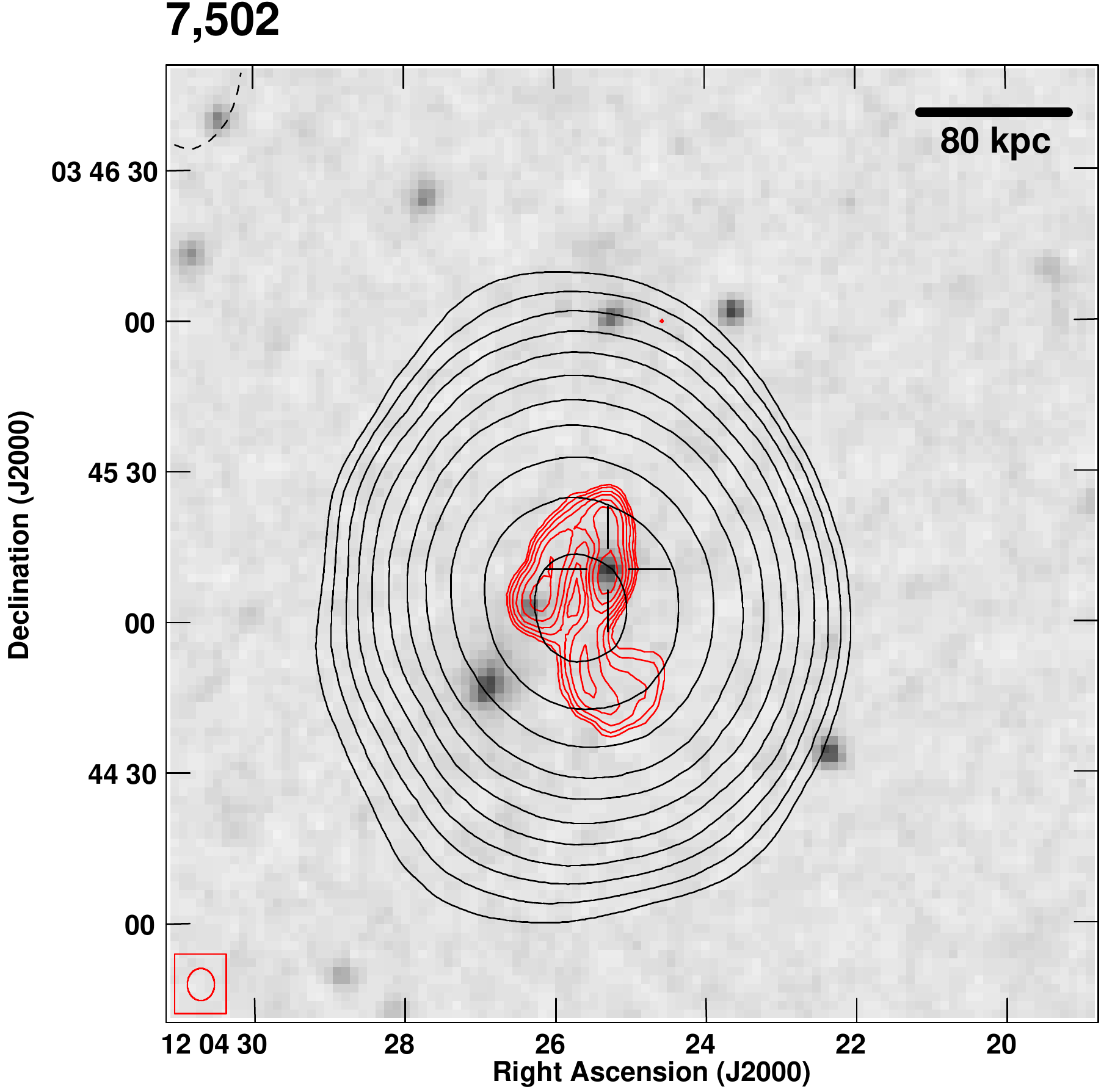}}
\hspace{0.2cm}{\includegraphics[width=0.3\textwidth]{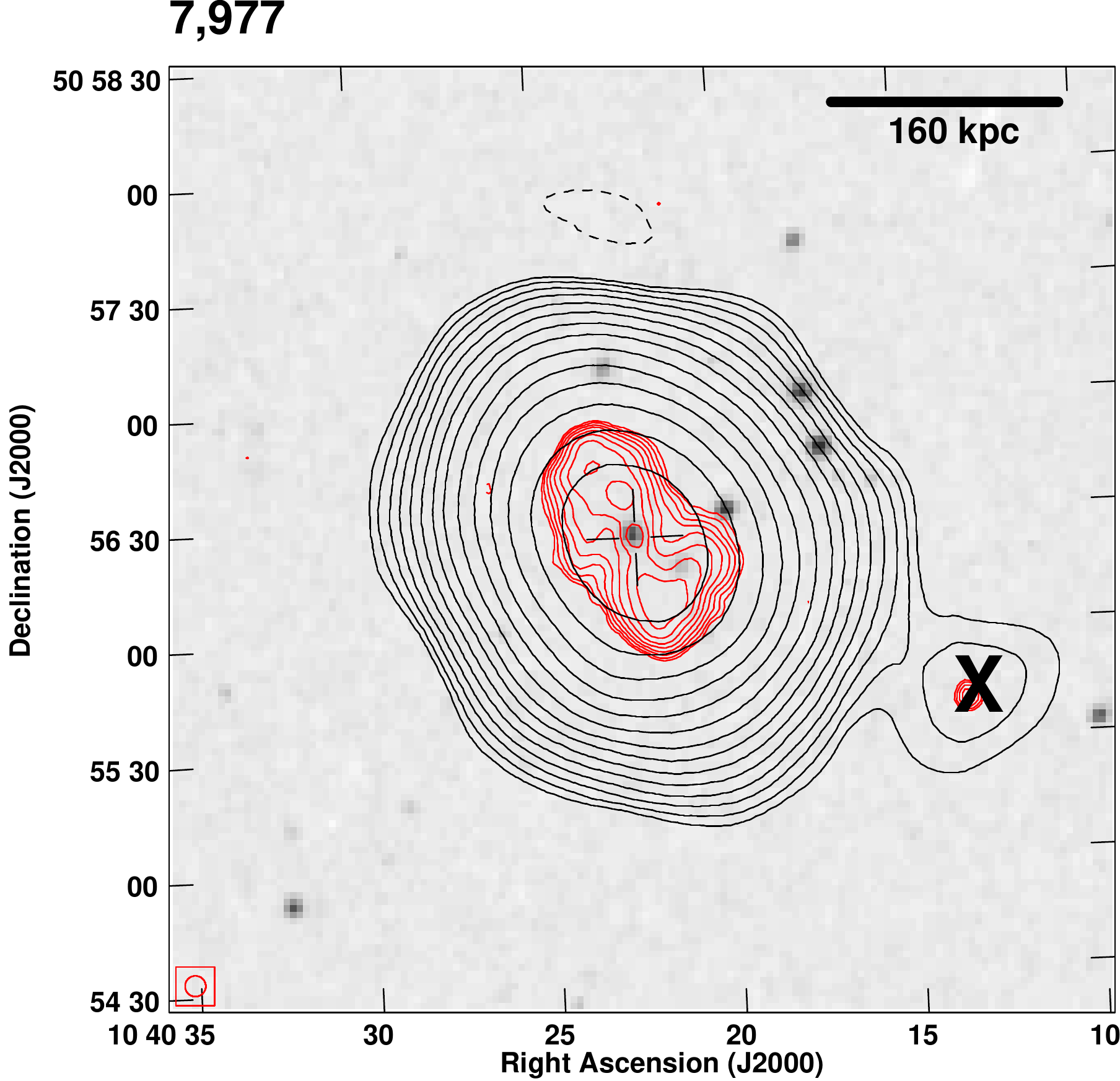}}
\hspace{0.2cm}{\includegraphics[width=0.3\textwidth]{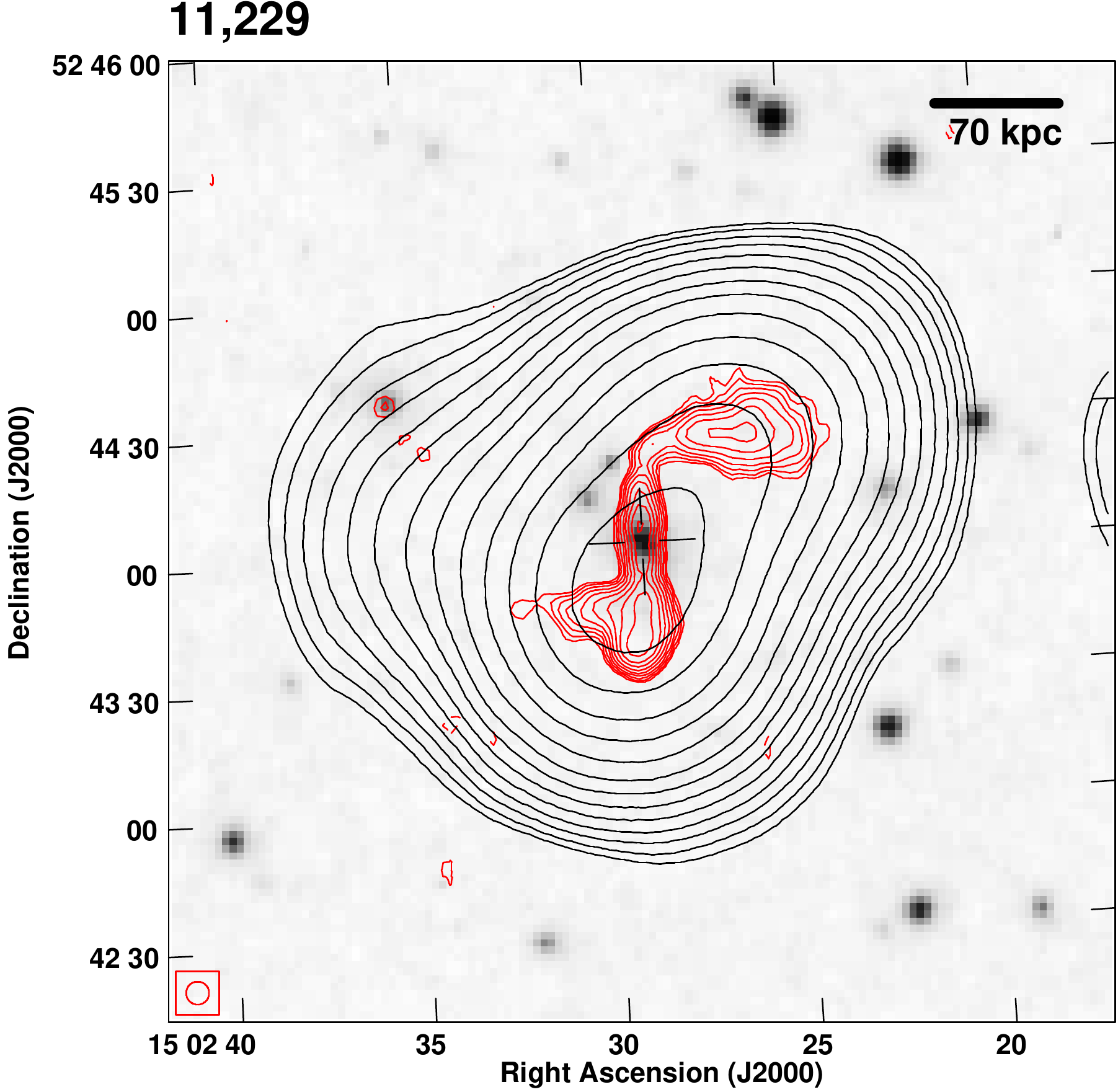}}
\hspace{0.2cm}{\includegraphics[width=0.3\textwidth]{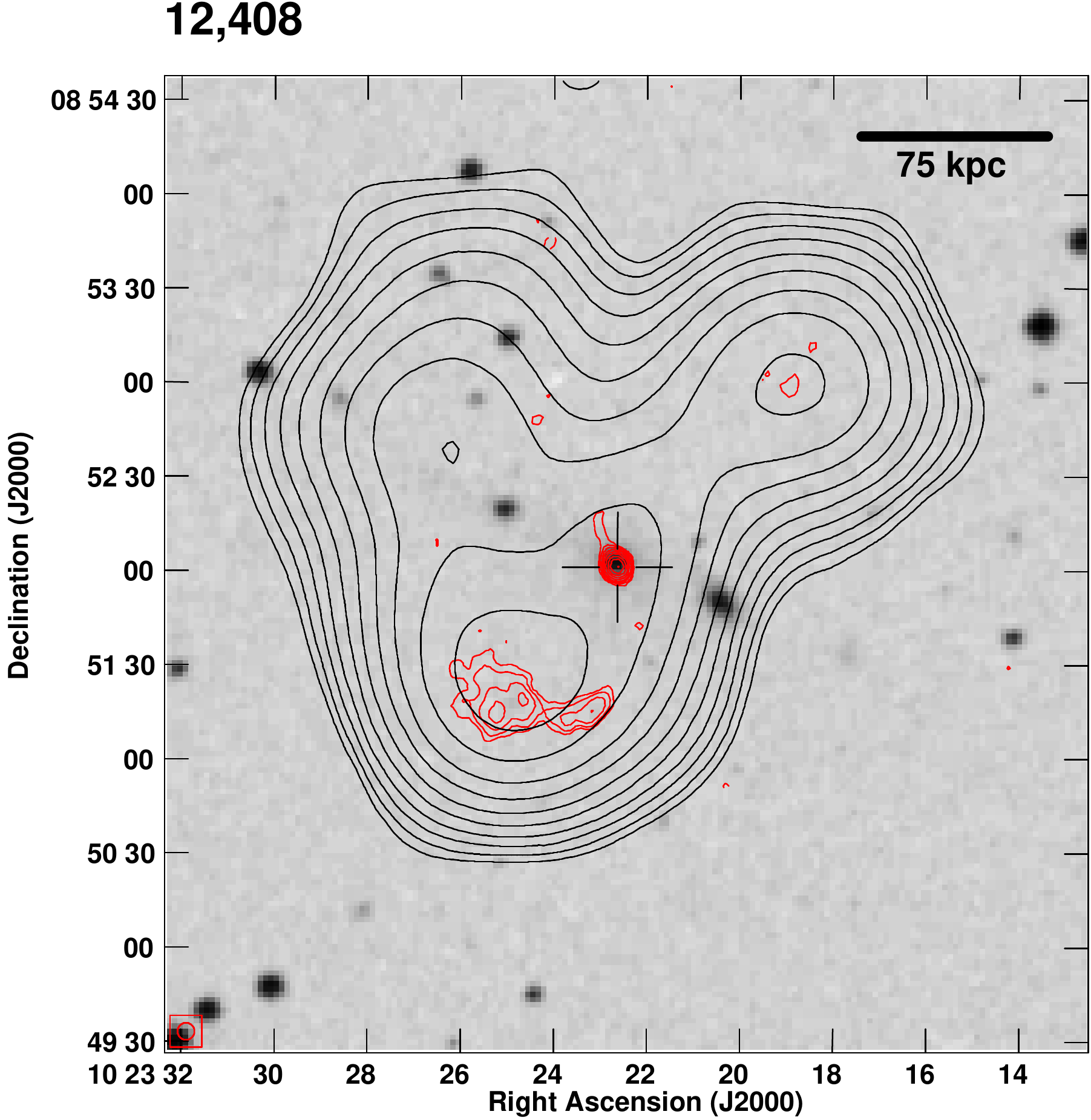}}
\hspace{0.2cm}{\includegraphics[width=0.3\textwidth]{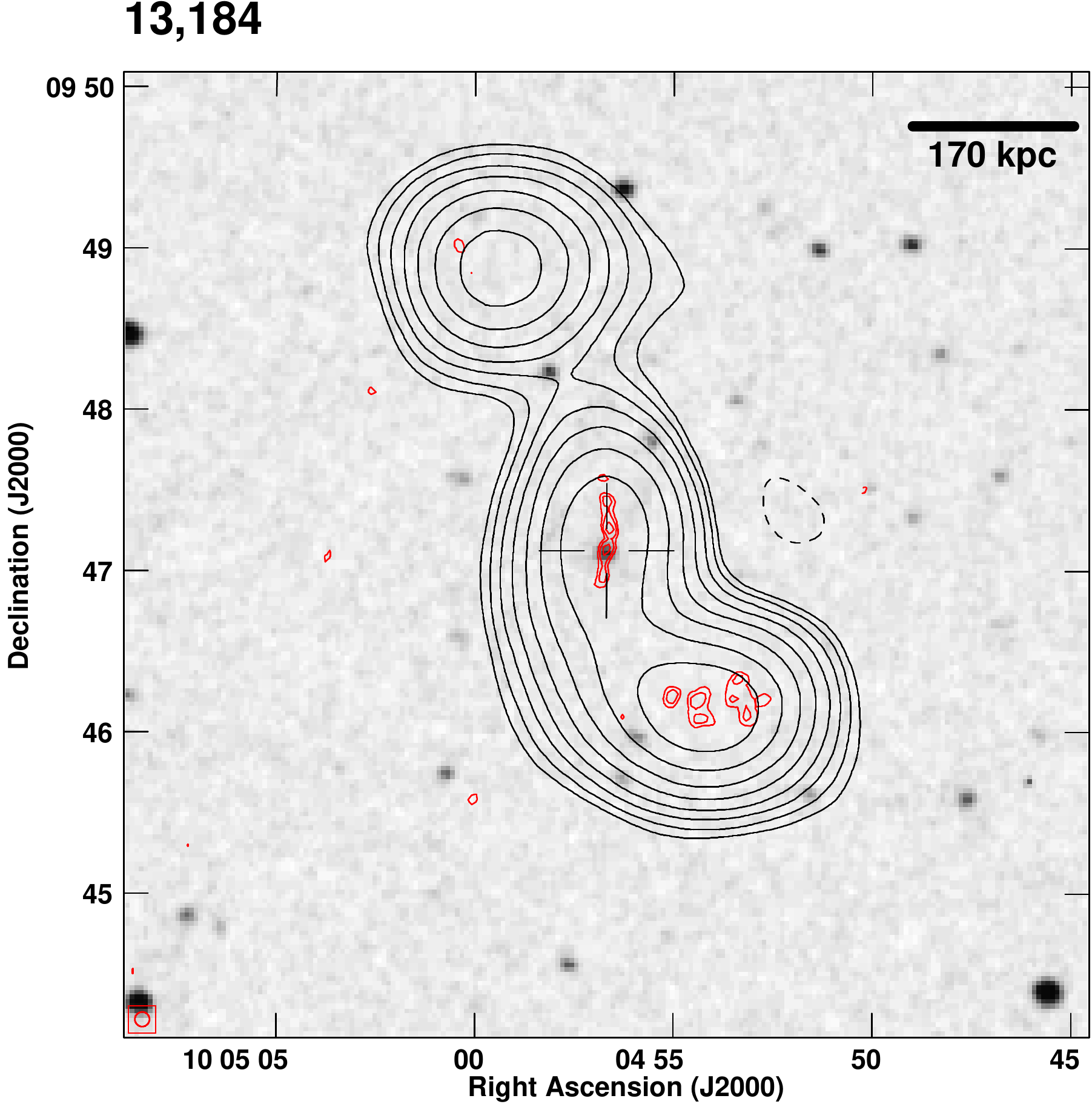}}
\hspace{0.8cm}{\includegraphics[width=0.3\textwidth]{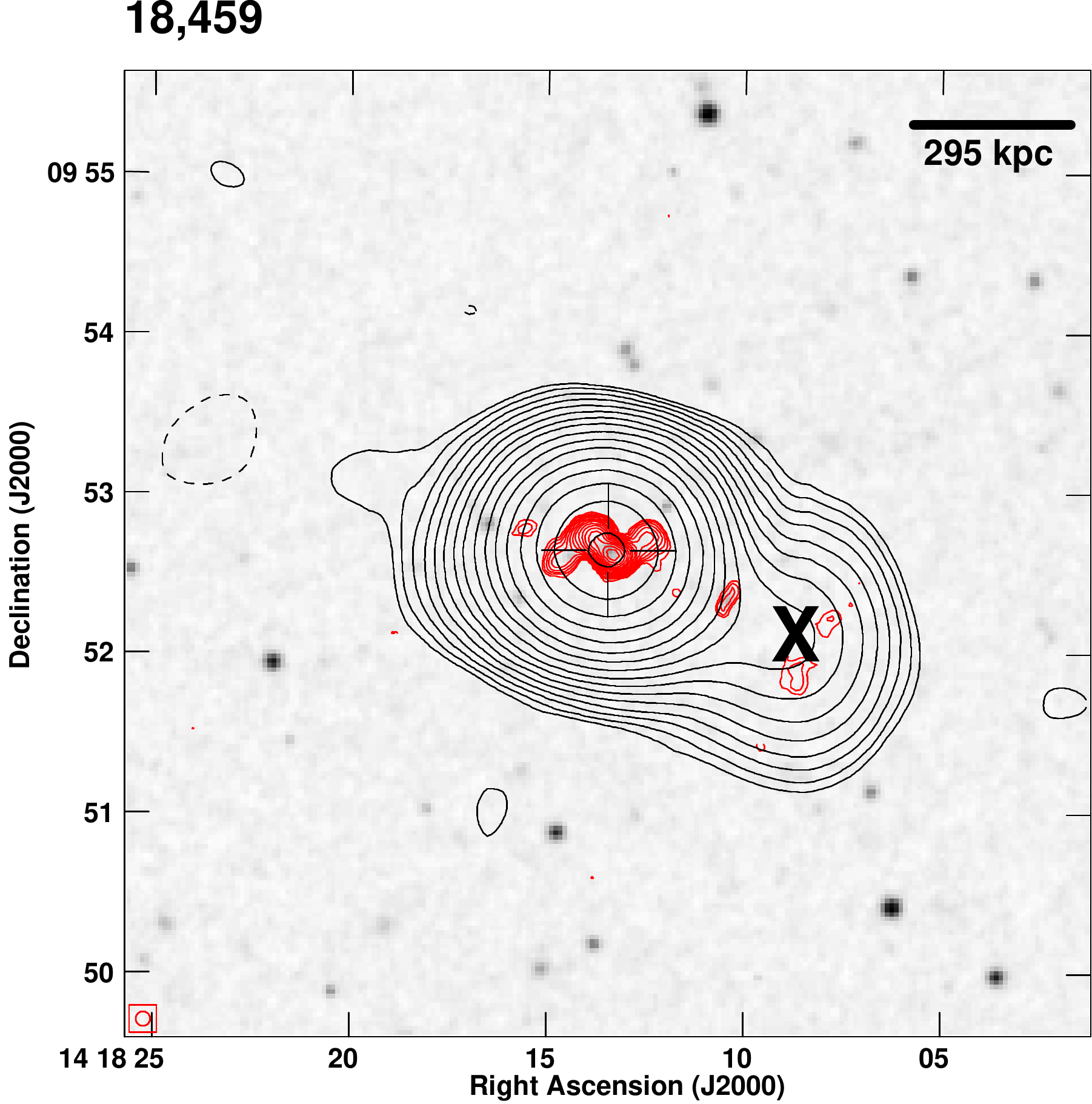}}
\hspace{0.8cm}{\includegraphics[width=0.3\textwidth]{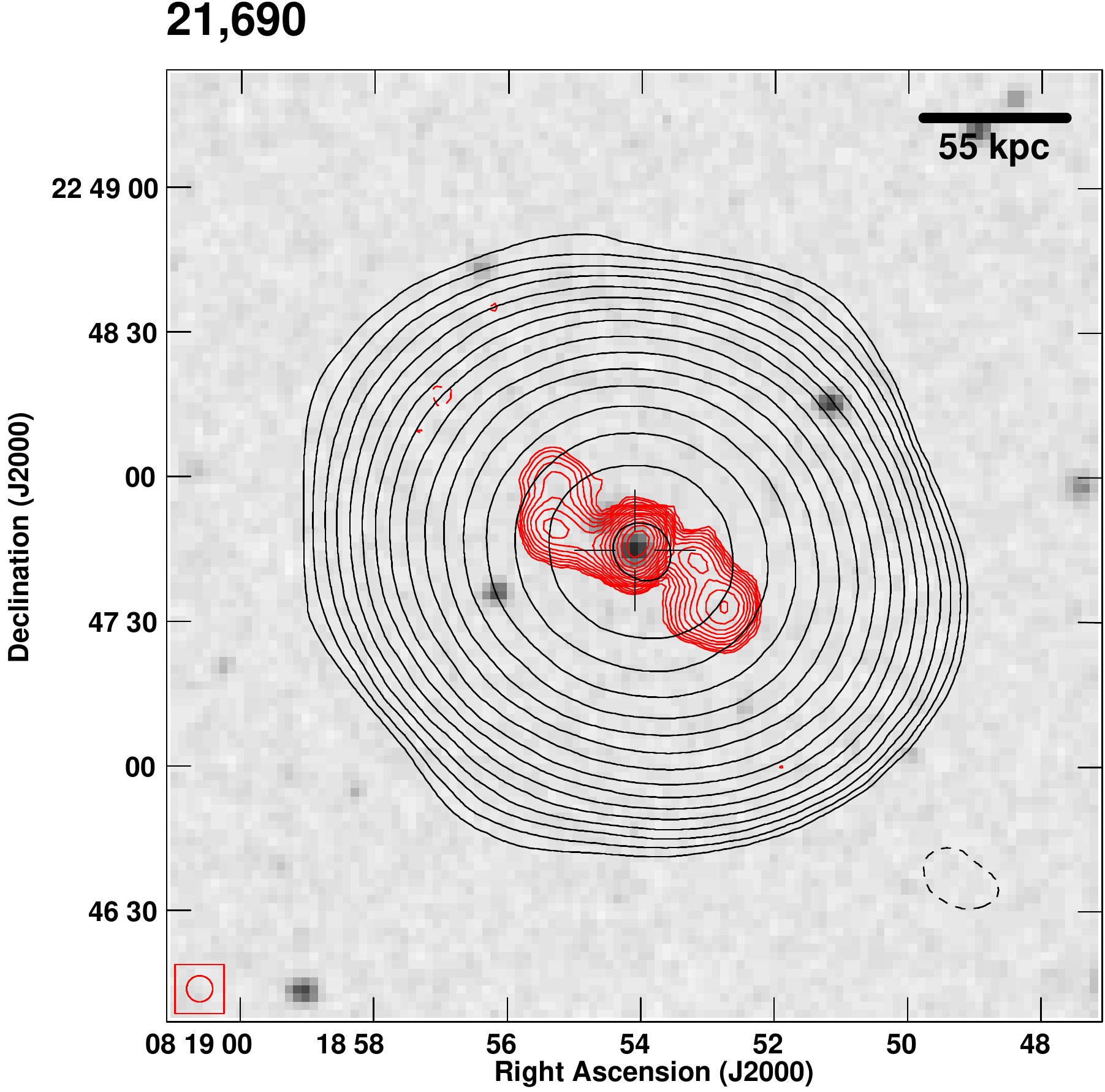}}
\caption{Newly classified Z--shaped and possible Z--shaped radio sources in ROGUE~I. The layout of the maps is the same as in Figure~\ref{fig:NewGiantsMaps}.}
\label{NewZMaps}
\end{figure*}

\renewcommand{\thefigure}{A.\arabic{figure} (Cont.)}

\begin{figure*}[htb!]
\ContinuedFloat
\hspace{0.0cm}{\includegraphics[width=0.3\textwidth]{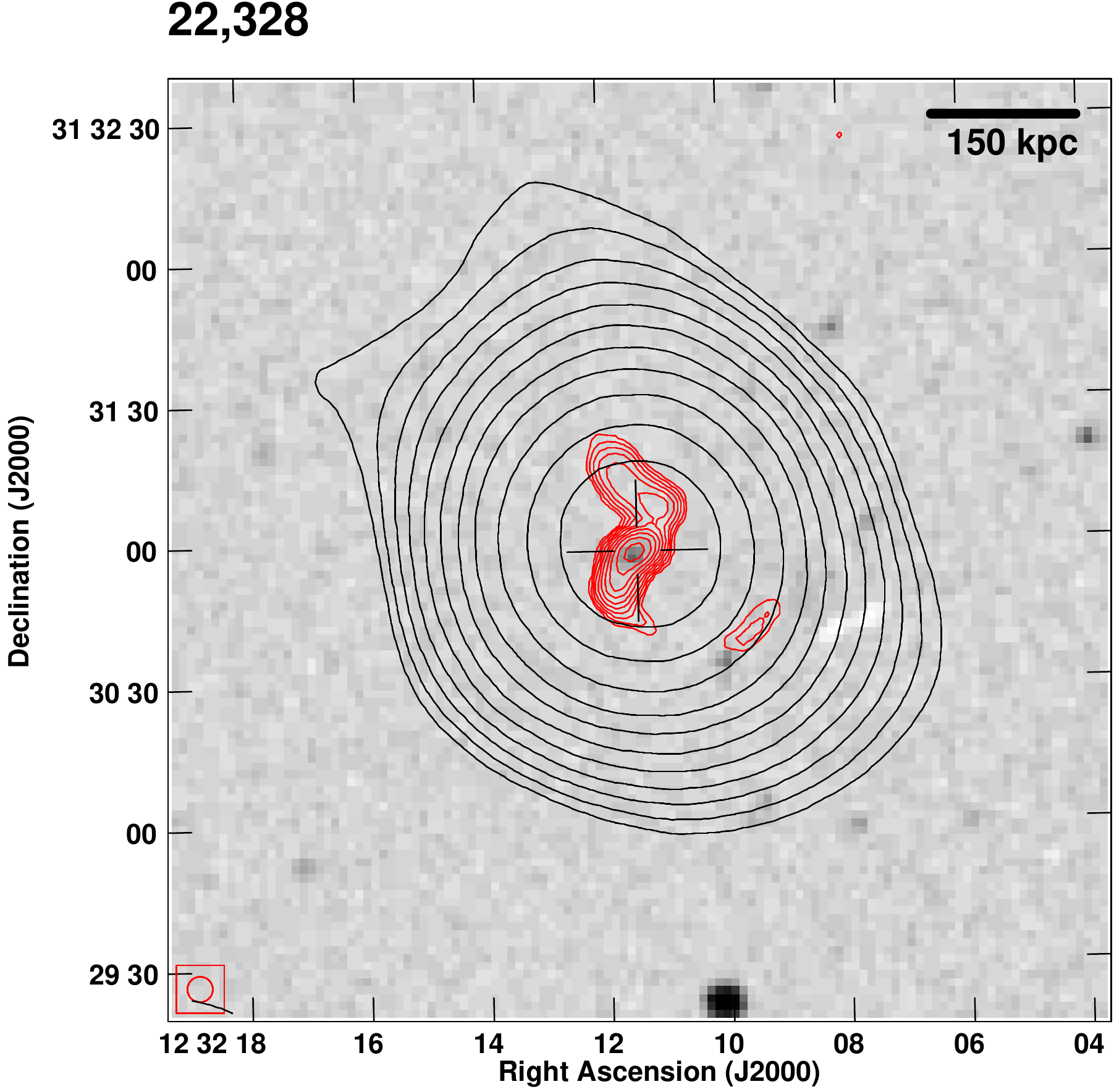}}
\hspace{0.2cm}{\includegraphics[width=0.3\textwidth]{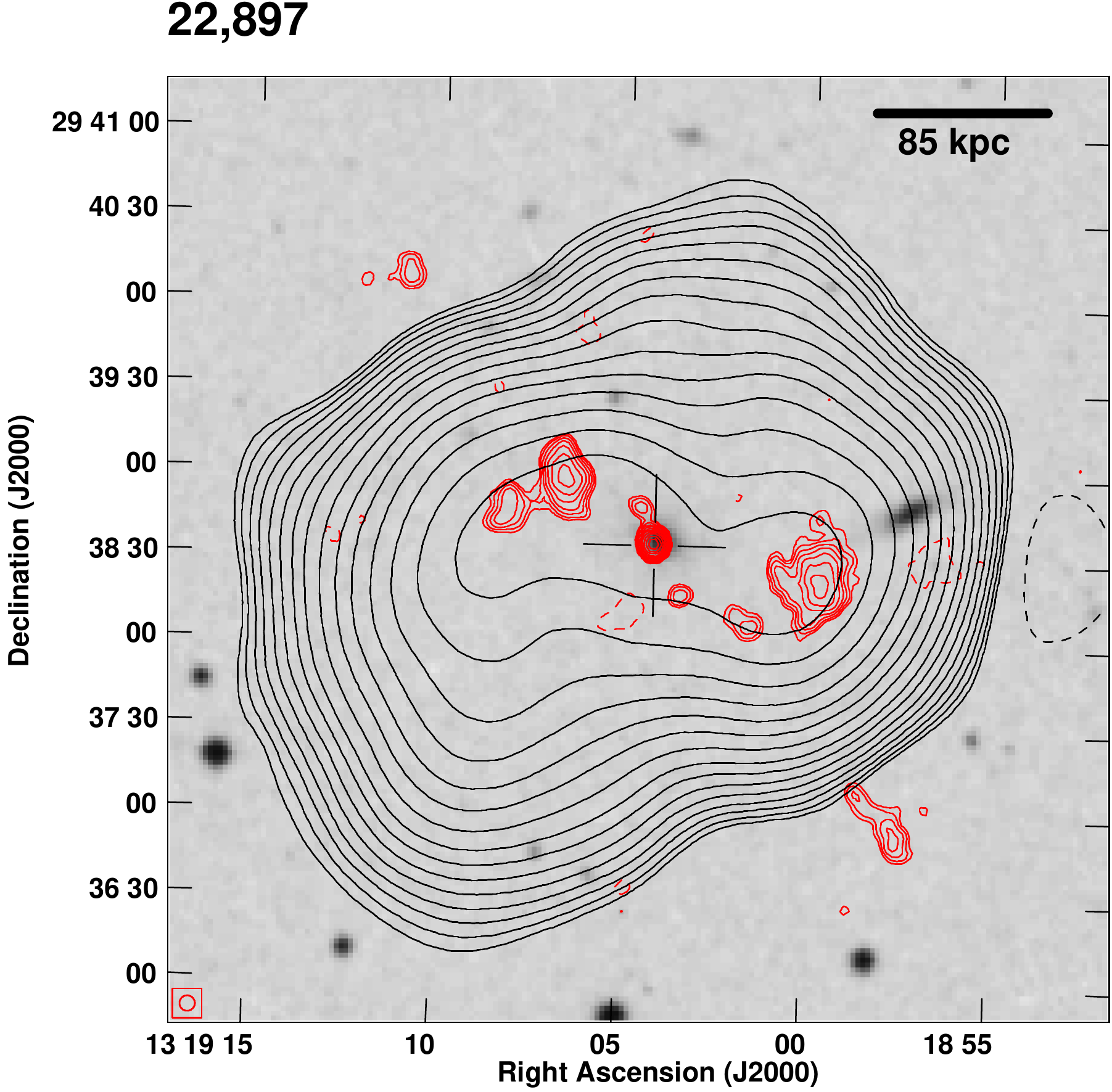}}
\hspace{0.2cm}{\includegraphics[width=0.3\textwidth]{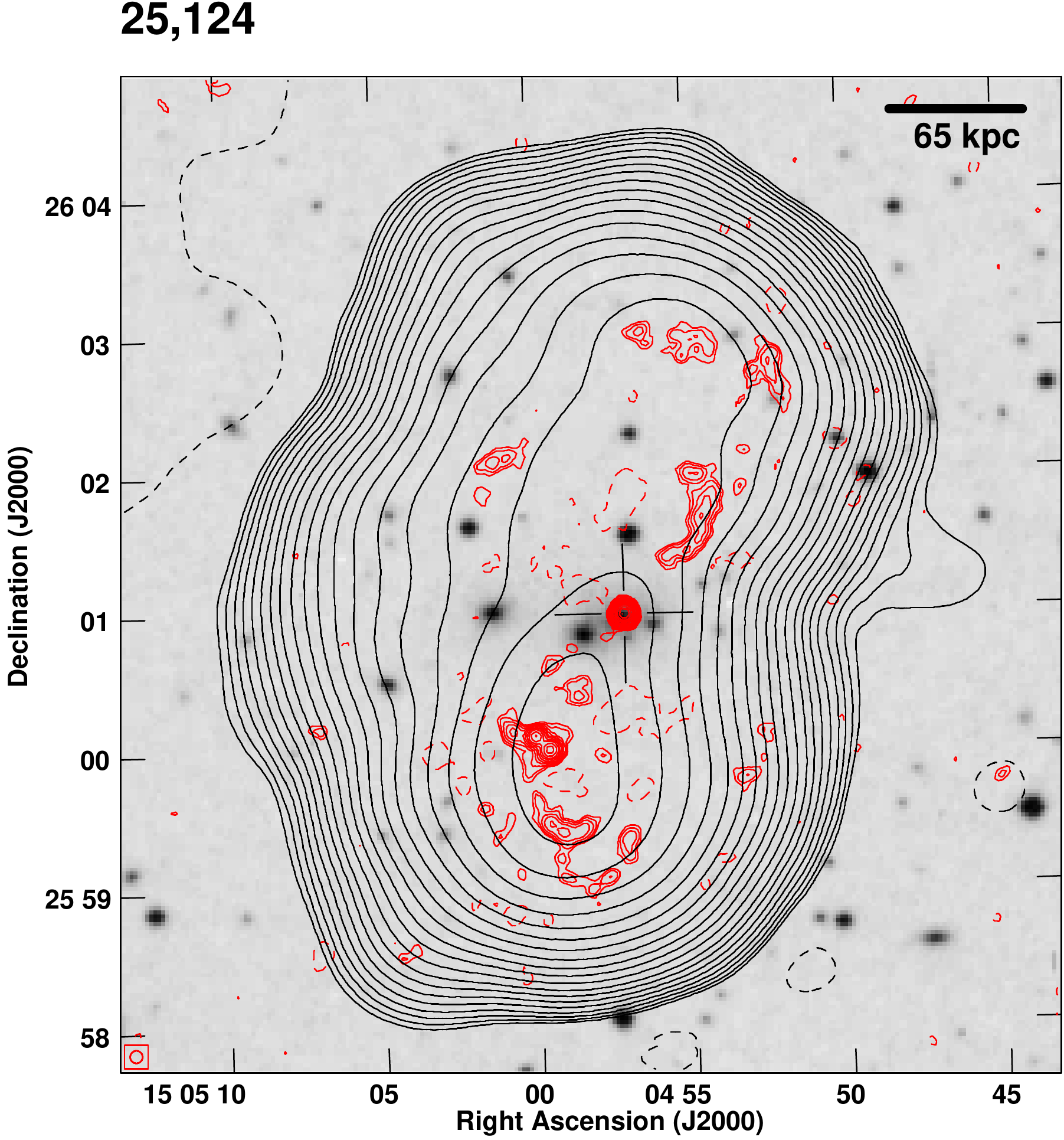}}
\hspace{0.2cm}{\includegraphics[width=0.3\textwidth]{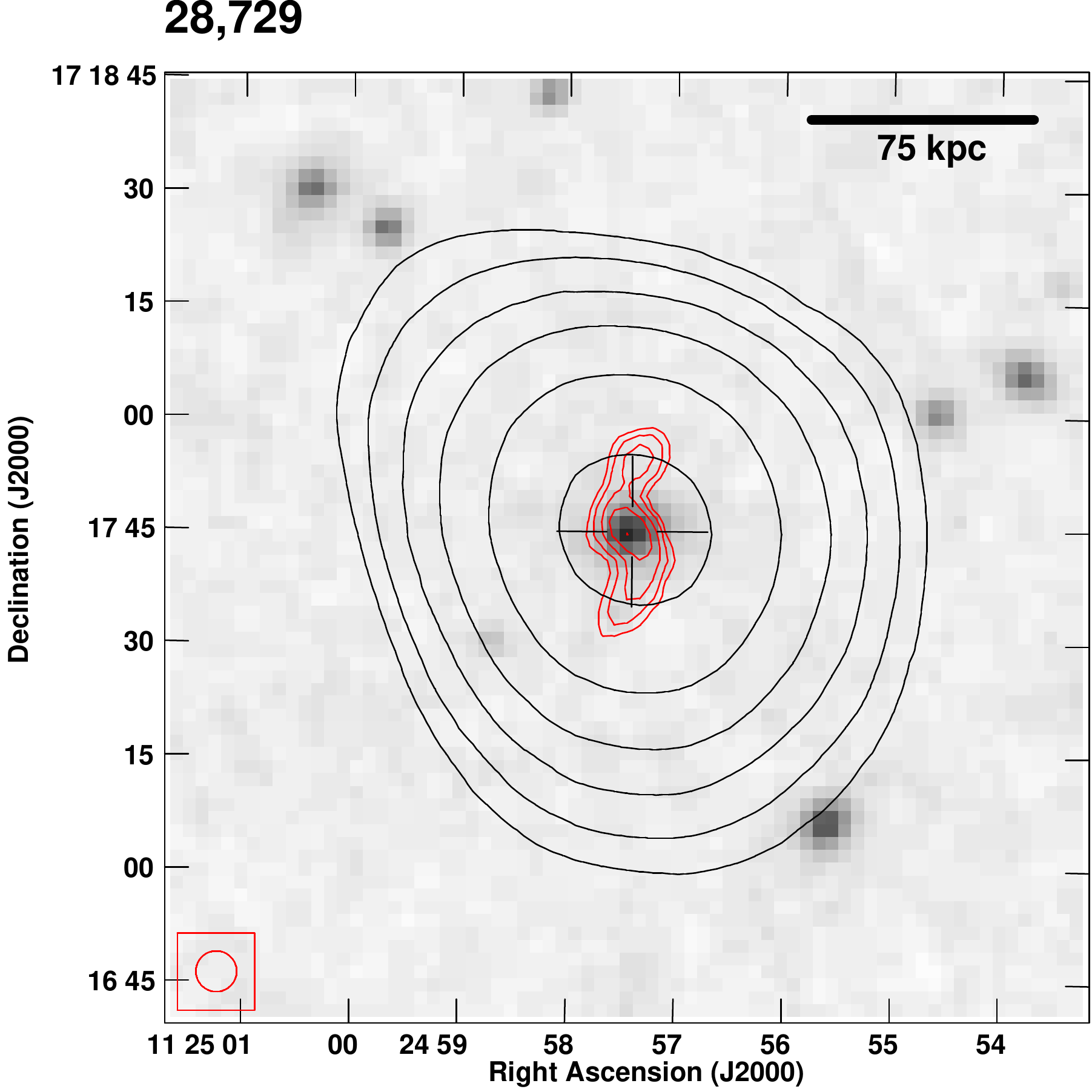}}
\hspace{0.0cm}{\includegraphics[width=0.3\textwidth]{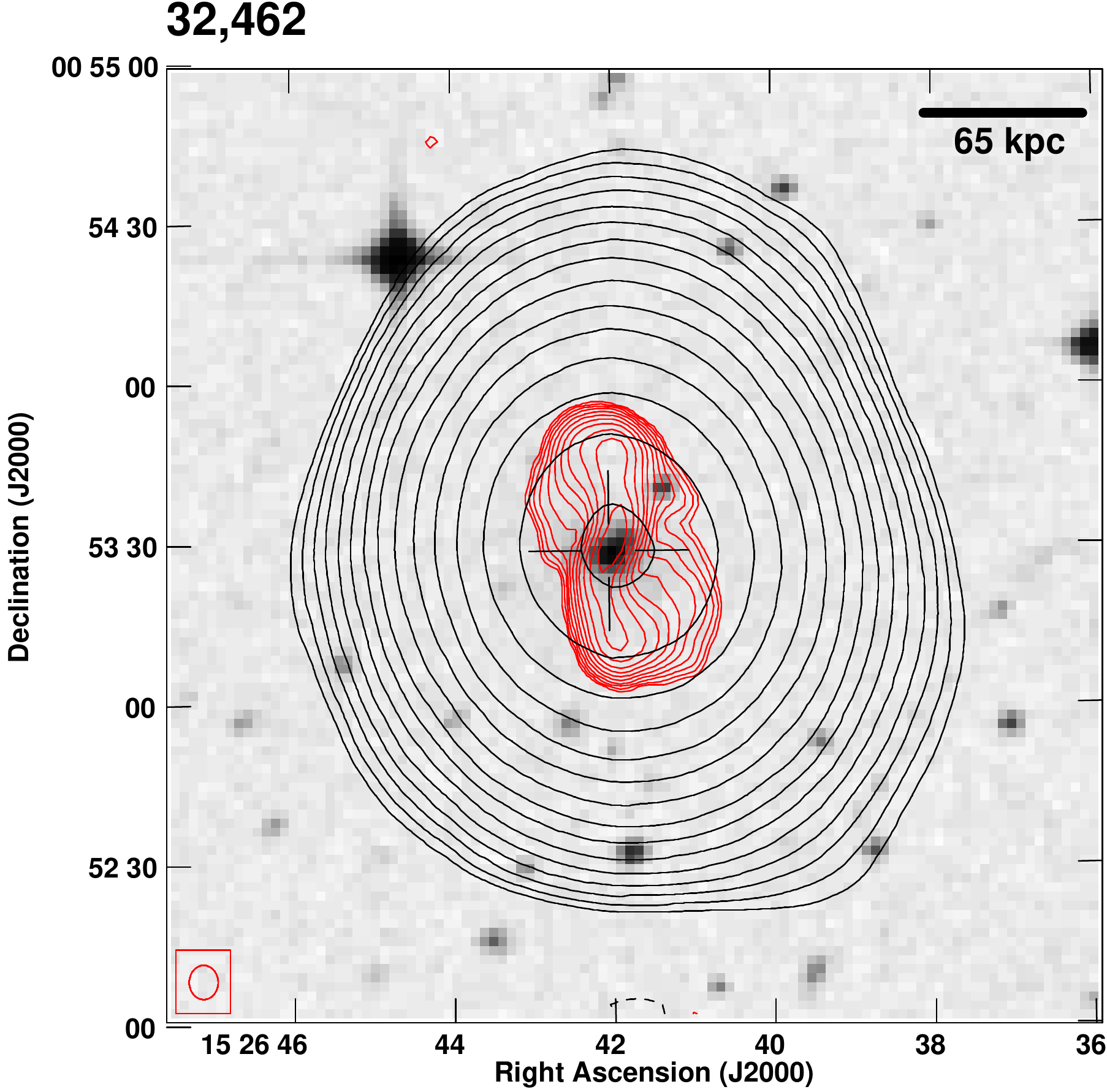}}
\hspace{0.2cm}{\includegraphics[width=0.3\textwidth]{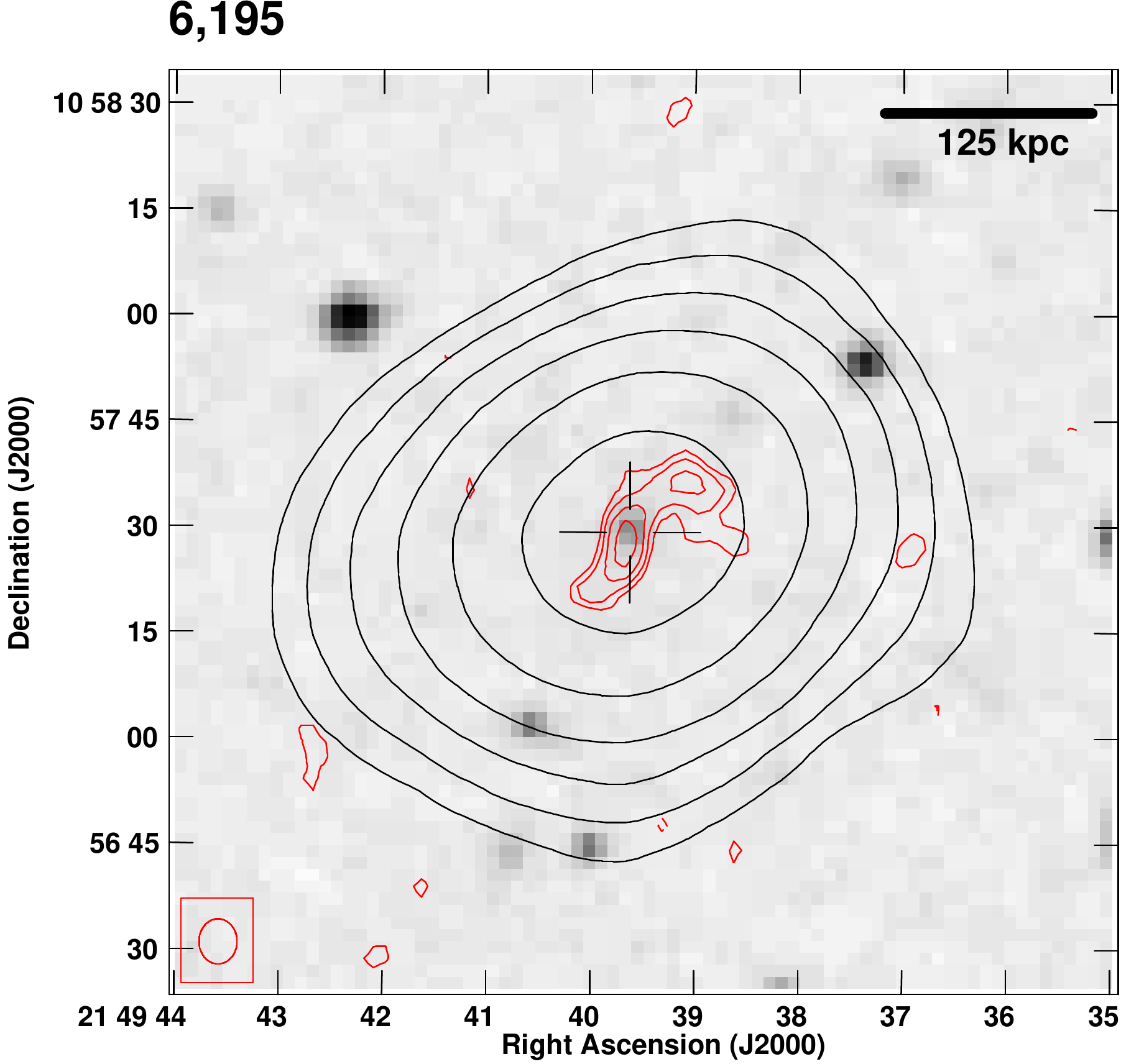}}
\hspace{0.2cm}{\includegraphics[width=0.3\textwidth]{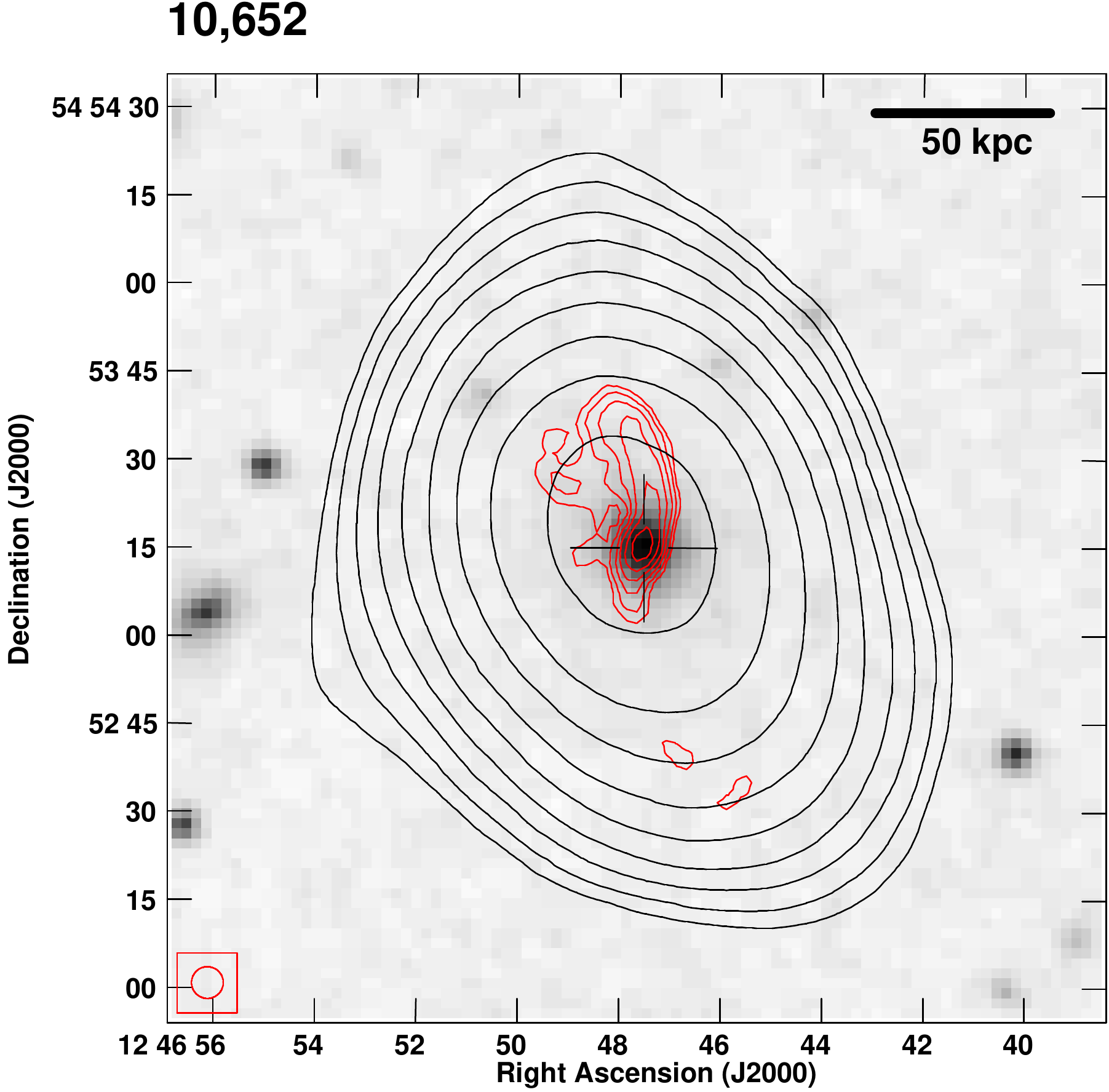}}
\hspace{0.2cm}{\includegraphics[width=0.3\textwidth]{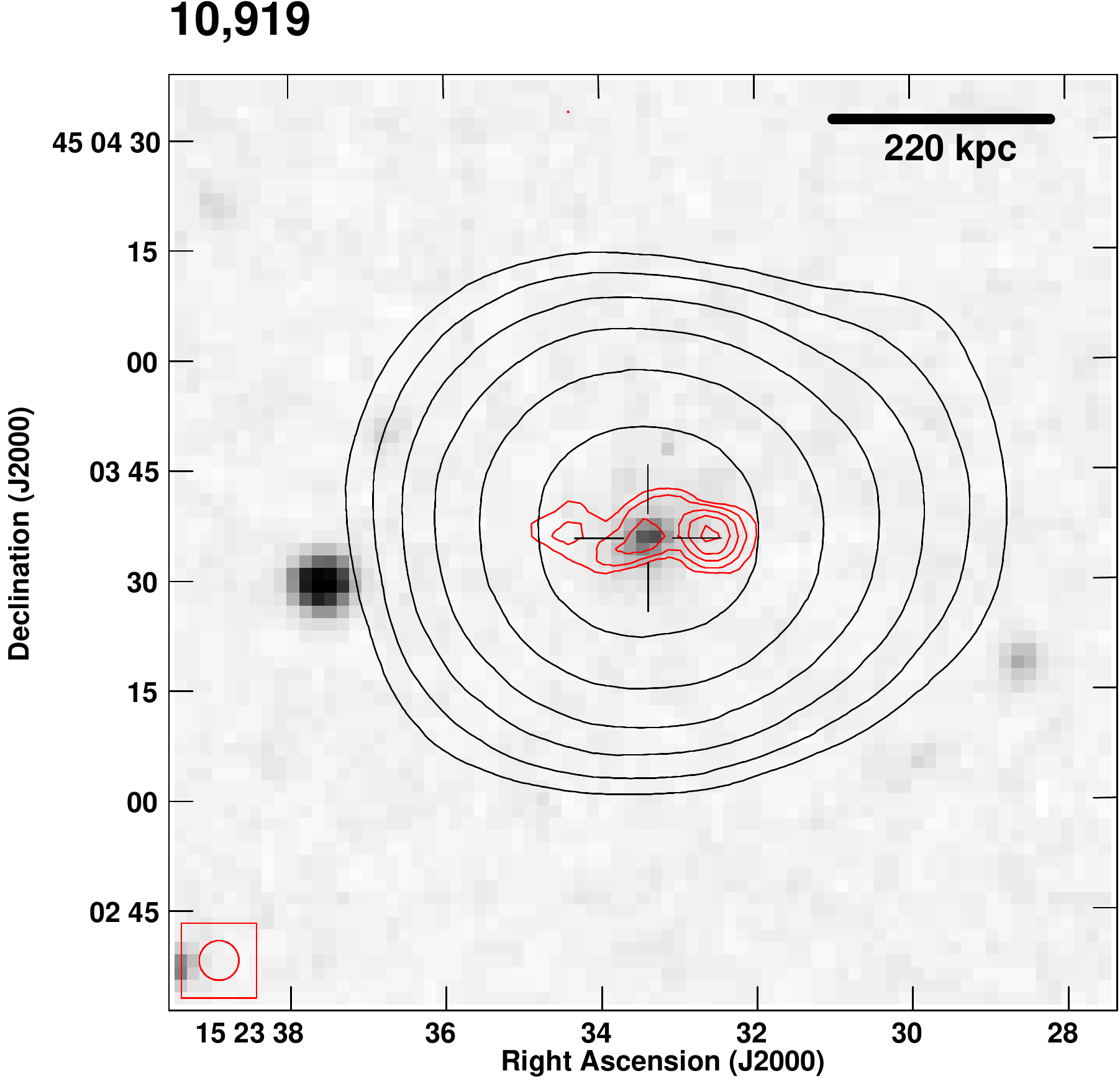}}
\hspace{0.2cm}{\includegraphics[width=0.3\textwidth]{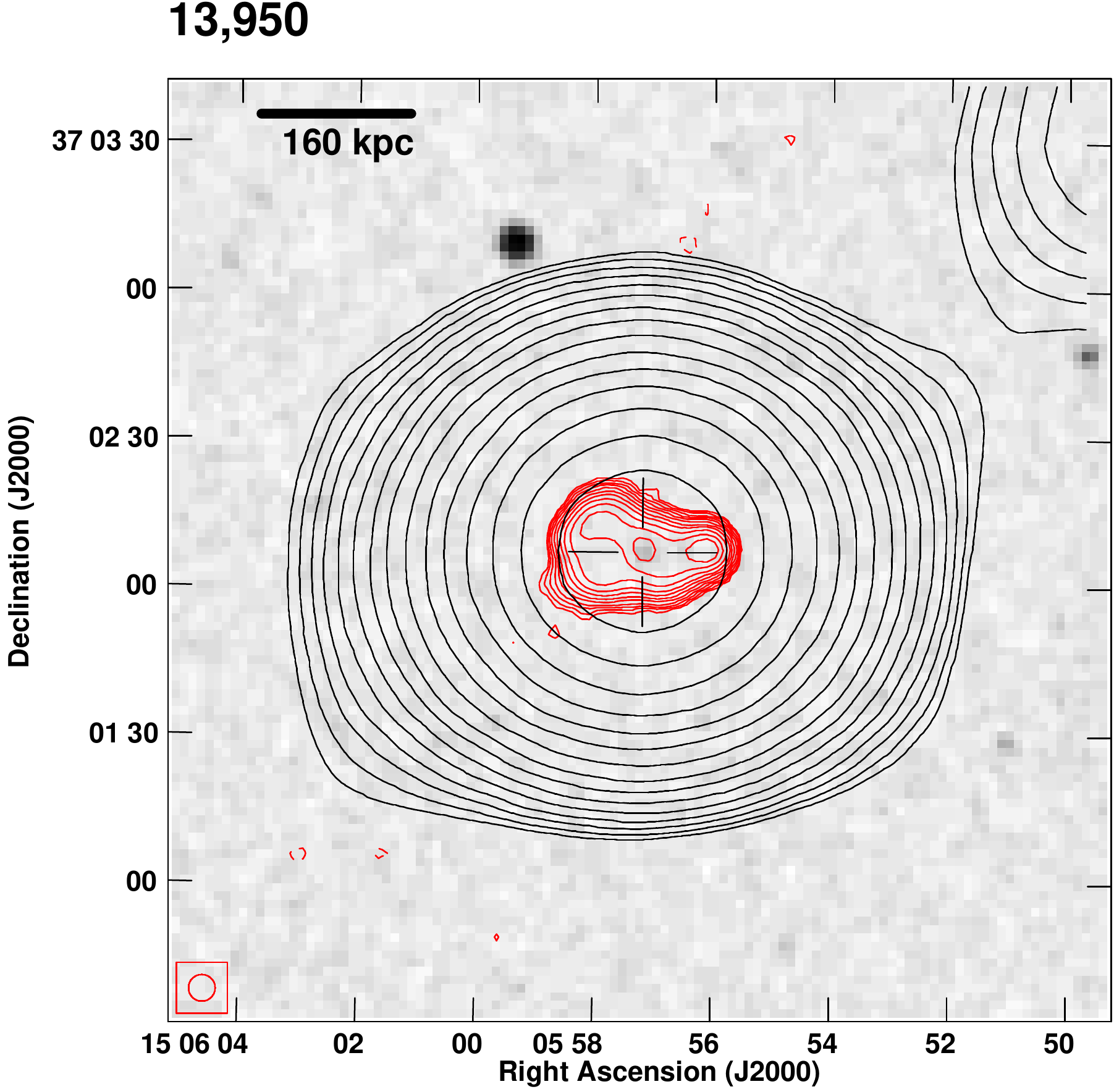}}
\hspace{0.2cm}{\includegraphics[width=0.3\textwidth]{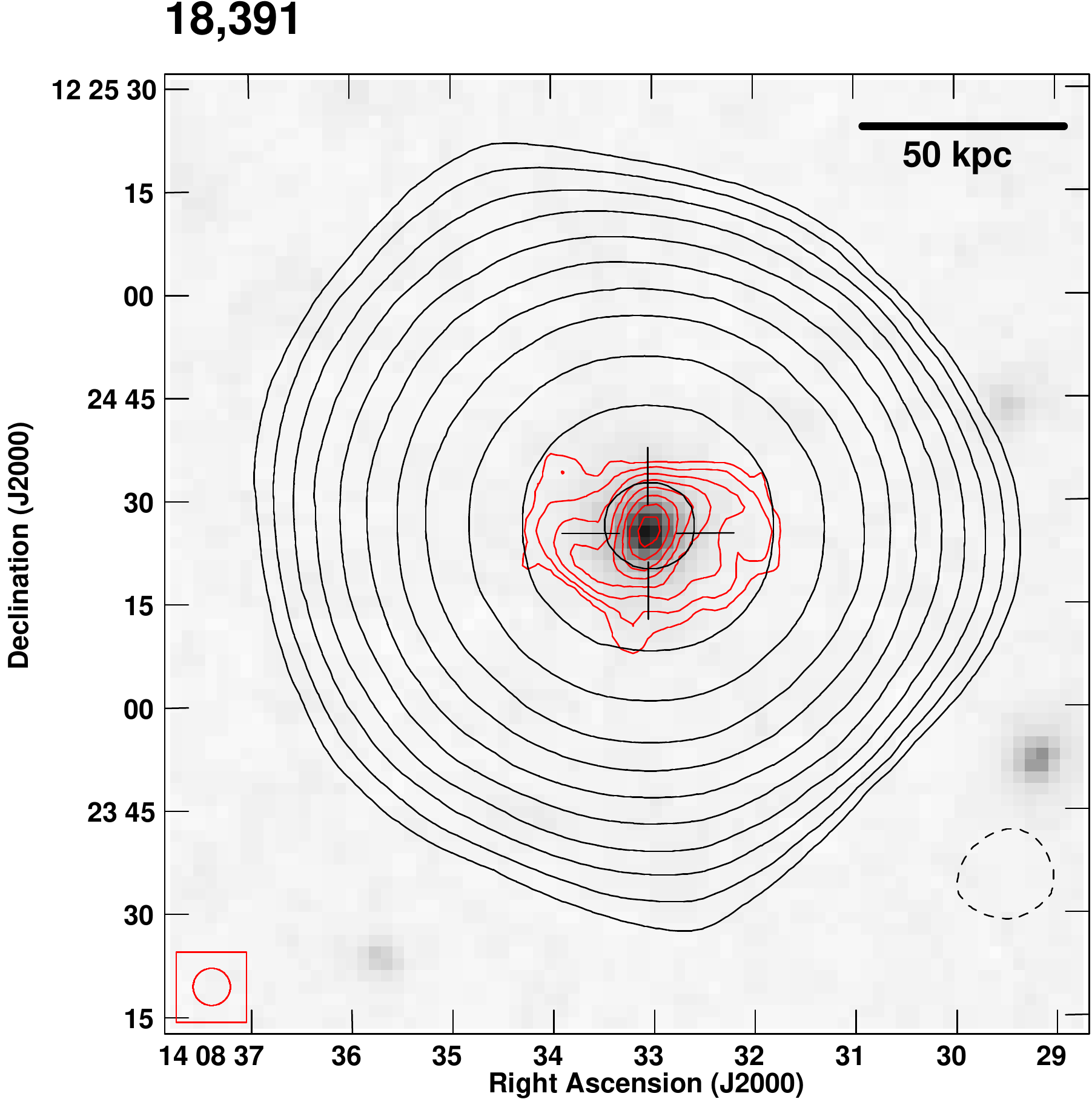}}
\hspace{0.8cm}{\includegraphics[width=0.3\textwidth]{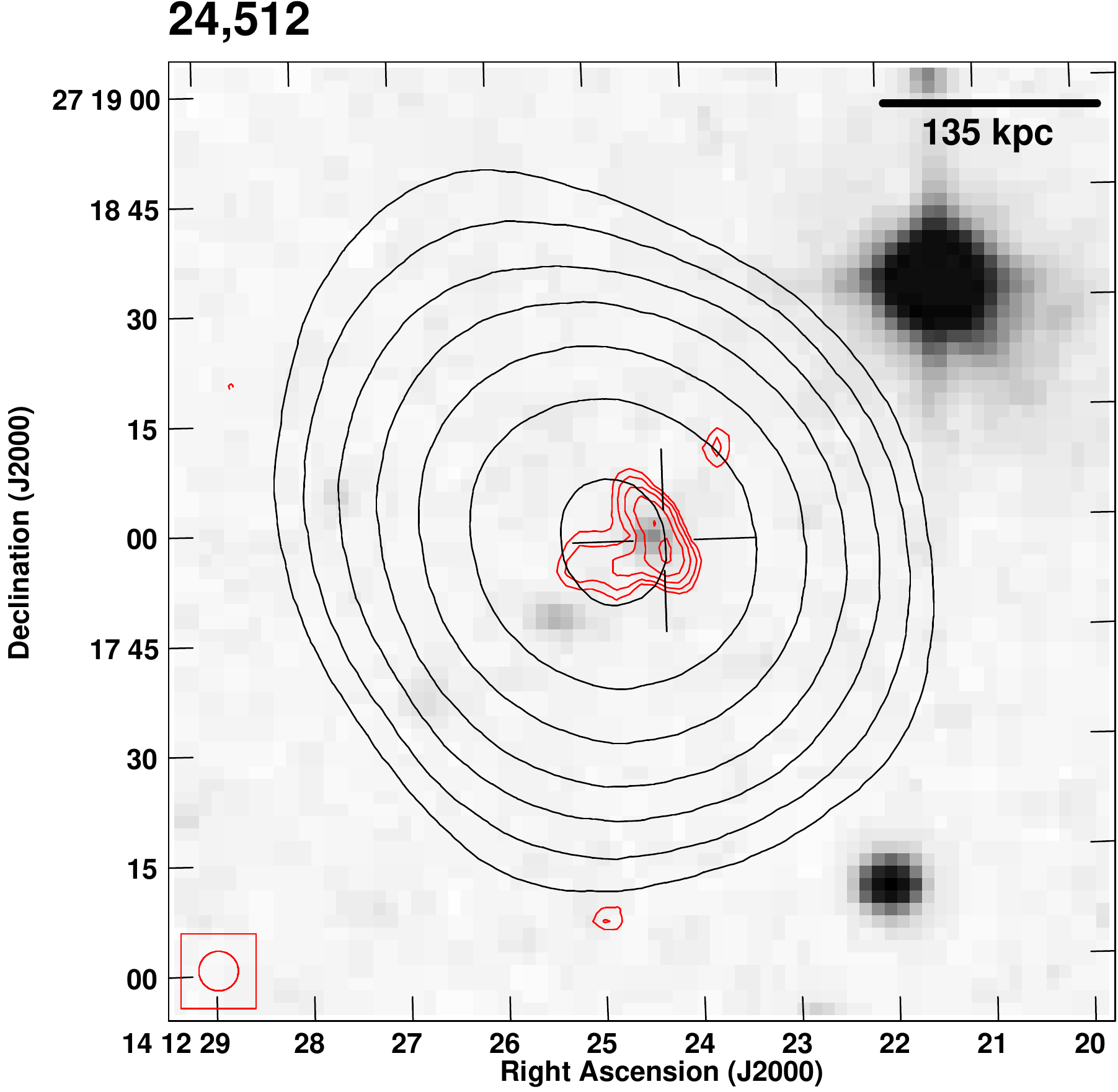}}
\hspace{0.8cm}{\includegraphics[width=0.3\textwidth]{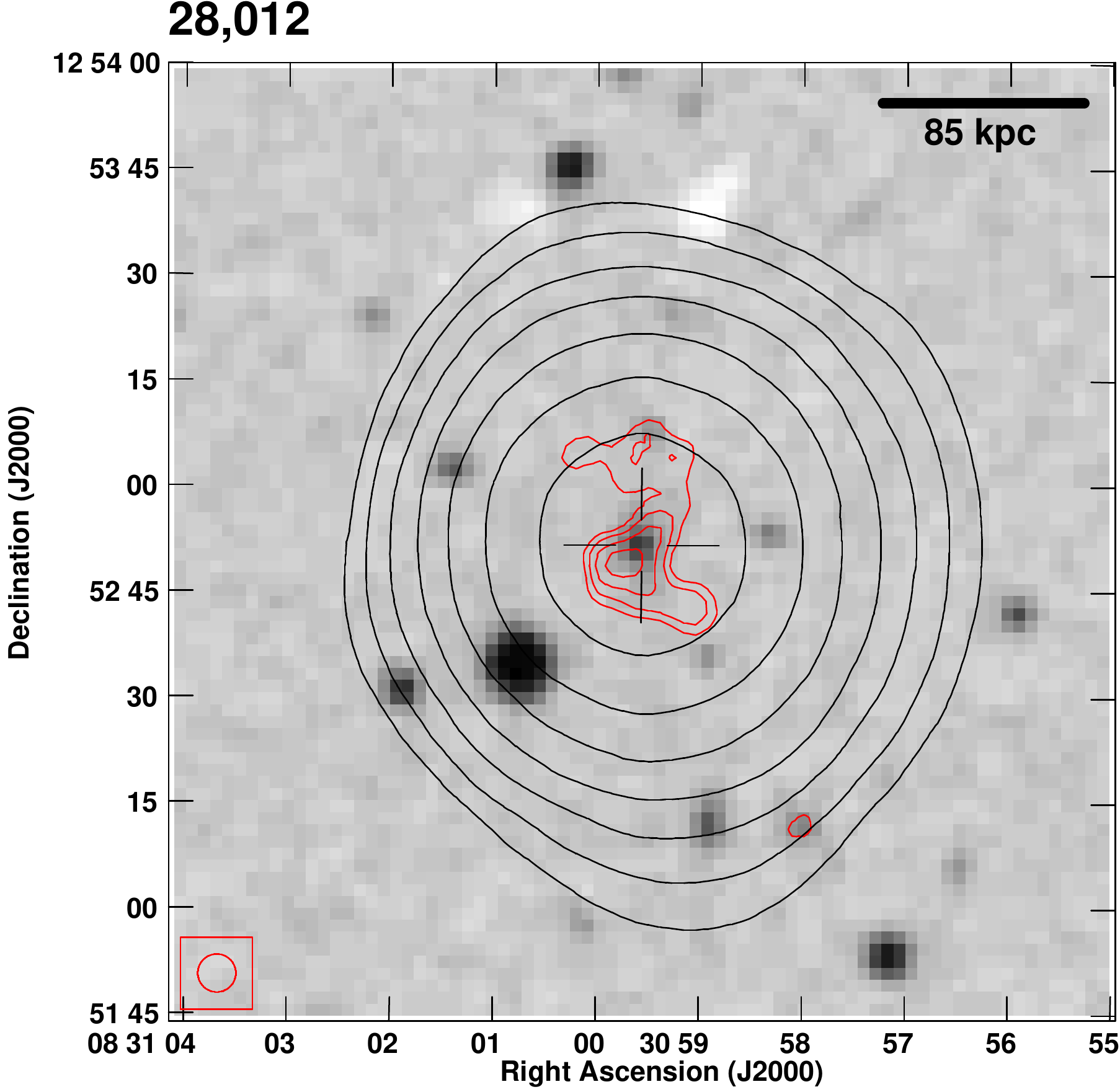}}
\caption{Newly classified Z--shaped and possible Z--shaped radio sources in ROGUE~I.}
\end{figure*}

\renewcommand{\thefigure}{A.\arabic{figure}}

\end{document}